\documentclass[12pt]{report} 
\usepackage{latexsym}
\usepackage{graphicx}
\usepackage{epsf}
\newcommand{\dessin}[4]{\begin{figure}{\refstepcounter{figure}\label{#3}}\end{figure}%
\vspace*{12pt}%
\begin{center}
{\hfill\strut \mbox{\includegraphics[scale=#1]%
{#2}}\strut\hfill}%
\vspace*{1.5pt}\\ %
\strut \\
{\hfill\strut{\hbox{\vbox{\raggedright{{\bf Fig. \ref{#3}}\ \sl
#4}}}\strut}\hfill\strut}\end{center}%
\strut \\ 
}%
\newcommand{\dessinbis}[4]{%
\vspace*{12pt}%
\begin{center}
{\hfill\strut \mbox{\includegraphics[scale=#1]%
{#2}}\strut\hfill}%
\vspace*{1.5pt}\\ %
\strut \\
{\hfill\strut{\hbox{\vbox{\raggedright{{\bf Fig. #3}\ \sl
#4}}}\strut}\hfill\strut}\end{center}%
\strut \\ 
}%
\newcommand\Dalamb{\mathchoice\sqr68\sqr68\sqr{4.2}6\sqr{3}6}

\def\mettresous#1\sous#2{\mathrel{\mathop{\kern0pt #2}\limits_{#1}}}
\def\sqr#1#2{{\vcenter{\vbox{\hrule height.#2pt
          \hbox{\vrule width.#2pt height#1pt \kern#1pt
           \vrule width.#2pt}
           \hrule height.#2pt}}}}
\def\square{\mathchoice\sqr68\sqr68\sqr{4.2}6\sqr{3}6}
\def\ket#1{|#1\rangle}
\def\bra#1{\langle #1|}
\def\braket#1#2{\mathrel{\langle #1|#2\rangle}}
\def\elematrice#1#2#3{\langle #1|#2|#3 \rangle}
\def\inoutexpect#1{\elematrice{0,\mbox{out}}{#1}{0,\mbox{in}}}
\def\inout{\langle 0,\mbox{out}|0,\mbox{in}\rangle}
\def\lrpartial{\mathrel{\partial\kern-.75em\raise1.75ex
\hbox{$\leftrightarrow$}}}
\def\lrD{\mathrel{{\cal D}\kern-.75em\raise1.75ex\hbox{$\leftrightarrow$}}}
\def\om{\omega	}
\def\la{\lambda}

\def\lr #1{\mathrel{#1\kern-.75em\raise1.75ex\hbox{$\leftrightarrow$}}}

\def\beq {\begin{equation}}
\def\feq {\end{equation}}
\def\beqa {\begin{eqnarray}}
\def\feqa {\end{eqnarray}}

\def\p {\prime}

\def\gd {\delta}
\def\ge {\epsilon}

\def\gL {\Lambda}

\newcommand{\scal}[2] {\mbox{$ < #1 \vert #2 > $}}    \newcommand{\ki}{\mbox{$ \ket{\psi_i} $}}
\newcommand{\bi}{\mbox{$ \bra{\psi_i} $}}  
\newcommand{\e}{\mbox{$ \epsilon $}}

\catcode`\@=11
\newcommand{\INTRO}[1]
{
\renewcommand{\thechapter}{\fnsymbol{chapter}}
\setcounter{chapter}{-1}
\renewcommand{\@chapapp}{}
\chapter{#1}
\renewcommand{\thechapter}{\arabic{chapter}}
\renewcommand{\@chapapp}{Chapitre}
}
\catcode`\@=12

\begin{document}

\begin{flushright}
ULB-TH 95/02\\
UMH-MG 95/01\\
LPTHENS\\
January 1995\\
\end{flushright}
\vskip 2.5 truecm
\centerline{\Huge{A Primer for Black Hole Quantum Physics}}
\vskip 1. truecm
\centerline{R. Brout\footnote{e-mail: smassar @
ulb.ac.be},\addtocounter{footnote}{-1} 
S. Massar\footnotemark $\/ ^{,}$\footnote{Boursier IISN}}
\centerline{Service de Physique Th\'eorique, Universit\'e Libre de Bruxelles,}
\centerline{Campus Plaine, C.P. 225, Bd du Triomphe, B-1050 Brussels, Belgium}
\vskip 5 truemm
\centerline{R. Parentani\footnote{
e-mail: parenta@physique.ens.fr}}
\centerline{Laboratoire de Physique Th\'eorique de l'Ecole
Normale Sup\'erieure\footnote{Unit\'e propre de recherche du C.N.R.S.
associ\'ee \`a l' Ecole
Normale Sup\'erieure et \`a l'Universit\'e de Paris Sud.}}
\centerline{24 rue Lhomond
75.231 Paris CEDEX 05, France}
\vskip 5 truemm
\centerline{Ph. Spindel\footnote{e-mail: spindel @ umh.ac.be}}
\centerline{M\'ecanique  et Gravitation, Universit\'e de Mons-Hainaut, 
Facult\'e des
Sciences,}
\centerline{15 avenue Maistriau, B-7000 Mons, Belgium}
\vfill
\newpage
\strut
\vskip 14truecm
\strut
\hfill
\begin{minipage}{6cm}{
\begin{flushright}
{\it And so long as we have this itch for explanations, must we not always
carry around with us this cumbersome but precious bag of clues called
History? }\\
\noindent --- G. Swift, {\it Waterland} (1983)
\end{flushright}
}\end{minipage}
\newpage

\centerline{\huge { A Primer for Black Hole Quantum Physics}}
\vskip 2.5truecm
{\bf Abstract: }The mechanisms which give rise to Hawking radiation are revealed by analyzing
in detail pair production in the presence of horizons. 
In preparation for the black hole problem, three preparatory problems are
dwelt with at length:  pair production in an external electric field,
thermalization of a uniformly accelerated detector and accelerated mirrors.
In the light of these examples, the black hole evaporation problem is then
presented.

The leitmotif is the singular behavior of modes on the
horizon which gives rise to a steady rate of production. Special emphasis is
put on how each produced particle contributes to the mean albeit arising
from a particular vacuum fluctuation. It is the mean which drives the
semiclassical
back reaction. This aspect is analyzed in more detail than heretofore and in
particular its drawbacks are emphasized. It is the semiclassical theory which
gives rise to Hawking's famous equation for the loss of mass of the black
hole due to evaporation $dM/dt \simeq -1/M^2$. Black hole thermodynamics is
derived from the evaporation process whereupon the reservoir character of the
black hole is manifest. The relation to the thermodynamics of the eternal
black hole through the Hartle--Hawking vacuum and the Killing identity are
displayed. 

It is through the analysis of the fluctuations of the field
configurations which give rise to a particular Hawking photon that the
dubious character of the semiclassical theory is manifest. The present
frontier of research revolves around this problem and is principally
concerned with the fact that one calls upon energy scales that are greater
than Planckian and the possibility of a non unitary evolution as well. These
last subjects are presented in qualitative fashion only, so that this review
stops at the threshold of quantum gravity. \newpage

\tableofcontents

\INTRO{Acknowledgments}

Out of little acorns big oaks grow. This review grew out
of a short series of lectures by R. Brout at the
University of Crete, Iraklion in May 1993. The topic was
``Vacuum Instability in the Presence of Horizons''. It has
grown into a rather complete introduction to quantum black
hole physics as we now see it, a half baked but intensly
interesting subject.

R. Brout would like to express his gratitude to Professor
Floratos and the other members of the physics faculty at
Crete, for their heartwarming hospitality.

We are beholden to many colleagues for the generous
offerings of their wisdom. They are too numerous to
enumerate, but we wish to cite in particular: R. Balbinot,
J. Beckenstein, A. Casher, P. Grove, J. Katz, A. Ottewill,
S. Popescu, T. Piran, G. Venturi and most of all F.
Englert for his interest, advice and fruitfull discussions.

\strut \vfill \pagebreak


\INTRO{Introduction}
\par This review is conceived as a pedagogical essay on black hole quantum
physics.
\par Black hole physics has a rich and varied history. Indeed a
 complete coverage of the subject would run parallel to a good part of the
development
of modern theoretical physics, embracing as it does general
relativity, quantum field theory and thermodynamics. The present review, a
primer that is
designed to bring the reader to the threshold of current research, covers
the middle ground.
It is a review of that aspect of the theory that dates from Hawking's seminal
 work on black hole evaporation up to but not including quantum gravity.
This latter
chapter has been the subject of intense investigation over the past few years
 and is still at a very speculative stage, certainly not yet in a 
state to receive consecration in a review article. What has emerged from
investigation
of present quantum field theory, in the presence of a black hole is the
awareness
that the problem cannot be confronted without at least some aspects that
emerge from the
(as yet unknown) quantum theory of gravity. It is our intention here to
develop the 
theory to the point where the reader will have a clear idea 
of the nature of the unsolved problems given the present state of our
ignorance. An optimist
would say that the solution to these problems leads to the discovery 
of the quantum theory of gravity. But even a minimalist will admit that a
good deal is to
 be learnt from black hole physics 
at its present stage of development. It is perhaps not too
much to hope that a careful presentation in terms 
of present concepts can illuminate the way to the unknown. Such is our endeavor.

\par There exist several situations in quantum field theory
which give rise to phenomena which are similar to black hole
evaporation.  Our own experience has been that it is
very helpful to analyze those examples in detail.
They furnish a body of information so as to constitute a useful
theoretical laboratory for the study of the more complex
issues which arise in the black hole problem.  To this end the
initial chapters of this review are devoted to these laboratory
exercises~: pair production in a static electric field, accelerated
systems which become spontaneously excited in Minkowski-space
and accelerated mirrors.  
 Black hole evaporation
is then presented in the light of these examples.
\par The common leitmotif that runs through all of this will
gradually emerge upon reading.  The two central elements are~:
\begin{enumerate}
\renewcommand{\labelenumi}{\arabic{enumi})}
\item The existence of a horizon in each case, a separation of
space-time into portions, which would have been in causal contact
in other circumstances, that are no longer so. Straight
trajectories become curved so as to approach or recede from
the horizon exponentially  slowly as seen by an external
observer.  The result of this law of approach is a
modification of the spectral decomposition of the 
quantum matter fields in such manner as to
give rise to a steady state of production. 
\item The
mechanism of production is the steady conversion of vacuum
fluctuations into particles.  A large part of the analysis
is devoted to the description of how a particular
fluctuation gives rise to a particular production event.
\end{enumerate} 
\par In this way, we reveal the mechanisms
at work 
in black hole evaporation.
These are understood in terms of usual quantum mechanics.
For the black hole problem they turn out to be
surprisingly complex in that they involve
a subtle interplay between the conversion of vacuum
fluctuations to physical particles and a restructuration
of the vacuum itself. Each gives its contribution to the
energy momentum tensor that provides the source of gravitational
feed-back.  Some of this richness is already present in the semi-classical
approximation 
wherein the source of Einstein's equations is given by the mean 
energy-momentum tensor.
But this approximation will certainly break 
down at later stages, obviously at
the Planckian stage, but possibly before, even much before.
 What we have set out to
do is display the structures at work in the hope that this
approach will sharpen the focus on the types of problems
which will have to be confronted at later stages.
The hope is that, from thence, quantum gravity will emerge.
But, as introduced in one BBC radio program, that was next week's news!

Considerable effort has been put into exhibiting the configurations of field
energy which contribute to the production mechanism. This is the content of
Section \ref{pair}, \ref{fluctmirr}, \ref{state} and \ref{VFHR}. Particularly
relevant for black hole physics is Section \ref{fluctmirr}. Of necessity, the
analysis is not without subtlety (though rigorous and mathematically
straightforward). The reader, who wants to get into the midsts of the
black hole problem immediatly, may choose to skip Sections \ref{fluctmirr} and
\ref{state} in a first reading. For his benefit Section \ref{VFHR} contains a
summary of all the necessary information related to the configurations of the
fluctuations. The reader may continue on from this point (and of course
then return if he wishes to the demonstrations and more detailed conceptual
discussion of Sections \ref{fluctmirr} and \ref{state}.

One word on presentation. Generally speaking, the type of physics discussed
is more conceptual than technical in character. As such the mathematical
formalism is usually quite simple, unencumbered by extensive algebraic
manipulation. Nevertheless, from time to time, the algebra does become
heavy. When we have judged 
that the details of the mathematics interrupts smooth reading, we begin
the topics under discussion with a resume of the results and relegate
details to a passage enclosed within square brackets, or to an appendix.

\strut \vfill \pagebreak


\chapter{Pair Production in a Static Electric Field}\label{ELEC}
\section{Qualitative Survey}\label{qualitative}
It was shown early on by Heisenberg and collaborators~\cite{HeEu} that the
vacuum of quantum field theory in the presence of a static
electric field is unstable against the creation of charged
pairs.  Subsequently using techniques of functional integration
in the context of the action formalism, Schwinger~\cite{Schw} showed in an
almost one line proof that the overlap function between vacuum
at the time the field was turned on with vacuum
at subsequent times decays exponentially fast.

After Hawking discovered black hole radiation~\cite{Hawk} several
authors~\cite{Mul}, \cite{Step},\cite{BPS},\cite{PaBr0}
 pointed out the analogy with this kind of pair production.
Though the analogy is not strict the formalisms which
are used to encode these two phenomena have many points in 
common
thereby making the electric pair production problem a 
fruitful exercise.
 To introduce the problem we first give a brief
sketch of the physics in terms of a given mode (solution
of the wave equation).
\par We shall work in one dimension with a charged scalar field.
This contains enough physics to make the exercise applicable
to the black hole problem. For definiteness we choose the gauge
$A_t =  Ex$, $A_x = 0$ (with the charge absorbed into the definition of $E$)
whereupon the Klein-Gordon equation
is 
\begin{eqnarray}
 \square \varphi = \left[ + \partial^2_x - \left( \partial_t - i Ex
\right)^2 \right] \varphi = m^2 \varphi \qquad .\label{boxphi}
\end{eqnarray}
Since the electric field is static, in this time independent gauge the operator
$\partial_t$
commutes with the d'Alembertian operator $\square $, and modes can be taken of the
form $\varphi_{\omega}(t,x)={\rm e}^{-i \omega t} f_\omega(x)$ where $f_\omega(x)$
 satisfies the equation
\begin{eqnarray}
\left[ \partial^2_x + \left( \omega + Ex \right)^2
\right] f_\omega (x) = m^2 f_\omega (x) \label{eqchi}
\end{eqnarray}
Upon dividing eq.~(\ref{eqchi}) by the factor $2m$, one comes
upon a non relativistic Schr\"odinger equation in the
presence of an upside down oscillator potential centred at the
point $x_c = - \omega/E$.  The eigenvalue for the effective energy,
 is $-m/2$. Therefore the solution tunnels
between two turning points 
$\left(x_c \pm m/E =x_c \pm a^{-1}\right)$ where $a$ is the classical
acceleration due to the field.
\par Take the case $E > 0$.  Then classically a particle  is 
uniformly accelerated
to the right, following the classical trajectory $(x-x_c)^2-(t-t_c)^2=a^{-2}$. So if
it comes in asymptotically from the right along its past horizon $(x-x_c)=-(t-t_c)$
it will turn around at $x=x_c+a^{-1}$ and $t=t_c$.  In quantum
mechanics this is translated  into the following description.
A wave packet centred for example around $\omega = 0$ (i.e. $x_c=0$)
will come  in from the right
asymptotically at the speed of light, slow down and turn around at
$x = a^{-1}$ and fly off to the right approaching once  more the
speed of light along its future horizon.  The incident  packet is
 localized around the classical orbit
$x^2 - \left(t-t_c\right)^2 = a^{-2}$.  Near the point $x =+a^{-1}$
there will be some amplitude  to tunnel through to
$x = - a^{-1}$ and then continue from that point on to
accelerate to the left.  Since $E > 0$, the tunneled particle
must therefore have been ``mesmerized" into an antiparticle
in the tunneling region.  This situation can only be met
by second quantizing $\varphi$ (confronting the Klein
paradox~\cite{Klein}).  The ``mesmerization" is simply reflecting the fact
that there is a probability amplitude to create 
a pair in the tunneling region.  Then the
above scattering
description must be amended to: the initial particle is scattered
as classically, but its final flux after the scattering is increased
by a factor $\alpha^2$ to accommodate  for the creation of an
antiparticle on the other side. The flux of this latter is
denoted by $|\beta|^2$.  Thus we see that $|\alpha|^2 -
|\beta|^2 = 1$ is the statement of charge conservation.
\par 
In Fig.~\ref{F1}.
\dessin{1.000}{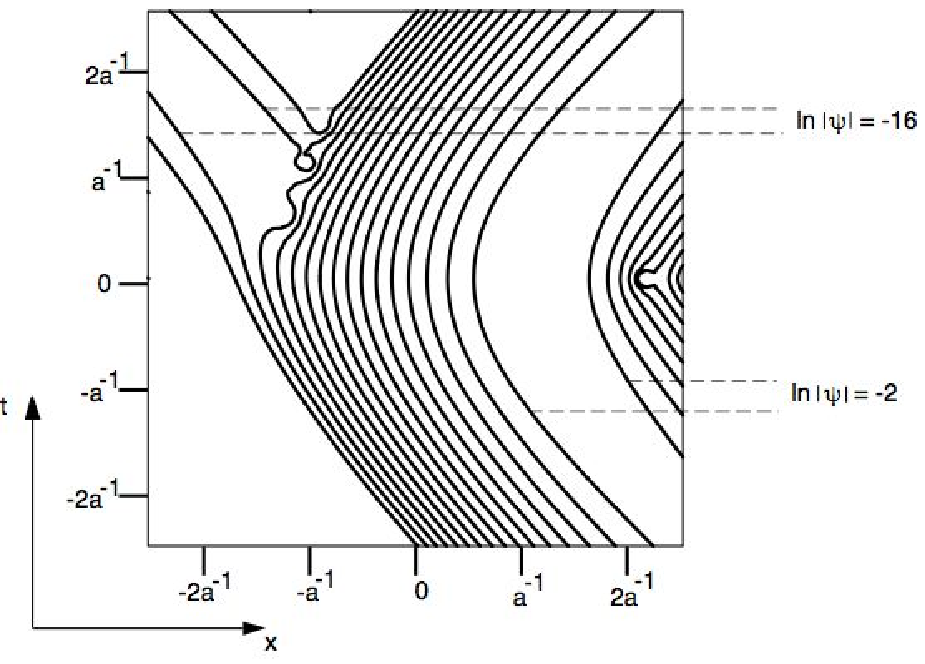}{F1}{The function $\ln \vert \psi(t,x)\vert$ is plotted for an incoming
wave packet solution of eq.~(1.1). It is a minimal packet as described
subsequently in the context of eq.~(1.54). We use the logarithm
because of the very small amplitude of the produced particle ($e^{-\pi m / 2 a}$
where we have taken $m/a = 9$). The particle (on the right) is centered on its
classical trajectory $x^2 - t^2 = a^{-2}$. The wave function of the antiparticle
behaves classically for $t>0$ and $x<-a^{-1}$ and vanishes in the past.}
we have plotted the modulus of a wave packet. One sees both the
classical path about which the packet is centered 
and the created pair.  
Note that the separation between the
turning points $\pm a^{-1}$, the length of the region from which the pair
emanates, coincides with
the distance $|\Delta x|$ necessary to have the electrostatic energy
compensate the rest mass $( E |\Delta x|=2m)$, so as to make
possible on mass shell propagation from these points outward.
It is the realization of this possibility, without energy cost, that causes
the vacuum instability.  Note also that since the amplitude of probability
for a vacuum fluctuation to have its particle-antiparticle
components separated by a space-like interval,
$|\Delta x|$, is $\exp (- m|\Delta x|)$, it may be anticipated that the
amplitude for production is
$\beta \sim \exp \left[-C(m/a)\right]$ where $C$
is some constant of $O(1)$ to be calculated by the formalism, and found to 
be equal to $\pi$. We have thus shown how a wave packet not only describes
 the classical
orbit but gives an idea of the location of the produced pair 
as well as its  probability amplitude.

\section{Mode Analysis}\label{mode}
We shall start with the analysis of eq.~(\ref{eqchi}) considered as the 
Schr\"odinger equation of a fictitious non-relativistic problem.
The solutions are well known \cite{WiWa}.  We will nevertheless 
present them
by a method~\cite{PaBr0}
 which is suitable for subsequent use in that it displays
the singularity of the solution along the horizons. It is this singular
behavior which in every case (pair production in an electric field,
accelerating observer and mirror, black hole) encodes the
production phenomena
 that is peculiar to quantum physics in the presence of horizons.

The function $f_\omega$, solution of eq.~(\ref{eqchi})
is the $x$
representation of 
a ket 
$\ket{\Xi}$, i.e. $f_\omega (x)= \braket{x}{\Xi}$.
 Go over to dimensionless variables centered at the origin by introducing
\begin{eqnarray}
\xi = \sqrt E \left(x + \omega/E \right) \label{2.3}
\end{eqnarray}
so as to write (\ref{eqchi}) in the suggestive form
\begin{eqnarray}
\frac{\pi^2 - \xi^2}{2} \langle \xi | \Xi \rangle = - \varepsilon
\langle \xi | \Xi \rangle \label{sugxichi}
\end{eqnarray}
where
\begin{eqnarray}
&& \varepsilon = m^2/2E = m/2a \label{2.5} \\
&& \left[\pi,\xi\right] = - i \quad \mbox{($\pi = - i \partial/\partial \xi$
in $\xi$-representation)} \nonumber
\end{eqnarray}
Now introduce the new canonical variables~\cite{BaVo}, analogous to
the standard annihilation and creation operators used to quantize the harmonic
oscillator, defined by \begin{eqnarray}
 \left. U \atop V \right\}& =& \frac{1}{\sqrt 2} 
\left[ \pi \mp \xi \right] \nonumber \\
 \left[ U,V \right]& =& - i \label{uvrep}
\end{eqnarray}
so that (\ref{sugxichi}) becomes
\begin{eqnarray}
u \frac{\partial}{\partial u} \langle u|\Xi \rangle =
\left(+ i \varepsilon - \frac 12 \right) \langle u | \Xi
\rangle \label{equchi}
\end{eqnarray}
We can go back to $\xi$-representation through
\begin{equation}
\langle \xi | \Xi \rangle = \int du \langle \xi | u \rangle
\langle u | \Xi  \rangle \label{xichi}
\end{equation}
where we have adopted the notation $U|u\rangle = u|u\rangle$, defining  
the $u$ representation whereupon
$V=i\partial _u$.  To find the kernel $\braket{\xi}{u}$ of
this unitary transformation  one must solve the conditions
\begin{eqnarray*}
&& \langle \xi \left| \frac{V-U}{\sqrt 2} \right| u \rangle =
\frac{1}{\sqrt 2}\left(-i \frac{\partial}{\partial u} - u \right)
\langle \xi | u \rangle = \xi \langle \xi | u \rangle       \\
&& \langle \xi \left| \frac{(\pi - \xi)}{\sqrt 2} \right| u
\rangle = \frac{1}{\sqrt 2} \left(- i \frac{\partial}{\partial \xi}
- \xi \right) \langle \xi |u \rangle = u \langle \xi | u \rangle
\end{eqnarray*}
to yield
\begin{equation}
\langle \xi | u \rangle  = {1\over \sqrt{2^{1/2}\pi}} \exp i [\xi^2/2+\sqrt{2}\xi u+u^2/2 ]
\label{xiu}
\end{equation}
where we have supplied the norm to make the transformation unitary.
\par\noindent The general solution of (\ref{equchi}) is
\begin{equation}
\lambda_{\varepsilon}(u)\equiv \langle u |  \Xi  \rangle  = 
A \ \Bigl[ \theta(u) { u^{i \varepsilon - 1/2}\over
\sqrt{2\pi}} \Bigr]
+ B \ \Bigl[ \theta(-u) {(-u)^{i \varepsilon - 1/2} \over
\sqrt{2\pi}} \Bigr]
\label{uchi}
\end{equation}
The constants $A$, $B$ follow from initial conditions. Consider the classical orbits
associated with eq.~(\ref{sugxichi}), described by the Hamiltonian problem 
$H= u v 
$. They are given by $ u=u_0 e^{-\tau} , v=v_0 e^{\tau}$ with $u_0v_0=-
\varepsilon$.
 At early times $\tau$ the orbits located near
$u=-\infty$, $v=0^+$ 
(i.e. incoming from $\xi = +
\infty$) are described by $A=0$ and the ones near $u=+\infty$, $v=0^-$ 
by $B=0$. These two sets of solutions
(labeled by $\varepsilon $) are orthogonal, and complete. In the effective
Schr\"odinger problem
 which we are discussing their norm (with respect to the measure $du$) is
such that $\int^{+\infty}_{-\infty} du \lambda_{\varepsilon}^{*} 
\lambda_{\varepsilon'}
=\delta(\varepsilon - \varepsilon')$. This fixes $|A|^2 +|B|^2=1$.
\par Inserting each of the 
above solutions and (\ref{xiu}) into (\ref{xichi}) gives
an integral representation for the scattered mode $\langle \xi |  \Xi  \rangle$.
Reference~\cite{WiWa}, identifies these as integral representations of  parabolic
cylinder functions, more succinctly, Whittaker functions $D_{\nu}(z)$ 
(with $\nu = - i \varepsilon -1/2$) and provides 
their properties and
connection formulae. For our purposes however
it is instructive to continue to work as ``quantum mechanics" 
(see ref.~\cite{PaBr0}) and show how to relate asymptotic
amplitudes to on-shell quanta. 
The effort is worthwhile because of remarkable analogs with the modes
that arise in black hole physics  (the reason being that the 
singular character of
the solution eq.
(\ref{uchi}) near the origin encodes the existence of a horizon). 

\par Take the solution (\ref{uchi}) with $B=0$.  We 
then must analyze
\begin{eqnarray} \langle \xi |  \Xi  \rangle &=& 
\int^\infty_0 {du \over 2^{3/4}\pi} 
u^{i\varepsilon-1/2}\exp i [\xi^2/2+\sqrt{2}\xi u+u^2/2 ] \label{xichiu}\\
\Bigl[&=&\frac{1}{2^{3/4}\pi}e^{-\varepsilon \pi /4}e^{i\pi/8}\Gamma
(1/2+i\varepsilon) D_{-i\varepsilon -1/2}(\sqrt{2}e^{-i\pi/4}\xi)\Bigr]\nonumber
\end{eqnarray} in the limit $\xi \to \pm \infty $.  The saddle points of the
integral lie at 
\begin{eqnarray}
u^{\star}= {-\sqrt{2}\xi \pm \sqrt{2\xi^2-4\varepsilon}\over 2} \label{usadpts}
\end{eqnarray}
For $\xi \to -\infty $, the saddle at $u^{\star}= - \sqrt{2} \xi + 
\varepsilon/(\sqrt{2}\xi)$
may be used to evaluate the integral \ref{usadpts} by the saddle point method since it lies well within
the limits of integration and one has therefore the reliable asymptotic 
estimate (W.K.B.
approximation) 
\begin{equation}
\langle \xi |  \Xi  \rangle_I =\frac{1}{\sqrt{2 \pi|\xi |}}
\exp {-i\left[\xi^2/2 -\varepsilon \ln(\sqrt{2}|\xi|)-\pi/4\right]}\label{asIxichi}
\end{equation}
Since the solution that we are describing ($B=0$ in eq.~(\ref{uchi}))
corresponds
 to an incoming particle coming from the left, hence with $u_{classical}\to 
\infty $ 
as $\xi \to - \infty $ for the incident beam, eq.~(\ref{asIxichi})
is to be identified with the incident wave function ($I$).
For $\xi \to + \infty $, the saddle at $u^{\star}=  - \sqrt{2} \xi + 
\varepsilon/(\sqrt{2}\xi)$
gives no contribution since it lies completely 
outside the limits of integration.

The other saddles at $u^{\star}= \varepsilon/(\sqrt{2}\xi)$
 cannot be used to evaluate the integral by saddle point integration since, 
for $\xi \to \pm \infty $, they lie 
at the edge of the integration domain.
However we expect that, since  classically,
$v \to + \infty $ for the transmitted solution 
and $v \to - \infty $ for the reflected one,
 one
might fruitfully exploit the $v$-representation.  This is indeed the case.  The
$v$-representation of $\ket{\Xi}$ is \begin{eqnarray}
\langle v |  \Xi  \rangle & =&\int^{+\infty}_{-\infty}\langle v | u\rangle 
         \langle u | \Xi\rangle \/ du \nonumber \\
&=& \int^{+\infty}_{0}\frac {e^{ivu}}{\sqrt{2\pi}}
       \frac { u^{i \varepsilon - 1/2}}{\sqrt{2\pi}} \/du \nonumber \\
& = &\ e^{i\pi/4}e^{-\varepsilon\pi/2}\frac{ \Gamma(1/2+i\varepsilon)}{\sqrt{2\pi}}
\\Bigl[ \theta(v) { v^{-i \varepsilon - 1/2}\over
\sqrt{2\pi}} \Bigr] \nonumber  \\
& & +e^{-i\pi/4}e^{\varepsilon\pi/2}
\frac{ \Gamma(1/2+i\varepsilon)}{\sqrt{2\pi}}
\Bigl[ \theta(-v) {(-v)^{-i \varepsilon - 1/2} \over
\sqrt{2\pi}} \Bigr]
  \nonumber \\
& \equiv &T \ \Bigl[ \theta(v) { v^{-i \varepsilon - 1/2}\over
\sqrt{2\pi}} \Bigr]
+ R \ \Bigl[ \theta(-v) {(-v)^{-i \varepsilon - 1/2} \over
\sqrt{2\pi}} \Bigr] \label{vchi}
\end{eqnarray}
This result follows from standard analysis,
most easily by deformation of the contour from the real axis to the 
 positive (resp. negative) imaginary axis
for positive (resp. negative) values of $v$. It is this
difference that
gives the different weights $R$, $T$
of the reflected to transmitted
amplitudes, 
the ratio of which is
\begin{eqnarray}
|\frac RT| = e^{\varepsilon \pi}\label{R/T}
\label{RTratio}
\end{eqnarray}
This exponential ratio is generic to all problems having horizons since the
behavior of the wave functions at the horizon gives rise to the singular form
$u^{i\epsilon}$, hence analytic continuation as exhibited in \ref{vchi}.

Since we have normed $\ket{\Xi}$, the unitarity condition $|R|^2+|T|^2=1$ is
fulfilled by eq.~(\ref{vchi}), as is easily checked thanks to
the formula $|\Gamma(1/2+i\varepsilon)|^2=\pi /
\mbox{\rm cosh}(\pi\varepsilon)$.

The identification of the waves 
multiplied by $R$ and $T$ as the reflected 
and transmitted waves,
thereby giving a
precise expression to the physics of the tunneling process, 
follows from
the asymptotic behavior of $\langle \xi |\Xi\rangle$. Indeed
\begin{eqnarray}
\langle \xi |\Xi\rangle &=&
\int^{+\infty}_{-\infty}\langle \xi |v\rangle \langle
v|\Xi\rangle dv
\nonumber  \\
&=&\  T \int^{+\infty}_{0}\frac {e^{i[-\xi^2/2+\sqrt{2}\xi
v-v^2/2+\pi/4]}}{2^{3/4}\pi}
v^{-i \varepsilon - 1/2}dv\nonumber  \\
& &+ R \int^{0}_{-\infty}\frac{e^{i[-\xi^2/2+\sqrt{2}\xi
v-v^2/2+\pi/4]}}{2^{3/4}\pi}
 (-v)^{-i \varepsilon - 1/2}dv\label{vxichi}  \\
\Bigl[&=&\frac{1}{2^{3/4}\sqrt{2\pi}}\frac{e^{\varepsilon \pi/4}e^{-i\pi/8}}
{\cosh{\pi\varepsilon}}
 \left(i e^{-\varepsilon\pi}D_{i\varepsilon
-1/2}(\sqrt{2}e^{-3i\pi/4}\xi)\right. \Bigr.\nonumber\\
&&\Bigl.\left.\quad\quad +D_{i\varepsilon
-1/2}(\sqrt{2}e^{i\pi/4}\xi)\right)\Bigr]\quad .
\end{eqnarray} 
The phase of $\braket{\xi}{v}$ (
$= -\xi^2/2+\sqrt{2}\xi v-v^2/2+\pi/4$) is fixed
by the relation $\braket{\xi}{v}=\int \braket{\xi}{u}\braket{u}{v} du$.
The saddle points 
are now located at 
\begin{eqnarray}
 v^{\star} =    
{\sqrt{2}\xi
\pm \sqrt{2\xi^2-\varepsilon}\over 2} \label{vsadpts} 
\end{eqnarray}
The saddles that go to $\pm \infty $ as $\xi \to \pm \infty $ are
responsible for the transmitted and reflected  solution respectively. For these
the integration can be estimated by a
saddle point approximation to give
\begin{eqnarray}
\langle \xi | \Xi \rangle_T & {\mettresous \xi \to + \infty
\sous =}
&T{e^{i[\xi^2/2-\varepsilon\ln\sqrt{2}\xi ]}\over \sqrt{2\pi\xi}}\qquad , 
\label{Txichi} \\
 \langle \xi | \Xi \rangle_R & {\mettresous \xi \to - \infty
\sous =}&R{e^{i[\xi^2/2-\varepsilon\ln|\sqrt{2}\xi |]}\over \sqrt{2\pi|\xi
|}}\qquad .
\label{Rxichi}
\end{eqnarray}

The other saddles ($v \to \pm 0$ as $\xi \to \pm \infty$)
are concerned with the incident
solution which has already been
analyzed in $u$-representation. This is fortunate in that  the width of these
saddles
extend outside the domains of integration
in eq.~(\ref{vxichi}). 
We call attention to the 
complementary r\^ oles of the saddle points in $u$ and $v$-representations 
in isolating the asymptotic incident wave on one hand and the transmitted and
reflected parts of the solution on the other.
As previously mentioned, this is due to the fact that the locus of saddle
points is along
classical trajectories.

\section{Vacuum Instability}\label{vacuum}
\par The above solution and notation describes the effective Schr\"odinger
wave function which solves
(\ref{eqchi}). We now analyze how it encodes vacuum instability and pair production.
 As we have announced the transmitted wave must
have opposite charge.  This can seen
by following the movement of physical wave packets, solution of the Klein-Gordon
equation (\ref{boxphi}).  These are of the form
\begin{eqnarray}
\int d\omega f(\omega - \omega_0) e^{-i \omega(t-t_0)}
\chi_{\varepsilon}(\sqrt{E}(x+\omega/E)) \qquad .\label{wavepakt}
\end{eqnarray}
Note the novelty of the construction  wherein the eigenvalue,
$\omega$, appears in the argument of the function $\chi_{\varepsilon}(\xi)
= \langle \xi | \Xi \rangle$ 
since $\xi = \sqrt E \left(x + \omega/E \right)$.  It is this
circumstance that allows for the correct charge and current
assignments of each  asymptotic part of the 
wave (\ref{asIxichi}), (\ref{Rxichi}),
(\ref{Txichi})~:
\begin{eqnarray}
e^{-i \omega t} \left\{ \begin{array}{ll}
1\ e^{-iE(x + \omega/E)^2/2}|x + \omega/E|^{im^2/2E-1/2} &\equiv {\cal I}\
(x\to -\infty ,t\to +\infty) \\
R e^{iE(x+\omega/E)^2/2}|x + \omega/E|^{-im^2/2E-1/2} &\equiv {\cal R}\
(x\to -\infty ,t\to -\infty)\\
T e^{iE(x+\omega/E)^2/2} (x + \omega/E)^{-im^2/2E-1/2}&\equiv {\cal T} \
(x\to +\infty ,t\to +\infty)
\end{array} \right.\\ \nonumber
\strut\label{asymptwave}
\end{eqnarray}
The motion of each branch is given by the group velocity
obtained by setting the derivative with respect to $\omega$ of the phases
equal to zero.  Thus, ${\cal I}$ corresponds to motion to the left 
at late times, 
whereas ${\cal R}$ describes motion 
 to the right at early times and
${\cal T}$ to the right, but at late times.  The r\^ oles of ${\cal I}$ and ${\cal R}$ have been
swapped once we consider the dynamics with respect to the physical time appearing in the 
Klein-Gordon equation (\ref{boxphi}) instead of the fictuous dynamics associated
to the solutions of the Schr\"odinger equation (\ref{equchi}).
\par Charge assignments follow from the sign of the charge density
 \begin{eqnarray}
J^t=-
J_t\equiv \varphi^*_{\omega} i {\lrD_t} \varphi_{\omega}=
\varphi^*_{\omega}(i \lr {\partial_t} +2 E x)\varphi_{\omega}=
2(\omega +Ex)\chi_{\varepsilon}^*\chi_{\varepsilon}
\label{jt}
\end{eqnarray}
  Thus
${\cal I}$ and ${\cal R}$ which live at large negative $x$ have negative
charge
whereas ${\cal T}$ has positive charge.   
In summary, for $E > 0$ the 
solution of eq.~(\ref{uchi}) with $B=0$ represents
an incoming anti-particle of amplitude $R$, and
outgoing reflected anti-particle of amplitude $1$ 
and a transmitted particle of amplitude $T$.   The
unitarity relation $|R|^2 + |T|^2 = 1$ is then best recast into
a form out of which one reads charge conservation rather than
conservation of probability.  Divide this relation by $|R|^2$ and recast
it into the form
\begin{eqnarray}
 &-|\alpha|^2 + |\beta|^2 = -1 \label{unitarity}& \nonumber\\
 &|\alpha|^2 = 1/|R|^2 \quad; \quad |\beta|^2 =
|T|^2/|R|^2 &\label{coefbog}
\end{eqnarray}
The right hand side then corresponds to incident flux of
unit negative charge coming from $x=-\infty$, $|\beta|^2$ is the outgoing 
 transmitted flux of
positive charge going to $+\infty$ and $(|\alpha|^2 - 1)$ is the increase of
flux of the reflected wave necessary to implement charge
conservation.
\par For each value of $\omega$, in addition to the above mode, there is
its parity conjugate, obtained from the transformation $(x+\omega /E)\to
-(x+\omega /E)$ and complex conjugation.
Clearly it represents the charge conjugate since one need only run through
its mirror image about the 
axis $x_c (=- \omega/E)$. It corresponds to the solution
of eq.~(\ref{sugxichi}) proportional to $\theta (v)$.
The modulus of a wave packet built of this latter mode 
is given in Fig.~\ref{F1}.
\par From these considerations, it follows that these basis functions for the
quantized 
field, whose quanta are single particles in the past (formed from wave packets),
lose this property in the future due the production. More  specifically let us
envisage the situation wherein $E$ is switched on during a time interval $0\leq
t\leq T$, over
a length $L$. Then the modes fall into two classes:
\begin{itemize}
\item{Class $I$:} Those that scatter $(\beta \neq 0)$ i.e. whose turning
points lie
within this space-time domain of area $LT$.
\item{Class $II$:} Those which are only deflected, so as to follow the
classical trajectory
before they reach the would-be turning point (in the sense of wave
packets). For these latter
$\beta =0$.
\end {itemize}
\par Indeed, we shall see subsequently  that, for modes in the first class,
if a
pair is formed, it will emerge from a region of area ${\cal O}(a^{-1}\times a^{-1})$
whereas
in the second class no pair is formed to within ``edge corrections". By
this, we mean 
for $L>>a^{-1}$ and $T>>a^{-1}$, no pair is formed except near the edges.
Thus to obtain
asymptotic results of ${\cal O}(L\times T)$ for the production of pairs, it is a
legitimate idealization to divide the modes into these two classes. For the
remainder
of this chapter, all considerations are devoted to those in the non trivial
$(\beta
\neq 0)$ class $I$. For these modes, the future fate of a single particle
mode in the
past is to become a many particle mode, hence not convenient 
to describe the result
of a counter experiment  in the future. Thus, in addition to the modes
previously
presented --~denoted by  ``in-modes"~-- we shall be obliged to introduce
``out-modes" as well. These latter will represent single-particle quanta
in the future of their turning points. 
\par It is in the framework of second quantized field theory that this discussion
makes sense. Therefore we begin by quantizing the field
operators in
the in-modes. The complete set is given by
\begin{eqnarray}
&& \varphi^{in}_{p,\omega}(t,x) =\frac{1}{(4 E)^{1/4} R^{*}} 
e^{-i\omega t}  \chi^{*}_{\varepsilon}(-x - \omega/E)
\label{inpphi} \\
&& \varphi^{in*}_{a,\omega}(t,x) =\frac{1}{(4 E)^{1/4} R}
 e^{-i \omega t} \chi_{\varepsilon} (x+\omega/E)
\label{inaphi}
\end{eqnarray}
Note that for completeness, $\omega$ should span the range $[-\infty, +\infty]$
 contrary to the case without $E$-field
where the energy has to be taken positive only.
The wave functions have been 
normed according to the Klein-Gordon scalar product
\begin{eqnarray}
\int dx \varphi^{in}_{p,\omega}(t,x) i{\lrD _t}\varphi^{in*}_{p,\omega}(t,x)
&=&+\delta(\omega-\omega^\prime)\nonumber\\
\int dx \varphi^{in}_{a,\omega}(t,x)i {\lrD _t}\varphi^{in*}_{a,\omega}(t,x)
&=&-\delta(\omega-\omega^\prime)\qquad.
\end{eqnarray}
Here the label $p$ designates particle and $a$ anti-particle (described
by the parity
conjugate as previously discussed). The reason for the complex conjugation
in the l.h.s of eq.~(\ref{inaphi}) is that
the function $\varphi^{in}_{a,\omega}$ is a solution of the charge conjugate 
field equation (here obtained by changing the sign of the charge, i.e.
$E\to -E$ in the field equation). Then $\varphi^{in}_{p,\omega}$ and
$\varphi^{in
*}_{a,\omega}$ obey the same equation and the second quantized field is
written 
in the in-basis,
\begin{eqnarray}
\Phi (t,x) & = & \int d\omega \bigl(a^{in}_\omega \varphi^{in}_{p,\omega}
(t,x) + b^{in\dagger }_\omega \varphi^{in*}_{a,\omega}(t,x)\bigr)
\label{qphi} 
\end{eqnarray}
with the usual commutation relations
\begin{equation}
 \left[ a^{in}_\omega,a^{in\dagger }_{\omega^\prime} \right] =
 \left[ b^{in}_\omega, b^{in\dagger }_{\omega^\prime} \right] =
\delta(\omega - \omega^\prime) \label{comrel}
\end{equation}
\par The out-basis is obtained simply by observing that if one
follows a single outgoing branch backwards in
time it will trace out the same backwards history
as that of one of the in-modes  when it moves forward in time. 
These modes at fixed $\omega$ are obtained by sending $t$ into
$-t$ and complex conjugating so that the new functions remain solutions of the 
 field equation (\ref{boxphi}).
\begin{eqnarray}
\varphi^{out}_{p,\omega}(t,x) & = & \varphi^{in *}_{p,\omega}(-t,x)\nonumber\\
\varphi^{out*}_{a,\omega} (t,x)& = &  \varphi^{in}_{a,\omega}(-t,x)
\label{outaphiB}\end{eqnarray}
\noindent Furthermore, since for each $\omega$,
the two sets of in-modes eqs~(\ref{inpphi},\ref{inaphi}) are complete,
these out-modes must be linear combinations of them.
In fact they are given (after  choices of phases) by
\begin{eqnarray}
\varphi^{out}_{p,\omega} & = & \alpha \ \varphi^{in}_{p,\omega}
- \beta ^{*} \ \varphi^{in*}_{a,\omega} \label{outpphi} \\
\varphi^{out*}_{a,\omega} & = & \alpha^{*} \ \varphi^{in*}_{a,\omega}
- \beta \ \varphi^{in}_{p,\omega} \label{outaphi}
\end{eqnarray}
where $\alpha$ and $\beta$
are the coefficients introduced in (\ref{unitarity}, \ref{coefbog}), now fixed precisely as: 
\begin{eqnarray}
\beta = T/R\ ,\quad \alpha = e^{i\pi/4} /R
\end{eqnarray}
 To understand this
result it suffices 
to follow the histories of the
$\varphi^{in}$'s as wave packets~\cite{BMPPS}.  In eq.~(\ref{outpphi}), for example,
$\varphi^{in}_{p,\omega}$ will give rise to a reflected wave of
amplitude $ \alpha^*$ on the right and a transmitted wave of amplitude $\beta^*$
on the left,
whereas $\varphi^{in^*}_{a,\omega}$ gives a transmitted wave
of amplitude $ \beta$ on the right and reflected wave of amplitude
$\alpha$ on the left.  The contributions on the left therefore
cancel and those on the right have total amplitude $|\alpha|^2 -
|\beta|^2\ (= 1)$ as required 
for a particle out-mode. 
The assiduous reader will check that the
right hand side
of eqs~(\ref{outpphi},\ref{outaphi}) are indeed the complex conjugates of
the $\chi_{\varepsilon}$
functions defined in the in-modes (all multiplied by ${\rm e}^{i\omega t}$).
 The easiest route to obtain these relations is to use  
eqs~(\ref{xichiu},\ref{vxichi}) in the context of integral
representations
for $\langle\xi|\Xi \rangle$:
\begin{eqnarray}
\chi_{\varepsilon}(\xi)=e^{\frac{i\pi}{4}} \left[ T
\chi^*_{\varepsilon}(-\xi)+R\chi^*_{\varepsilon}(\xi) \right]
\end{eqnarray} 
 and the correct identification of these wave
functions as 
used in second quantization. 

\par One develops $\Phi$  in the out basis
by inverting  eqs~(\ref{outpphi}, \ref{outaphi}),
\begin{eqnarray}
\varphi^{in}_{p,\omega} & = & \alpha^* \ \varphi^{out}_{p,\omega}
+ \beta ^{*} \ \varphi^{out*}_{a,\omega} \label{inpout} \nonumber \\
\varphi^{in*}_{a,\omega} & = & \alpha \ \varphi^{out*}_{a,\omega}
+ \beta \ \varphi^{out}_{p,\omega}\qquad , \label{inaout}
\end{eqnarray}
 and substituting into eq.\ref{qphi} to give
\begin{eqnarray}
\Phi = \int d\omega \bigl(a^{out}_\omega \varphi^{out}_{p,\omega} +
b^{out\dagger }_\omega \varphi^{out^*}_{a,\omega}\bigr)\label{phiout}
\end{eqnarray}
where
\begin{eqnarray}
a^{out}_{\omega} & = & \alpha^*\ a^{in}_\omega + \beta\ b^{in \ \dagger }_{\omega}
\nonumber \\
b^{out \ \dagger }_{\omega} & = & \alpha \ b^{in \ \dagger }_\omega +
 \beta ^* a^{in}_{\omega}\
\label{bogtransf}
\end{eqnarray}
These equations (\ref{bogtransf}) define a Bogoljubov 
transformation~\cite{BD}
wherein eq.~(\ref{unitarity}) ensures that $a^{out}$ and $b^{out}$ obey
the correct commutation relation.  Historically it came up
when Bogoljubov~\cite{Bogo} 
noted that the free particle Bose-Einstein ground
state was unstable against creation of pairs of equal and opposite momentum
once interparticle interactions were introduced.   The interested reader
will find a brief account of Bogoljubov's considerations in 
Appendix~\ref{appbog}.
\par From (\ref{bogtransf}) it is clear that the in-vacuum $
\vert 0,\mbox{in} \rangle $ (i.e. the
state annihilated by in-annihilation operators) will contain out-particles,
ie. we set up this Heisenberg state at early times and count the mean number of
out particles that are realized from it at some later time.
\begin{equation} 
\langle
n_{\omega}\rangle = 
\elematrice{0,\mbox{in}}{a^{out\dagger }_{\omega}a_{\omega}^{out}}{0,\mbox{in}}
=|\beta|^2 
 \qquad .\label{nout}
\end{equation}

One can express the in-vacuum state 
as the linear combination~\cite{KaUm}
of out states since the Bogoljubov transformation is unitary.
\begin{eqnarray}
|0, \mbox{in} \rangle & = & N^{-1/2} \exp \left[\left(\frac {\beta}
{ \alpha}\right) \sum_\omega
a^{out\ \dagger }_{\omega} b^{out\ \dagger }_{\omega}\right]
\ |0,\mbox{out} \rangle \label{outvac}
\end{eqnarray}
This result is  obtained by setting $|0, \mbox{in} \rangle =
 f(a^{out\ \dagger },b^{out\ \dagger }) |0,\mbox{out} \rangle $ and 
imposing that $0= a^{in} |0, \mbox{in} \rangle =
(\alpha a^{out} - \beta b^{out\ \dagger } ) 
f |0,\mbox{out} \rangle = 
(\alpha [ a^{out}, f] - \beta b^{out\ \dagger } f)|0,\mbox{out} \rangle$. This in turn
implies that $ \alpha (\partial f / \partial a^{out\ \dagger }) -
 \beta b^{out\ \dagger } =0$, hence that $f$ has the form 
written in \ref{outvac}.
Note that the creation operators $a^{out\ \dagger }_{\omega}, b^{out\ \dagger
}_{\omega}$ appear only as a product. Hence to each produced particle in mode
$p,\om$, there corresponds one and only one antiparticle in mode $a,\om$.
  From what has been
described these particles are born in pairs of opposite values of their
conserved quantum numbers such as charge and energy . Thus the energy of 
the states containing pairs spontaneously created is equal to the energy of
 the in-vacuum. This is readily seen by expressing the hamiltonian of the 
field in terms of the $a^{out}, b^{out}$ operators. One finds
\begin{eqnarray}
H= \int ^{\infty}_{-\infty} d\omega \omega 
(a^{out\dagger }_{\omega}a_{\omega}^{out}
- b^{out\dagger }_{\omega}b_{\omega}^{out}) 
\label{hamiltE}
\end{eqnarray}

To compute the normalization factor $N$ in equation \ref{outvac} we have 
\begin{eqnarray} \langle 0, \mbox{in} |0, \mbox{in} \rangle & = & 1  =  
N^{-1} \prod_\omega\sum_n \left|\frac \beta \alpha \right|^{2n}
=N^{-1} \prod_\omega \frac{1}{1-|\beta/\alpha|^2} = N^{-1}
\prod_\omega |\alpha|^2 \nonumber \\
N & = & \prod_\omega |\alpha|^2 \label{norme}
\end{eqnarray}
\noindent From this it follows
that
\begin{equation}
|\langle 0,\mbox{out}|0,\mbox{in} \rangle|^2 = {1\over N} = \prod_\omega
\frac{1}{|\alpha|^2} = \exp - \sum_\omega \ln (1 + |\beta|^2) \label{inout}
\label{Z}
\end{equation}
thereby delivering  the probability to find no pairs at future times.

To complete the calculation
we must now give a meaning to the $\sum_\omega$ i.e.
we must count the number of modes in the non trivial class (those
which turn around in the interval $0 \leq t \leq T$ and $0 \leq x \leq L$).
\par The density of orthogonal modes of frequency
$\omega$ in the interval $0 \leq t \leq T$ is
$\frac{T}{2\pi} d\omega$.  To calculate $\int d\omega$, we note
that the scattering centers of non-trivial modes are at the points
$x = -\omega/E$ 
where $0 \leq x \leq L$.  Therefore the total number of
non trivial modes is $(T/2\pi)(EL)$.  
The values of $\omega$ which contribute to $\int d\omega$ can also be 
obtained from classical mechanics. Indeed the classical equation 
of motion is
$m d^2t /d\tau^2=E\ dx/d\tau $ which
integrates to $mdt/d\tau=Ex + \omega $ in the region $ 0\leq x\leq
L$. 
The turning point is given by 
$dt/d\tau = 1$ (i.e. by $\omega = - E( x - a^{-1})$ ). Therefore
 the modes that have turning points in the region $ 0\leq x\leq
L$ have energies in the interval $-EL \leq \omega \leq 0$, 
so that for $L>>a^{-1}$ the number of modes in class $I$ 
is $T/{2\pi} \int_{-EL}^{0} d\omega =  ETL/2\pi$ \footnote{It is amusing that
it is precisely the same consideration that affords
a very simple proof of the chiral anomaly of Fermi fields in 2 dimensions
in the presence of a magnetic field $H$. 
 In its Minkowski version, replace $H$ by $E$ and chirality
$(\gamma_5 = \gamma_1\gamma_2)$ by velocity $(= \gamma_0 \gamma_1)$.  Then
the change in velocity of the vacuum due to $E$ ($= \int dx 
\overline \psi \gamma_0 \gamma_1 \psi)$ is $(1/2\pi) \int Edxdt$. See for 
example~\cite{Nien}}.
 Of course the same result can be obtained
in the gauge $A_x =- Et$, $A_t = 0$.  
\par From this discussion
 a physical picture of production emerges
wherein a certain class of vacuum fluctuations (class $I$) in the past
contains the potentiality of making pairs. This comes about because 
these fluctuations fall into resonance with a state  which contains pairs
of out-quanta. The pair production is possible because of the degeneracy of
states of zero energy (see eq.~(\ref{hamiltE})).
 In each case
that we review in this article the same physics is repeated.  Only the
details of the mechanism of conversion from vacuum to  physically
propagating states
changes from problem to problem. Moreover
the selection of those fluctuations that are predestined to become
physical particles 
is such as to give rise to a steady rate of
production, the common ground being the constancy of acceleration.
\par The Schwinger formula~\cite{Schw} now follows from (\ref{inout})
\begin{eqnarray}
\left|\langle 0, \mbox{out} | 0,\mbox{in} \rangle\right|^2 = \exp -
\frac{ELT}{2\pi} \ln \left(1 + e^{-m^2\pi/E} \right) 
\label{schwfor}
\end{eqnarray}
where we have used (\ref{R/T}) 
$|\beta|^2 = |T/R|^2$.  Actually Schwinger
worked out the case of $3+1$ dimensions.  This is obtained by replacing
$m^2$ by the transverse mass squared $(=m^2 + k^2_{\perp})$ and integrating
over $k_{\perp}(\int d^2 k_\perp/2\pi)$ in the exponential so as to
replace $ETL/2\pi$ by $E^2TV/4\pi^2$).  For the interested reader
Appendix~\ref{functint} contains a modified version of Schwinger's derivation 
obtained by functional integration.  
The mode by mode analysis
given here is obviously far more detailed in revealing physical
mechanisms and hence of more pedagogical value for the black hole
problem. 

The physical interpretation of eq.~(\ref{schwfor}) is 
illuminating when one recalls that the mean number of quanta produced in
the mode  $\omega$ is $|\beta|^2 \  =\langle
n_{\omega} \rangle$ (c.f. eq.~(\ref{nout}). The argument of 
the exponential in eq.~(\ref{schwfor}) is thus $\sum_{\omega}\ln (1+\langle 
n_{\omega} \rangle )$, as in 
a partition function.

It is interesting to remark that the population ratio of produced pairs of
two charged fields one of mass $m$ and the other of mass $m + \Delta m$
is given, in the case $\Delta m << m$ and $ m^2 /E << 1$, by
\begin{eqnarray}
\frac{\langle n \rangle_{m+ \Delta m}}{\langle n \rangle_{m}} 
= e^{- 2\pi \Delta m /a}
\label{popth}
\end{eqnarray}
i.e. the ratio of the mean number of created pairs with neighbouring masses
is Bolzmannian with temperature $a/2 \pi$. That the physics of accelerated 
systems is associated with this temperature is the subject of the next
chapter. From \ref{popth} and Section \ref{ACCEL}, we can say that particles are born in
equilibrium.

\section{Pair Production as the Source of Back Reaction}\label{pair}

Following~\cite{BMPPS} let 
us examine more closely the physical situation that arises 
when the $E$ field is  turned off after
a finite time lapse ($E\neq 0$; in the interval $0\leq t \leq T$), wherein
the (Heisenberg)
state is the vacuum $|0,\mbox{in}\rangle$. This state, at time $T$, the content of
which is then
 measured by a counter system (i.e. a measuring device sensitive to
out-quanta), is
expressed in eq.~(\ref{outvac}). It appears as a linear superposition of
 different
outcomes according
 to the number and nature (mode number) of produced pairs.
For example eq.~(\ref{schwfor}) gives the 
probability to find no pairs and eq.~(\ref{nout}) gives the mean number of pairs in
a mode. These pairs produce a non vanishing expectation value for the
electric current at intermediate time $t$:
\begin{eqnarray}
\langle J_{\mu}(t,x) \rangle &\equiv&\elematrice{0,\mbox{in}}
{J_{\mu}(t,x)}{0,\mbox{in}}\nonumber \\
&=&\sum_{m}A_{m}^{*}\elematrice{m,\mbox{out}}{J_{\mu}(t,x)}
{0,\mbox{in}}\
\label{expectj}
\end{eqnarray}
where the second equality results from 
the insertion of  a complete set of states in the out-basis. The coefficients
$A_{m}$ ($=\langle m ,\mbox{out}|\mbox{in}\rangle)$ 
are the probability amplitudes to find the system in the state
$\ket{m,\mbox{out}}$ at time
$T$. For a system with many pairs, the state $\ket{m,\mbox{out}}$, is very
complicated due to
interactions among these pairs. The whole development of the previous
sections has been
devoted to the production of non interacting pairs and we shall continue to
work in this
approximation valid for sufficiently small times $T$ or large masses $m$.
More precisely we require 
\begin{equation}
\frac {ET}{2\pi}e^{-\pi m^2/E}[a^{-1}+T]<<1\qquad .\label{cond}
\end {equation}
This inequality ensures negligible overlap among the pairs produced in time
$T$, hence
negligible interaction\footnote {Recall that the pairs are one-dimensional
dipoles}.
Equation (\ref{cond}) comes about by multiplying the density of pairs
($=(ET/2\pi)e^{-\pi m^2/E}$) by their mean separation ($=a^{-1}+T$) since they are
produced at
separation $a^{-1}$ whereupon the members of the pair fly apart at the
speed of light.
\par The opposite limit of large density, where mean field approximation is
valid has been the
subject of detailed investigation~\cite{CoMo}
 where the back reaction has been solved.
Unfortunately
many body effects which give rise to plasma oscillations are an important
element of this
development. Hence it is not a relevant laboratory for the study of black
hole radiation.
Therefore we shall continue with the approximation of 
non interacting pairs, which turns out to
be more 
relevant to the study of black hole evaporation.
When eq.~(\ref{cond}) is valid, the relative fluctuations
of density of pairs 
are large, so the mean can no longer be used. It
is no longer appropriate
for calculating the effect of back reaction on the production. 
For this reason
it becomes of interest to take apart eq.~(\ref{expectj}) term by term.

For the case of
negligible
interaction the normalized states $\ket{m,\mbox{out}}$ are direct products
over pair states~:
\begin{equation}
\ket{m,\mbox{out}}=\prod_{\omega}\frac {\left[
a^{out \dagger }_{p,\omega}
b^{out \dagger }_{a,\omega}\right]^{n_{\omega}}}{n_{\omega}!}
\ket{0,\mbox{out}} \qquad .\label{mout}
\end {equation}
Since $J_{\mu}$ 
 is quadratic in the field operator $\Phi$, it contains
creation and
annihilation operators in a 
double sum over modes. It is easy to check that the diagonality
of the Bogoljubov transformation
(\ref{bogtransf}) 
not mixing different values of $\omega$,
 reduces the double sum to a single one. Thus
the in-vacuum expectation value
 $\langle J_{\mu} \rangle$ in eq.~(\ref
{expectj}) is expressible as a single sum over $\omega$. It is given by
\begin{eqnarray}
\langle J_{\mu}(t,x)\rangle &=&
\frac {\inoutexpect{J_{\mu}(t,x)}}{\inout}\nonumber\\
& &-\ \sum_{\omega}
\left(|\frac {\beta}{\alpha}|^2\right) \left( \frac {\alpha
^{*}}{\beta ^{*}}
\varphi_{p,\omega}^{in*}(t,x)\lrD _{\mu} \varphi_{a,\omega}^{in*}(t,x)
\right) \qquad ,
\label{injin}
\end{eqnarray}
which may be found by either hard work from eq.~(\ref{outvac}) 
or using the
identity for the
Feynman propagator:
 \begin{eqnarray}
G_{F}(t,x;t',x')&=& 
\frac {\inoutexpect{\Phi(t,x)\Phi^{\dagger }(t',x')}}{\inout}\nonumber \\
&=& \elematrice{0,\mbox{in}}{\Phi(t,x)\Phi^{\dagger }
(t',x')}{0,\mbox{in}}
+ \sum _{\omega}\frac {\beta}{\alpha} \varphi_{a,\omega}^{in*}(t,x)
\varphi_{p,\omega}^{in*}(t',x')\nonumber \\
\label{feynman}
\end{eqnarray}
easily obtained by expressing $\Phi^{\dagger }$ 
in the in-basis, 
 and expanding the out-vacuum state $\bra{0,\mbox{out}}$ up to the term 
quadratic in the in-annihilation operators acting on the in-vacuum~:
\begin{equation}
\langle 0, \mbox{out}| =  N^{-1/2}\langle 0,\mbox{in}|\exp \left[\left(\frac {\beta}
{ \alpha}\right) \sum_\omega
a^{in}_{\omega,p} b^{in}_{\omega,a}\right] \nonumber
\end{equation}
\par The first term on the r.h.s. of eq.~(\ref{injin}) is the current
contained in the state $\ket{0,
\mbox{in}}$ which arises when no pairs are produced. Hence it should
vanish. It does. There
are two proofs. 
The first one is related to 
subtraction problems
in general. This method will be presented in Section \ref{Tmunuren} 
(see ref.~\cite{BMPPS} for an explicit treatment in the 
case of a background electric
field).  The other
is to prove that upon
expanding $J_{\mu}$ into modes, each term is
separately zero. We leave this as an exercise
to the reader (see also ref.~\cite{BMPPS}). This latter proof 
is completely  satisfactory since the summation over the alternating series
is absolutely convergent due to the finite domain of space-time $L \times T$
in which the $E$ field is non vanishing.

\par The second term on the r.h.s. of eq.~(\ref{injin}) is 
the mean current, expressed as a weighted sum over
the contribution of the current carried by pairs arising from the mode
$\omega$. The
weight $|\beta /\alpha|^2$ is the probability that at least one pair be
produced
in this mode (since $1/|\alpha|^2$ is the probability that none be
produced). 
The c-number in parenthesis which multiplies this weight 
will be called a weak-value. It is related to what one
measures in an
experiment in which the presence of a pair is detected, thereby
making
contact with the weak measurement theory of Aharonov 
et al.~\cite{Ahar}. 
In our situation, the post-selection (i.e. the detection of a localized pair)
is physically performed by registering the click
in a counter.
More generally, whether or not the pairs 
are treated as independent,
one may write $\langle J_{\mu} \rangle$ as 
(see eq.~(\ref{expectj})
\begin{equation}
\langle J_{\mu}(t,x)\rangle =\sum_{m}P_{m} \left[{
 \elematrice{m,\mbox{out}}{J_{\mu}(t,x)}{0,\mbox{in}}
\over A_{m}} \right]
\qquad . \label{pexpectj}
\end{equation}
where $P_{m}=|A_{m}|^2$ is the probability 
to find the system in the state $\ket{m,\mbox{out}}$.
In other words the
expectation
value of $J_{\mu}$ is a weighted sum of non diagonal matrix
elements (weak values) in exactly the same way as usual conditional
probabilities. The physical relevance of these matrix elements as well as the
concepts of post selection and weak value are discussed in more detail at
the end of this section and in Appendix \ref{weak}.

For the nonce, we shall enquire into the 
properties of these matrix elements
using well localized wave packets
 To this end a Gaussian envelop
turns out to be most convenient for the following reason. First consider the
asymptotic regions.  For simplicity we take a mode which is centered on the
space-time origin.  Then,
see eqs~(\ref{asymptwave},\ref{inpphi}), we obtain :
 \begin{eqnarray}
\lim_{t \to -\infty}
\psi^{in}_p & \simeq & \int d\omega e^{-\omega^2/2\sigma^2}
e^{-i \omega t} e^{-iE(x+\omega/E)^2/2} \nonumber \\
& \simeq & e^{-iEx^2/2} e^{-(x+t)^2/2 \Sigma^2_+}
\nonumber \\
\lim_{t \to +\infty} \psi^{in}_p & \simeq &
\int d\omega e^{-\omega^2/2\sigma^2} e^{-i \omega t}
\left[\alpha e^{+iE(x+\omega/E)^2/2} +
\beta ^* e^{-iE(x+\omega/E)^2/2} \right] \nonumber \\
& \simeq & \alpha e^{+iEx^2/2} e^{-(x-t)^2/2\Sigma^2_-}+
\beta ^* e^{-i Ex^2/2} e^{-(x+t)^2/2 \Sigma^2_+} \label{asymptwp}
\end{eqnarray}
where
\begin{eqnarray}
\Sigma^2_{\pm} = \left[ \frac{1}{\sigma^2} \pm \frac{i}{E} \right]
\label{Sigma}
\end{eqnarray}
and inessential phases factors have not been taken into account.
The asymptotic widths are given by $\left[{\rm Re} (1/\Sigma^2_{\pm})
\right]^{-1/2}$.  If ${\rm Im} \sigma \not= 0$ the two outgoing
branches in (\ref{asymptwp}) will have different widths. Charge symmetry
then dictates the choice ${\rm Im} \sigma = 0$ whereupon once sees
that the width $(= \left[(E^2 + \sigma^4)/E^2 \sigma^2 \right]^{1/2})$
is minimized by the choice $\sigma^2 = E$. In what follows, wave packets of such width
 will be qualified as minimal.  They are not only optimally
localized, but they constitute a good approximation to a complete
orthogonal set.  This follows from the fact that their width is
$\sim E^{-1/2}$ about the classical orbit so that the number of such 
(asymptotically) non overlapping packets is ${\cal O}(ELT)$ as is required for
the correct count of non trivial modes.  With a little tinkering
on their size and shape they could be shaped into a
rigorous complete orthogonal set~\cite{Meye}, but for our purposes the minimal
packet is sufficient to show how the back reaction field emerges from
a single pair.
\par The subsequent analysis is facilitated by using the integral
representation (\ref{xichiu}) which from (\ref{xiu}) is a Gaussian
transform of $\langle u|\chi \rangle$.  Hence with $\xi$ given
by (\ref{2.3}) the Gaussian integral over $\omega$ is trivial and one
finds once more that the packet is nothing but another Whittaker function.
With $\sigma^2 = E$, one obtains
\begin{eqnarray}
\psi_a^{in}(t,x)&=&\int_{-\infty}^{+\infty}e^{-\omega^2/2E}
e^{-i\omega t}\braket{\xi}{\Xi}d\omega \nonumber \\
&=&\frac{e^{i\pi /4}\Gamma(1/2+i\varepsilon)}{2\pi}
e^{-(x^2+t^2+i2xt)E/4}
D_{-i\varepsilon -1/2}[
e^{i\pi/4}\sqrt{E/2}(t-ix)]\mbox{\hfill}
 \nonumber \\
\label{gausswpkt}
\end{eqnarray}
and $\psi_p^{in}$ is given by the same superposition in terms of
$\chi^*_{\varepsilon}$ see eq.~(\ref{inpphi}).
The algebraic details are given in ref.~\cite{BMPPS}. 
We have plotted 
in Fig.~\ref{F1} the (logaritm) of the modulus  of $\psi_p^{in}$.
\par The current~(\ref{injin}) has been written as a sum over modes. this can
be rewritten as a sum over packets. The contribution to this sum from 
$\psi_p^{in}$ and $\psi_a^{in}$ is
\begin{equation} \vert \alpha \vert^2 J_{\mu}^{\psi}(t,x) = 
\frac{\alpha^*}{\beta^*} \psi_p^{in *}(t,x) i \lrD_{\mu} \psi_a^{in *}(t,x)
\label{jpsi}
\end{equation}
(The factor $\vert\alpha\vert^2$ on the l.h.s. is introduced here for
 subsequent physical interpretation, see eq.~(\ref{jmatr})). We emphasize that
this current is a product of two localized wave packets. It is a contribution to
the non-vanishing second term of eq.~(\ref{injin}) when the set of modes is
recast into a set of packets.

In Fig.~\ref{F4} we have plotted the real and the
imaginary part of $J_{\mu}^{\psi}$.
Note in particular that this product vanishes in the remote past
since we 
start from vacuum at early times.
Indeed each factor ($\psi_p^{in}$ and $\psi_a^{in}$ respectively) of 
this product
is a minimal packet centered
 on the space-time origin and they do not overlap before they 
reach their common
turning point.  Being in-modes, one comes
in from the left only and the other from the right only. 
This is at is should be,
the current produced by the pair should not exist before it is produced.
\dessin{.800}{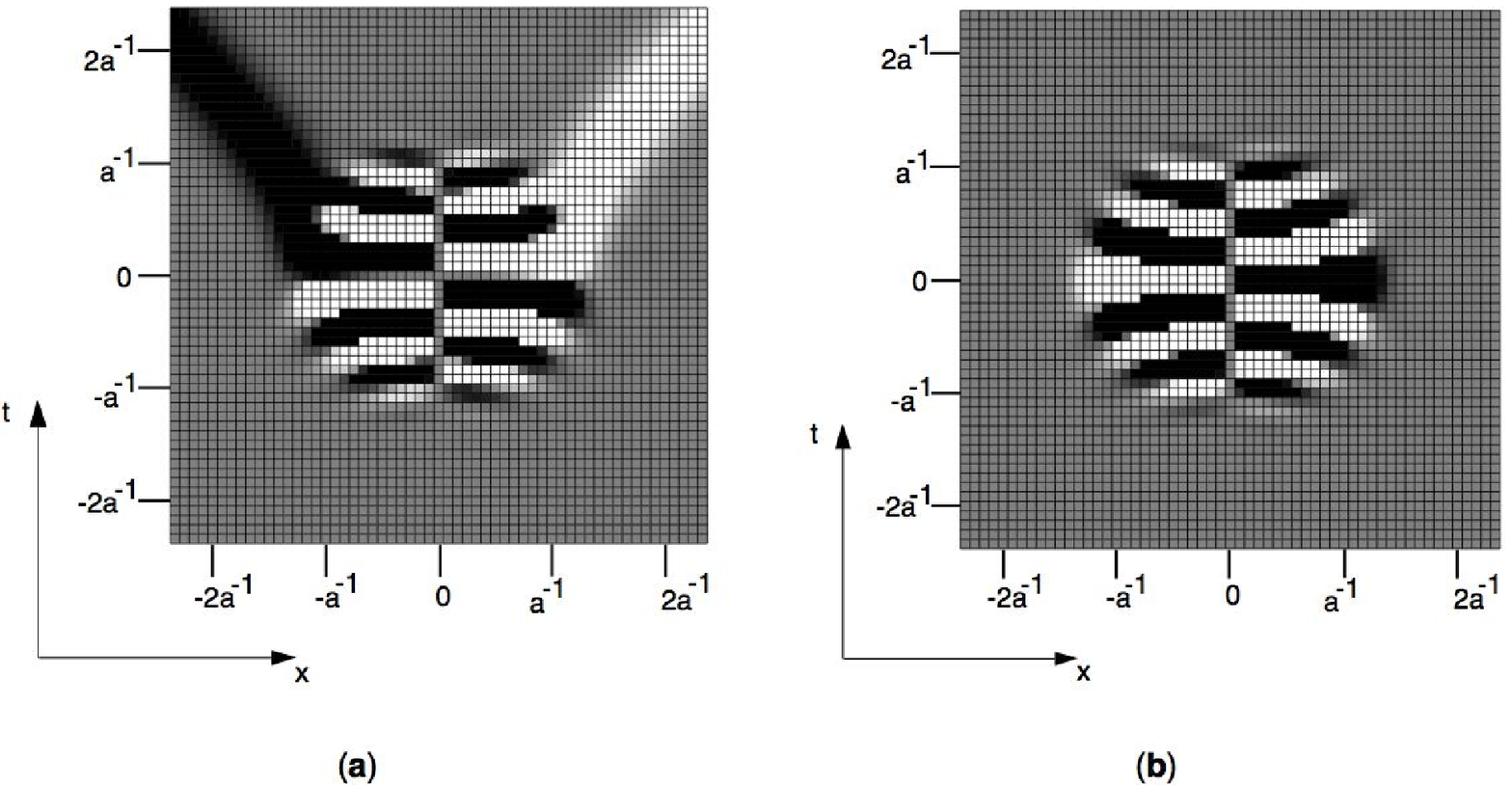}{F4}
{The real (fig. {\bf a}) and imaginary (fig. {\bf b}) parts of
$J^{\psi}_0$ are plotted for a minimal wave packet 
where we have taken $m/a = 9$. In the figure
zero is grey, positive is white and negative is black. Within the
production region $\Delta t \times \Delta x = 2 a^{-1}
\times 2 a^{-1}$ the real and imaginary parts are
comparable and oscillate. Outside the production region,
the quanta are on mass shell and propagate classically: the
imaginary part vanishes and the real part takes its
classical value.
} 

Moreover, in the far future,
 from the inverse of eqs.  (\ref{outpphi}, \ref{outaphi}) taken together
with the vanishing overlap of $\psi^{out}_p$and $\psi^{out}_a$, 
(where $\psi^{out}_p$ is the same wave packet as  $\psi^{in}_p$ but made of
out-modes) 
we find
\begin{eqnarray}
\lim_{t\to +\infty} & |\alpha|^2 & \left [ \frac{\psi^{in *}_p(x)
i \lrD_{x} \psi^{in *}_a(x)}{\alpha
\beta^*} \right]
 = \nonumber \\
& |\alpha|^2 
& \left[ \psi^{out *}_p(x) i \lrD_{x} \psi^{out}_p(x)
 + \psi^{out}_a(x) i \lrD_{x} \psi^{out *}_a(x) 
\right]
\label{2.47}
\end{eqnarray}
This once more is at it should be.  The future current is the sum of currents
carried by the particle and anti-particle separately according to one's
classical expectation.  Remark on the importance of the
denominator in eq.~(\ref{injin})  (or more generally in eq.~(\ref{pexpectj})),
wherein $\langle m, {\mbox out}\vert$  is given by
$\langle 1, \psi^{out} \vert$ the out state which contains only the
pair described by $\psi^{out}_p$ and $\psi^{out}_a$. Hence it is given by
\begin{equation}
\frac{\langle 1, \psi^{out} \vert J_{\mu} \vert 0, {\mbox{in}} \rangle}
{\langle 1, \psi^{out} \vert0, {\mbox{in}} \rangle} =
\frac {1}{\alpha \beta^*} \psi_p^{in *}(t,x) i \lrD_{\mu} \psi_a^{in *}(t,x)
= J_{\mu}^{\psi}
\label{jmatr}
\end{equation}
It is the denominator $\alpha
\beta^*$ that garantees that the current carried by the detected out pair is 
unity. 
The multiplicative factor $|\alpha|^2$ in eq.~(\ref{2.47}) is there
to account for the
fact that more than one pair may be produced in the packet. It is equal to
the ''induced emission"
factor  $\langle n+1\rangle$ where  $\langle n\rangle \equiv |\beta|^2$ is
the mean density number of pairs
produced. It is not present in eq.~(\ref{jmatr}) since in that case only
one out pair characterizes the out-state.

 \par Having understood that asymptopia goes according
to rights, we note further features seen in Fig.~\ref{F4}
\begin{enumerate}
\renewcommand{\labelenumi}{\arabic{enumi})}
\item The region of production lies within a circle 
of radius $2/a$,
\item Within this region the contribution to $J_{\mu}$ is complex and it
oscillates.
\end{enumerate}
This is the quantum region.  The oscillations occur because for $x$ and
$t \ll a^{-1}$ the modes oscillate with frequencies of ${\cal O} (m)$.
This is seen most clearly in the gauge where $A_t = 0$, $A_x=-Et$.  Then
the packets are built out of $\chi_{-\varepsilon} (t + k/E)\/\exp(ikx)$ 
 and these\footnote{Which are the analytic continuation from $m^2$ to $-m^2$ of the functions
introduced in eq.~(\ref{sugxichi}).}
oscillate near the origin with frequency $\sqrt{m^2 + k^2}$.  The 
amplitudes of these oscillations is very strong near the origin
(like $\exp E(a^2 - t^2 -x^2)$) and then fade out as the particles
settle down to get on to their mass shell.  We are
seeing how the transients work themselves out so as
to arrive at the completion of a quantum event.

\par  Note however that in the sum for $\langle J_{\mu} \rangle $
eq.~(\ref{injin}), the imaginary
part vanishes identically and
furthermore, since $E$ is constant in the box $\langle J_{\mu}(x)\rangle$
is homogeneous, so that
the oscillations drop out as well. 
Thus one should question the physical relevance of these complex 
oscillations. 
We therefore conclude this section with a few remarks on this count, including
a short discussion of the back reaction problem.

The simplest way to put them in evidence in a 
physical amplitude is by considering the change in the probability to find
the $\psi$ pair upon slightly modifying the electric field, i.e. by
replacing $E$ by $E+\delta E(x,t)$. Since the change in the action 
is given by $\delta S = \int dt dx J_{\mu} \delta A^{\mu}$, the change of
the probability to find the $\psi$ pair is given, in first order in $
\delta E(x,t)$ by
\begin{eqnarray}
P_{E+\delta E}
&=& \vert \langle 1, \psi^{out} \vert e^{i\delta S} \vert 0, {\mbox in} 
\rangle \vert ^2 \nonumber \\
&=& \vert \langle 1, \psi^{out} \vert ( 1 + i \delta S ) \vert 0, {\mbox in} 
\rangle \vert ^2 \nonumber \\
&=& P_{E} \left( 1 + 2 \int dt dx \delta A^{\mu}(t,x) {\mbox {Im}}
\left[ \frac{\langle 1, \psi^{out} \vert J_{\mu} \vert 0, {\mbox{in}} \rangle}
{\langle 1, \psi^{out} \vert0, {\mbox{in}} \rangle} \right] \right)
\label{PEnew}
\end{eqnarray}
(We recall that $P_{E} = \vert \langle 1, \psi^{out} \vert 0, {\mbox in}
\rangle
\vert ^2= \vert \beta /\alpha \vert ^2 /N$ where $N$ is 
given by eq.~(\ref{norme})). 
Hence it is the imaginary part of the weak current (given in eq.~(\ref{jmatr}))
which furnishes
the change of the probability when one compares two neighboring
$E$-external fields. But given that each created pair carries an electric field
these weak values will also govern the modification of multi-pair
production due to current-current interaction in
the usual
perturbative approach.
One way to proceed is to include the interaction term in the hamiltonian
$\frac 12 \int\int dx dx' J_t(x) v(x-x')J_t(x')$ where $v$ is the Coulomb
potential. 
It can be shown that once one pair is produced, the electric field which acts on
fluctuations to give rise to subsequent pairs is of the form $E+\int dx'
J_t^{\psi}(x')$ where $J_t^{\psi}(x')$ is the  matrix element of the charge
density due to the pair which has been created. The self interaction of a pair
as it is produced is given by a more complicated loop correction.
The statement of the back
reaction problem is thus given in terms of a self consistent highly
non trivial problem.
 One will then be led to treat the complete wave function of the system
as a function of the
configuration of both the matter and electric field. This will then give a
generalization of the
usual semi-classical theory $\left[\partial_{\mu}F^{\mu
\nu}=\elematrice{\mbox{in}}{J^{\nu}}{\mbox{in}}\right.$, wherein $F^{\mu
\nu}$ on the l.h.s.
is classical, but modified due to the mean of $\left. J^{\nu}\right]$, to a
full quantum
Heisenberg equation among operators $\partial_{\mu}F^{\mu
\nu}=J^{\nu}$.
\par The hard job then remains.  How do these individual production
acts affect subsequent production.  This is the true feed-back problem
and it has not been solved.
Current research is now under way to find
an approximate solution
which is better than the mean field 
approximation (see for instance ref.~\cite{CHKMPA}) wherein fluctuations
and correlations among pairs are taken into account.

Another way to understand the status of the matrix elements eq.~(\ref{jmatr})
and the relevance of the complex oscillations 
is by appealing to the formalism developed by Aharanov et al~\cite{Ahar}.
 In this formalism, it is shown how those matrix elements 
give the result of a ``weak measurement''.
 This
development contains two
elements that we briefly expose. The interested reader will consult 
Appendix \ref{weak}
as well.
\begin{itemize}
\item A weak detector, a quantum 
device in weak interaction with the system, where
weak
means that one can correctly work
in first order in the interaction. The 
detector wave function is
spread out 
so that first order perturbation theory is valid.
Then one picks 
up not only 
the detector's displacement (in position)  
as in the von Neuwmann  model  of
measurement~\cite{Vonn}, but 
one finds also that the imaginary part of the matrix element ( here of
$J_{\mu}$) imparts
 momentum to the detector.
For the case we discuss here the detector could be for
example a test
charge whose path is deflected by the external $E$ field, modified by the
back reaction field (induced by the weak value of the current) 
due to the selected pair, i.e. the modification of the $E$ field is given 
by Gauss' law: $\Delta E(x,t)= \int^x dx' J_t^{\psi} (x',t)$, where
$J_t^{\psi} (x',t)
$ is the weak value of the current. Its statistical interpretation in the
context of measurement theory is given below. \item A post selection, the
specification of the final state of the system, here the state
$\ket{m,\mbox{out}}$. In the context of our wave packet development 
with
no
interaction among the
pairs a possible out state is 
the one
post-selected by the click of a localized counter.
One selects, thus, that part of the wave function $\ket {\mbox{0,in}}$ which causes the
counter to click.
 Physically this post selection consists 
of registering all results 
of the weak detector
(as one usually does to establish mean values, in the manner prescribed by
the
Copenhagen interpretation of quantum mechanics)
and then to keep only the results 
when the particular localized click is registered
 (as one does in constructing  conditional 
probabilities).
\end{itemize}


\chapter{Accelerating Systems}\label{ACCEL}
\section{The Accelerated Detector}\label{general}
A system in constant acceleration through empty Minkowski space, coupled 
to a field whose state is quantum vacuum in Minkowski
space, will heat up and its internal degrees of freedom will become thermally
distributed with a temperature given by
\begin{eqnarray}
T = \beta^{-1} = a/2\pi \qquad .\label{temp}
\end{eqnarray}
More succinctly : an accelerated detector perceives the vacuum
as a thermal bath.  This remarkable observation is due to Unruh~\cite{Unru1},
who thereby gave substance to an equally remarkable observation by
Fulling~\cite{Full} who showed that 
quantization of a field in Rindler coordinates~\cite{Rind} 
is inequivalent to the usual
Minkowski quantization. 

The
definition of Rindler coordinates $\rho, \tau$ in $1+1$ dimensions is
\begin{eqnarray}
t=\rho\ \mbox{sinh} a \tau\quad , \quad
x=\rho\ \mbox{cosh} a \tau \nonumber\\
\rho > 0 \quad , \quad - \infty < \tau < +\infty
\label{coor}\end{eqnarray}
hence applicable to the quadrant $t>0$, $x>0$, hereafter referred to as the
right quadrant (R). 
These coordinates are naturally associated to a uniformly  accelerated system
with  acceleration $a$ in the sense that its trajectory is 
$\rho =a^{-1}= const $; its proper time is $\tau$. In addition the
Minkowski metric expressed in coordinates $\rho, \tau$ 
\begin{equation}
ds^2=-dt^2+dx^2=-\rho^2 a^2 d\tau^2 + d \rho^2
\label{coor2}\end{equation}
is static, translations in $\tau$ (boosts) being Lorentz
transformations.

\dessin{1.000}{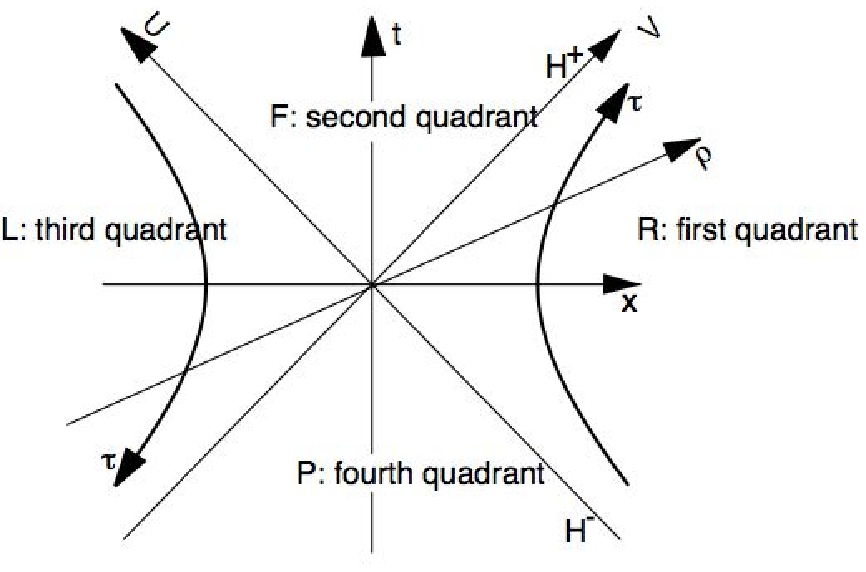}{Mink}{The coordinate system ($\rho,\tau$), the trajectory of a
uniformly accelerated detector, the four quadrants of Minkowski space 
and the horizons $H_\pm$ (U=0,V=0) which separate them.} 
The
remarkable result, now stated with more precision is that Minkowski vacuum is
described in R (or alternatively in the left quadrant (L) wherein $t= - \rho\ 
\mbox{sinh} a \tau$, $ x=-\rho\ \mbox{cosh} a \tau$, see Fig.\ref{Mink}.) by a
thermal bath in the quantization scheme based on the coordinates $\rho,\tau$
defined by eq.~(\ref{coor}). The connection between the remarks of Unruh and
Fulling is through the proper local dynamics of the accelerator when it is
coupled to the radiation field so as to undergo transitions. Conventional
emission and absorption is then given in terms of Rindler quanta

The easiest way to see this phenomenon is by looking at rates of
absorption and emission of an accelerated two level detector.  One may
for example take a two level ion whose center of mass is described by the wave
function in the W.K.B. approximation to the minimal wave packet of 
Section \ref{pair}.
The type of idealization envisioned is to let the mass of the ion
tend to infinity at fixed acceleration, $a$, i.e. $M \to \infty $,
$E \to \infty $, $E/M \to a$.  The packet then describes a $\delta$ function
along a classical orbit whose center is used to define the origin of 
space-time
\begin{eqnarray}
&x^\mu x_\mu = -t^2 + x^2 = a^{-2}&\nonumber\\ 
& \left\{ \begin{array}{c}t=a^{-1} \mbox{sinh}a \tau\\
x=a^{-1} \mbox{cosh}a \tau
\end{array} \right.&
\label{acctraj}
\end{eqnarray}
(Acceleration is 
 constant in the sense that $a_\mu a^\mu = a^2$ where 
$a^\mu =
d^2 x^\mu/d\tau^2 = a^2 x^\mu
$ with $\tau$  the proper
time on the path of the accelerator, i.e. $u_\mu u^\mu = 
(dx^\mu/d\tau) (dx_\mu/d\tau) =-1$.) The two levels are separated
by the mass difference $\Delta M$, taken to be finite.

 We take the example of a massless scalar field (here chosen
hermitian) in $1 + 1$ dimension, coupled to the two level accelerating
detector~\cite{Unru1},\cite{Davi},\cite{Dewi}. In the interaction
representation, the  interacting hamiltonian of the system is
\begin{eqnarray}
H &=&   g \left[ e^{i \Delta M \tau}\sigma_+ +
e^{-i \Delta M \tau}\sigma_-
\right]
 \phi (x^\mu(\tau))
\nonumber \\
&=& g \int {d k \over \sqrt{
4 \pi \omega}} \left[ e^{i \Delta M \tau}\sigma_+ +
e^{-i \Delta M \tau}\sigma_-
\right]
\left[ e^{-i k_\mu x^\mu(\tau)}a_k
+e^{i k_\mu x^\mu(\tau)}a_k^\dagger \right]
\label{ham}
\end{eqnarray}
where $\sigma_{\pm}$ are the operators which send the system from
ground (excited) state to excited (ground) state and
where $k_0=\omega =\vert k \vert$ ,$k_1 =k$ and $a_k^\dagger$ ($a_k$) are the
creation (destruction) operators of field quanta of momentum $k$.  

 Standard golden rule physics gives in lowest order perturbation
theory the following formula for the rates expressed in terms of the proper 
time $\tau$, of absorption (--) and
emission (+)
\begin{eqnarray}
R_{\mp} = g^2 \int^{+\infty}_{-\infty}\! d\Delta\tau\ e^{\mp i \Delta M
\Delta\tau } W(\Delta\tau  - i\epsilon) \label{rates}
\label{Rone}
\end{eqnarray}
where $W$ is the Wightman function between two points on the orbit
separated by proper time $\Delta\tau $. 
For the reader unfamiliar with this
formula we sketch a brief derivation

  Take the case of absorption, then
we begin at $\tau = 0$ with the atom in its ground state and
the radiation field in vacuum.  The amplitude to find it in the
excited state is the matrix element of $-i \int^\tau_0
H_{int}(\tau^\prime)d\tau^\prime$. The probability is
obtained by squaring the amplitudes and by integrating over all $k$ and is
therefore equal to
\begin{eqnarray}
g^2 \int^\tau_0 \! d\tau_2 \int^\tau_0 \! d\tau_1\ e^{-i \Delta M(\tau_2 -
\tau_1)} \left[ \elematrice{0_M}{\varphi \left(x^{\mu} (\tau_2)\right)
\varphi \left( x^{\mu }( \tau_1)\right) }{0_M} \right] 
\label{Rtwo}
\end{eqnarray}
where $\ket{0_M}$ is Minkowski vacuum.
For emission change the phase $e^{-i \Delta M  (\tau_2 -
\tau_1)}$ to
$e^{+i\Delta M (\tau_2 -
\tau_1)}$.  The quantity in brackets is called the
Wightman function $W(\tau_2,\tau_1)$ and the positivity of the Minkowski
frequencies ($\omega$) encoded in
its decomposition into annihilation and creation operators impose an
integration prescription in the complex plane 
given by $i\epsilon$ in eq.~(\ref{Rone})

  The Wightman function for the
case of a massless field in $1+1$ dimension is $-(4\pi)^{-1} \ln
\left[(\Delta t-i\epsilon)^2 - \Delta x^2) \right]$ where for constant
acceleration eq.~(\ref{acctraj}), we have
\begin{eqnarray} (\Delta
t-i\epsilon)^2 - \Delta x^2 & = &
a^{-2} \left[ \left(\mbox{sinh} a\tau_2 - \mbox{sinh}a \tau_1  
- i\epsilon\right)^2 -
\left(\mbox{cosh} a\tau_2 - \mbox{cosh} a \tau_1\right)^2 \right]
\nonumber \\ & = & 
4a^{-2}
\mbox{sinh}^2 a (\Delta\tau - i \epsilon)/2 \label{Rthree} \end{eqnarray}
Therefore for this case, as well as for the case where the
path is inertial ($u^\mu = \mbox{const})$, the
integrand in eq.~(\ref{Rtwo}) is a function of $\tau_2 - \tau_1$
only\footnote{\label{pgftnote1} The origin of
this miracle is that both kinds of trajectories are orbits of the Lorentz
group. 
In the accelerated case $\tau$ translations are boosts.}.  Take $\tau$
in eq.~(\ref{Rtwo}) such that $\tau >>  {\rm Max} (\Delta M^{-1},a^{-1})$.
[The  usual condition to establish the golden rule is $\tau >> \Delta M
^{-1}$ (for spontaneous emission) and $\tau >> \beta$ (for absorption of
photons from a thermal bath). In the case of uniform acceleration this
implies  $ \tau >> a^{-1}$]
 One may change
variables to $(\tau_2 + \tau_1)/2$ and $\tau_2 - \tau_1 = \Delta \tau$. 
Since the integral comes from the finite region around $\tau_2 - \tau_1 =
O(\Delta M^{-1})$, the integral over the latter may  have its limits extended
to $\pm \infty $. (For $(\tau_2 + \tau_1)/2 \leq \Delta M$  a
small error is made which does not contribute to the rate.) Then 
integration over
$(\tau_2 + \tau_1)/2$ gives $\tau$ thereby yielding a rate formula per unit
propertime. The rate is given by eq.~(\ref{Rone}).  The  reader unfamiliar
with this formalism will check out the usual golden rule in terms of density  of
states for the inertial case.  Also he will relate eq.~(\ref{rates}) to the
imaginary part of the self energy. \par Substituting for $W$ in eq.~(\ref{Rone})
we have \begin{eqnarray} R_{\mp} & = & -g^2 \frac{1}{4\pi}
\int^{+\infty}_{-\infty}\! d\tau \  e^{\mp i \Delta M\tau} \ln
\left[\mbox{sinh}^2 \left(a \frac{\tau}{2} - i \epsilon \right) \right]
\nonumber \\ &=& \pm ig^2 \frac{a}{4\pi}  \int^{+ \infty}_{- \infty} \! d\tau \ 
\frac{e^{\mp i \Delta M \tau}}{\Delta M} \frac{\mbox{cosh} \frac{a}{2}
\tau}{\mbox{sinh} \left( \frac{a}{2} \tau - i \epsilon \right)} \label{Rfour}
\end{eqnarray}
where we have integrated by parts accompanied by the usual prescription of
setting infinite oscillating functions to zero.  We
remark that in $3+1$ dimensions the same type of formula obtains with $W$ given
by  $a^2 \mbox{sinh}^{-2}(  a\tau/2 - i \varepsilon) $.
\par From eq.~(\ref{Rfour}), it is seen that for the minus sign (case
of absorption), complexifying $\tau$ and extending the  domain of integration
to a closed contour in the lower half complex plane picks up poles at
$a \tau = 2 \pi ni$ ($n=-1,-2,\ldots$). For the plus sign (emission)
the contributing poles lie in the upper half plane, at $a\tau =
+i\epsilon$ and $a \tau =  2\pi ni$ ($n=1,2,\ldots$) whence the ratio
\begin{eqnarray}
\frac{R_-}{R_+} = 
\frac{\sum^\infty_{n=1} e^{-\beta \Delta M n}}{\sum^\infty_{n=0}
e^{-\beta \Delta M n}} = e^{-\beta \Delta M} \label{Rfive}
\end{eqnarray}
with $\beta = 2\pi/a$. Since $R_-$ is proportional
to $\langle n \rangle$ and $R_+$ to $\langle n \rangle + 1$ one obtains,
following Einstein's famous argument, $\langle n \rangle = (e^{\beta \Delta M}
- 1 )^{-1}$ as in a thermal bath. Explicit evaluation yields 
\begin{eqnarray} R_- &=& {g^2 \over \Delta M}(e^{\beta \Delta M}
-1)^{-1}\qquad ,\nonumber\\ R_+ &=& {g^2 \over \Delta M}(1-e^{-\beta \Delta
M})^{-1} \label{Rfivea}
\end{eqnarray}
Note that these rates
 coincide exactly with the rates given in an inertial thermal bath for
the model considered. This occurs for massless fields
in 2 and 4 dimensions only. 

Equation~(\ref{Rfive}), in fact, results from a general property of the
propagator $W$ appearing in eq.~(\ref{Rone}) related to the periodicity of the
orbit eq.~(\ref{acctraj}) when  $\tau$ is imaginary, hence transforming the
orbit to a circle. To see how this is related to thermal properties we first
present the thermal Wightman function for a general system 
wherein $H$ may include interaction of the $\phi$ field with itself
\begin{eqnarray}
W_{\rm thermal} &=& {\rm tr}\left [ e^{-\beta H} \phi(t_1) \phi(t_2) \right ]
\label{wwone}\\
&=& {\rm tr} \left [ e^{-\beta H} \phi(\Delta t) \phi(0)\right ]
\label{wwtwo}\\
&=& {\rm tr}\left [ e^{-H (\beta - i \Delta t) } \phi(0) 
e^{-i H \Delta t }\phi(0) \right ] \label{wwthree}
\end{eqnarray}
where eqs.(\ref{wwtwo}) and (\ref{wwthree}) follow from $\phi(t) = e^{i H t}
\phi(0) e^{-i H t}$. Because of the positivity of the
spectrum of $H$ (i.e. $E_n - E_0 \geq 0$), $W_{\rm thermal}(\Delta \tau)$ is
defined  by eq.~(\ref{wwthree}) in the strip in the complex plane
\begin{equation}
-\beta +\epsilon < {\rm Im} \Delta \tau <  - \epsilon \label{wwfour}
\end{equation}
In this strip one has the identity
\begin{equation}
W_{\rm thermal}(t) = W_{\rm thermal} (-i \beta -t)
\label{wwfive}
\end{equation}
obtained by using the cyclic invariance of the trace. Under general
conditions $W_{\rm thermal}$ is analytic in the strip. [For a general
discussion of thermal Green's functions see ref.~\cite{FuRu}].

Analyticity and eq.~(\ref{wwfive}) then imply eq.~(\ref{Rfive}). This is derived
by integrating $\oint \! dz\ e^{i \Delta M z} W_{\rm thermal}(z) =0$ over the
contour surrounding the strip (\ref{wwfour}), and we assume that there is no
contribution from the ends at ${\rm Re} z = \pm \infty$. The integral along ${\rm Im}
z= -\epsilon$ is then $- R_+$ and along ${\rm Im}
z= -\beta +\epsilon$
 is $e^{\beta \Delta M } R_-$ thereby recovering eq.~(\ref{Rfive}).

The point of all this is to note that $W$ for the accelerating system enjoys
the property (\ref{wwfive}) since it is a function of $( \Delta x^{\mu})^2 
= - 4 a^{-2}
{\rm sinh}^2 a \Delta \tau /2$ and this is true not only for a free field but
in general. Thus  the ratio of the accelerating detector rates of
absorption and emission is given by eq.~(\ref{Rfive}) simply in consequence
of the imaginary periodicity of the orbit. We emphasize that it 
is not necessary that
$W(\Delta \tau)$ be equal to an inertial thermal propagator eq.~(\ref{wwone})
and indeed for free field theory one may check that apart from $d=2$ and
$d=4$ it is not since the density of Rindler modes of energy $ \Delta M$
differs from the inertial one. 
However eq.~(\ref{wwfive}) and therefore eq.~(\ref{Rfive})
always obtains.

\section{Quantization in Rindler Coordinates}\label{quantization}
We now show how it is possible to interpret these results in terms of
annihilation and creation of Rindler quanta (Rindlerons) in situ. Rindler
coordinates are defined by eq.~(\ref{coor}) in the first quadrant and it is in
this quadrant where we situate the accelerator. We make this point somewhat more
explicit. The transformation eq.~(\ref{coor}) is analogous to euclidean polar
coordinates, with the r\^ole of the angle being played by $a \tau$
(hence Im$\tau$ has period $2\pi/a$). However
whereas euclidean space is completely covered by polar coordinates, Minkowski
space is only partially covered in one quadrant: $x>0$, $x> \vert t \vert$
designated in Fig. \ref{Mink} by R.

The d'Alembertian $-\partial_t^2 + \partial_x^2$ is 
$[-\rho^{-2} a^{-2} \partial^2_\tau + \rho^{-1} \partial_\rho \rho
\partial_\rho]$ so that for massless particles the modes in these coordinates
are solutions of 
\begin{equation}
\left[
-{1 \over a^2} {\partial^2 \over (\partial \tau)^2} + 
{\partial^2 \over \left[ \partial ({\ln a \rho})\right]^2} \right] \phi =0
\label{eqone}
\end{equation}

In R, a complete set of eigenmodes 
of $i \partial_\tau$, solutions of eq.~(\ref{eqone}), is
\begin{eqnarray}
\begin{array}{c}
\varphi_{\la,R}(u) = \theta(-U){e^{-i \la u} /\sqrt{4 \pi \la}}
  =\theta(-U) {(-aU)^{i \la /a}   /\sqrt{4 \pi \la}}
 \\
\tilde \varphi_{\la,R} (v)= \theta(V) {e^{-i \la v} / \sqrt{4 \pi \la}}
= \theta(V) {(aV)^{-i \la /a }/\sqrt{4 \pi \la}}
\end{array}
\quad ,\ \la > 0\nonumber\\
\label{onefive}
\end{eqnarray}
taken together with their complex conjugates. Here
\begin{eqnarray}
\left.
\begin{array}{r}
u\\v
\end{array}
\right\} &=& \tau \mp a^{-1} \ln a \rho \label{onesix}\\
\left.
\begin{array}{r}
U\\V
\end{array}
\right\} &=& t \mp x = \left\{  
\begin{array}{l}
-a^{-1}e^{-au}\\a^{-1}e^{av}
\end{array}\right.
\label{onesixb}
\end{eqnarray}
The modes are normalized according to the Klein Gordon norm which for
$u$ modes can be written as
\begin{equation}
\int_{-\infty}^{+\infty} du\ 
\varphi_{\la^\prime,R}^*\  i \! \lr {\partial_u} \varphi_{\la,R}
= \delta(\la-\la^\prime)
\end{equation}
and similarily for $v$. We note that even though $u,v$ are defined 
in  R  only by eq.~(\ref{onesix}),
by trivial
extension $\varphi_{\la,R}(u)$ is also defined in P since $u$ is finite on the
past horizon $V=0$. Similarly  $\tilde \varphi_{\la,R}(v)$ is defined as well in
F as well as R. 

In L, the corresponding complete set of modes of positive Rindler frequency is
\begin{eqnarray}
\begin{array}{c}
\varphi_{\la,L}(u_L) = \theta(U) {e^{+i \la u_L} / \sqrt{4 \pi \la}}
 =\theta(U) {(aU)^{-i \la /a}   /\sqrt{4 \pi \la}} \\
\tilde \varphi_{\la,L} (v_L)= \theta(-V){e^{+i \la v_L} / \sqrt{4 \pi \la}}
=\theta(-V) {(-aV)^{+i \la /a }/\sqrt{4 \pi \la}}
\end{array}
\quad ,\ \la > 0\nonumber\\
\label{oneeight}
\end{eqnarray}
taken together with their complex conjugates where in L we have
\begin{eqnarray}
&&\begin{array}{r}
t=-\rho \mbox{sinh} a \tau \\
x=-\rho \mbox{cosh} a \tau
\end{array}
  \label{onenine}\end{eqnarray}
and $u_L,v_L$ in L are also given in terms of $\tau, \rho$ by eq.~(\ref{onesix}).
Hence in L 
\begin{eqnarray}
\left.
\begin{array}{r}
U\\V
\end{array}
\right\} &=& t \mp x = \left\{  
\begin{array}{l}
a^{-1}e^{-au_L}\\ -a^{-1}e^{av_L}
\end{array}\right. \quad .
\label{twonulab}
\end{eqnarray}
 Once more, by extension
$\varphi_{\la,L}(u_L)$ is valid in L and F and
$\tilde \varphi_{\la,L}(v_L)$  is valid in L and
P. Note that in L, $dt/d\tau <0$ whereas in R, $dt/d\tau
>0$. Therefore the modes (\ref{oneeight}) have Rindler frequency opposite
to the usual one.  The convenience of the convention (\ref{onenine}) is that a
given boost is represented by the same displacement in $\tau$ in both L and
R. 
In addition the mapping of R into L by reflection through $x=0,\ t=0$ is
very simply described by the analytic continuation $\tau \to \tau + i \pi /
a$,  $u \to u + i \pi / a=u_L$, $v \to v + i \pi / a=v_L$ which maps (\ref{coor})
into (\ref{onenine}) and (\ref{onesixb}) into (\ref{twonulab}) respectively. If
one performs this inversion twice one obtains that R is invariant under 
$\tau \to \tau + 2i \pi / a$ which is the essential ingredient used to obtain
the thermal properties in the discussion following eq.~(\ref{Rfivea}).

These sets of modes are the Rindler versions of the Minkowski light-like modes
\begin{eqnarray}
\xi_\om (U) &=& {e^{-i \om U} / \sqrt{4 \pi \om}}\nonumber\\
\tilde \xi_\om (V) &=& {e^{-i \om V} / \sqrt{4 \pi \om}}
\label{twonul}
\end{eqnarray}
and their conjugates.

The advantage of the use of these light-like variables is that the right
movers (functions of $u$ or $U$) and left movers (functions of $v$ or $V$)
do not mix under Bogoljubov transformations (e.g. $\varphi_{\la,R}(u)$ is a
linear combination of $\xi_\om (U)$ only). For the remainder of this section
we shall work with the right movers only. The left follow suit.

We have two representations of the right moving part of the field
\begin{eqnarray}
\phi(U) &=& \int_0 ^\infty d \om \left[ a_\om \xi_\om(U) +
\hbox{h.c.} \right]
\nonumber\\
&=& \theta( -U)
\int_0 ^\infty d \la \left[ c_{\la,R} \varphi_{\la , R}(U) + \hbox{h.c.}
\right]\nonumber\\
&\ & \ + \theta(U)
\int_0 ^\infty d \la \left[ c_{\la,L} \varphi_{\la , L}(U) + \hbox{h.c.}
\right]
\label{twoone}
\end{eqnarray}
To find the Bogoljubov transformation in R we express $\xi_\om (U)$ as
linear combination of $\varphi_{\la,R}(U)$ 
\begin{equation}
\theta(-U)\xi_\om (U)
=
\int_0^\infty \! d \la \ \left[
\alpha_{\la \om}^R \varphi_{\la , R}(U) + \beta_{\la \om}^R \varphi^*_{\la ,
R}(U) \right]
\label{twotwoa}\end{equation}
\begin{eqnarray}
\alpha_{\la \om}^R &=& 
\int_{- \infty}^{0}
\! dU\ \varphi_{\la,R}^*(U)\  i \! \lr {\partial_U} 
\xi_\om (U) \nonumber\\ 
&=& 
\int_{- \infty}^{+\infty}
\! du\ \varphi_{\la,R}^*(u)\  i \! \lr {\partial_u} 
\xi_\om (U(u)) \nonumber\\ 
&=& {1 \over 2 \pi} { \sqrt{ \lambda} \over \sqrt{\omega}}
\int_{-\infty}^0 dU 
(-a U )^{- i \lambda / a -1}
e^{- i \omega U} \nonumber\\
&=& 
{1 \over 2 \pi a} \sqrt{\la \over \om} 
\Gamma(-i \la/a)  
\left( {a \over \om} \right)^{-i \la/a}
e^{\pi \la / 2a}\nonumber\\
\mbox{and}\quad
\beta_{\la \om}^R  &=& \int_{- \infty}^{+ \infty}
\! du\  \varphi_{\la,R}(u) (-i\! \lr{ \partial_u}) 
\xi_\om (U(u)) \nonumber\\ &=& 
{1 \over 2 \pi a} \sqrt{\la \over \om}
\Gamma(i \la/a) 
\left( {a \over \om} \right)^{i \la/a}
e^{-\pi \la / 2a} 
\label{twotwo}
\end{eqnarray}

In terms of operators eq.~(\ref{twoone}) gives the Bogoljubov
transformation \begin{equation}
c_{\la , R} = \int_0^\infty d \om \left[
\alpha_{\la \om}^R  a_\om +  \beta_{\la \om}^{R *} a_\om^\dagger \right]
\label{twothree}
\end{equation}
\noindent From the explicit values (\ref{twotwo}), or more simply from their
integral representations, or more formally from the canonical commutation
relations of $c_{\la,R}$ and 
$c_{\la,R}^\dagger$ one checks out the completeness of $\xi_\om$
\begin{eqnarray}
\int_0^\infty\! d \om\ 
\left[ \alpha_{\la \om}^R \alpha_{\la^\prime \om}^{R *} -
\beta_{\la \om}^{R *} \beta_{\la^\prime \om}^{R }\right]
&=& \delta(\la - \la^\prime) \quad ,\nonumber\\
\int_0^\infty\! d \om\ 
\left[ \alpha_{\la \om}^R \beta_{\la^\prime \om}^{R *} -
\beta_{\la \om}^{R *} \alpha_{\la^\prime \om}^{R }\right]
&=& 0 \quad .
\label{twofour}
\end{eqnarray}
Of course, the set $\varphi_{\la,R}$ is complete only in $R$. For complete
completeness its complement $\varphi_{\la,L}$, as expressed in
eq.~(\ref{twoone}), is required. Then the operators $a_\om$ may be expressed as
linear combinations of $c_{\la,R}$ and $c_{\la,L}$. So the totality of
Bogoljubov transformations is eq.~(\ref{twothree}), its analog in $L$ (wherein
one has the simple rule 
$\alpha_{\la \om}^L =\alpha_{\la \om}^{R *}$ and 
$\beta_{\la \om}^L =\beta_{\la \om}^{R *}$), and the inverse given by
\begin{equation}
a_\om = \int_0^\infty \! d \la \ 
\left[ \alpha_{\la \om}^{R *} c_{\la,R}
+ \alpha_{\la \om}^{L *} c_{\la,L} - 
\beta_{\la \om}^R c_{\la,R}^\dagger
-\beta_{\la \om}^L c_{\la,L}^\dagger\right] \quad .
\label{twofoura}
\end{equation}

To calculate the Rindler content of Minkowski vacuum  we use
eq.~(\ref{twothree})  to obtain the mean number of Rindler particles in
Minkowski vacuum \begin{eqnarray} \elematrice {0_M}{ c_{\la R}^\dagger
c_{\la^\prime R} }{0_M} &=& \int_0^\infty d \om \beta_{\la \om}^R
\beta_{\la^\prime \om}^{R *} =\delta (\la - \la^\prime) n(\la) \nonumber\\
n(\la) &=& (e^{\beta	\la} -1)^{-1}
\label{twoseven}
\end{eqnarray}
where we have used \ref{twotwo} and $| \Gamma(ix)|^2 =  \pi  / x
\mbox{sinh} \pi x$ and $n(\la)$ is the Planck distribution.
 One finds also
\begin{eqnarray}
\elematrice {0_M}{ c_{\la R}  c_{\la^\prime R} }{0_M} = 
\int_0^\infty d \om \alpha_{\la \om}^R \beta_{\la^\prime \om}^{R *}=0
\label{twosev}
\end{eqnarray}
In this way one understands the thermal propagator $W$ of eq.~(\ref{Rone}).
Indeed since $W$ has been calculated only in R, the accelerator
perceives Minkowski vacuum in R (to which it is confined) as a thermal bath (of
Rindlerons since in the proper frame of the accelerator it is they which are
in resonance with the excitations of the accelerator).

\section{Unruh Modes}\label{unruh}
There is an interesting and powerful technique of handling the Bogoljubov
transformation \ref{twotwo} which is due to Unruh~\cite{Unru1}. It couples
together Rindler modes which lie in R and L respectively
($\varphi_{\la,R}$ and  $\varphi_{\la,L}$) so as to yield
 a density matrix description for what happens in one quadrant when
one traces over the states in the other. This technique makes contact with
the physics of pair production in a constant electric field 
and finds important use in the black hole problem. It is
implemented by inverting \ref{twotwoa} (using \ref{twofour}) to give
\begin{equation} \varphi_{\la,R}(U) = \int_0^\infty d \om 
\left[
\alpha_{\om \la}^{R *} \xi_\om (U) -
\beta_{\om \la}^{R *} \xi_\om^* (U)\right ]
\label{twoeight}
\end{equation}
Though the inversion has been
carried out in R, eq.~(\ref{twoeight}) can be extended into L (since the
$\xi_\om (U) $ are defined there as well). It will be checked out below
that this is perfectly consistent since we will find automatically that in
this continuation the right hand side of eq.~(\ref{twoeight}) vanishes in L.
We rewrite eq.~(\ref{twoeight}) in the form 
\begin{equation}
\varphi_{\la,R} = \alpha_\la \hat \varphi_\la	(U) - \beta_\la
\hat \varphi_{ - \la}^*	(U)
\label{twonine}
\end{equation}
where 
\begin{eqnarray}
\hat \varphi_\la	(U) &=&
{1 \over \alpha_\la}\int_0^\infty \! d \om \ 
\alpha_{\om \la}^{R *}\ \xi_\om (U) \nonumber\\
\hat\varphi_{ - \la}^*	(U) &=&{1 \over \beta_\la}
\int_0^\infty \! d \om \ 
\beta_{\om \la}^{R *}\ \xi_\om^* (U)
\label{threenul}
\end{eqnarray}
and
\begin{eqnarray}
\alpha_\la &=& {1 \over \sqrt{1 - e^{- \beta \la}}}
\nonumber\\
\beta_\la &=& {1 \over \sqrt{ e^{ \beta \la} -1}}\label{threeone}
\end{eqnarray}
so that the modes $\hat \varphi_\la$ are normed in the usual way, i.e.
$ \int_{-\infty}^{+\infty} dU \hat \varphi_\la^*(U) \ i\! \lr {\partial_U} \hat
\varphi_{\la^\prime}(U) = \delta(\la - \la^\prime)$. Substituting
eq.~(\ref{twotwo}) into eq.~(\ref{threenul}) gives 
\begin{eqnarray}
\hat \varphi_\la &=&
{1 \over \alpha_\la} {1 \over 2 \pi a} \sqrt{ \vert \la \vert \over 4 \pi }
e^{\pi \la / 2 a} \Gamma(i \la /a )
  \int_0^\infty \! { d \om \over \om} 
\left( \om \over a \right)^{-i \la /a} 
e^{-i \om U} \nonumber\\
&\ &\nonumber\\
\ &=& \left.\left \{ 
\begin{array}{c}
\alpha_\la \theta(-U) { (-aU)^{i\la/a} /\sqrt{4 \pi \la}} + 
\beta_\la \theta(U) { (aU)^{i\la/a} / \sqrt{4 \pi \la}} 
\quad \la >0
\\ 
 \beta_{\vert \la \vert}\theta(-U) { (-aU)^{i\la/a} /\sqrt{4 \pi \la}}
+ \alpha_{\vert \la \vert} { \theta(U) (aU)^{i\la/a} / \sqrt{4 \pi \la}}
\quad \la <0
\end{array} \right.\right.
\nonumber\\
&\ &\nonumber\\
\ &=& \left.\left \{ 
\begin{array}{c}
\alpha_\la \varphi_{\la,R} + \beta_\la \varphi_{\la,L}^* \quad \la >0 \\
\beta_{\vert \la \vert}\varphi^*_{{\vert \la \vert},R} + 
\alpha_{\vert \la \vert} \varphi_{{\vert \la \vert},L} \quad
\la < 0
\end{array} \right.\right.
\label{xxx}
\end{eqnarray}
We shall call the modes, $\hat \varphi_\la$, Unruh modes. As announced the
linear combination (\ref{twonine}) does vanish in $L$. 

Unruh modes enjoy the following
properties:

1) They are eigenfunctions of $i aU \partial_U $ with eigenvalue $ \la$ in both
$R$ and $L$.

2) $\hat \varphi_\la$ are manifestly positive Minkowski frequency modes for
both signs of $\la$ (c.f. eqs~(\ref{threenul})). Together with their
conjugates they form a complete orthogonal set just as plane waves.
This is proven trivially by direct computation.

3) They are linear combinations of the Rindler modes $\varphi_\la^R$ and
$\varphi_\la^L$ given by the Bogoljubov linear combination (\ref{xxx}) (we
remind the reader once more that the mode $U^{i\la/a}\theta(U) , \la >0$ is
a negative frequency mode in $L$ (eq.~(\ref{oneeight})).

An independent and interesting derivation of eq.~(\ref{xxx}) is obtained by
appeal to analyticity in the lower half $U$ plane. This is because the Minkowski
modes eq.~(\ref{twonul}) are analytically extendable in the lower half plane 
for positive frequencies only. This 
expresses the stability of the ground state (Minkowski vacuum) of the theory. 
Eigenfunctions of $i\partial_u$ considered as functions of $U$ obey the
differential equation \begin{equation}
-i U \partial_U \chi_\la =  {\la \over a} \chi_\la
\label{threefive}
\end{equation}
having solutions
\begin{eqnarray}
\chi_\la &=& 
A \ \Bigl[ \theta(-U) {(-U)^{i \la /a} \over \sqrt{4 \pi \la}} 
\Bigr]
+ B \ \Bigl[ \theta(U) { U^{i \la /a} \over \sqrt{4 \pi \la}} \Bigr]
\nonumber\\ 
&=& A \varphi_{\la,R} + B \varphi_{\la,L}^*
\label{threesix} \end{eqnarray}
(Note the similarity to eq.~(\ref{uchi}). This is no accident. Any problem with
an exponential approach to a horizon in the classical theory will be reflected
in such a singular differential equation when expressed in terms of global
coordinates.)

To determine $A$ and $B$ in eq.~(\ref{threesix}) we require that $\chi_\la$ be
a positive frequency Minkowski mode i.e. an Unruh function. Therefore set $U$
equal to $\vert U \vert e^{i \theta}$ with $- \pi \leq \theta \leq 0$ so as to
continue the function $U^{i \la/a}$ analytically in the lower half complex
plane to go from $L$ to $R$. The boundaries at $\theta= - \pi , 0$ then give the
ratio $ B  /  A  = e^{- \pi \la /a} = e^{- \beta \la /2}$
(compare with eq.~(\ref{RTratio}).
Normalization then gives $\chi_\la = \hat \varphi_\la$. This technique has
been fruitfully used by Unruh~\cite{Unru1} and Damour and Ruffini~\cite{DaRu}
in the black hole problem.

The Unruh modes can then be synthetically written as
\begin{equation}
\hat \varphi_\lambda = {1 \over \sqrt{4 \pi \lambda (e^{\beta \lambda}-1)}}
\left( { U - i \epsilon  \over a} \right ) ^{i \lambda / a}
\label{Unm}
\end{equation}
with  the $i \epsilon$ encoding how the function should be continued in the
complex plane ($  U - i \epsilon  = U$ for $U>0$ and
$ U - i \epsilon  = \vert U \vert e^{-i\pi}$ for $U<0$). This expression for
$\hat \varphi_\lambda$ has the added luster that for small but finite
$\epsilon$ the high frequency behavior of the modes in the vicinity of $U=0$
has been regularized~\cite{Pare}. This will be seen to have important
consequences  when evaluating energy densities near and on the horizon, see
 Section \ref{state}. 

The quantized field $\phi$ can be decomposed in Unruh modes
everywhere according to \begin{equation}
\phi = \int_{-\infty}^{+\infty} \!{ d \la}
\ \left[ \hat a_\la \hat \varphi_\la + \hbox{h.c.} \right ]\quad .
\label{threeoneb}
\end{equation}
The bogoljuobov transformation among 
Rindler and Unruh annihilation and
creation operators is
\begin{equation}
\begin{array}{c}
\hat a_\la = \alpha_\la c_{\la,R} - \beta_\la c^\dagger_{\la,L}\\
\hat a_{-\la}^\dagger = \alpha_\la c_{\la,L}^\dagger - \beta_\la
c_{\la,R}
\end{array} \quad , \ \la>0
\label{threetwo}
\end{equation}
where we have used eq.~(\ref{xxx}).

In this way Minkowski vacuum can be viewed as a linear combination of pairs
of L and R Rindlerons. Defining $\ket{0_{Rindler}}$ as the direct product 
$\ket{0_R} \ket{0_L}$ of Rindler vacuum in the left and right quadrant
($c_{\la,L}\ket{0_L} =0$ and $c_{\la,R}\ket{0_R} =0$), we have following 
eq.~(\ref{outvac})
\begin{equation}
\ket{0_M} = Z^{-1/2} \exp \left [ \sum_{\la=0}^\infty {\beta_\la \over
\alpha_\la} c_{\la,R}^\dagger c_{\la,L}^\dagger \right]\ket{0_{Rindler}}
\label{threethree}
\end{equation}
where $Z$ is given by (see eq.~(\ref{norme}))
\begin{equation}
Z= \prod _\la \vert \alpha_\la \vert^2 
= \exp \left[{L \over 2 \pi} \int_0^{\infty}\! d\la \ \ln (1 + \beta_\la^2)
\right]=\exp \left[ L{\pi \over 12} {1 \over \beta}\right]
\label{partf}
\end{equation}
which is the partition function of a massless gaz in 1+1 dimensions in 
a  volume of size $L$.
Thus for an operator $O_R$ localized in $R$ one then has
\begin{eqnarray}
 \langle O_R \rangle &=& Z^{-1} 
\bra{0_{Rindler}} \exp \left[{ \sum_{\la=0}^\infty {\beta_\la \over
\alpha_\la} c_{\la,R} c_{\la,L}} \right] O_R
\exp \left [{ \sum_{\la=0}^\infty {\beta_\la \over \alpha_\la}
c_{\la,R}^\dagger c_{\la,L}^\dagger} \right] \ket{0_{Rindler}}
\nonumber\\
&=& Z^{-1}\sum_{ \{ n_\la \} }
\left( \beta_{\la} \over \alpha_{\la} \right)^{2 n_{\la}}
 \elematrice{\{ n_{\la} \}}{ O_R }{\{ n_{\la} \}}\nonumber\\
&=& Z^{-1} \sum_{ \{ n_\la \} }
e^{-\beta \sum_\la  \la n_\la  } 
\elematrice{\{ n_\la \}}{ O_R }{\{ n_\la \}}
\nonumber\\
&=& { {\rm tr} e^{-\beta H_R} O_R \over {\rm tr} e^{-\beta H_R}}
\label{threefour}
\end{eqnarray}
where $H_R$ is the Rindler hamiltonian generator of $\tau$ translations
restricted to $R$.
So single quadrant operators have their means given by a thermal density
matrix. 
Note that the Rindler hamiltonian (the generator of boosts)
is equal to $-i aU \partial_U $  and is given
in terms of the operators $c_{\la,R}, c_{\la,L}$ by
\begin{eqnarray}
H_{Rindler} &=& H_R - H_L \nonumber\\
&=& \theta(-U) \int_{-\infty}^{+\infty} \! du\ T_{uu}
-  \theta(+U) \int_{-\infty}^{+\infty} \! du_L\ T_{u_Lu_L}
\nonumber \\
&=& \int^{+\infty}_0 d\la \la (c_{\la,R}^{\dagger} c_{\la,R} - 
c_{\la,L}^{\dagger} c_{\la,L} )
\label{hamr}
\end{eqnarray}
and possess therefore the same structure as the hamiltonian in the $E$-field
given in eq.~(\ref{hamiltE}). Thus the  pairs of
Rindlerons in eq.~(\ref{threethree}) have zero Rindler energy. This degeneracy
allows for the  creation of pairs of Rindler quanta.

Of course everything that has been derived for $u$ modes applies equally for $v$
modes where the r\^ole of P and F are interchanged hence the past and
future horizons $H_\pm$ (see fig. \ref{Mink}). Also one can  perform the
analysis for fermions and include the effects of  a
mass and of higher dimensionality (see for instance~\cite{Spin}, \cite{SCD}).

It is more interesting (and relevant for blackholology as well) to inquire into
the physical condition of the radiation due to its interaction with the
accelerator.
 We shall therefore in the next section delve more
intimately into the details of the transition amplitude.

\section{Spontaneous Emission of Photons by an Accelerated
Detector}\label{spontaneous} 

We here dissect the rate formula eq.~(\ref{Rfivea}) by
displaying the transitions of the atom as resonance phenomena with Doppler
shifted Minkowski photons~\cite{PaBr}. This will introduce in a natural way the
conversion of a particular Minkowski vacuum fluctuation into an on mass shell
quantum. Similar resonance phenomena constitute the dynamical origin of
particle emission from a non inertial mirror (Section \ref{mirro}) and of black
hole radiation. 

We shall see that in this formulation the rate formula comes
about by a somewhat different mechanism from that of usual golden rule analysis,
i.e. in terms of density of states. It is rather a consequence of the steady
process in which photons are brought into resonance through the ever--changing
Doppler shift occasioned by the accelerator, much like the analysis  of Section
\ref{vacuum} (paragraph following equation (\ref{inout})), wherein different
$\omega$ values brought different vacuum fluctuations into resonance with real
pairs giving rise to a time sequence of produced pairs.

We focus for definiteness on right
movers. In lowest order in $g$, the amplitude for a transition in time
$\tau$ is \begin{eqnarray}
A_\mp (\om, \tau) &=& -i \bra{\pm} \elematrice{0_M}{ a_\om \int_0^\tau d\tau^\p
H_{int}(\tau^\p) }{0_M}\ket{\mp}\nonumber\\
&=& {-i g \over \sqrt{4 \pi \om}} \int_0^\tau d\tau^\prime
e^{\pm i \Delta M \tau^\prime} e^{-i \om a^{-1} e^{-a \tau^\prime}}
\label{threefourb}
\end{eqnarray}
where  the plus (minus) subscript of $A$ corresponds to  
spontaneous deexcitation
(excitation) and the ket $\ket{\pm}$ means excited (ground) state. Begin
with the former, spontaneous deexcitation. The integrand presents 
a saddle point at $\tau^*$ given by
\begin{equation}
\Delta M = \om e^{-a \tau^*(\om) }
\quad .\label{threefiveb}
\end{equation}
If $\om$ is such that $\tau^*(\om) (=a^{-1} \ln \om / \Delta M)$
lies well within the domain of integration in eq.~(\ref{threefourb}) (i.e. the
gaussian width $ =(\Delta M a)^{-1/2}$ does not overlap the limits of
integration), and furthermore if higher derivatives of the exponential are
relatively small (i.e. $ (\Delta M / a) >>1$), then the saddle point
estimate is valid and one finds \begin{equation}
A_+(\om, \tau) \simeq {-i g \over \sqrt{4 \pi \om}} \sqrt{2 \pi
\over -i \Delta M a}
e^{- i \Delta M a^{-1}} \left ( {\Delta M \over \om} \right)^{i \Delta M
a^{-1}} \quad .\label{threesixb}
\end{equation}
The crucial inequality for legitimization of eq.~(\ref{threesixb}) is 
that $\om$ should be bounded by
\begin{equation}
0 \leq a^{-1} \ln {\om \over \Delta M} \leq \tau
\label{threeseven}
\end{equation}
but we shall shortly loosen up on the condition $
(\Delta M / a) >>1$.

Once more there is, as for the electric case, a division into class I, the
frequencies $\om$
which resonate, i.e. satisfy eq.~(\ref{threeseven}) and class II modes which do
not resonate, i.e. lie outside the range (\ref{threeseven}) and for which
$A_+(\om,\tau)\simeq 0$.  And once again as in Section \ref{vacuum} this must be
qualified by "apart from edge effects". Clearly $\tau$ has to be sufficiently
large for these asymptotic estimates to make sense, as in all "golden rule" type
estimates.

For spontaneous excitation the saddle point condition has a  minus sign
on the r.h.s. of eq.~(\ref{threefiveb}) so that the saddle point lies at
\begin{eqnarray}
{\rm Re} \tau^*(\om) &=&  a^{-1} \ln {\om / \Delta M}\nonumber\\
{\rm Im} \tau^*(\om) &=& \pi/a\label{threeeight}
\end{eqnarray}
whereupon  one has
\begin{eqnarray}
A_-(\om,\tau) &=& e^{- \beta \Delta M /2} {-i g \over \sqrt{4 \pi \om}}\sqrt{ 2
\pi \over i \Delta M a} e ^{ i \Delta M a^{-1}}
\left ( {\Delta M \over \om} \right)^{i \Delta M
a^{-1}} \nonumber\\
&=& - A^*_+(\om,\tau) e^{- \beta \Delta M /2}
\label{threenine}
\end{eqnarray}
with $\beta = 2 \pi / a$. Equation~(\ref{threenine}) is valid only for class I
modes; for class II, $A_-(\om,\tau)\simeq 0$ as well.  
Squaring these formulae one
recovers the thermal ratio of rates gven in eq.~(\ref{Rfive}). 
Integrating $\om$ over the
bounds set by eq.~(\ref{threeseven}) and dividing by $\tau$, one recovers half
of $R_\pm$ given by eq.~(\ref{Rfivea}) provided one limits oneself to the
contribution from the leading poles in the integrand (\ref{Rfour}) [ $\tau = -
i \epsilon$ for emission and $2 \pi i / a$ for absorption, consistent with the
condition $\Delta M / a >>1$]. The other half is due to the left movers. Note
how the rate formula comes about, e.g. 
\begin{eqnarray}
P_{-}(\tau)&=&\int_{\Delta M}^{\Delta M e^{a\tau}}\!d\om
\vert A_-(\om,\tau)\vert^2\nonumber\\
&=&{g^2 e^{-\beta \Delta M}
\over 2 \Delta M a}
\int_{\Delta M}^{\Delta M e^{a\tau}}\!{d\om\over \om}\nonumber\\
&=&{g^2 e^{-\beta \Delta M}
\over 2 \Delta M a}
\int_0^\tau \!d\tau^\p
\label{threenineb}
\end{eqnarray}
The integral over $\om$ is $\int d \ln \om$
hence an integral over saddle times ${\rm Re} \tau ^* (\om)$. Neglecting edge
effects this is equal to $\tau$. The physical interpretation is clear. At each
time $\tau^*(\om)$, the Minkowski frequency $\om$ enters into resonance because
of the changing Doppler shift occasioned by the acceleration ($\om_{\rm Res} =
\Delta M e^{a \tau^* }$).

It is interesting to try to interpret the complex saddle encountered in
eq.~(\ref{threeeight}) (the case of absorption).  In
the complex $\tau$ plane the contour has to be deformed from ${\rm Im} \tau =0$
so as to go through ${\rm Im} \tau =\pi / a$ (as in the construction of Unruh
modes--paragraph after eq.~(\ref{threefive}). In both cases the voyage in
the complex plane encodes the positivity of Minkowski frequencies). Reference to
eq.~(\ref{onesixb}) then indicates that there has been a voyage from $U$ to
$-U$ (or $x\to -x$, $t\to -t$), hence from a point in the quadrant R to its
inversion in the third quadrant L where $\tau$ runs backwards (i.e. 
$d\tau / dt < 0$) hence where Rindler energy conservation is satisfied
since the frequency $\om$ appears as carrying negative Rindler energy
(since it is given by $i aU \partial_U$ see eq.~(\ref{hamr}). 
In  Sections \ref{state} and \ref{weakacc}
it will be shown by more detailed considerations that in fact the Minkowski
photon which is emitted when the atom is excited arises from a vacuum
fluctuation one part of which exists in the quadrants $L$ and $F$, the other
part lives in $P$ and $R$. This latter has crossed the past horizon so as to be
absorbed by the atom and the former continues out to infinity in quadrant $F$,
i.e. it arises in $L$ and radiates from there into $F$. It is this effective
emission act in $L$ which is encoded in the the rough Born approximation saddle
point integral of saddle point eq.~(\ref{threeeight}). The reader at this point
can pick up some flavor for the true state of affairs by peeking ahead at
Fig.\ref{excit}.

One  can offer oneself the luxury of completing this type of analysis to get the
full expressions for $R_\pm$ (i.e. by relaxing the condition 
$\Delta M/ a >> 1$) 
by noting that for any value of $\Delta M / a$ the
contribution to the integrand from a given $\om$ is dominated by the region
(gaussian width) around $\tau^*(\om)$. So provided 
one maintains eq.~(\ref{threeseven})
one can extend the $\tau^\prime$ integration over the whole real
axis. 
In this limit the transition amplitudes eqs~(\ref{threefourb}) and
(\ref{threenine})
 become 
\begin{eqnarray}
A_+ &=& -ig \sqrt{{\pi\over \Delta M}} \alpha_{\Delta M \omega}^{R *}\nonumber\\
A_- &=& -ig
\sqrt{{\pi\over \Delta M}} \beta_{\Delta M \omega}^{R *}
\label{extrab}\end{eqnarray}
thereby giving a dynamical
content to the Bogoljubov coefficients (\ref{twotwo}).
Squaring and integrating
over $\om$ within the bounds limited by eq.~(\ref{threeseven}) leads to
the exact rates given in eqs~(\ref{threenine}) multiplied by the interval
$\tau$. This 
is how the Golden rule comes about in this
approach, successive resonances with the Doppler shifted frequencies of
the proper atomic frequency (here $= \Delta M$).

For the skeptical reader we now present a more rigorous analysis of this
argument as it plays an important role in the black hole problem as well. At
the same time we shall clearly exhibit the difference between the class I and
class II modes.

[
The amplitude to emit a photon of frequency $\omega$ 
in the interval $(-\tau/2, \tau/2)$ is 
\begin{eqnarray}
A_+(\omega, \tau) &=&
{-i g\over \sqrt{4 \pi \omega}}
\left( {\omega \over a} \right)^{-i \Delta M/a}
\int_{\omega a^{-1} e^{-a \tau/2}}^{\omega a^{-1} e^{a
\tau/2}}
\! \!dx\  e^{-ix} x^{i \Delta M/ a -1}
\label{jazz00}\\
&=&{-i g\over \sqrt{4 \pi \omega}}
\left( {\omega \over a} \right)^{-i \Delta M/a}
{e^{-\Delta M \pi / 2 a} \over a^2}
\left[ \gamma(i\Delta M /a , i \omega a^{-1} e^{a
\tau/2} ) \right.\nonumber\\&\ &\quad\quad\quad
\left. -\gamma(i\Delta M /a , i \omega a^{-1} e^{-a
\tau/2} )  \right]
\label{jazz01}
\end{eqnarray}
 where the interval is
taken symmetric around $\tau=0$ for mathematical
convenience; the physics is unmodified by this
translation in $\tau$ and where 
$\gamma$ is the incomplete gamma function~\cite{WiWa} which for large
and small values of its argument takes the form
\begin{eqnarray}
\gamma(i\mu , ix) &\mettresous{x \to \infty}\sous{\simeq}& 
\Gamma(i \mu) - i e^{- \pi \mu / 2}
x^{i \mu} e^{-ix} x^{-1}
\label{jazz02}
\\
\gamma(i\mu , ix) &\mettresous{x \to 0}\sous{\simeq}&  - i e^{- \pi \mu / 2}
x^{i \mu}
\label{jazz03}
\end{eqnarray}

Before analyzing equation (\ref{jazz01}) we discuss the unphysical infra-red
divergence which arises as $\om\to 0$. This divergence exists in the inertial
case as well. Indeed as $\om \to 0$ one has $A_+=2 i g \sin(\Delta M \tau /2 )
/\sqrt{4 \pi \omega}$ independent of the acceleration. It is unphysical
because in one dimension the coupling of the massless field to the atom
becomes strong as $\om \to 0$. Therefore the perturbation theory which has
been used is inadequate. No doubt a resummation of all terms is possible to
give the true infra red physics (like the Bloch Nordziek~\cite{BochN} theory
in QED). But this is irrelevant to our purpose, since these infra red photons
do not contribute to the rate.

One way to get rid of the problem is to subtract off the inertial amplitude. We
shall do the computation in another manner so as to obtain an infra red finite
answer. This is accomplished by adiabatically 
switching on and off the coupling to the field by a 
function $f(\tau)$.
 Then the amplitude
for $\om \to 0$ will decrease as a function of the period $T$ of switch on and
off. To see this we remark that 

1) for $\om \to
0$, $A_+$ is given by the Fourier transform of $f$ 
($A_+(\omega \simeq 0, \tau) \simeq \int d\tau e^{-i \Delta
M \tau} f(\tau) / \sqrt{\om}$). 

2) If the interval $\tau$
during which $f$ is constant is equal to $\tau = 2 \pi k /
\Delta M$ with $k$ an integer then $A_+(\om\simeq 0,\tau)$ is
independent of $k$. 

3) Therefore we can take $k=0$. 

4) If $f$
can be differentiated $n$ times, the Fourier transform of
$f$ decreases for large $\Delta M$ as $\Delta M ^{-n}$. 

5)
Since the only dimensional parameter is $T$, the Fourier
transform of $f$  must be of the form   $T^{-n+1} \Delta
M^{-n}$. 

\noindent The correct procedure to calculate
the rate of excitation consists in first taking the
adiabatic limit $T \to \infty$ (with however the condition
$T << \tau$ ) and only then
performing the integral over $\omega$. We shall show that
with this order of operations the concept of resonant
frequency described in the main text appears. 

To perform thus the adiabatic limit
it proves rather more convenient to introduce a time average of the amplitude
\begin{equation}
\bar A_+(\om,\tau) = \int\! d \tau^\prime
\ g(\tau^\prime) A_+(\om, \tau^\prime)\label{jazzxx}
\end{equation} 
where $g(\tau^\prime)$ is a $n$ times
differentiable function, centered on $\tau^\prime=\tau$, with width
$T$, and normalized such that $\int \! d\tau\ g(\tau) =1$.
One verifies by permuting the integral over $\tau$ in
eq.~(\ref{jazzxx}) and the integral over $x$ in the definition of
$A_+$ (eq.~(\ref{jazz00}) that the averaging of $A_+$ is
equivalent to the introduction of a switch function
$f(\tau)$ with the same regularity as $g$.

We now permute the series expansions eqs~(\ref{jazz02}) and
(\ref{jazz03}) with the averaging in eq.~(\ref{jazzxx}) to
obtain that $\bar A_+ $ is given by a similar
expression to eq.~(\ref{jazz01}) with $\gamma$ replaced by
$\bar \gamma $ where 
(see eq.~(\ref{jazz03}))
\begin{eqnarray}
\bar  \gamma (i\Delta M/a , i \omega a^{-1}
e^{\pm a \tau / 2})
&\mettresous{\frac{\omega 
e^{\pm a \tau / 2}}{a} \to \infty}\sous{\simeq}&
 \Gamma ( i\Delta M / a) + \nonumber\\
&\ &\!\!\!\!+ O\left({e^{- \pi \Delta M / 2a}
\over \omega a^{-1}
e^{\pm a \tau / 2} } \right)
\label{jazz04}
\\
\bar  \gamma (i\Delta M/a , i \omega a^{-1}
e^{\pm a \tau / 2})
&\mettresous{\frac{\omega 
e^{\pm a \tau / 2}}{a} \to 0}\sous{\simeq} &
 O\left(e^{- \pi \Delta M / 2a}
{a e^{\pm i\Delta M \tau/2} \over T^{n-1} \Delta M^n}\right) \quad 
\label{jazz05}
\end{eqnarray}

We are now
in a position to rederive  the results obtained in the main
text. Three cases are to be considered according to the
values of $\omega$:

1) $\omega < a e^{- a \tau / 2} $. Then both $\gamma$
functions in eq.~(\ref{jazz01}) are given by eq.~(\ref{jazz05}),
hence their difference yields \hfil \\ $\bar A_+ (\omega, \tau)
\simeq O[a e^{-\pi \Delta M/2a} {\rm sin} (\Delta M \tau / 2) / (T^{n-1}\Delta
M^n)]$.

2) $a e^{- a \tau / 2}  < \omega < a e^{a \tau / 2} $.
Then the first $\gamma$ function in eq.~(\ref{jazz01}) is given
by eq.~(\ref{jazz04}) and the second by eq.~(\ref{jazz05}, hence
their difference is $\Gamma(i \Delta M /a)+ O[a\/e^{-\pi \Delta M/2a}/T^{n-1}
\Delta M^n]$
and $\bar A_+(\omega , \tau)$ is given by eq.~(\ref{extrab}) as announced.

3) $\omega > a^{a\tau / 2}$. Then both gamma functions are
given by eq.~(\ref{jazz04}) and $A_+$ vanishes once more.

In Fig.\ref{rate} we have plotted a numerical calculation of
$\vert \gamma(i\Delta M /a , i \omega a^{-1} e^{a
\tau/2} ) 
-\gamma(i\Delta M /a , i \omega a^{-1} e^{-a
\tau/2} ) \vert^2$ as a function of $\tau$ and $\ln
\omega$. The plateau of this function when 
$-a \tau/ 2 < \ln \omega /a < a \tau /2$ is clearly
apparent. The oscillations on the plateau are due to the fact that the
adiabatic limit has not been taken  in
this figure. \dessin{1.000}{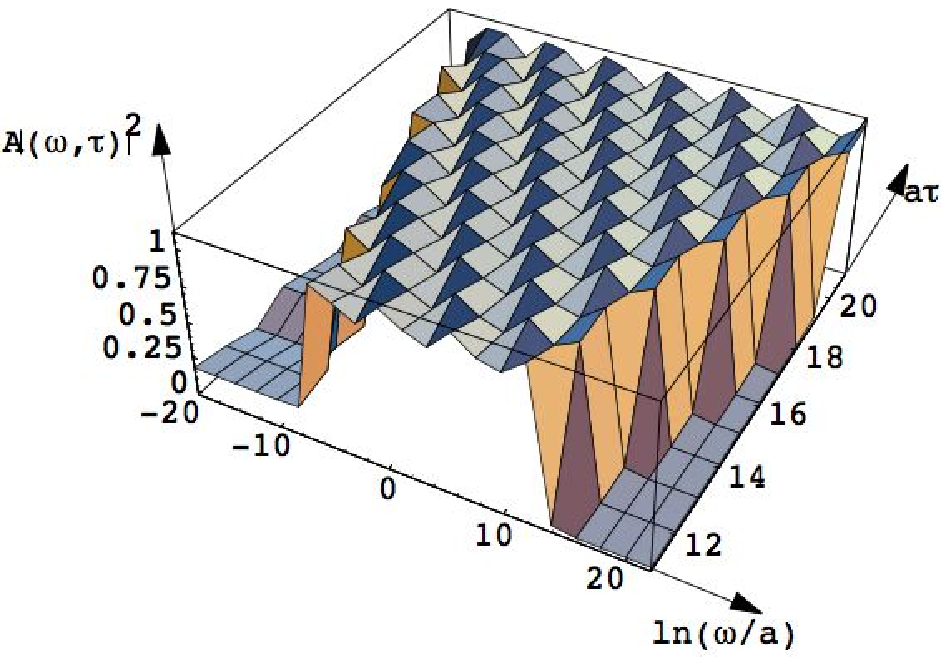}{rate}{The square of the amplitude $A(\tau , \omega)$ to
have made a transition in time $\tau$ by emitting a photon of
Minkowski frequency $\omega$. plotted as a funtion of
$\tau$ and $\ln \omega$.  The plateau that arrises when
the resonance condition $-a\tau/2 < \ln \omega/\Delta M <
a \tau/2$ is satisfied is clearly seen.
} 

In this way we have confirmed that 
eq.~(\ref{threefourb}) arises from the saddle point region to
yield the result (\ref{extrab}). Hence that the successive
resonances do build up to yield a rate.]


\section{The Accelerating Mirror}\label{mirro}

\subsection{General Description}

A highly instructive chapter in the physics of accelerating
systems is that of the accelerating mirror\cite{FuDa}, \cite{DaFu}, \cite{BD} in
that the analogy to the production of Hawking radiation is remarkably close.

We have shown that a uniformly accelerating system with
 internal degrees of freedom displays thermal properties
due to the exponentially changing Doppler shift (eq.~(\ref{threefiveb})) which
relates the inertial frequencies to its local resonant frequency. Similarly a
non inertial mirror scatters  modes with the changing Doppler shift associated
with its trajectory, hence giving rise to physical
particles. If the mirror trajectory is such that the
Doppler shift is identical to eq.~(\ref{threefiveb}) the particles
emitted by the mirror will be thermally distributed as well.
We write this Doppler shift as
\begin{equation}
\omega e^{-aU} = k
\label{mirone}
\end{equation}
where the energy difference $\Delta M$ is now replaced by
the frequency of the produced particle $k$, and
$\omega$ is, as before, the frequency of the Minkowski
vacuum fluctuation in resonance with the produced quantum which is reflected
from the point $U$ on the mirror. In the next paragraph the order of magnitude of
$k$ is  taken to be of $O(a/ 2 \pi)$ whereas $\omega$ varies strongly. We now
determine the mirror trajectory which leads to this Doppler shift.

The general solution of $\square \phi = \partial^2 /
\partial_U\partial_V \phi =0$ is 
\begin{equation}
\phi = F(V) + G(U)
\label{mirtwo}
\end{equation}
The reflection condition is that $\phi$ vanish on the
mirror hence the solutions take the form
\begin{equation}
\phi = F(V) - F(V_m(U))
\label{mirthree}
\end{equation}
where the mirror trajectory is expressed as
\begin{equation}
V=V_m(U)
\label{mirfour}
\end{equation}

To determine the Doppler shift associated with this
trajectory we take an incoming Minkowski mode: $e^{- i
\omega V} /\sqrt{4 \pi \omega}$. The reflected wave is then
$-
e^{- i
\omega V_m(U)} /\sqrt{4 \pi \omega}$. This mode should be decomposed into the
inertial outgoing modes  $e^{- i
k U} /\sqrt{4 \pi k}$ so as to determine its particle content. The
scattering amplitude (the Bogoljubov coefficient $\alpha_{\omega k}$) 
is
the overlap of the modes
\begin{equation}
\alpha_{k \omega } = 
\int dU { e ^{i k U}\over \sqrt{4 \pi k}} i {\lrpartial}_U
{e^{- i \omega V_m(U)} \over\sqrt{4 \pi \omega}}
\label{mirfive}
\end{equation}
The resonance condition given by the stationary phase of
the integrand is
\begin{equation}
k=\omega {d V_m \over dU}
\label{mirsix}
\end{equation}
Hence in order to recover eq.~(\ref{mirone}) we must take as trajectory for the
mirror
  \begin{equation}
V_m(U)= -{1 \over a} e^{-a U}
\label{mirseven}
\end{equation}
where we have set to zero an irrelevant integration
constant.  This trajectory tends exponentially fast towards
the asymptote $V=0$ which is the last reflected ray. It plays the role of a horizon in
this problem (see Fig~\ref{ACCEL1}). It is by construction 
that this mirror trajectory has the same expression as an
inertial trajectory expressed in Rindler coordinates $u,v$ as it approaches the
future horizon $u=\infty$ (see reference \cite{BD} for a mapping 
of the one problem into the other by
a conformal transformation). \dessin{1.000}{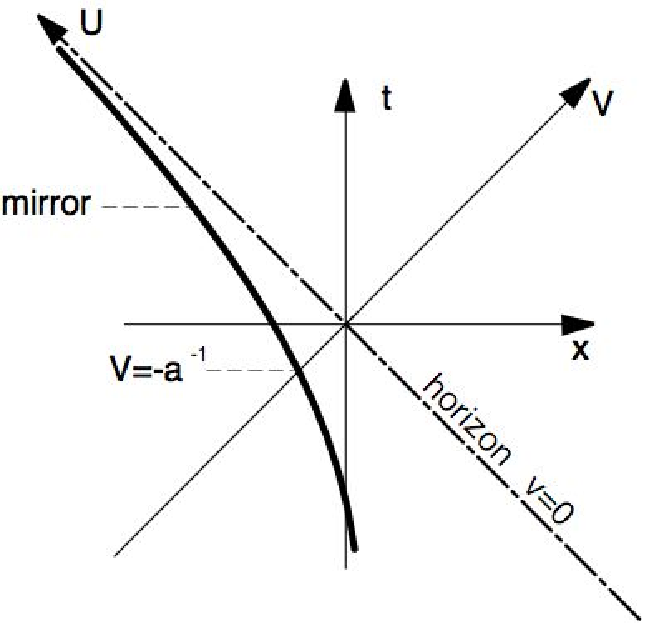}{ACCEL1}{The mirror trajectory $V_m(U) = - a^{-1} e^{-aU}$ which
gives rise to a steady thermal flux.} The mirror therefore follows a very
different trajectory from that of the uniformly accelerated detector.

In order to describe the particles emitted by the mirror it
is necessary to work in the second quantized context. As in
the electric field problem one introduces two bases. The
initial one is given by
\begin{equation}
\varphi^{in}_\omega = {1 \over \sqrt{4 \pi \omega}}(e^{-i
\omega V} - e^{
-i \omega a^{-1} e^{-aU}})\qquad \omega > 0
\label{mireight}
\end{equation}
and corresponds on the past null infinity surface  ${\cal I^-}$ $(U=-\infty)$
 to the usual Minkowski basis. The in-vacuum is the state annihilated by the
destruction operators $a_\omega^{in}$ associated with these modes:
\begin{equation}
\phi= \int_0^\infty \!d \omega\ ( a^{in}_\omega
\varphi^{in}_\omega + {\rm h.c.} ) \label{mirnine}
\end{equation}
The final basis corresponds  to the usual
Minkowski basis on the future null infinity surface ${\cal I^+}$ $(V=+\infty)$,
hence given by \begin{equation} \varphi_{k}^{out L} = {1 \over \sqrt{4 \pi
k}}(\vert aV \vert^{i k /a}\theta(-V) - e^{-i kU} )\qquad k > 0 \label{mirten}
\end{equation}
 The $V$ part of
$\varphi^{out L}$ is identical to the left 
Rindler modes expressed in Minkowski
coordinates (see eq.~(\ref{oneeight})). Similarly the $U$ part of
$\varphi^{in}$ is the same function as a Minkowski mode when it is expressed in
Rindler coordinates. Hence the mathematics as well as the physics of Sections
\ref{general} to \ref{spontaneous} apply to obtain the number and type of
particles emitted from the mirror. Indeed the alpha Bogoljubov
coefficient~(\ref{mirfive}) and the beta Bogoljubov coefficient 
\begin{equation}
\beta_{k \omega  } = 
\int^{\infty}_{-\infty} dU { e ^{-i k U}\over \sqrt{4 \pi k}} i {\lrpartial}_U
{e^{- i \omega V_m(U)} \over\sqrt{4 \pi \omega}}
\label{mirfiveB}
\end{equation}
are identical to the Bogoljubov coefficients obtained in eq.~(\ref{twotwo}).
In particular the ratio $\vert \beta_{\om k}/ \alpha_{\om k}
\vert = e^{-  \pi k / a}$
obtains and implies a constant rate of particle production in a thermal spectrum
of produced particles at temperature $T = a/ 2
\pi$. 

The mechanism wherein one finds a constant rate of
emission  is the same as in Section \ref{spontaneous}.
It arises from the successive  resonances of in-modes 
of varying energy $\omega$ with the emitted
photons of energy $k$.  To a photon in a wave packet of energy
$k$ emitted around the value $U=U_0$ corresponds
one and only one $V$-mode. Its frequency is $\omega = {k}e^{a
U_0}$ as in equation eq.~(\ref{threefiveb}). 
And the number of photons of energy $k$ emitted during a certain
$\Delta U$ lapse is given by the integral
of 
$ \vert \beta_{\om k} \vert^2 = (1 / \omega) n(k)$
(  where  $ n(k)=(e^{k/T} -1)^{-1}$)
over the resonant frequencies $\om$.
This integral yields $\Delta U  n(k)$, i.e. a thermal flux times 
the time lapse, as in eq.~(\ref{threenineb}).

We now calculate the energy momentum carried by these quanta. In the
next subsection the mean energy will be obtained by two different
techniques. In the third subsection we shall describe the fluctuations of
the energy density: we shall obtain the energy momentum correlated
to the observation of an outgoing particle around a particular
value of $U$. We wish to emphasize at this point that we are now seriously
trespassing into the domain of the black hole problem. The next two sections
contain a great deal of the essential physics of black hole evaporation.

\subsection{The Mean Energy Momentum Tensor} \label{Mean}

To start, from eq.~(\ref{mireight}) we have
\begin{equation}
\langle 0_{in} \vert T_{VV} \vert 0_{in} \rangle
\vert_{U=-\infty} =0
\label{eq11}
\end{equation}
The surface ${\cal I}^-$  ($U=-\infty$) is where
Minkowski vacuum is laid down. It is the Cauchy surface
from which emanates the modes~(\ref{mireight}). This defines the
Heisenberg state $\ket{0_{in}}$ which is annihilated by the
operators $a^{in}_\omega$. From eq.~(\ref{mireight}) the modes $\varphi
^{in}_\omega$ are pure $V$-like on ${\cal I}^-$ (in the sense of broad wave
packets) and are Minkowski modes. Thus eq.~(\ref{eq11}) is true in the
usual sense of normal ordering. Furthermore since
$\partial_U T_{VV}=0$ for massless fields in Minkowski
space it follows that eq.~(\ref{eq11}) remains true for all points
$(U,V)$ which lie to the right of the mirror's
trajectory~(\ref{mirfour}):
\begin{equation} 
\langle 0_{in} \vert T_{VV} \vert
0_{in} \rangle \vert_{V>V_m(U)} =0 .
\label{eq11B}
\end{equation}
More interesting is $\langle T_{UU} \rangle_{in}$. We
calculate it in two different ways: 

1) by mode analysis 

2)
more synthetically through use of Green's functions. 

\noindent In
the first method we have
\begin{eqnarray}
\langle 0_{in} \vert T_{UU} \vert 0_{in} \rangle &-& 
\langle 0_{out} \vert T_{UU} \vert 0_{out}\rangle
= \nonumber \\ 
\int_0^\infty\! d \omega \ \partial_U \varphi^{in}_\omega
\partial_U \varphi^{in *}_\omega \ &-& \ 
\int_0^\infty\! d k \ \partial_U \varphi^{out L}_k \partial_U
\varphi^{out L *}_k \label{TRBB}
\end{eqnarray}
wherein we have implemented the prescription of normal ordering
by making a subtraction of the value of $\langle T_{UU}
\rangle$ in Minkowski vacuum. Indeed by definition the $U$ part of the out modes
eq.~(\ref{mirten})  is Minkowski in character. We
 use the Bogoljubov coefficients~(\ref{mirfive}) and (\ref{mirfiveB}) to express
$\varphi^{in}_\om$ in terms of $\varphi^{out  L}_k$ 
\begin{equation}
\varphi^{in}_\om =\int_0^\infty\!dk\ 
\alpha_{k \om } \varphi^{out}_k +
\beta_{k \om } \varphi^{out *}_k
\label{TRB}\end{equation}
Thus obtaining \begin{eqnarray}
\langle 0_{in} \vert T_{UU} \vert 0_{in} \rangle -
\langle 0_{out} \vert T_{UU} \vert 0_{out}\rangle
= \nonumber\\
\int_0^\infty\!dk
\int_0^\infty \! dk^\p
\left[
\left(
\int_0^\infty\! d\om
\alpha_{k\om}\alpha^{*}_{k^\p\om}\right)
\partial_U \varphi^{out  L}_k \partial_U
\varphi^{out L *}_{k^\p} \right.\ +\nonumber\\
\quad \left(
\int_0^\infty\! d\om
\beta_{k\om}\beta^{*}_{k^\p\om}\right)
\partial_U \varphi^{out L *}_k \partial_U
\varphi^{out  L}_{k^\p}\ + \nonumber\\
\quad \left(
\int_0^\infty\! d\om
\alpha_{k\om}\beta^{*}_{k^\p\om}\right)
\partial_U \varphi^{out  L}_k \partial_U
\varphi^{out  L}_{k^\p}\ + \nonumber\\
\quad\left.\left(
\int_0^\infty\! d\om
\beta_{k\om}\alpha^{*}_{k^\p\om}\right)
\partial_U \varphi^{out L *}_k \partial_U
\varphi^{out L *}_{k^\p}\right]
\nonumber\\
-\ \int_0^\infty\! d k \ \partial_U 
\varphi^{out  L}_k \partial_U
\varphi^{out L *}_{k^\p}
\label{TRR}
\end{eqnarray}
The unitary
relations~(\ref{twofour}) simplify the result to the form
\begin{eqnarray}
&=& \int_0^\infty\!dk
\int_0^\infty \! dk^\p\ 
2 {\rm Re} \left[
\left(
\int_0^\infty\! d\om
\beta_{k\om}\beta^{*}_{k^\p\om}\right)
\partial_U\varphi_{k}^{out L *} \partial_U
\varphi_{k^\p}^{out  L} \right.
\nonumber\\
&\ & \quad 
\left. +
\left(
\int_0^\infty\! d\om
\alpha_{k\om}\beta^{*}_{k^\p\om}\right)
\partial_U\varphi_{k}^{out  L} \partial_U
\varphi_{k^\p}^{out  L} \right]
\label{TRRR}
\end{eqnarray}
We now recall eqs.~(\ref{twoseven}, \ref{twosev})
\beqa
\int_0^\infty\! d\om
\beta_{k\om}\beta^{*}_{k^\p\om}&=&
{{e^{-\pi(k+k^\p)/(2a)}\Gamma(ik/a)\Gamma(-ik^\p/a) \sqrt{kk^\p}}\over {(2 \pi
a)^{2}}} \int_0^\infty\! (d\om /\om) (a/\om)^{i(k-k^\p)/a}\nonumber\\
&=&\delta (k - k^\p) n(k)\label{TRRRBBB}
\feqa 
where
\beq
n(k)={{e^{-\pi k/a}|\Gamma(ik/a)|^2 k}\over {(2 \pi a)}}={1\over {(e^{2\pi
k/a}-1)}}
\feq
One also verifies  that
\beq
 \int_0^\infty\! d\om \beta_{k\om}\alpha^{*}_{k^\p\om}=
{{e^{-\pi(k-k^\p)/(2a)}\Gamma(ik/a)\Gamma(ik^\p/a) \sqrt{kk^\p}}\over {(2 \pi
a)^{2}}} \int_0^\infty\! (d\om /\om) (a/\om)^{i(k+k^\p)/a}
\feq
is proportional to $\delta(k+ k^\p)$
hence does not contribute to eq.~(\ref{TRRR}). The vanishing of this interference
term is the expression in the particular case we are considering  of the general
theorem proven in eq.~(\ref{threefour})  that expectation values of operators
restricted to one quadrant are given by their 
average in a thermal density matrix (see also eqs~(\ref{twoseven},
\ref{twosev})).
This theorem is applicable here since the $U$
 part of each mode arises from the part of the
mode to the left of the horizon ($V<0$).

The final answer is thus \begin{eqnarray}
\langle 0_{in} \vert T_{UU} \vert 0_{in} \rangle -
\langle 0_{out} \vert T_{UU} \vert 0_{out}\rangle
= \int_0^\infty \! dk\ 2\ n(k)\ \partial_U\varphi_{k}^{out  L}
 \partial_U \varphi_{k}^{out L *}\nonumber \\
={1 \over 2 \pi} \int_0^\infty\! dk \ k \ n(k)
= {\pi \over 12} \left({a \over 2 \pi }\right)^2\label{TRRRR}
\end{eqnarray} 
which is a thermal flux in one dimension with $T= a/ 2 \pi$ as
announced.

The other technique to calculate the flux 
relies only on the trajectory
of the mirror eq.~(\ref{mirseven}) \cite{DFU}. We now pass it in review.

 In usual global null coordinates the metric of Minkowski 
space reads as
\begin{equation}
ds^2 =- dUdV
\label{A7}
\end{equation}
and the modes  used to quantize about the usual Minkowski vacuum are
 $e^{-i \omega U}$ and $e^{-i \omega V}$. The
$U$ part of the reflected wave in eq.~(\ref{mireight}) is in terms
of modes $e^{-i\omega f(U)}$ with $f(U) = -a^{-1} e^{-aU}$.
If we adopt $f(U)$ in place of $U$ as coordinate we have
\begin{equation} ds^2 =- {dU\over df} dfdV
\label{A8}
\end{equation}
The $U$ part of the Green's function in usual Minkowski
vacuum is $-{1 \over 4 \pi}\ln \vert U - U^\prime\vert$
whereas in the modes $e^{-i\omega f(U)}$ it is 
$-{1 \over 4 \pi}\ln \vert f(U) - f(U^\prime ) \vert$.
Thus the difference in $\langle T_{UU} \rangle = 
\langle \partial_U \phi  \partial_U \phi \rangle$ between
the two is
\begin{eqnarray}
\Delta \langle T_{UU} \rangle &=&
-{1 \over 4\pi} \lim_{U \to U^\prime}
\partial_U\partial_{U^\prime} \left[
\ln \vert f(U) - f(U^\prime ) \vert
-
\ln \vert U - U^\prime  \vert\right]\nonumber\\
&=& {1 \over 12 \pi}
\left[ f^{\prime 1/2}\partial_U^2 f^{\prime -1/2}
\right]\label{A9}
\end{eqnarray}
where $f^{\prime} = df/dU$ and where
the second line is obtained by expanding $f$ to third order in $U-U^\p$. 
In the present case $f(U) = V_m(U)$ which yields
\begin{eqnarray}
\langle 0_{in} \vert T_{UU} \vert 0_{in} \rangle -
\langle 0_{out} \vert T_{UU} \vert 0_{out} \rangle
&=&{a^2 \over 48 \pi} = {\pi \over 12 }T^2\quad (T= a/ 2
\pi) \label{A9B}
\end{eqnarray}
as required. The second line of eq.~(\ref{A9})
is a remarkably elegant formula which relates the
average energy momentum  to the coordinate transformation
$f(U)$ only. It does not refer to the presence of the
mirror. It suffices to refer to the non inertial coordinates used in the
expression~(\ref{A8}) of the metric. We shall see in Section (\ref{Tmunuren})
that there is a natural generalization to curved space which is of
important use in the black hole problem. 

Note that the complete in-in Green's
function involves terms in $V,V^\prime$ and mixed terms
$U,V^\prime$ due to the linear combinations which appear in eqs~(\ref{mireight})
and (\ref{mirten}). The term in $V,V^\prime$ gives $\langle T_{VV} \rangle$
unchanged from Minkowski vacuum, eq.~(\ref{eq11B}). The $U,V$ term gives $U,V$
energy-energy  correlations
but no contribution to $\langle T_{\mu\nu} \rangle$ itself
since the trace ($4\ \langle T_{UV} \rangle = m^2 \langle \phi^2 \rangle $)
vanishes identically for a massless field in Minkowski space
(see Section (\ref{Tmunuren}) and ref. \cite{BD})).

\subsection{The Fluctuations around the Mean}\label{fluctmirr}

In preparation for black hole physics, we now ask a more detailed question,
concerning fluctuations.  We wish to display the field configuration in vacuum
that is responsible for the emission of a particular photon whose wave function is
localized around $U=U_0$. [ Recall that the
pure state Minkowski vacuum is a linear superposition of field configurations
(it is the product of the ground states of the field oscillators: $\ket{0_M} =
\Pi e^{-\vert \phi_n(x) \vert^2 \omega_n}$ 
in $\phi$ representation)]. In particular we shall describe the energy
distributions which characterize this fluctuation (we consider this
quantity since in the black hole problem the energy momentum is the
source of the back reaction). .
At the same time we will also answer the complementary
question, to wit: what is the energy distribution when no photon is observed
around $U_0$ (since we know the average). The calculation is straightforward
but somewhat lengthy. We therefore sketch below the main points and then go on
to the proofs. 

The Heisenberg state $\ket{0_{in}}$ is a linear superposition of out
states. It is a straightforward exercise to describe the contribution
of each final state to the energy on ${\cal I}^+$  (in order to do this
in a local manner it is necessary to use localized wave packets). In
addition to the positive energy density on ${\cal I}^+$ associated to
the production of a particle at $U=U_0$, there is
 correlated to it a
``partner'' which is a bump of field that propagates along the
mirror (hence with $V>0$).
Before reflection
there is also a bump  in the region $V<0$.
This is the ``ancestor'' of the produced 
photon. The total energy carried by this
pair of bumps vanishes as behooves a vacuum fluctuation. It does so in rather
subtle fashion. It is positive definite for $V>0$ whereas in the region $V<0$,
there is a positive energy bump, mirror image of the former as well as an
oscillating broader  piece which is negative. The sum of all these contributions
vanishes on $\cal{I}^-$ and therefore for all $U$
 until the wave packet starts reflecting. 

If two (or in general n) photons are produced around $U_0$ then the ancestors
carry twice (n times) the energy if one photon is produced. If no photons are
produced there is also a vacuum fluctuation which is proportional to minus
 the energy if one photon is produced. The coefficient is such that  upon
averaging the energy over the production of zero, one, two ... photons one
recovers the mean, 
(zero for $\langle T_{VV}\rangle$, see eq.~(\ref{eq11B}) and the
thermal flux for $T_{UU}$ (see eq.~(\ref{TRRRR})).

To see all of this we first display the pair. This stands in strong analogy to
Section \ref{unruh}. We then consider the non-diagonal matrix elements of $
T_{\mu \nu}$ associated to the produced photon at $U_0$.

To display the pair we introduce the analog of the
Unruh modes (Section \ref{unruh}) since they diagonalize the Bogoljubov
coefficients. An out mode (eq.~(\ref{mirten})) is proportional to
$\theta(-V)$. By adding to it a piece proportional to $\theta(+V)$ one can
obtain a purely positive frequency mode on ${\cal I}^-$. Writing only the $V$ 
part of the modes we have
\begin{equation}
\hat \varphi_k = {1 \over \sqrt{4 \pi k}} \left(
\alpha_k \vert a V \vert^{i k/a} \theta(-V) +
\beta_k ( a V )^{i k/a} \theta(+V)
\right)
\quad k>0
\label{Ta}
\end{equation}
The $U$ part is given by the reflection condition (\ref{mirthree}).
Analyticity in the lower half of the complex $V$ plane (i.e.
positive frequency $\om$ in eq.~(\ref{mireight})) 
fixes the ratio (see eqs~(\ref{threefive}) et seq.)
\begin{equation}
 { \beta_k \over \alpha_k} = e^{-\pi k/a}
\label{Taa}
\end{equation}
and the normalization of $\hat \varphi_k$ with the Klein Gordon scalar product fixes
$\alpha_k^2 -\beta_k^2 =1$. To obtain a complete orthogonal set of positive
frequency modes one must include the modes for $k<0$
\begin{equation}
\hat \varphi_k = {1 \over \sqrt{4 \pi \vert k\vert}} \left(
\beta_{\vert k\vert} \vert a V \vert^{i k/a} \theta(-V) +
\alpha_{\vert k\vert} ( a V )^{i k/a} \theta(+V)
\right)
\quad k<0
\label{Tb}
\end{equation}

\par From eqs~(\ref{Ta}) and (\ref{Tb}) it is seen that the set of
``Rindler'' type modes given by eq.~(\ref{mirten}) and the mode given by
\begin{equation}
\varphi_k^{out  R} = \theta(+V)
{1 \over \sqrt{4 \pi k}} ( a V )^{-i k/a} 
\label{Tc}
\end{equation}
constitute a complete orthonormal set. It is to be noted that the
modes~(\ref{Tc}) do not have a $U$ part since they don't reflect (when the mirror
follows eq.~(\ref{mirseven}) forever).

The quantum number $k$ in eqs~(\ref{Ta}) to (\ref{Tc}) is not the usual
Minkowski energy. Rather it is the eigenvalue of the boost operator $i a V
\partial_V$ whereas the  energy $\om$ 
is the eigenvalue of $i\partial_V$. (We shall
call it therefore ``Rindler energy'' see eq.~(\ref{hamr}).)
 However upon reflection the time dependent Doppler shift
(\ref{mirone}) gives $k$ the meaning of Minkowski energy for out modes. 
It is then the
eigenvalue of $i\partial_U$. 

Equations~(\ref{Ta},\ref{Tb}) therefore are to be read as the Bogoljubov
transformation ( written in terms of eqs~(\ref{mirten},\ref{Tc})):
\begin{equation} 
\left. \begin{array}{rclc}
\hat \varphi_k &=& \alpha_k \varphi_k^{out  L} + \beta_k
\varphi_k^{out R *}&\ \\
\hat \varphi_{-k} &=& \beta_{k} 
\varphi_{k}^{out L *} +
\alpha_{k} \varphi_{k}^{out R }&\
\end{array}\right\}k>0
\label{Td}
\end{equation}
or in terms of operators:
\begin{equation}
\left.
\begin{array}{rclc}
\hat a_k &=& \alpha_k a_k^{out L} - \beta_k 
a_k^{out R \dagger}&\\
\hat a_{-k} &=&- \beta_k a^{out L \dagger}_k +
\alpha_{k} a_k^{out R }&\ 
\end{array}\right\}k>0
\label{Tdoplaboum}
\end{equation}
The creation of particles of energy $k$ by the moving mirror gives
physical content to the modes~(\ref{Tc}) as the partners of
the created particles since the Minkowski 
vacuum on ${\cal I}^-$ ($\ket{0_{in}}$)
can be expressed (see eq.~(\ref{threethree})) as
\begin{equation}
\ket{0_{in}} = {1 \over \sqrt{Z}}
\prod_{k>0} \exp^{{\beta_k \over \alpha_k} 
a_k^{out L \dagger}a_k^{out  R  \dagger}} \ket{0_{out}}
\label{Te}
\end{equation}
where $\ket{0_{out}} =\ket{0_{out L}}\otimes\ket{0_{out R}}
$
with $a_k^{out  L}\ket{0_{out L}} =0$ and  $a_k^{out  R}\ket{0_{out R}} =0$. 
In this way one sees that to each produced
$\varphi_k^{out  L}$ particle corresponds a partner 
$\varphi_k^{out  R}$ living on the other side of the horizon
($V>0$) with the opposite Rindler energy. 
 Upon tracing over the states of different
$R$-quanta in eq.~(\ref{Te}) one obtains a
thermal density matrix,  with
temperature $T= a/ 2 \pi$, defined  on the subspace of  $L$-quanta states.
This is exactly what was seen upon computing the flux on ${\cal{I}}^+$ in
eqs~(\ref{TRBB} $\to$ \ref{TRRRR}).

As stated, to each particle created on ${\cal{I}}^+$ there corresponds  a 
bump of something on the other side ($V>0$) which is correlated to it. To
understand that 
this
correlated bump is really present, consider a charged
field and a measurement that reveals the production after reflection of a
quantum of positive charge. Then the correlated bump
necessarily has unit negative charge. 

To exhibit the correlations between the produced particle
and its partner configuration we consider 

a packet localized around the line
$U_0$, having mean energy $k_0$. This packet issues from
a vacuum fluctuation which is propagating in the $V$ 
direction and which (from the reflection condition) is
centered around $V = - a^{-1} e^{-a U_0}$. From 
the perfect symmetry ($V \to - V$) between $\varphi_k^{out  L}$ and $\varphi_k^{out  R}$, the ``partner fluctuation'' is centered
around $V= + a^{-1} e^{-a U_0}$. 
 This particular
configuration of the field gives one of the contributions
to $\langle 0_{in} \vert T_{\mu\nu}\vert 0_{in}\rangle$.
Our object therefore is to decompose this in-vacuum expectation
value into its component parts, these latter constituting a
complete set of post-selected photons, i.e. out states. Thus we introduce a
complete set of localized out-states (wave packets), considering the 
tensor products of all possible
states of arbitrary numbers of right and left quanta, we write (as in
eq.~(\ref{expectj}))
\begin{equation} 
\langle 0_{in} \vert T_{\mu\nu}\vert
0_{in}\rangle   = \sum_{ \begin{array}{c}
\{n_k^{out  L} \}\\
\{ n_{k^\p}^{out  R}\} 
\end{array}
}
\langle 0_{in}
\vert \{n_k^{out  L} \}\{ n_{k^\p}^{out  R}\} \rangle
\langle \{n_k^{out  L} \}\{ n_{k^\p}^{out  R}\}
 \vert T_{\mu\nu}\vert 0_{in}\rangle 
\label{Tf}
\end{equation}
Let us consider that piece of this expression wherein the post
selected states are of the form 
\begin{equation}
\left(\int_0^\infty\! dk \ c_k a^{out L
\dagger}_k\right) \ket{0_{out L}}\ket{\{ n_{k^\p}^{out  R}\}}
\label{Tg}
\end{equation}
i.e. where we have specified that the $out L$ state factor contains one
particle in the packet 
\begin{equation}
\psi = \int_0^\infty \! dk \ c_k^*
\varphi_k^{out  L}
\label{TgB}\end{equation}
 (with the normalization $\int dk \vert
c_k \vert^2 =1$) but the number or type of $R$-quanta in $ \ket{\{
n_{k^\p}^{out  R}\}}$ not prescribed. This
partial specification is appropriate in the present situation wherein only the $
out\ L$ quanta are realized on-shell \cite{MaPa}.
 However we shall soon prove that, due to
the correlations in the pure state $\ket{0_{in}}$ the 
configuration in $R$ is automatically
specified as well.

It is convenient to define the projection
operator  \begin{equation}
\Pi = {\rm I}_{out R} \otimes \int_0^\infty \! dk\
c_k a_k^{out L \dagger} \ket{0_{out L}}\bra{0_{out L}}
\int_0^\infty \! dk\
c_k^* a_k^{out  L}
\label{Th}
\end{equation}
which projects onto states of the form (\ref{Tg}) since ${\rm I}_{out R}$ is
the identity operator in the $V>0$ region.  One may then rewrite the
decomposition (\ref{Tf}) in such manner as to isolate the contribution from
$\Pi$: \begin{eqnarray}
\bra{0_{in}}  T_{\mu\nu} \ket{0_{in}} &=&   
\bra{0_{in}} \Pi T_{\mu\nu} \ket{0_{in}}
+ \bra{0_{in}} (I -\Pi) T_{\mu\nu} \ket{0_{in}} 
\label{Ti}
\end{eqnarray}
We rewrite the first term on the right hand side of eq.~(\ref{Ti})
as
\begin{eqnarray}
\bra{0_{in}} \Pi T_{\mu\nu} \ket{0_{in}} &=&   
\bra{0_{in}} \Pi  \ket{0_{in}}
\left[ { \bra{0_{in}} \Pi T_{\mu\nu} \ket{0_{in}} \over  
\bra{0_{in}} \Pi  \ket{0_{in}} }\right]
\label{Tj}
\end{eqnarray}
so as to express it as the product of the probability of
being in the state~(\ref{Tg}) times the weak value of $
T_{\mu\nu}$ in that state (see eq.~(\ref{pexpectj})).
In the sequel it is this non diagonal matrix element (weak value) 
\begin{equation}
\langle T_{\mu\nu} \rangle_w ={ \bra{0_{in}} \Pi T_{\mu\nu}
\ket{0_{in}} \over   \bra{0_{in}} \Pi  \ket{0_{in}} }
\label{Tk}
\end{equation}
which we shall analyze,
since from the above analysis this is the value of 
$T_{\mu\nu}$ which corresponds to the final state $\Pi \ket{0_{in}}$
which we now construct explicitly.

We first check that the specification of the $out\ L$ 
quantum uniquely fixes the configuration
in R to be its partner. To this end we 
change basis from the waves $\varphi_k^{out  L}$ to a
complete set of wave packets labeled by $i$. The matrix of this change of basis is the
unitary matrix $\gamma_{ik}$. We shall take the wave packet
$i=0$ to be that specified in eq.~(\ref{Tg}), i.e.
$\gamma_{0k}=c_k$. We now rewrite the argument of the
exponential in eq.~(\ref{Te}) so as to isolate the creation
operators of the wave packets labeled by $i$:
\begin{equation}
\int\! dk\ {\beta_k\over \alpha_k} a_k^{out R\dagger}a_k^{out
L\dagger} =\int\! dk\int\! dk^\p \sum_i
{\beta_k\over \alpha_k} a_k^{out R\dagger}
\gamma_{ik}^*\gamma_{ik^\p} a_{k^\p}^{out L\dagger}\label{labeli}
\end{equation}
Hence eq.~(\ref{Te}) becomes
\begin{eqnarray}
\ket{0_{in}}&=&
{1 \over \sqrt{Z}}
\exp \left(\int\! dk^\p c_{k^\p}a_{k^\p}^{out L\dagger}
\int\! dk\ c_k^*{\beta_k\over \alpha_k} a_k^{out R\dagger}
\right)\nonumber\\
&\ & \otimes\prod_{i\neq 0}
\exp \left(\int\! dk^\p \gamma_{ik^\p} a_{k^\p}^{out L\dagger}
\int\! dk\ \gamma_{ik}^*{\beta_k\over \alpha_k} a_k^{out
R\dagger} \right)\ket{0_{out}}\ .
\label{Tl}
\end{eqnarray}
Note that we have arranged this construction so as to put into evidence the
combination that creates the observed wave packet 
($\int\! dk^\p c_{k^\p}a_{k^\p}^{out L\dagger}$). This  construction
shows clearly
the asymmetry 
 between the particle and its partner wave functions induced by the presence
of ${\beta_k / \alpha_k}$.
This will be crucial in what follows.
 
Since by construction all the
states created by the operators $\int\! dk^\p \gamma_{ik^\p} a_{k^\p}^{out
L\dagger}$ ($i\neq 0$) are orthogonal to the states involving $\int\! dk^\p
c_{k^\p}a_{k^\p}^{out L\dagger}$, the state 
$\Pi \ket{0_{in}}$ is easily found to be 
\begin{equation}
\Pi \ket{0_{in}} =
{1 \over \sqrt{Z}}
\int\! dk^\p c_{k^\p}a_{k^\p}^{out L\dagger}\ket{0_{out R}}
\int\! dk\ c_k^*{\beta_k\over \alpha_k} a_k^{out R\dagger}
\ket{0_{out L}}\ .
\label{Tm}\end{equation}
In this way we see that in the projection $\Pi$ of eq.~(\ref{Th}) onto
$\ket{0_{in}}$ there is an implied specification
of the partner. This EPR \cite{epr} effect results from the global structure of
the Heisenberg state $\ket{0_{in}}$.

All is now ready for the evaluation of eq.~(\ref{Tk}). In order to simplify the
notation we shall calculate $\phi(x)\phi(x^\p)$ rather than $T_{\mu\nu}$. The
latter is obtained by taking derivatives with respect to $x,x^\p$ and then the
coincidence  limit. By expressing the $out$ operators which appear on the
right hand side of eq.~(\ref{Tm}) in terms of $in$ operators and writing
$\phi(x)\phi(x^\p)$ in terms of the $in$ basis a straightforward calculation
yields 
 \begin{eqnarray} \langle
\phi(x) \phi(x^\p) \rangle_w &=& {\bra{0_{in}} \Pi \phi(x) \phi(x^\p)
 \ket{0_{in}}\over   \bra{0_{in}} \Pi  \ket{0_{in}} }\nonumber\\
 &=& {
\bra{0_{out}} \phi(x) \phi(x^\p)
\ket{0_{in}}
\over
\bra{0_{out}} 0_{in}\rangle }
\nonumber\\
&\ &+ \ 
\Bigl[{
\left(\int_0^\infty \! dk\ 
(c^*_k/\alpha_k)  \hat\varphi^*_k(x)\right)\left(
\int_0^\infty \! dk^\p\ 
c_{k^\p} (\beta_{k^\p} / \alpha_{k^\p}^2 )
 \hat\varphi^*_{-k^\p}(x^\p)\right)
\over 
\int_0^\infty \! dk\ 
\vert c_k \vert^2 ( \beta_k /\alpha_k)^2
}\nonumber \\
&&\quad +\ (x)\leftrightarrow(x^\p)\Bigr]\qquad .
\label{Tn}
\end{eqnarray}
 
To derive eq.~(\ref{Tn}) we use eq.~(\ref{Tdoplaboum}) to obtain the sequence of
equalities: 
 \begin{eqnarray}
\bra{0_{out}} a^R_pa^L_q\phi \phi\ket{0_{in}} &=&
\bra{0_{out}}{1\over \alpha_p}\hat  a_{-p}\ {1\over \alpha_q} (\hat a_q+\beta 
\hat a^{R\dagger}_q)
\phi \phi\ket{0_{in}}\nonumber \\
&=& 
{\beta_q\over \alpha_q}\delta (p-q)
\bra{0_{out}}\phi \phi\ket{0_{in}} +
\bra{0_{out}}{1\over {\alpha_p \alpha_q}}\hat a_{-p}
\hat a_q \phi \phi\ket{0_{in}}
\nonumber \\
\label{333}
\end{eqnarray}
Expanding $\phi$ in $\hat \varphi_k$ and making the packet construction 
indicated in 
eqs~(\ref{Tl}) and (\ref{Tm}) yields the two terms in eq.~(\ref{Tn}) when the
denominator  $ \bra{0_{in}} \Pi  \ket{0_{in}}$ is taken into
account. 
This denominator gives the probability to find the state eq.~(\ref{Tm})
on ${\cal{I}}^+$. It is given by
\begin{eqnarray}
 \bra{0_{in}} \Pi  \ket{0_{in}}&=& 
\vert \bra{0_{out}} 0_{in}\rangle \vert^2
\int_0^{\infty}\!dk\ \vert c_k \vert^2 (\beta_k/\alpha_k)^2
\nonumber \\
&=&{1 \over Z}\int_0^{\infty}\!dk\ \vert c_k \vert^2 (\beta_k/\alpha_k)^2.
\end{eqnarray}
(as for the expression for $P_E$ given after eq.~(\ref{PEnew})).

The first term in eq.~(\ref{Tn})
is background. It is evaluated  by expressing $\phi(x)$ in terms of
$out $-modes and $\phi(x^\p)$ in terms of $in$-modes and one obtains
\begin{eqnarray}
{
\bra{0_{out}} \phi(x) \phi(x^\p)
\ket{0_{in}}
\over
\bra{0_{out}} 0_{in}\rangle }
&=& \bra{0_{in}} \phi(x) \phi(x^\p)
\ket{0_{in}}\nonumber\\
&\ &-  \int_0^\infty  dk {\beta_k \over \alpha_k}
[\hat \varphi^*_{k}(x)\hat\varphi^*_{-k}(x^\p)+(x
\leftrightarrow x^\p)]
\label{mir201} \end{eqnarray} 

The first term gives the mean value of $T_{\mu \nu}$ calculated in 
the previous subsection.

The
second term is equal to $-2\int_0^\infty \beta_k^2(|\varphi^L_k|^2+|\varphi^R_k|
^2) dk$. Taking derivatives with respect to $U$ and the coincidence limit,
gives $-\int_0^\infty k n(k)\!dk$ which is the negative Rindler energy density
in Rindler vacuum. Indeed, in Rindler vacuum (paragraph after 
eq.~(\ref{threetwo})),
using Rindler coordinates~(\ref{onesixb}), one has
\begin{eqnarray}
\bra{0_{Rindler}} T_{uu}  \ket{0_{Rindler}}=-(\pi/12) T^2
\label{trindl}
\end{eqnarray}
This can be interpreted as the removal of the thermal distribution
of Rindler quanta present in Minkowski vacuum. In the present case it
corresponds to the removal of the energy of all the produced quanta. 

Putting the two contributions together,
we thus have
\begin{eqnarray} {
\bra{0_{out}} T_{UU}\ket{0_{in}}
\over
\bra{0_{out}} 0_{in}\rangle }=
{
\bra{0_{out}} \partial_{U}\phi \partial_{U}\phi
\ket{0_{in}}
\over
\bra{0_{out}} 0_{in}\rangle }
&=& 0\label{iii}\\
{
\bra{0_{out}} T_{VV}
\ket{0_{in}}
\over
\bra{0_{out}} 0_{in}\rangle }=
{
\bra{0_{out}} \partial_{V}\phi \partial_{V}\phi
\ket{0_{in}}
\over
\bra{0_{out}} 0_{in}\rangle }
&=& -{\pi \over 12} {1 \over (2 \pi a)^2} {1 \over ( a
V )^2}
\label{iiii}\end{eqnarray}
since $dU/dV= 1/aV$ see eq.~(\ref{mirseven}).
The absence of outgoing flux in the in-out matrix element~(\ref{iii})
is natural since it corresponds to a state with no out-particle produced.
This specification implies that before reflection the vacuum fluctuations
leading to the production of out-particles be absent, hence to
the negative Rindler energy~(\ref{iiii}).
We note that we have neglected to treat properly the singularity
at $V=0$. This shall be  analyzed  subsequently.

We now consider the second term of eq.~(\ref{Tn}) (hereafter denoted 
$\langle T_{\mu\nu}\rangle_{\psi}$)
which contains
the contribution of the selected particle $\psi$ eq.~(\ref{TgB}). 
Its contribution to $\langle T_{UU}\rangle_w$ is
\begin{equation}
\langle T_{UU}\rangle_{\psi} =2{
 \left(\int_0^\infty \! dk\ 
c_k (\beta_k^2 /\alpha_k^2) \partial_U \varphi^{out  L}_k\right)\left(
\int_0^\infty \! dk^\p\ 
c_{k^\p}^* 
\partial_U \varphi^{out L *}_{k^\p}\right)
\over 
\int_0^\infty \! dk\ 
\vert c_k \vert^2 ( \beta_k^2/ \alpha_k^2)
}
\label{To}
\end{equation}
To calculate the energy of the particle we shall take a
gaussian packet $c_k = e^{-ik U_0} e^{-\Delta^2
(k-k_0)^2/2}$ where the phase factor $e^{-ik U_0}$ locates the
produced particle around $U=U_0$, its energy being
approximatively $k_0$. Then one verifies by saddle point
integration that $\langle T_{UU}\rangle_{\psi}$ is also
located around $U=U_0$
and carries also the energy 
\begin{equation}
\int_{-\infty}^{+\infty}\!dU
\langle T_{UU}\rangle_{\psi} = {{\left(\int_0^\infty\!dk\ k \vert c_k
\vert^2 ( \beta_k^2/ \alpha_k^2) \right)}\over{\left(
\int_0^\infty\!dk\ \vert c_k
\vert^2 ( \beta_k^2/ \alpha_k^2) \right)}}\simeq k_0
\label{ToB}\end{equation}
As in the electric field, once the selected particle is on mass shell
the $\psi$ part of the weak value behaves classically.

We now consider the contribution of $\psi$ to $\langle T_{VV}\rangle_{w} $.
Before reflection ($U<U_0$),
there is a piece both for $V<0$ (which upon reflection becomes 
$\langle T_{UU}\rangle_{\psi}$) and for $V>0$ (the partner contribution):
\begin{eqnarray}
\langle T_{VV}\rangle_{\psi} &=&2\theta(-V){
\left(\int_0^\infty \! dk\ 
c_k ( \beta_k^2/ \alpha_k^2) \partial_V \varphi^{out  L}_k\right)\left(
\int_0^\infty \! dk^\p\ 
c_{k^\p}^* 
\partial_V \varphi^{out L *}_{k^\p}\right)
\over 
\int_0^\infty \! dk\ 
\vert c_k \vert^2( \beta_k^2/ \alpha_k^2)
} \ +\nonumber\\
&\ & 2 \theta(+V){\left(
\int_0^\infty \! dk\ 
c_k( \beta_k/ \alpha_k)\partial_V \varphi^{out  R}_k\right)\left(
\int_0^\infty \! dk^\p\ 
c_{k^\p}^* ( \beta_{k^\p}/ \alpha_{k^\p})
\partial_V \varphi^{out R *}_{k^\p}\right)
\over 
\int_0^\infty \! dk\ 
\vert c_k \vert^2 ( \beta_k^2/ \alpha_k^2)
}\nonumber\\
&&\label{Tp}
\end{eqnarray}
After reflection, for $U>U_0$, only the $\theta(+V)$ piece remains
since the $\theta(-V)$ is reflected and gives $\langle T_{UU}\rangle_{\psi}$.
To grasp the energy content of the particle and partner let
us first check that each ``Rindler'' piece (proportional to
$\theta(+V)$ and $\theta(-V)$ respectively) carries the same
``Rindler'' energy (i.e. $i aV\partial_V$) 
approximatively equal to $k_0$.
To this
end we introduce $v=a^{-1}\ln \vert a V \vert$ and 
analyse the ``Rindler'' flux $T_{vv}$.  
One has: $\int_{-\infty}^{+\infty}\! dv \  \langle T_{vv} \rangle =
\int_0^{\pm \infty}\! dV \ aV
\langle T_{VV} \rangle$ with the $\pm$ corresponding to the
$\theta(\pm V)$ pieces respectively. In both cases one finds
\begin{equation}
\int_{-\infty}^{+\infty}\! dv \  \langle T_{vv} \rangle =
\left(\int_0^\infty\!dk\ k \vert c_k
\vert^2 ( \beta_k^2/ \alpha_k^2)\right)/\left(
\int_0^\infty\!dk\ \vert c_k
\vert^2 ( \beta_k^2/ \alpha_k^2) \right)\simeq k_0
\label{Tq}
\end{equation}
For the  $\theta(-V)$ piece this simply follows from
eq.~(\ref{ToB}) since $v=U$ upon reflection.
For the  $\theta(+V)$ piece it follows from the orthogonality
rules enjoyed by the $\varphi^{out R}$ modes.

Thus $\langle T_{vv}\rangle_w$ is 
simply obtained by adding the ``Rindler energy''
density of the $\psi$-Rindler 
pair to the background, i.e. the first term of eq.~(\ref{Tn})
 which we have explained gives the
Rindler vacuum see eq.~(\ref{iiii}).
 
However $\langle T_{VV}\rangle_\psi$ is quite asymmetric and its
analysis is more subtle. This asymmetry is already present in 
eq.~(\ref{Tm}) where the partner wave function is given explicitly.
 The source of
the asymmetry in the formalism comes from the use of packets
which is necessary to exhibit correlations in space-time. Indeed the
detection of a particle by a counter requires a description in terms of
a broad packet for this outgoing particle. Once this is done the
partner's packet becomes complicated: the convolution of the particle
Fourrier components with the energy dependence of the ratio
$\beta / \alpha$, see eq.~(\ref{Tm}). One loses therefore the
simplicity of the Bogoljubov transformation between Unruh and
Rindler modes (eq.~(\ref{Td})). Hence  the  two terms on the right hand
side of eq.~(\ref{Tp}) are not symmetric with respect to $V=0$.

Continuing with the configuration before reflection
we emphasize that 
\begin{eqnarray}
\int_{-\infty}^{+\infty}
\!dV\ \langle T_{VV} \rangle_{\psi} &=&0\label{vanish}\\
\int_{-\infty}^{+\infty} \!dV\ { \bra{0_{out}} T_{VV}\ket{0_{in}}
\over
\bra{0_{out}} 0_{in}\rangle } &=& 0 \label{vanishB}
\end{eqnarray}
  for $U\leq U_0$. To see this, recall that 
$\int_{-\infty}^{+\infty}
\!dV\  T_{VV} \ket{0_{in}}=0$ on $\cal{I}^-$ in virtue of normal ordering (i.e. 
subtraction of the zero point energy of each mode, here the energy carried by
each orthogonal packet). Thus $\int_{-\infty}^{+\infty}\!dV\  T_{VV} \ket{0_{in}}$
vanishes mode by mode on $\cal{I}^-$.
 
Equation~(\ref{vanishB}) implies that the energy distribution in 
eq.~(\ref{iiii}) has a positive singular contribution at $V=0$ which exactly
compensates the negative energy density for $V\neq 0$
\cite{Pare}, \cite{MaPa}. This singularity will come up once again in Sections
\ref{state} and \ref{VFHR} where it will play a critical role in ensuring the
consistency of the theory.

Let us now decompose $\int_{-\infty}^{+\infty} \!dV\ \langle
T_{VV} \rangle_{\psi}$ into the contribution from the particle $\int_{-\infty}^0
\!dV\  \langle T_{VV} \rangle_{
\psi}$ and from the partner  $\int_0^{+\infty}
\langle T_{VV} \rangle_{\psi}$. 

Starting with the latter one
sees that the integral is positive since the integrand is
(as is seen by inspection of the second term in the right hand side of
eq.~(\ref{Tp})). One estimates  $\int_{0}^{+\infty}\!dV\ 
\langle T_{VV} \rangle_{\psi}
 \simeq k_0 e^{a U_0}$ (since we know that
$\int_0^{+\infty} \!dV\ a V \langle T_{VV} \rangle_{\psi}\simeq k_0$
and it may be checked by stationary phase that the main contribution comes from $V_0\simeq
a^{-1} e^{-a U_0}$). This relation between ``Rindler'' and Minkowski
energy is exactly that given by the Doppler shift equation~(\ref{mirsix}). 

These properties imply that $\int_{-\infty}^0\!dV\ 
\langle T_{VV} \rangle_{\psi}$ is negative and $\simeq - k_0
e^{a U_0}$. But we have seen that the ``Rindler'' energy 
$\int_{-\infty}^0\!dV\  \vert a V \vert\langle T_{VV} \rangle_{\psi}\simeq k_0$
 is positive. How is that possible? The answer is that 
$\theta(-V)
\langle T_{VV} \rangle_{\psi}$ is not positive definite and
contains strong negative oscillations for small values of $V$. When
evaluating 
$\int_{-\infty}^0\!dV\ 
(- a V ) \langle T_{VV} \rangle_{\psi}$ (or upon
reflection, $\int_{-\infty}^{+\infty} dU \langle
T_{UU} \rangle_{\psi}$) these oscillations  give   negligible contributions. But
in the Minkowski energy the oscillations near $V=0$ play a dominant
r\^ ole ( because of the weight factor $|V|^{-1}$ in going from Rindler to 
Minkowski energy ) in
such fashion as to make the Minkowski energy of particle plus partner vanish.
We shall discuss similar oscillatory effects in quantitative detail in the
next section.

In conclusion the  post selection of an outgoing photon of
frequency $k_0$ entails a distribution of $T_{VV}$ which
exists on both sides of the horizon $V=0$. The partner
piece, localized around the line $V_0=+ a^{-1} e^{-a U_0}$
propagates forever and has Minkowski energy equal to the
Doppler shifted value given by $k_0e^{a U_0}$. The particle
piece propagates out to $U=U_0$ and is then reflected.
Before reflection it carries energy equal and opposite to
that of the partner. This energy has a positive piece equal
to $k_0e^{aU_0}$ centered around $V=-a^{-1}e^{-a U_0}$ and a
broader oscillating piece which is net negative in such
fashion that the sum of all contributions vanishes. After
reflection this piece has gotten converted into $T_{UU}$. 
The energy carried in
the reflected wave is almost all in the center positive
piece (see eq.~(\ref{To})). The oscillating negative piece is diffuse and
negligible. For $U>U_0$ the partner continues alone
carrying a net positive energy always equal to
$k_0e^{aU_0}$. However it is not a quantum in the usual
sense. It does not manifest itself on the average since the field 
configurations, on the average, are still Minkowski (by causality).
 To pick up the effect of the
partner requires an EPR correlation type experiment.
 What is important is that
the fluctuations in play when an outgoing particle is
produced have energies which blow up exponentially and which
hugs the horizon at exponentially small distances. It is this
circumstance which constitutes a major hiatus in the more
realistic case of black hole collapse.

Had we post selected the absence of an outgoing photon we would come upon 
an ``anti-partner'' whose energy is negative being the weighted sum of the
energies of all the negative energy corresponding to the absence of of
$1,2,\dots,n,\dots$  photons. As previously calculated (eq.~(\ref{iiii}) this is
precisely the energy of the Rindler vacuum (i.e. the absence of the average
thermal energy ). The sum of all weak values is of 
course the net average as given in  Section (\ref{Mean}).

Perhaps a more physical way to exhibit the correlations between the
emitted $U$ photons and the $V$ partners is to decelerate the
mirror after a while. As seen in Fig~\ref{ACCELiii},
\dessin{1.000}{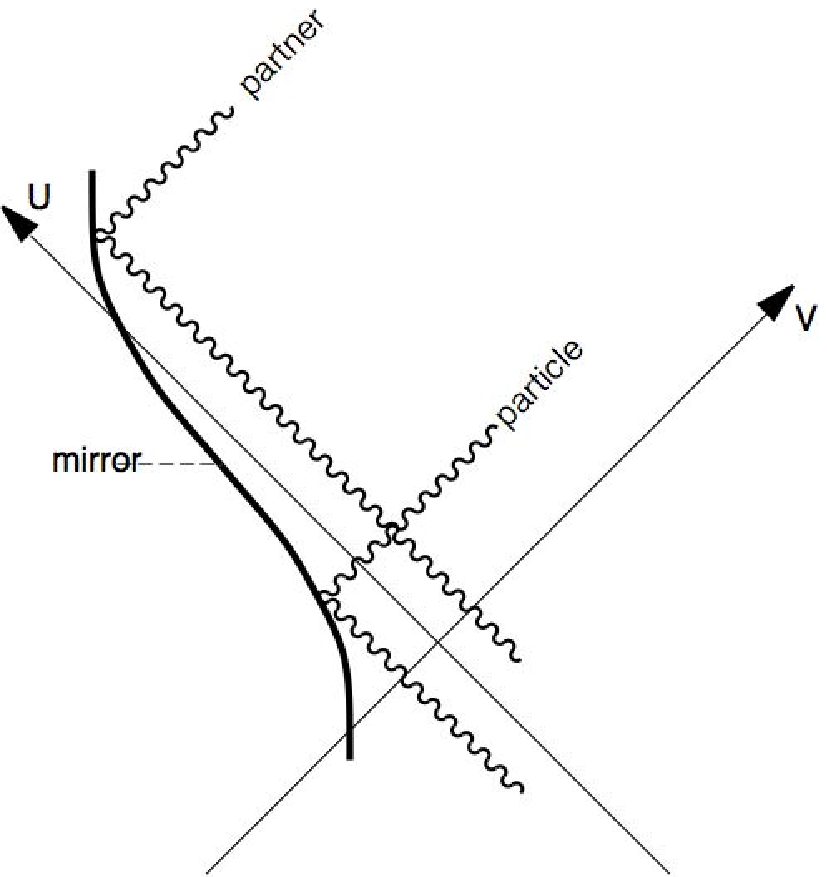}{ACCELiii}{
The trajectory of a mirror which decelerates after a
while. The classical trajectories (stationnary phase) of
a pair of particles are indicated by wavy lines. It is
apparent that the deceleration of the mirror allows the
partners to be realized on shell.}
$\varphi_{k}^{out  R}$ will then be reflected at late
times, transformed into a real quantum and can therefore 
be observed, in particular in coincidence
with $\varphi_{k}^{out  L}$. 
One could choose a trajectory which starts
from rest $V-U=const$ and come back to rest  after having followed 
the trajectory eq.~(\ref{mirseven}) for a while. 
The main
point is that after the mirror becomes inertial again the system goes back to
vacuum plus the radiation that goes out to infinity, always a pure state.
This was  pointed out by Carlitz and Willey in 
\cite{Carlitz} 
who suggested possible applicability to
the black hole problem (this was also mentioned in \cite{PaBr} and discussed in
\cite{Wil}).


\section{The  
Energy Emitted by the Accelerated Detector}\label{state}

\subsection{Introduction and Qualitative Description}

Though not directly related to  black hole physics we include this section in
our review, not only because it is of interest in itself, given the rather
stormy history of the problem, but also because one comes upon concepts that
arise in the black hole problem as well.

We shall present in detail the energy emitted by the
accelerated detector. The first subsection is qualitative and shows how the
paradoxes which have been raised (see for instance and in
historical order
\cite{UnWa}, \cite{Grow},
\cite{RSG}, \cite{Unru2}, \cite{MPB}, \cite{AuMu}, \cite{MaPa}) can be
rationalized. In the second subsection the energy emitted is computed in
perturbation theory and particular attention is paid to how it
is correlated to the final state of the atom. In the third
subsection the same decomposition of final states is used to
display the vacuum fluctuations which induce the transitions
giving rise to these final states.

It has been seen in Section \ref{spontaneous} that in perturbation theory the leading
order in $H_{int}$ corresponds to photon {\it emission} in both cases,
excitation and deexcitation, as it should be since one perturbs Minkowski
vacuum. Nevertheless as Grove pointed out \cite{Grow}, when equilibrium is
reached there is no net change of the state of the radiation in the quadrant,
$R$, of the accelerator except for transient effects. We refer to this as
Grove's theorem. His argument is the following. The accelerator feels the effect
of a thermal bath. 
So first consider the inertial two level system in thermal equilibrium. The
principal ingredients which guarantee the absence of net energy flux to or
from the atom is the time independence of the Hamiltonian and the
stationarity of the 
state of the atom (in the thermodynamic sense), so that each
photon which is absorbed  is re-emitted  with the
same energy, i.e. energy conservation results from time translational
symmetry. Equilibrium is maintained through the implementation of the
Einstein conditions.

This argument is immediately applicable to the accelerator since the
fact that $a={\rm constant}$ implies that his physics is translationally
symmetric in his proper time (ie. invariance under boosts). Since Minkowski vacuum is
also an eigenstate of the boost operator, the implication is that the total eigenvalue
of $\partial /\partial \tau$ is conserved. The dynamical realization is the time
averaged conservation of the energy of the totality of Rindler quanta which are
absorbed and emitted, in strict analogy to thermal equilibrium.
This implies no net energy flux.

The above considerations result in a dilemma since, as we said,
both excitation and deexcitation of the atom leads to emission
of a Minkowski photon (see for instance the
amplitude eq. (\ref{threefourb}) wherein the first order interaction of the atom
with the field always results in the creation of a photon of frequency
$\omega$).
The resolution of the dilemma will be shown to lie in a
global treatment of the radiation field which also takes into account the
transients due to switching on and off the detector. These transients play a r\^ole
similar to the oscillations  encountered near $V=0$ in equation (\ref{vanish})
and (\ref{vanishB}) which ensure the global vanishing of the energy.

The above general considerations have been verified in an exact model, that
of an accelerating harmonic oscillator coupled linearly to the radiation
field \cite{RSG}, \cite{Unru2},  \cite{MPB}. It is noteworthy that in this work
Heisenberg equations of motion have been integrated to give a long time steady
state solution wherein initial conditions become irrelevant. This is what makes
the analogy to thermal equilibrium possible.
Rather than describe the exact oscillator system we
shall continue with the two level atom in perturbation theory (since this
is more relevant for the understanding of some corresponding problems which
come up in the black hole problem).

A detailed picture of the steady state emerges from the following
consideration. Focus on the ground state of the accelerator which excites
by absorbing a Rindleron coming in from its left. Then the field
configurations  to its right is depleted of this Rindleron. Since this
Rindleron carried positive energy, its removal  can be described as the
emission of  negative energy to the right. In equilibrium there is also to be
considered the process of deexcitation corresponding to the emission of
positive energy   to the right. The Einstein relation eq. (\ref{Rfive})
guarantees that the two cancel. 

A nice way to express the physics is to appeal to exact
eigenstates of the photon field, scattering states. Their energy (Minkowski
energy for the inertial atom, Rindler energy for the accelerating one) is
the same as that of the free states and their number is also the same since
the scattering matrix is unitary.  Therefore the average energy of the
radiation field of any ( mixed) state is unperturbed by the scatterer.

We now discuss qualitatively the transient behavior in both the Rindler
and Minkowski representation of the radiation field by calculating the
energy density emitted in terms of the mean energy momentum tensor, $\langle
T_{\mu \nu} \rangle$. Take, for example, right movers. The relevant energy
density is  $\langle T_{UU} \rangle = \langle (\partial \phi / \partial U)^2
\rangle $.

By Grove's theorem, one should have $\langle T_{uu} \rangle$, the energy
density measured by a co--accelerator, equal to zero in the absence of
transients. (Once more we normalize the energy so that the expectation values of
$T_{UU}$ and $T_{VV}$ vanish in Minkowski vacuum, therefore of 
$T_{uu}$ and $T_{vv}$ as well). Now consider the transients, with switch
on (off) time at $\tau_i$ ($\tau_f$) modeled by some function $f(\tau)$
which vanishes outside the interval $(\tau_i , \tau_f )$ and is equal to $1$
inside the interval except for a time $\Delta \tau$ during which $f$ passes
smoothly from $0$ to $1$. We choose $\tau_f - \tau_i >> {\rm Max} (\Delta M
^{-1}, a^{-1})$ which is the time necessary to establish the Golden rule.
In addition we take $\tau_f - \tau_i >> \Delta \tau$. The compensation
mechanism based, as it is, on translational invariance in time is then no
longer operative. Insofar as $T_{uu}$ is concerned this will introduce minor
effects, but these become dramatic for $T_{UU}$ owing to the exponential
character of the Doppler shift measured by the inertial observer. Indeed the
total Minkowski energy emitted is \begin{equation} E_M = \int_{ - \infty}^{0}
dU <T_{UU}> =  \int_{ - \infty}^{+ \infty} du <T_{uu}> {du \over dU}
\label{fournul} \end{equation}
where $du / dU = e^{au}$. The first integral in eq. (\ref{fournul}) is limited
to the domain $U<0$ because the accelerator lives in $R$ thereby confining
right movers (i.e. those which contribute to $T_{UU}$ in the integrand) to
the quadrants $P$ and $R$. To pick up the total energy emitted, one
must integrate along the surface $V=V_0$ with $V_0 > a^{-1}e^{a\tau_f }$ so 
that all emitted right movers cross 
the surface (see Fig.~(\ref{excit})). From what has been said 
the integrand vanishes except
at the endpoints $u_i$ ($= \tau_i $) and $u_f$ ($= \tau_f $) whereupon eq
\ref{fournul} integrates to a form \begin{equation} E_M = C_f e^{a \tau_f} -
C_i e^{a \tau_i} \label{fourone}
\end{equation}
where $C_i$ and $C_f$ depend on the exact form of the switching function as
it turns on and off the interaction.

 Most surprisingly eq. (\ref{fourone}) can be approximately written as an
integral over the naive rates of absorption and of emission of Rindler
photons taking into account that each transition, be it excitation or
deexcitation, is accompanied by the emission of a Doppler shifted Minkowski
photon, according to the resonance condition $\om(\tau) = \Delta M e^{a\tau}$
(eq; ( \ref{threefiveb}) )
i.e.
\begin{eqnarray} E_M &=& \int_{\tau_i}^{\tau_f}\! d\tau \ f(\tau)( R_+ p_- +
R_- p_+ ) \Delta M e^{a\tau }\nonumber\\
&=&( R_+ p_- +
R_- p_+ ) \Delta M ( e^{a\tau_f } - e^{a \tau_i}) +( C^\prime_f e^{a \tau_f} 
-C^\prime_i e^{a \tau_i})
\label{fourtwo}
\end{eqnarray}
Here $R_\pm$ are rates of excitation (deexcitation) and $p_\pm$ are
probabilities of being in excited (ground) states (Einstein's condition
eq. ( \ref{Rfive}) is $R_+ p_- = R_- p_+ =$constant). The constants
$C^\prime_i$ and $C^\prime_f$
 depend on the form of the function $f$ near the endpoints. So the
integral eq. (\ref{fourtwo}) is of the same form as eq. (\ref{fourone}). In the
subsequent development we shall see how it is that the energy density
$<T_{UU}>$ which appears in eq. (\ref{fournul}) can be decomposed into a positive
steady piece and an interference term. The positive piece has an integral of the
form eq. (\ref{fourtwo}) and should be interpreted in a similar way. The
interference term carries no net energy, however it plays an essential role in
ensuring that causality be respected. In the interval $(\tau_i,\tau_f)$ far from
the transients the interference term is negative and exactly compensates the
positive piece  so as
to recover Grove's theorem in the region where it should hold, this being
characterized in good approximation by translational symmetry. This
detailed method proceeds event by event so as to provide a description of
the radiation field in all four quadrants and its correlations to the
detector. 

[This state of affairs wherein transients have a global content which
depends on the whole history also occurs in the problem of the classical
electromagnetic field emitted by a uniformly accelerated charge in 3
dimensional Minkowski space. Here also one can argue, in a way strongly
reminiscent of Grove's result, that no radiation should be emitted. To wit,
the equivalence principle asserts that a uniformly accelerated charge is
equivalent to a static charge in a uniform gravitational field. In a
static situation no energy should be emitted (this is confirmed by everyday
experience at the earth's surface) hence none should be emitted in the
accelerated case. On the  other hand the charge should radiate at a constant
rate given by Larmor's formula \begin{equation} {dE \over d\tau} = {2 \over 3}
\left( e^2 \over 4 \pi \right) \dot u^2 u^0 \label{Bone}
\end{equation}
where ${dE \over d\tau}$ is the energy radiated per unit proper time of the
particle, $\dot u^2$ is the square of the acceleration of the particle. 

This dilemma was resolved by Boulware \cite{Boul2} who showed that both
results are compatible. The crucial remark is that the equivalence principle
is valid only "inside" the right quadrant (where the static Rindler
coordinates  eq. (\ref{coor}) are valid) and indeed inside R the field is the
Coulomb field of the charge with no radiation part. However Maxwell's equations
imply that along the past horizon $U=0$ there is a delta like singularity in the
field. This singularity can be interpreted as the infinitely blue shifted
transient which occurred when the charge was set into acceleration. This
singularity along the past horizon is unobservable by a coaccelerator, but 
in the future quadrant it propagates and gives rise to a
 flux of radiation at exactly the rate  eq. (\ref{Bone}). Hence once more
transients in the accelerated frame acquire a global content for the
Minkowski observer.]

\subsection{The energy emitted at $O(g^2)$}

We first consider the 
energy emitted during spontaneous excitation of the atom.
The coupled state of atom and field is
\begin{eqnarray}
\ket{\psi_-} &=& { \rm T} e^{-i\int H_{\rm int} d\tau^\prime} \ket{0_M}\ket{-}
\nonumber\\
&=& \ket{0_M}\ket{-} -ig \int_{\tau_i}^{\tau_f} \! d\tau^\prime
e^{i \Delta M \tau^\prime} \phi(\tau^\prime) \ket{0_M}\ket{+} \nonumber\\
&\ &\ \  -g^2 \int_{\tau_i}^{\tau_f} \! d\tau_2 \int_{\tau_i}^{\tau_2} \!
d\tau_1 e^{-i \Delta M \tau_2} \phi(\tau_2)
e^{i \Delta M \tau_1} \phi(\tau_1) \ket{0_M}\ket{-} + O(g^3)
\nonumber\\
\label{fourthree}
\end{eqnarray}
where ${\rm T}$ is the time ordering operator. $\ket{+}$ ($\ket{-}$)
 refer to excited
(ground) state of the atom respectively. The mean energy momentum is
(considering once more only right movers and therefore $T_{UU}$)
\begin{equation}
\langle T_{UU} \rangle _- = \elematrice{\psi_-}{ T_{UU} }{\psi_-}
\label{fourfour}\end{equation}
The mean $T_{UU}$ eq. (\ref{fourfour}) is dissected into its constituent parts by
considering the final state of the atom  \cite{UnWa},  \cite{Grow},
 \cite{MaPa} (i.e. by making a post selection similar to eq. (\ref{Ti}) where
the final state of the radiation field was specified).
This is
realized by inserting into  eq. (\ref{fourfour}) the projectors $\Pi_+$ ($\Pi_-
=1-\Pi_+$) onto the
excited (ground) state of the atom:
\begin{eqnarray}
\langle T_{UU} \rangle _- =
\elematrice{\psi_-} {\Pi_+ T_{UU} }{\psi_-}  
+ \elematrice{\psi_-} { \Pi_- T_{UU}
}{\psi_-}
\label{fourfive}
\end{eqnarray}
We have
\begin{equation} 
\elematrice{\psi_-} {\Pi_+ T_{UU} }{\psi_-}
=
g^2 \elematrice {0_M} { \int_{\tau_i}^{\tau_f} \! d\tau_1
e^{ -i \Delta M \tau_1} \phi(\tau_1) T_{UU}
\int_{\tau_i}^{\tau_f} \! d\tau_2
e^{i \Delta M \tau_2} \phi(\tau_2) }{0_M}
\label{foursix}
\end{equation}
Equation (\ref{foursix}) is the energy emitted due to excitation of the atom
as it would be calculated in lowest order perturbation theory (
$O(g)$ in the wave function, $O(g^2)$ in the energy). But, to $O(g^2)$,
one requires the correction to the wave function as well, thereby leading
to interference terms. These are in the $\Pi_-$ term of eq.
\ref{fourfive} corresponding to emission and reabsorption in the wave
function. This term is 
\begin{equation} 
\elematrice{\psi_-} {\Pi_- T_{UU} }{\psi_-}
=
-2 g^2 {\rm Re} \elematrice {0_M} { \int_{\tau_i}^{\tau_f} \! d\tau_2
e^{ -i \Delta M \tau_2} \phi(\tau_2) 
\int_{\tau_2}^{\tau_f} \! d\tau_1
e^{i \Delta M \tau_1} \phi(\tau_1)  T_{UU}}{0_M}
\label{fourseven}
\end{equation}

It is convenient to reexpress  eq. (\ref{fourseven}) as
 \begin{eqnarray}
\elematrice{\psi_-} {\Pi_- T_{UU} }{\psi_-}
&=& 
- g^2 {\rm Re} [ D(U) + {\cal F}(U) ]
\label{fff}\end{eqnarray}
where
\begin{eqnarray}
 D(U)&=&
 \elematrice {0_M} { \int_{\tau_i}^{\tau_f} \! d\tau_2
e^{ -i \Delta M \tau_2} \phi(\tau_2) 
\int_{\tau_i}^{\tau_f} \! d\tau_1
e^{i \Delta M \tau_1} \phi(\tau_1)  T_{UU}}{0_M} \nonumber\\
{\cal F}(U)&=& 
\elematrice {0_M} { \left[ \int_{\tau_i}^{\tau_f} \! d\tau_2 
\int_{\tau_i}^{\tau_f} \! d\tau_1 \epsilon(\tau_2 - \tau_1)
e^{ -i \Delta M \tau_2} \phi(\tau_2) 
e^{i \Delta M \tau_1} \phi(\tau_1)  , T_{UU}\right]}{0_M}\nonumber\\
\label{foursevena}
\end{eqnarray}
Complications from time ordering are no longer present in $D(U)$. The term
${\cal F}(U)$ plays no important role in the physics. Indeed a detailed
analysis shows that it enjoys the following properties: 

1) It is smaller
than $D(U)$ by a factor $1/\Delta M 
(\tau_f - \tau_i)$ except in the transitory regime 
where it is comparable to $D$. 

2) It vanishes in the non
causal domain $U>0$. 

3) The integral $\int dU {\cal F}(U)$ vanishes (ie.
it does not contribute to $E_M = \int dU T_{UU}$). 

4) The integral $\int du
(dU / du)^2 {\cal F}(U(u))$ vanishes also
(ie. It does not contribute to the
Rindler energy $E_R = \int du T_{uu}$). 

All these properties are also valid
when one introduces a switch on and off 
function $f(\tau)$ as in equation  eq. (\ref{fiveseven})
below. Hence from now on we shall drop the term ${\cal F}(U)$.

To execute the calculation, and so check out 
Grove's theorem at this order, we need to evaluate 
$\elematrice{\psi_-} {\Pi_+ T_{UU} }{\psi_-}$  and $D(U)$ in the limit
$\tau_f - \tau_i \to \infty$. Begin with
 eq. (\ref{foursix}) and expand $\phi(\tau_2)$ and $\phi(\tau_1)$ in Unruh modes
(eq. (\ref{threetwo})). These are the most convenient because their vacuum
is Minkowski. Then
$\phi(\tau_2)$ (and $\phi(\tau_1)$) contain the creation of an Unruh mode in
$R$. Since $\Delta M >0$, the part of $\phi(\tau_2)$ which contributes to the
integral (i.e. resonates with the factor $e^{i \Delta M \tau_2}$) is that which
corresponds to the annihilation of a Rindler mode in $R$, hence carrying a
factor $\beta_\la$ in $\hat \varphi_\la$ (eq. (\ref{threetwo})). The result is
thus (compare with eq. (\ref{extrab}))
\begin{equation}
\int d\tau e^{i\Delta M \tau } \phi (\tau)\ket{0_M} = \beta_{\Delta M}
 \sqrt{\pi/\Delta
M} \hat a^\dagger_{\Delta M} \ket{0_M}. 
\end{equation}
This Unruh creation operator must
be contracted with the corresponding Unruh annihilation operator appearing in
$T_{UU}$ to give
\begin{eqnarray} \elematrice{\psi_-}
{\Pi_+ T_{UU} }{\psi_-} &=& {\pi \over \Delta M} 2 g^2\beta_{\Delta_M}^2
\partial_{U} \hat \varphi_{-\Delta M}(U)
\partial_{U} \hat \varphi_{-\Delta M}^*(U)
\nonumber\\
&=& {g^2   \over 2} {1 \over (aU)^2}\left(\theta(-U) 
\beta_{\Delta_M}^4 +\theta(U)  \beta_{\Delta M}^2\alpha_{\Delta M}^2 
\right)
\label{fourninea} \end{eqnarray}

The calculation of  eq. (\ref{fourseven}) proceeds along similar lines and yields
\begin{eqnarray}
\elematrice{\psi_-} {\Pi_- T_{UU} }{\psi_-}&=& -
{\pi \over \Delta M} 2 g^2\beta_{\Delta_M}^2 {\rm Re} \left[
\partial_{U} \hat \varphi_{\Delta M}^*(U)
\partial_{U} \hat \varphi_{-\Delta M}^*(U)\right]
\nonumber\\
&=& 
-{g^2   \over 2} {1 \over (aU)^2}
 \beta_{\Delta M}^2\alpha_{\Delta M}^2 
\label{fournineb}
\end{eqnarray}

The important minus sign is the same as comes up in the verification of
unitarity wherein to $O(g^2)$ the interference term  cancels
the direct term in 
\begin{eqnarray}
1&=& \braket{\psi}{\psi}\ =\ \braket{\psi_0 + i\delta \psi}
{\psi_0 -i \delta \psi}
\nonumber\\
&=&
\braket{\psi_0 }
{\psi_0 } + \braket{ \delta \psi}
{\delta \psi} +  2 {\rm Im} \braket{\psi_0 }
{ \delta \psi}\ =\ \braket{\psi_0 }
{\psi_0}
\label{unit}
\end{eqnarray}
This minus sign is in fact essential to satisfy causality which requires
that all expectation values of $T_{UU}$ vanish for $U>0$. The photons
emitted cannot affect the region $U>0$. Indeed there is a rigorous
theorem stating that $\elematrice{\psi_-}{T_{UU}}{\psi_-} = 0$ for $U>0$
which we now prove \cite{UnWa}:
\begin{eqnarray}
\langle T_{UU}(U>0) \rangle _-
&=& \bra{-}\elematrice{0_M}{Te^{i \int d\tau H_{int}} T_{UU}(U>0)
Te^{-i \int d\tau H_{int}} }{0_M}\ket{-}\nonumber\\
&=& \bra{-}\elematrice{0_M}{ T_{UU}(U>0)T e^{i \int d\tau H_{int}}
Te^{-i \int d\tau H_{int}} }{0_M}\ket{-}\nonumber\\
&=& \bra{-}\elematrice{0_M}{ T_{UU}(U>0) }{0_M}\ket{-} \nonumber\\
&=& 0
\label{fivesix}
\end{eqnarray}
where we have used the commutativity of $\phi(\tau)$ appearing in $H_{int}$
(i.e. on the accelerating trajectory) with $\phi(U)$ for $U>0$ appearing in
$T_{UU}$. The same proof applies immediately if the initial state is
$\ket{\psi_+}= \ket{0_M} \ket{+}$.
[This proof may be also generalized to massive fields and interacting
fields for $T_{UU}(t,x)$ with $(t,x)$ in $L$ since then
it is space like separated from $\phi(\tau)$ but in general we will have 
$\langle T_{UU} (t,x) \rangle \neq 0$ for $(t,x)$ in $F$.]

The sum of  eqs. (\ref{fourninea}) and \ref{fournineb} gives a negative result to
the average energy density
\begin{equation}
 \langle T_{UU} \rangle_- = -{g^2 \over 2}\beta^2_{\Delta M} 
{\theta (-U) \over (a U)^2}
\label{fournine}
\end{equation}
in accord with the expectation that absorption of a Rindleron diminishes
the energy density. More precisely the total reduction of Rindler energy in
the interval $(\tau_i, \tau_f)$ at $O(g^2)$ due to absorption of right movers
is 
\begin{eqnarray}
\int \langle {T_{uu}} \rangle_- du &=&
\int \langle T_{UU} \rangle_- ({dU \over du})^2 du
=-{g^2 \over 2} \beta_{\Delta M}^2 (\tau_f - \tau_i)
\nonumber\\
&=& - {R_+\over 2} {\Delta M} (\tau_f - \tau_i)
\label{sixfive}
\end{eqnarray}
where we have used eq. (\ref{Rfivea}). Dividing  eq. (\ref{sixfive}) by the
probability of excitation $P_+= R_+(\tau_f-\tau_i)$ yields the energy emitted in
U modes if the atom is found excited
\begin{equation}
{1\over P_+}\int \langle {T_{uu}} \rangle_- du = -{\Delta M / 2}
\label{enmode}\end{equation}

Thus one obtains a steady absorption of energy exactly as in the usual golden rule. The factor
$1/2$ in  eq. (\ref{sixfive}) arises because we have taken into account right
movers only. We have set the integral $\int_{- \infty}^{+\infty} du = \tau_f -
\tau_i$. This very reasonable result can be obtained rigorously by going to the
limit $\tau_f - \tau_i \to \infty$ in a more controlled manner (see 
\cite{MaPa}).

The same procedure as that discussed in the paragraph preceding equation
 eq. (\ref{fourninea}) applies to the energy emitted if the initial state is
$\ket{\psi_+}$. One finds
\begin{equation}   
\elematrice{\psi_+}{T_{UU}}{\psi_+} = {g^2 \over 2
 } \alpha_{\Delta M}^2 {\theta (-U) \over (a U)^2}
\label{sixsix}
\end{equation}
whence
\begin{eqnarray}
 \int \elematrice{\psi_+}{T_{uu}}{\psi_+} du = 
{R_-\over 2} {\Delta M} (\tau_f - \tau_i) \label{sixseven}
\end{eqnarray}

We can now check out Grove's theorem for the equilibrium
situation. In the inside region ($\tau_i < \tau < \tau_f$)
one has at thermal equilibrium
\begin{equation}  
p_+/ p_- =R_+ / R_- = e^{- \beta \Delta M} = \beta_{\Delta M}^2 /
\alpha_{\Delta M}^2
\end{equation}
where $p_-$ ($p_+$) is the probability to find the atom in the
ground (excited) state.
This guarantees no net flux:
\begin{equation}  
p_+ \langle T_{uu} \rangle_+ + 
p_- \langle T_{uu} \rangle_- =0 
\end{equation}

It is important to remark that the separate contributions to 
$\langle T_{UU} \rangle _-$ given by eq. (\ref{fourninea}) and
eq.~(\ref{fournineb}) are each non causal. Each has $\theta(+U)$ contributions
which cancel in the sum. These non causal contributions occur because in
the calculation of eq.~(\ref{foursix}) and
eq.~(\ref{fourseven}) we have introduced the projectors $\Pi_+$ and $\Pi_-$ (the
$\Pi_+$ contribution was already anticipated in the simple saddle point
approximation (Section \ref{spontaneous}).
They encode the correlations in Minkowski vacuum
between $L$ and $R$ Rindlerons (recall eq.~(\ref{threethree})). Thus in the 
description wherein an atom excites (deexcites) by means of Rindleron
absorption (emission) its transitions correlate to the presence (absence) of
the corresponding space like separated Rindler quantum in the other quadrant
($U>0$). This is summarized in Fig.~(\ref{excit}) which is presented at the end of
this section.

Instrumental in the explicit realization of Grove's theorem is the negativity
of $\langle T_{uu} \rangle_-$ and hence of $\langle T_{UU} \rangle_-$ in the
 region where translational symmetry is valid\footnote{
After this manuscript was completed, further research on this
subject was done. 
It has now been proven  \cite{Par} 
that for a detector of finite mass $M$ (but nevertheless
with $M/a >>1$) the interference term becomes 
negligible after a few transitions,
thereby reinstating the validity of the naive Born approximation for the energy
emitted. This occurs because, when recoil 
is taken into account, the detector shifts its
orbit from $\rho=a^{-1}$ to an orbit characterized by a new horizon (ie. 
the center of
the hyperbola which describes the detector's trajectory shifts). Thus one loses the
translational invariance in $\tau$ (boost invariance) and Grove's theorem is no longer
applicable. More formally, the term eq.~(\ref{fournineb}) exists because the
atom which has emitted a photon and then reabsorbed it interferes with the atom
which has not made any transitions. When 
the atom recoils these two amplitudes  no longer interfere destructively.
Recoil induces decoherence. This decoherence occurs
after a logarithmically short time 
$ \tau \simeq a \ln ( M / \Delta M) $. This very short time is a manifestation of the exponentially
large frequencies which resonate with the accelerated atom 
at early times (see eq.~(\ref{threefiveb})). 
The same exponential rise of frequencies in the Hawking radiation
is a source of anguish when the gravitational back-reaction of these
frequencies is considered. More on this in Section \ref{troub}.}  (recall
we have made the computation in the limit $\tau_f-\tau_i \to \infty$). But 
one has
$\int_{-\infty}^{+\infty}  dU \langle T_{UU} \rangle_- > 0$ 
since to order $g^2$ the contribution to it from eq.~(\ref{fournineb}) must in
fact vanish  on the basis of a
rigorous theorem 
\begin{eqnarray}
\int_{-\infty}^{+ \infty} \! dU \ \elematrice{\psi_-}{\Pi_-
T_{UU}}{\psi_-} =0
\label{sixnine}
\end{eqnarray}
This is because 
$\int_{-\infty}^{+ \infty} \! dU \ 
T_{UU}$
is the total energy operator carried by right movers; hence it
annihilates the vacuum. (Recall we are normalizing the vacuum energy to
zero).
Furthermore the integral 
$\int_{-\infty}^{+ \infty} \! dU \
\elematrice{\psi_-}{\Pi_+T_{UU}}{\psi_-}$ is strictly positive (it 
is the expectation value of a positive definite operator:
the hamiltonian, see eq.~(\ref{foursix})).

Thus the sum of the two is positive, as stated.
Therefore there is  positive energy density
which does not appear in eq.~(\ref{fournine}) and which
can only come  from transient behavior at the end points.

In order to study the details of  the energy distribution one must introduce
explicitly a switching function, for instance by  taking the
interaction Hamiltonian to be of the form \begin{equation}
H_{int} (\tau)= f(\tau) e^{-i \Delta M \tau} \sigma_- \phi(\tau) +
\hbox{h.c.} \label{fiveseven}
\end{equation}
where $f(\tau)$ controls the switching on and off of the interaction.
In order for $f(\tau)$ not to induce spurious switch on and off effects it
should have a sufficiently long plateau $\Delta \tau$ that the golden rule can
establish itself ($\Delta \tau >> \Delta  M^{-1}$ and $\Delta \tau >> a^{-1}$)
(see discussion after eq.~(\ref{Rthree})).
Furtherore in  \cite{MaPa}
it is shown that if $f(\tau)$ obeys the condition
\begin{equation}
\int d\tau e^{a\vert \tau \vert} f(\tau) < \infty
\quad \Leftrightarrow \quad
\int dt f(\tau(t)) < \infty
\label{fiveeight}
\end{equation}
then the energy fluxes are regular and no singularities appear on the
horizons.

When $f(\tau)$ does not obey eq.~(\ref{fiveeight}) the
transients do not appear in finite Rindler time. Rather they are 
found on the horizons
$U=0$ and $V=0$ where they give rise to  singular energy
fluxes. This situation is analyzed by
regulating  the Bogoljubov coefficients eq.~(\ref{twotwo}) which
amounts to replacing $U$ by $U-i \epsilon$ as in eq.~(\ref{Unm}). When
this is done one finds that the two terms eq.~(\ref{fourninea}) and
eq.~(\ref{fournineb}) take the form
 \begin{eqnarray}
\elematrice{\psi_-} { \Pi_+ T_{UU} }{\psi_-} &=&
{g^2 \over 2 a^2 } \beta^2_{\Delta M} \alpha^2_{\Delta M} 
(U-i\epsilon)^{-i\Delta M / a -1}(U+i\epsilon)^{i\Delta M / a -1}\nonumber\\
&=& {g^2 \over 2 a^2 } \beta^2_{\Delta M} {1 \over U^2 + \epsilon^2}
\left( \alpha^2_{\Delta M} \theta(U) + \beta^2_{\Delta M} \theta( -U)\right)
\label{fivefoura}
\end{eqnarray}
and
 \begin{eqnarray}
\elematrice{\psi_-} { \Pi_- T_{UU} }{\psi_-}&=&
-{g^2 \over 2 a^2 } \beta^2_{\Delta M}
\alpha^2_{\Delta M}  {\rm Re }\left[
(U+i\epsilon)^{-i\Delta M / a
-1}(U+i\epsilon)^{i\Delta M / a -1}\right]\nonumber\\ 
&=& -{g^2 \over 2 a^2 }
\beta^2_{\Delta M} \alpha^2_{\Delta M}  {\rm Re }\left[ 
{1 \over( U + i\epsilon )^2}\right]
\label{fivefourb}
\end{eqnarray}
Thus for $\vert U \vert > \epsilon$, eq.~(\ref{fournine}) remains valid whereas
for  $\vert U \vert < \epsilon$ all terms are positive. The total energy emitted
is obtained by integrating over $U$ (by contour integration). The contribution
from eq.~(\ref{fivefourb}) vanishes as required whereas eq.~(\ref{fivefoura})
gives a positive contribution equal to $(g^2/ 2 a^2) (\pi /\epsilon)
\beta^2_{\Delta M} \alpha^2_{\Delta M}$ ie. total positive energy is radiated,
one part of which is concentrated along the horizon and is positive, and the
rest is in the detectors quadrant. In this way both causality and the positivity
of energy of Minkowski excitations are respected.

  A similar analysis is possible for
$\elematrice{\psi_+} { T_{UU} }{\psi_+}$. In the equilibrium situation 
the terms from $\elematrice{\psi_-} { T_{UU} }{\psi_-}$ and 
$\elematrice{\psi_+} { T_{UU} }{\psi_+}$ combine in such a way that they give zero
everywhere except for a singular positive energy flux on the horizon $-\epsilon < U <
\epsilon$. 

This completes the formal proof of the qualitative discussion
presented in the preceding subsection. We thus have shown
that as
 in the accelerated charge
problem considered by Boulware this singular flux can be 
interpreted as  infinitely
blueshifted transients which occurred when the atom was set into
acceleration at $\tau = a^{-1} \ln a \epsilon$. When eq.~(\ref{fiveeight}) is satisfied
complicated expressions arise wherein the role of $\epsilon$ is played
by  $\Delta\tau^{-1}$. As these functions are not very
interesting to display here, we reserve a more interesting spot
in Section \ref{weakacc} to put them on exhibition (see
Fig.~(\ref{excit})).

It is now clear how eq.~(\ref{fourtwo}) makes sense. The total contribution to 
$\int_{-\infty}^{+\infty} \! dU \ \elematrice{\psi_-} { T_{UU} }{\psi_-}$
is from the $\Pi_+$ contribution only
and hence can be expressed as coming
almost entirely from
the wrong quadrant. 
Thus for $\beta^2 <<1$ one may write the $\Pi_+$ contribution as an integral
over $u_L$ (see eq.~(\ref{twonulab})) thereby obtaining 
the absorption part
of eq.~(\ref{fourtwo}). The emission part (ie. the contribution of
$\ket{\psi_+}$) is obtained in similar manner to yield the sum
eq.~(\ref{fourtwo}). Causality is verified when one analyzes the complete
distribution of energy, taking into account the $\Pi_-$ interference term. The
mean energy density radiated is then found only in the transients.

\subsection{The Vacuum Fluctuations Correlated to the Excitations of the Atom}
\label{weakacc}

In the previous subsection, we have analyzed the mean energy radiated by the
atom and we saw how it can be decomposed into two contributions. These
correspond to the energy emitted by the atom when it is found  excited or not
excited at $t=\infty$. We now address the
question: what configurations of energy-momentum 
were
present in Minkowski vacuum which
give rise to spontaneous excitation of the two level atom? 
They certainly carry zero
total energy but locally should
have a positive Rindler energy density to excite the
atom.

We shall use the weak-value formalism since
it is shown in Appendix \ref{weak} that it
provides a framework 
to investigate on ${\cal{I^+}}$ or on ${\cal{I^-}}$ 
the nature of the field configurations
correlated to the excitation of the atom.

Before the atom interacts with the radiation, we have that in the mean
\begin{equation}
\bra{-} \elematrice{0_M}{T_{UU}({\cal{I^-}})}{0_M } \ket{-}=0
\end{equation}
where 
$T_{UU}({\cal{I^-}})$ is the Heisenberg operator, $T_{UU}$, evaluated on the surface
${\cal I}^-$, ie. on a surface which is temporally situated before the time $\tau_i$
when the atom begins to interact with the field. 

In this mean appear two
cancelling contributions according to the class of final states
considered: contributions for which the atom excites in the period
$(\tau_i, \tau_f)$ where both $\tau_i >t$ and $\tau_f >t$ and that for
which the atom does not excite in this same time interval. Thus we must
project $\ket{0_M}\ket{-}$, the Schr\"odinger state at times $\tau< \tau_i$,
into the various outcomes that are realized at later times. We carry this out by inserting
 at $t=+\infty$ the
projectors $\Pi_+$ and $\Pi_-$ (see eq.~(\ref{fourfive})) \begin{eqnarray}
0 &=& \bra{-} \elematrice{0_M}{T_{UU}({\cal{I^-}})}{0_M}
\ket{-}
 = \bra{-} \elematrice{0_M}{ {\rm T}e^{i \int_{\tau_i}^{\tau_f} d \tau
H_{int} } {\rm T}e^{-i\int_{\tau_i}^{\tau_f} d \tau
H_{int} }T_{UU}}{0_M} \ket{-}\nonumber\\
 &=& \bra{-} \elematrice{0_M}{ {\rm T}e^{i \int_{\tau_i}^{\tau_f} d \tau
H_{int} } ( \Pi_+ + \Pi_-){\rm T} e^{-i\int_{\tau_i}^{\tau_f} d \tau
H_{int} }T_{UU}}{0_M} \ket{-} \label{wseven}
\end{eqnarray}

Let us examine the $\Pi_+$
contribution
\begin{eqnarray}
\bra{-} \elematrice{0_M}{{\rm T} e^{i \int_{\tau_i}^{\tau_f} d \tau
H_{int} }  \Pi_+ {\rm T}e^{-i\int_{\tau_i}^{\tau_f} d \tau
H_{int} }T_{UU}}{0_M} \ket{-}
\label{wtt}
\end{eqnarray}
where $\Pi_+$ is the Heisenberg operator which projects onto the excited
state at $\tau=\tau_f$. This contribution
is anticipated to be positive since it is the contribution to the mean
$\elematrice{0_M}{ T_{UU}({\cal{I^-}})}{0_M}$ that results in excitation. 
The correct normalization which appears in the weak value formalism 
(see eqs.~(\ref{pexpectj})  and (\ref{Tj})) is to
rewrite
eq.~(\ref{wtt}) 
 as $P_+ \langle T_{UU}({\cal{I^-}}) \rangle_{w+}$ where $P_+$ is 
the probability for excitation in the interval $(\tau_i,
\tau_f)$:
\begin{eqnarray}
P_+ &=&  
 \elematrice{\psi_-}{   \Pi_+ }{\psi_-}  \nonumber\\
\langle T_{UU}({\cal{I^-}}) \rangle_{w+} &=&  {1 \over P_+}
\bra{\psi_-}\Pi_+ T_{UU}({\cal{I^-}}) \ket{\psi_-}
\label{wts}
\end{eqnarray}
In this rewriting $\langle T_{UU} ({\cal{I^-}})\rangle_{w+}$  
is 
what
has been identified in Section \ref{pair} 
and \ref{mirro} and Appendix \ref{weak} 
with the energy of the field configuration which 
gives rise to excitation of the atom.

To order $g^2$ we find
 \begin{eqnarray}
\langle T_{UU}({\cal{I^-}}) \rangle_{w+} &=&  
{ 
\bra{-} 
\elematrice{0_M}{
\left( 1 + i g \int_{\tau_i}^{\tau_f}
e^{i\Delta M \tau} \phi(\tau) \sigma_+ \right) \Pi_+\left(
\hbox{h.c.} \right) T_{UU}
}{0_M} 
\ket{-}
\over  
\bra{-} 
\elematrice{0_M}{
\left( 1 + i g \int_{\tau_i}^{\tau_f}\! d \tau\ 
e^{i\Delta M \tau} \phi(\tau) \sigma_+ \right) \Pi_+\left(
\hbox{h.c.} \right) 
}{0_M} 
\ket{-} }
\nonumber
\\
&=& {  
\elematrice{0_M}{  
\int_{\tau_i}^{\tau_f}\! d\tau\ 
e^{i\Delta M \tau} \phi(\tau)  
\int_{\tau_i}^{\tau_f}\! d\tau^\prime\  
e^{-i\Delta M \tau^\prime} \phi(\tau^\prime) 
T_{UU}
}{0_M} 
\over 
\elematrice{0_M}{  \int_{\tau_i}^{\tau_f}\! d\tau\ 
e^{i\Delta M \tau} \phi(\tau) 
\int_{\tau_i}^{\tau_f}\! d\tau^\prime\  
e^{-i\Delta M \tau^\prime} \phi(\tau^\prime) 
}{0_M} } \label{weight}
\end{eqnarray}
As in eq.~(\ref{foursix}) the $g^2$ term in the wave function does not
contribute to matrix elements when $\Pi_+$ is inserted.

It is instructive once more to consider 
the resonant piece (ie. $\tau_f -\tau_i \to \infty$) of 
$\langle T_{UU}({\cal{I^-}}) \rangle_{w+}$
 even though we know from the previous subsection that the transients
are not correctly described in this approximation. 
Thus we replace the double integral in $T_{UU}$ by a simple integral and
take all Rindler frequencies to be equal to the resonant frequency
$\Delta M$ and obtain 
\begin{eqnarray}
\langle T_{UU} ({\cal{I^-}})\rangle_{w+} &=&  2{\alpha_{\Delta M}\over \beta_{\Delta M}}
\partial_U \hat  \varphi^*_{-\Delta M}
\partial_U \hat  \varphi^*_{\Delta M} \nonumber\\
 &=& 
{\Delta M \over 2 \pi}{1 \over (aU)^2} \alpha_{\Delta M}^2
\label{woneone}
\end{eqnarray}
Where we have used eq.~(\ref{xxx}). 
Notice that $\langle T_{UU}({\cal{I^-}})\rangle_{w +}$ is non vanishing both for $U>0$ and
$U<0$. 
It can be interpreted in $R$ by appealing to the
isomorphism with the thermal bath. Since we have post-selected that the
atom will get excited  necessarily there was a Rindler quanta in $R$ which will 
excite the atom. The weak value therefore contains the Rindler energy of this
particle. The factor $\alpha_{\Delta M}^2= n(\Delta M) + 1$ rather than $1$
takes into account that in a thermal bath their may be more than one quantum.
The two level atom is sensitive to this since it responds to the 
 mean number of quanta.

In $L$, the partner of the Rindleron in $R$ appears with the same Rindler energy
$\Delta M$ because Minkowski vacuum is filled with correlated Rindlerons
in R and L (eq.~(\ref{threethree})) of zero total Rindler energy
eq.~(\ref{hamr}). But since we are in Minkowski vacuum and the interaction has
not yet occurred, we necessarily have \begin{eqnarray}
\int_{-\infty}^{+\infty} dU \langle T_{UU}({\cal{I^-}}) \rangle_{w+} &=&
0\label{wonetwo}
\end{eqnarray}
because  the operator $\int_{-\infty}^{+\infty} dU T_{UU}$ annihilates  
Minkowski vacuum.  Hence $\langle T_{UU}({\cal{I^-}})
\rangle_{w+}$
 carries no integrated energy as behooves a vacuum fluctuation
(see eqs.~(\ref{vanish}) and (\ref{vanishB})).

Upon integrating eq.~(\ref{woneone}) a contradiction
seems to arise as in the previous section. The transients have been
incorrectly taken into account. A correct handling of the ultraviolet
Minkowski frequencies (eq.~(\ref{Unm})) solves this problem and 
in the limit $\tau_f - \tau_i \to \infty$, the weak value is the distribution
\begin{equation} \langle T_{UU}({\cal{I^-}}) \rangle_{w+} = {\Delta M \over 2\pi}{1 \over
a^2 (U+i\epsilon)^2} \alpha_{\Delta M}^2 \label{wonethree}
\end{equation}
with a regularized energy content on the horizon $U=0$ which restores the
property eq.~(\ref{wonetwo}) (verified by closing the contour in the complex
plane whereupon the double pole at $U=-i\epsilon$ has zero residue).

We now consider the model eq.~(\ref{fiveseven}) so as to have smooth $T_{UU}$'s
(functions rather than distributions). 
The manner in which eq.~(\ref{wonetwo}) is realized is a subtle interplay of
several effects which we now summarize.

We first obtain that for $U>0$ the weak value is real and positive since
\begin{eqnarray}
\langle T_{UU}(({\cal{I^-}};U>0) \rangle_{w+} &=&{ \elematrice{0_M}{
\int \! d\tau H_{int}(\tau) \int\! d\tau^\prime H_{int}(\tau^\prime)
T_{UU} }{0_M} \over \elematrice{0_M}{
\int \! d\tau H_{int}(\tau) \int\! d\tau^\prime H_{int}(\tau^\prime)}
{0_M}}\nonumber\\
\ &=& {
\elematrice{0_M}{
\int \! d\tau H_{int}(\tau)  
T_{UU}
\int\! d\tau^\prime H_{int}(\tau^\prime)
 }{0_M} \over \elematrice{0_M}{
\int \! d\tau H_{int}(\tau) \int\! d\tau^\prime H_{int}(\tau^\prime)}
{0_M}}
\label{wonefour}
\end{eqnarray}
which is manifestly real. We have used the causality development of 
eq.~(\ref{fivesix}) to make the necessary commutation. Moreover 
eq.~(\ref{wonefour})
is positive since it is the expectation value of $T_{UU}$ in a one particle
state. 

Since the integral over all $U$ vanishes, the integral over $U<0$
must yield a negative result which exactly compensates the integral over
$U>0$. On the other hand the Rindler energy 
$\int_{-\infty}^{+\infty}\! du \ T_{uu}({\cal{I^-}};U<0)$ is
positive (although the integrand contains small end effects) since it
describes the energy of the Rindleron which shall be absorbed by the atom.
The net negative Minkowski energy on the right ($(U<0$) 
once more finds its origin in the transients which
are small in the Rindler description but large in the Minkowski
description due to the enormous Doppler shift $du/ dU = e^{a u}$.
Furthermore for $U<0$, the weak value is not real. Nevertheless
the complexity
appears only in the transients (in eq.~(\ref{weight}) at $U=0$). This imaginary
part also integrates to zero by virtue of eq.~(\ref{wonetwo}). In Section
\ref{VFHR} we shall dwell somewhat further on the physical content of the
imaginary part of $\langle T_{\mu\nu} ({\cal{I^-}})\rangle _{w +}$
(see eq.~(\ref{physrela})).

The reader may check that the energy fluctuations correlated
to the transitions of the atom posess all the properties of the 
fluctuations correlated to the production of a specific photon 
by a mirror as studied in the previous section. However whereas in the mirror
the post selection was an abstract construction, we have here shown that it is
realized operationally by coupling an external system to the radiation.

Putting together the analysis of this subsection  and the
preceding one  we have obtained a  description in terms of
energy density of the field configurations in Minkowski vacuum which make an
accelerated two level atom excite 
(eq.~(\ref{weight})) and the resulting field configuration after excitation
(eq.~(\ref{fourninea})). This is depicted shematicaly in Fig.~(\ref{excit}a)
for an atom which begins in the ground  state around time $\tau_i$ and
 excites in the 
interval between $\tau_i$ and
$\tau_f$. If no excitation occurs one has the complement of eq.~(\ref{weight})
and (\ref{fourninea}). This is depicted in Fig.~(\ref{excit}b). The mean
energy momentum is recovered when summing over excitation and no
excitation and is depicted in Fig.~(\ref{excit}c).

For simplicity of drawing it is the Rindler energy density $T_{uu}$ 
which is drawn in every
case. This is because the rapid fluctuations of the Minkowski energy 
in the vicinity of the
horizon makes it virtualy impossible to represent in a comprehensible 
diagram.
In all three figures the trajectory of the atom is drawn as a 
hyperbola between the points
$\tau_i$ and $\tau_f$. Energy fluxes for U--modes are 
represented by dotted lines. These flow
from the past energy configuration (drawn in the SO 
corner of the diagram) to the future energy
configuration (drawn in the NE corner).
\dessin{1.000}{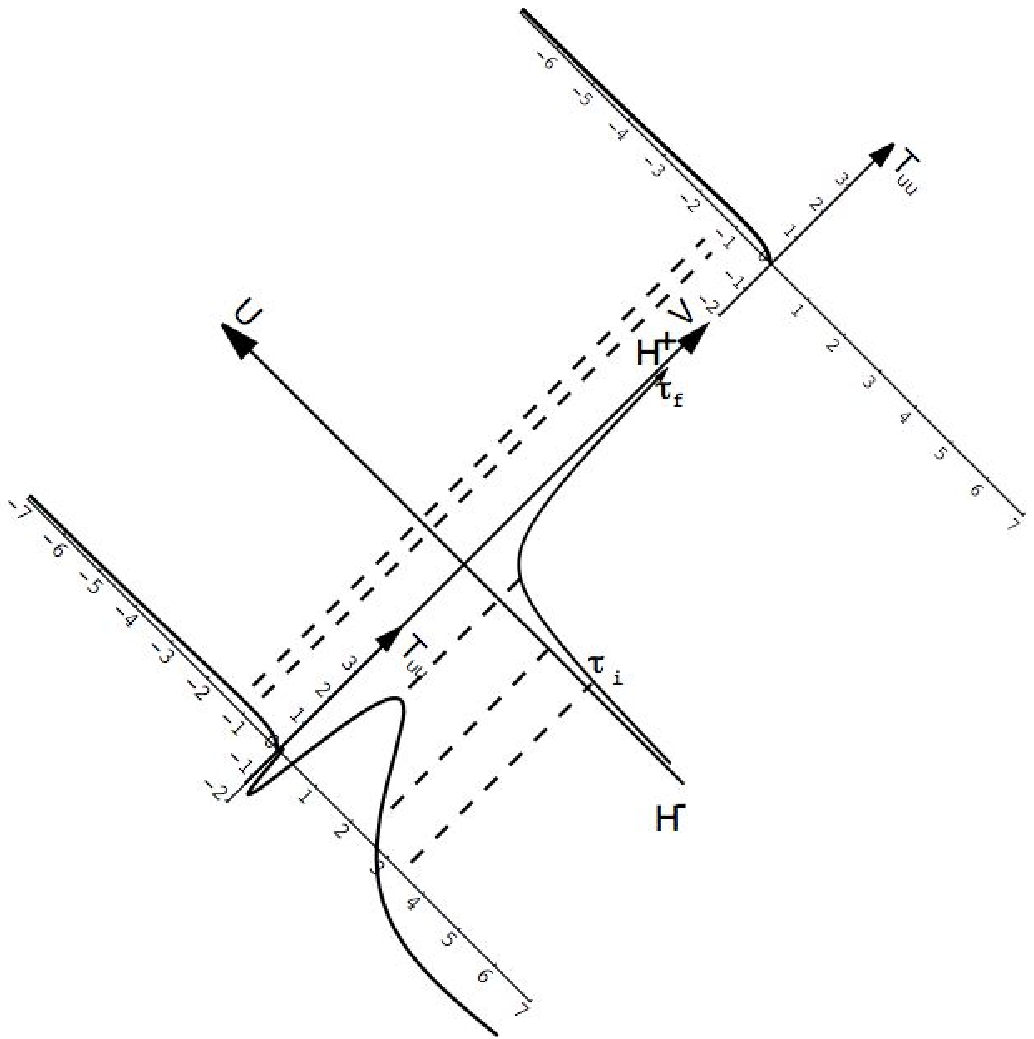}{excit}
{Birds'eye view of the energy of the field 
configurations both in the past
and in the future of a two level atom which begins in the ground 
state around time $\tau_i$ and
which either does (Fig. {\bf a}) or does not (Fig. {\bf b}) excite in the 
interval between $\tau_i$ and
$\tau_f$. Fig. {\bf c} represents the mean effect.}

In Fig (a) the atom has become excited as is witnessed 
by the absorption of the positive Rindler
peak in the middle of the configuration for $U<0$. In
consequence this peak has disappeared in
the future and there only remains the partner in the 
region $U>0$. (In fact the energy in the
future for $U<0$ is not quite zero but is order 
$e^{-\beta \Delta M}$ taken to be 
$e^{-\beta \Delta M}\simeq 10^{-5}$ in this case 
so that this is in fact too small to
draw).\\
Fig (b) is the complementary case in which the 
atom has not become excited. In order to make the
drawing visible the scale of the axes has been changed by a factor $\simeq
10^{-5}$.\\
\dessinbis{1.000}{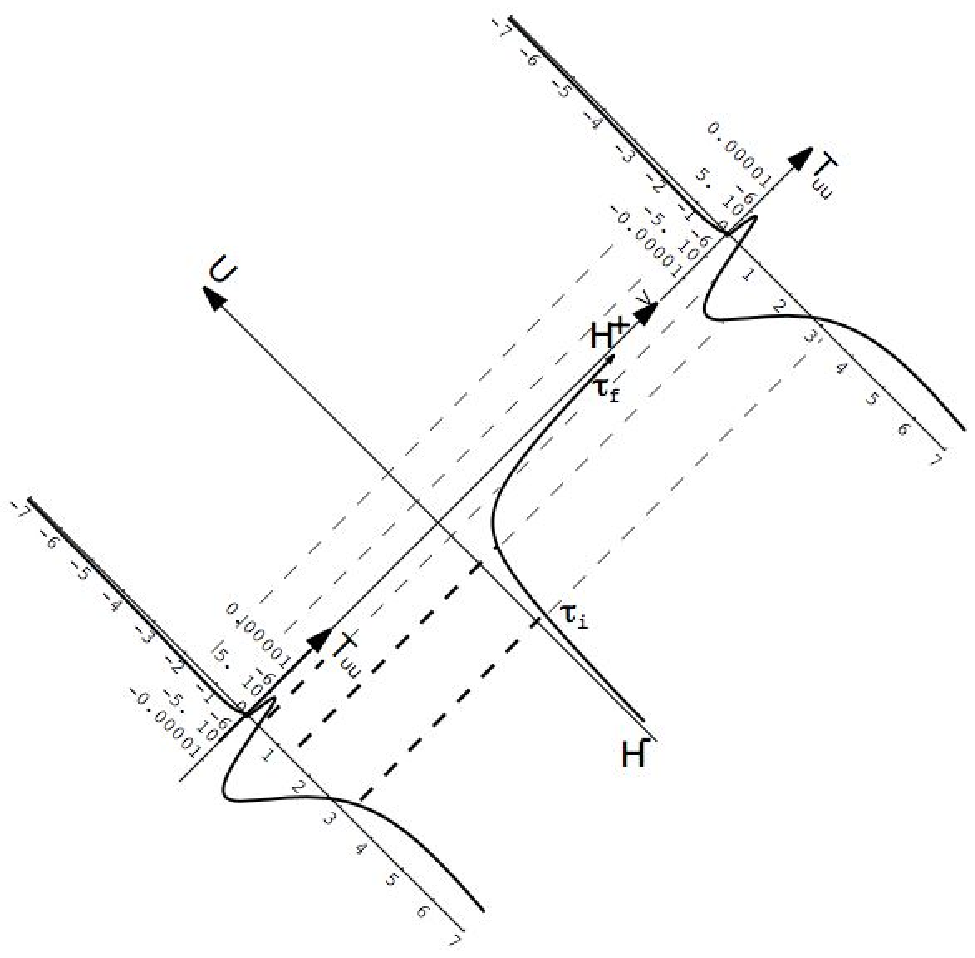}{2.5 b}{}
 The mean effect (fig (c)) therefore having mean energy zero in the
past gives rise to a negative Rindler energy in the future (illustrated by the
central negative dip in the future). As emphasized in the text there are
transient effects which carry positive energy and which contrive to render the
mean emitted Minkowski energy positive. It is further to be
 noted that causality
is respected in that the partner contributions ($U>0$) have cancelled between
fig (a) and (b). 
\dessinbis{1.000}{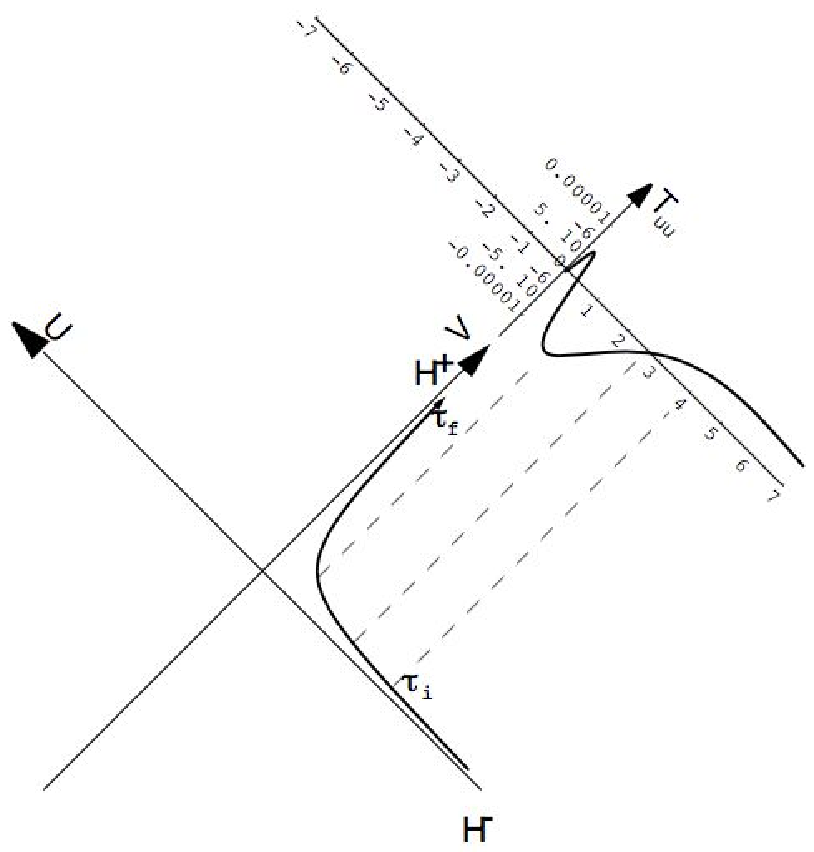}{2.5 c}{}

\chapter{ Black Hole Evaporation}\label{BLACK}

\section {Kinematics}\label{kinematics} 

The reader is referred (for instance) to
references
\cite{MTW},\cite{Houches},\cite{NATO},\cite{Chan},\cite{NoFr} for an introduction
to classical black hole physics. For the conceptual issues raised by quantum
mechanics in the presence of horizons it suffices to work with the Schwarzschild
black hole. Not that the Kerr hole 
does not give rise to interesting effects, but
its complications appear out of context in the present review. By Schwarzschild
black hole we include, and indeed mostly 
discuss, the incipient black hole wherein
the star's matter has fallen into a region which is asymptotically 
close to its
Schwarzschild radius, its future event horizon at $ r_S = 2M. $ 
Throughout we shall
take the Planck mass equal to unity. Thus $r$ is measured in Planckian
distances and $M$ in Planck masses. For a
star the size of the sun where $ M = 1.1\  10^{57} $ proton
masses = $0.9\ 10^{38}$ Planck masses, we have $ r_S = 1.8 \ 10^ {38} $
Planck distances.

Outside the star the  metric is Schwarzschild

\begin{eqnarray} 
ds^2& =&- (1 - {2M \over r}) dt^2 + (1 -{2M \over r})^{-1} dr^2 + r^2
d \Omega ^2  \label {1.1} \\\nonumber
& =& (1 - {2M \over r}) (-dt^2 + dr ^{ *2}) + r^2 d \Omega
^2 \nonumber\\
 & =& -(1- {2M \over r}) du\ dv + r ^2 d \Omega^2 \label {1.2} 
\end {eqnarray}
where $ r ^{*}$ is the tortoise coordinate 
\begin{eqnarray}
 r^* = r + 2M \ln \bigl\vert { r-2M \over 2M}\bigr\vert  \label {1.3} 
\end{eqnarray}
and 
\begin{equation}
d\Omega^2 = d \theta^2 + \sin^2 \theta\ d \varphi^2\ . \label{1.3B}
\end{equation}
Thus radial light rays follow $u$ or $v= constant$ where
\begin{eqnarray} 
\left.\begin{array}{r}
u \\ v
\end{array} \right\} = t \mp r ^* \ .\label {1.4}
\end{eqnarray}

In addition to the Schwarzschild coordinates displayed in eqs (\ref{1.1}, 
\ref {1.2}) we
shall refer to both Kruskal \cite{Krus},\cite{MTW}, and advanced  Eddington
Finkelstein \cite{Eddi},\cite{Fink},\cite{MTW} coordinates. In the first
Schwarzschild quadrant (R),  in which $r> 2M$, Kruskal coordinates are defined by 
\begin{equation}
\begin{array}{rcl} 
V&=& 4M e^{v/4M}  \\
 U& = &- 4M e^{-u/4M} 
\end{array}
\quad \hbox{in R} \label{1.5} 
\end{equation}
The relation between the Kruskal $U,V$ and the Schwarzschild $u,v$ is thus
identical to the relation, eq.~(\ref{onesixb}), 
between Minkowski and Rindler light-like coordinates. This
isomorphism will play a crucial role in the understanding of the various
properties of the Hawking radiation. 
In Kruskal coordinates the metric reads
\begin{equation}
ds^2 =-{2 M \over r} e^{-r/2M}  dUdV  +
r^2d \Omega ^2 \label{1.6} 
\end{equation}
where $r$ is given implicitly by $UV = (4M)^2 (1 - r/2M) e^{r/2M}$, 
and radial light cones are surfaces of constant $U$ or $V$.

Were the complete space given by the analytic extension of the geometry 
whose metric is
eq.~(\ref {1.6})
there would be four quadrants (see Fig.~(\ref{krusk})) separated 
by horizons $U=0$ and
$V=0$ the first of which, R, is coordinatized by the Schwarzschild 
coordinates $u$,
$v$ of  eq.~(\ref {1.4}). The other three 
are coordinatized by Schwarzschild local
coordinates in a manner analogous to the coordinatization 
Minkowski space into four
Rindler quadrants. 
\dessin{1.000}{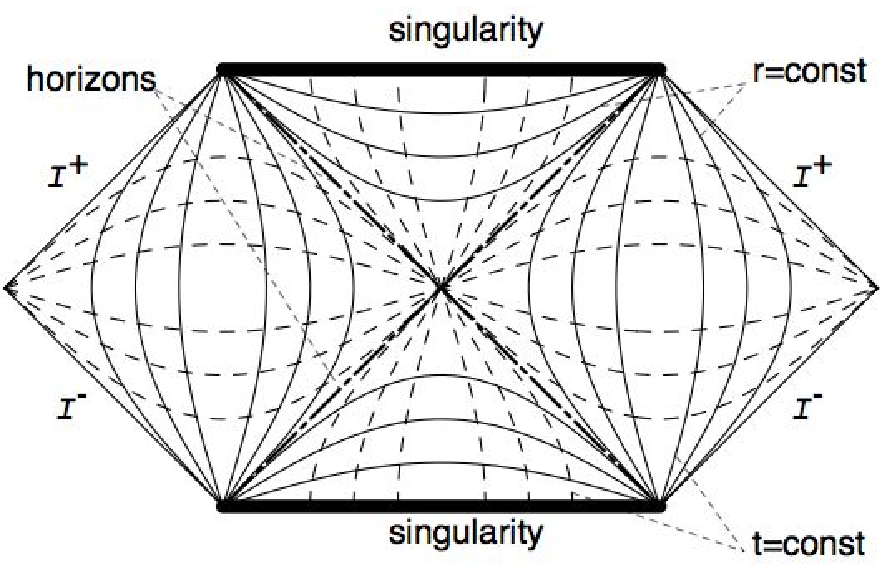}{krusk}{Penrose diagram of the maximal analytic extension 
of Schwarzschild
space. The horizons, the singularities as well as $r=const$ and $t=const$ 
lines have been represented.}

For the collapsing black hole the outside space is confined to two of
these quadrants R  and
F (see Fig.~(\ref{collapse})).

 \noindent Whereas Kruskal coordinates can
be used to describe both of them, the equations  which relate $t$ and $r$ to $U$ and $V$ through
eq.~(\ref{1.3}) to eq.~(\ref{1.5}) are  good only for the quadrant R. In the F quadrant, one may
introduce $u_F$ given by
 $U = 4M e^{u_F/4M}
$. The relation between $V$ and $v$ is still given by eq.~(\ref{1.5}) since 
$v$ is finite on the future horizon $U=0$.
Then, with $t,r$ given in terms of $u_F,v$ as in 
eq.~(\ref {1.4}) 
the metric is once more eq.~(\ref{1.1}). Note that  in R,  $t$
and $ r$ are time-like and space-like variables respectively whereas in F,
where $r-2M<0$, 
 $t$ is space-like and $r$ time like.
Finally, we introduce the advanced  Eddington-Finkelstein set $v$, $r$ 
that covers both R and F
and will be found convenient when we study the back reaction.
 In regions exterior to the
star one has
\begin {eqnarray}
 ds^2 =-( 1 - {2M \over r}) dv^2 + 2dvdr +r^2 d\Omega^2\label {1.7} 
\end{eqnarray}
\dessin{0.8000}{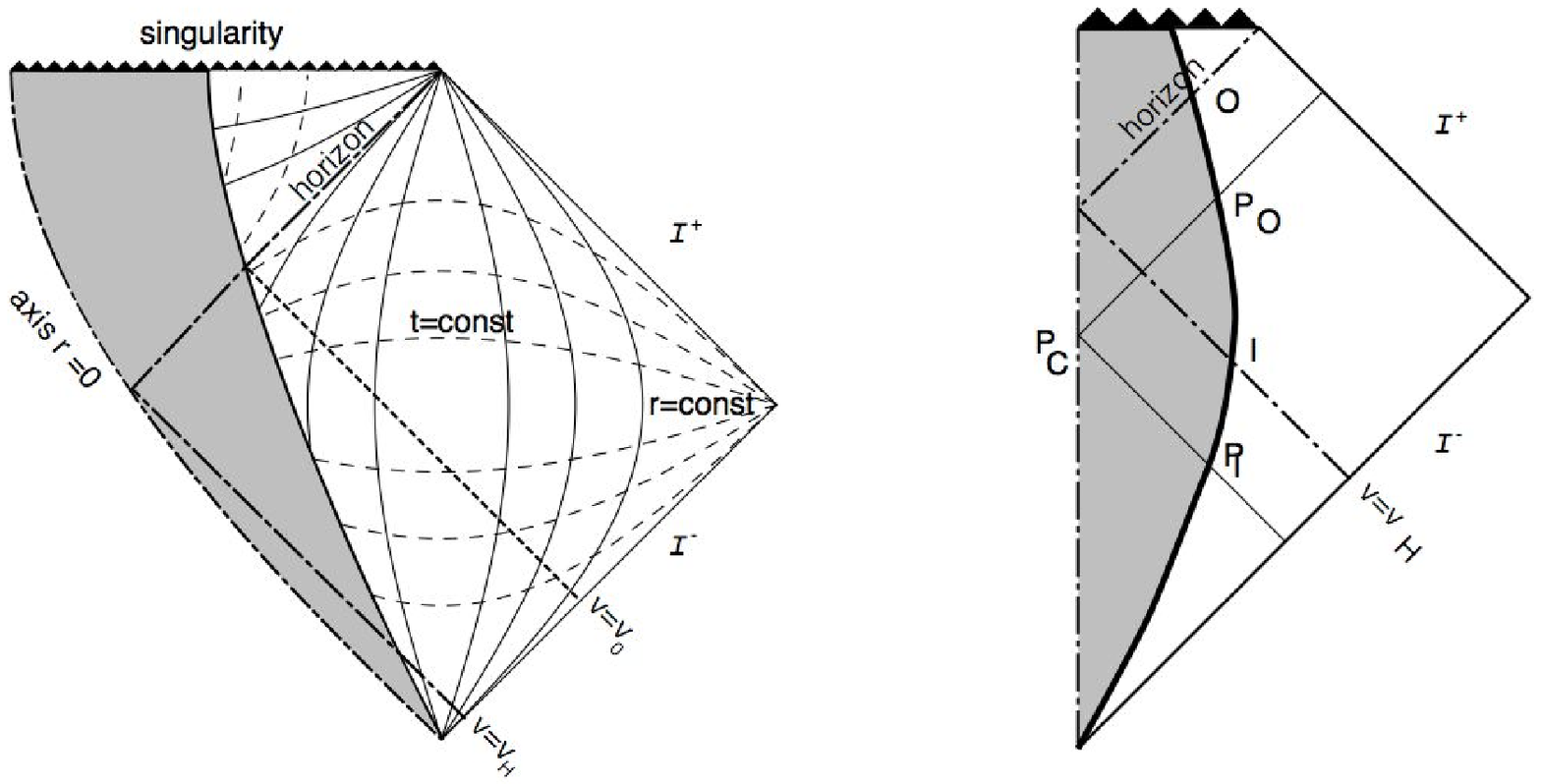}{collapse}{{\bf a}The exterior of the star (in white) is a 
recopy of the relevant part of the
complete Penrose diagram Fig. (3.1) whereas the grey  part 
representing the inside of the star is
obtained by extending ingoing and outgoing light cones through 
the two regions. The locus of the
apexes of these cones being the line $r=0$. As in all 
Penrose diagrams radial light rays are
represented by straight lines at $45^{\circ}$ degrees In te
to the vertical.\hfill
{\bf b} The same Penrose diagram as in {\bf a} redrawn by availing 
ourselves of further reparametrisation 
freedom of
radial null rays
 inherent to the construction in such manner that the line $r=0$ 
is drawn as a straight line. 
This is
the Penrose diagram which is usualy found in standard texts. The points
labeled $P_I$, $P_C$, $P_O$, $I$ and $O$ refer to the discussion of 
Hawking radiation in Appendix D}
In the relevant quadrants the domain of variation of the variables are:
\begin{itemize}
\item Schwarzschild in R
\begin {eqnarray}
&- \infty  < u<  + \infty  \  , \ - \infty  <  v <   + \infty 
&\nonumber\\
&\hbox{also } 
-\infty < t < + \infty \ ,  \  2M < r < \infty \ 
( \hbox{or } - \infty < r^* < + \infty )&
\label{1.9}
\end {eqnarray}
\item Kruskal in R
\begin {eqnarray}
-\infty < U <  0 \  , \ 
0  < V <  \infty \label {1.10}
\end {eqnarray}
\item Kruskal in F
\begin {eqnarray}
0 < U <  \infty \  , \ 
0  < V <  \infty  \  \hbox{and} \  UV < 1 \label {1.11}
\end {eqnarray}
\item{  Eddington Finkelstein in R and F}
\begin {eqnarray}
0 < r <  \infty \  , \ 
-\infty  < v <  \infty \label {1.12}
\end {eqnarray}
\end{itemize}

So much for the geometry of Schwarzschild space, i.e. the exterior of
the star. We must now describe the interior of the star. A
convenient idealization, first used by Unruh \cite{Unru1}, is the model of a
collapsing shell wherein the interior is empty, hence described by
flat space. We shall pursue this case explicitly and show how, in
Appendix \ref{sphradd} the essential result of the analysis emerges from the
general case.

Inside the space is Minkowski so that
\begin{equation}
ds^2 =-dT^2 + dr^2 + r^2 d\Omega^2 =-d {\cal{U}}  d {\cal{V}}  + r^2 d\Omega^2
\label{E.1}
\end{equation}
where $ {\cal{U}}  , {\cal{V}}  = T \mp r$.
We have set
the spatial coordinate $r$ to be the same in both regions so 
that the area of spheres
(SO(3)  orbits) is $4 \pi r^2$ everywhere and for all times.

In Fig.~(\ref{collapse}) we have drawn
 the Penrose diagram which shows the salient
features of the collapsing geometry. 
In this diagram we do not take into
account the loss of mass due to the Hawking radiation, hence it is only
useful to describe the early stages of evaporation. The heavy line
traces out the trajectory of the surface of the star $R_{st}(v)$. We
parametrize the trajectory in terms of the  
Eddington Finkelstein time $v$ rather than
the Schwarzschild time $t$, so as to cover its whole history. At 
$v= v_0$ the shell reaches its horizon 
$H$, the light like
surface given by $r=2M$. One may extend this light 
like surface $H$ into the
interior of the star as indicated. 

At still later times the shell 
reaches the space-like singularity $r=0$. This is a 
3-surface not to be
confused with the 1-dimensional time-like line $r=0$ which is the axis of
 SO(3)  symmetry of the whole collapsing geometry. This is brought out in Fig.~(\ref{tridcoll}).
  This line of symmetry is the locus of the vertices of light cones which trace out
the paths of spherical waves in the journey from ${\cal I}^-$ to
 ${\cal I}^+$ (light like past and future infinity). In Fig.~(\ref{collapse})
these cones are
represented by
lines reflected off $r=0$. 
This picture is blown
up into 3 dimensions in Fig.~(\ref{tridcoll}) wherein spheres 
are represented by
circles. 
\dessin{1.000}{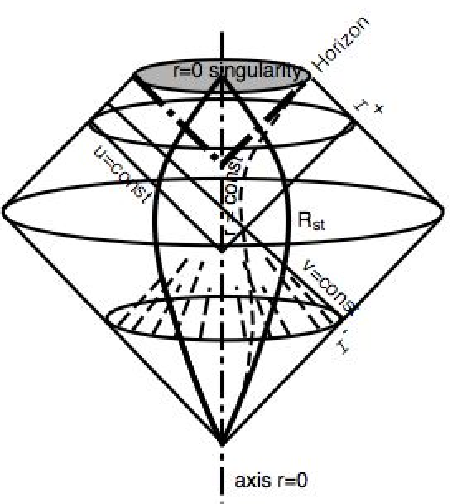}{tridcoll}{The Penrose diagram Fig. (3. 2 b) for a collapsing star blown up to $2+1$
dimensions. We have represented the horizon, the singularity, the surface of
the star, a typical light cone. The dotted line represents a $r=const$ line.
}
We have also represented the horizon light cone, $H$,
 its extension into the star and its backward
history as a light cone that terminates on  ${\cal I}^-$. A
spherical flash of light on this cone is the last flash that can be
emitted from  ${\cal I}^-$ so as to arrive at  ${\cal I}^+$. 
It will be shown that Hawking
radiation is concerned with the combined Doppler and gravitational
red shifts which spherical light cones, traveling just before $H$,
experience in their trip from   ${\cal I}^-$ to  ${\cal I}^+$.

Therefore, in order to compute those combined effects,
we need to match the internal coordinate system $ 
{\cal{U}} ,  {\cal{V}} $ to the external Schwarzschild set $u,v$.
We now turn to this task.

Since $u=$const. defines a future half light
cone, it can be extended back in time and enter the star with apex
at $r=0$; similarly for $v$. Thus the light cones labeled outside
the star by the Schwarzschild coordinates $u,v$ can be used to
coordinatize points inside the star. The same future half light cone
is described either by a constant value of $ {\cal{U}} $ or $u$ 
(and similarly for the past half null cones in terms of $ {\cal{V}} $ and $v$). In consequence $
{\cal{U}} $ is a function of $u$ only and $ {\cal{V}} $ a function of $v$ only.

The radial coordinate $r$ has the same meaning
inside and outside the star.
Thus on the shell the displacement in $r$ is the same in both systems:
\begin{eqnarray} 2 dr = d{\cal{V}} - d {\cal{U}}  = (1 - {2M \over
R_{st}})(dv - du) \label{E.2}
\end{eqnarray}
where differentials are along the trajectory. A further
condition is that intervals of proper time on the shell be the
same in both systems. Thus 
\begin{eqnarray} d{\cal{V}} d {\cal{U}}  = (1 - {2M \over
R_{st}})dv du \label{E.3}
\end{eqnarray}

These equations together with a trajectory for the star's surface
$R_{st}(v)$ are sufficient to solve for 
$ {\cal{U}} (u)$ and $ {\cal{V}} (v)$. We
rewrite them as
\begin{equation}
{ d {\cal{V}}  \over dv} - {d {\cal{U}}  \over du} {du \over dv}\!\bigr\vert_{st}
= (1 - {2M \over R_{st}})(1 - {du \over dv}\!\bigl\vert_{st})
\label{E.2B}\end{equation}
\begin{equation}
{ d {\cal{V}}  \over dv}  {d {\cal{U}}  \over du}
= (1 - {2M \over R_{st}})\label{E.3B}\end{equation}

When the shell is far from its horizon ($R_{st}>>2M$)
$ {\cal{U}} (u)$ and $ {\cal{V}} (v)$ are slowly varying functions 
the exact form of
which is irrelevant for our purpose. On the other hand when the shell 
approaches its horizon asymptotically these functions acquire
a universal behavior characterized by $M$ only. 
As it is precisely this domain of the variables which is
relevant to describe the steady state Hawking evaporation we shall
consider this case. 

To characterize the shell's trajectory near the horizon is very
simple. Suppose $R_{st}$ crosses the horizon at some  
Eddington Finkelstein advanced
time $v_0$ with finite acceleration (one verifies that this is indeed the case 
when the shell follows a geodesic). 
Then to first order we have
\begin{eqnarray}
2M - R_{st}(v) = k (v - v_0) \quad (k>0)\label{E.4}
\end{eqnarray}
Here $v(=t + r^*)$ is equal to $t+R^*_{st}$ on the surface so this
same equation gives $R_{st}(t)$
\begin{eqnarray}
-{R_{st}(t)-2M \over k}=  R^*_{st}(t) + t - v_0 \label{E.5}
\end{eqnarray} 
Near the horizon $R^*_{st} \simeq 2M + 2M \ln (R_{st}/2M - 1) $ so
we can solve iteratively to give
\begin{eqnarray}
R_{st}(t)-2M =  2M e^{(v_0 -2M -t)/2M} + O(e^{-t/M})= A e^{-t/2M} +
O(e^{-t/M}) \label{E.6} \end{eqnarray}
where $A$ is a positive constant.
Moreover to the same approximation we may replace $t$ by
$(u+v_0)/2$ so as to yield 
\begin{eqnarray}
R_{st}(u)-2M =  A^\prime e^{ -u/4M}  + O(e^{-u/2M})
\label{E.7} \end{eqnarray}
where $A^\prime$ is another positive constant. The relation between $u$ and
$v$ on the surface is thus
\begin{eqnarray}
v- v_0 &=& -{A^\prime\over k} e^{ -u/4M} + O(e^{-u/2M})
\label{E.8b}\end{eqnarray}
whence asymptotically
\begin{eqnarray}
{dv \over du}\!\vert_{st} &=& {A^\prime\over 4 M  k} e^{ -u/4M}
=  {R_{st}(u) - 2M \over 4M k}
\label{E.8}\end{eqnarray}  

Inserting eq.~(\ref {E.8}) into eq.~(\ref {E.2B}) and eq.~(\ref {E.3B}) one sees
that the first term on the r.h.s. of eq.~(\ref {E.2B}) is 
negligible with respect to $du/dv$. 
Further we have in the limit $v\to v_0$, $d {\cal{V}}  / dv  \vert_{v_0} =
\lambda$ where $\lambda$ is some positive constant. It then follows
that 
\begin{equation}
{d {\cal{U}}  \over du} = B e^{-u / 4M} + O (e^{-u /2M})
= - {{\cal{U}} \over 4M} + O ( {\cal{U}}^2 )  
\label{E.9}
\end{equation}
where $B$ is another positive constant. These constants will
be discussed in Appendix \ref{sphradd}, see also \cite{EMP}.
 Their value is irrelevant for the
calculation of the asymptotic Hawking radiation. The upshot is that
near the horizon
\begin{eqnarray}
 {\cal{V}} - 4M &=& \lambda (v - v_0)\nonumber\\
 {\cal{U}}  &=& B({-4M} e^{-u/4M})  + O (e^{-u /2M})
\label{E.10}
\end{eqnarray} 
Very important is the fact that ${\cal{U}}$ tends exponentially fast to
Kruskal $U$ 
(defined in eq.
(\ref{1.5}))
and
that ${\cal{V}}$ does not but rather is lineary related 
to the Schwarzschild $v$.
Note also that 
\begin{equation}  { d {\cal{U}}  \over du } = (B/A') (R_{st} -2M)
\label{g00}
\end{equation}
i.e. is proportional to  $g_{00}$, rather
than the usual $\sqrt{g_{00}}$. 
A direct calculation reveals that the factor $
R_{st} -2M$ is a composite
of the static red shift and a Doppler shift due to the retreating surface (see Appendix
\ref{sphradd}). 

In eqs (\ref{E.9}, \ref{E.10}) we have set $ {\cal{V}}  = 4M $ 
and $ {\cal{U}} =0$ at the point where the
star crosses the horizon $H$. The light cone $H$ which shall
generate the horizon is
 given by $ {\cal{U}} =0$ after it reaches $r=0$ and therefore 
by $ {\cal{V}} =0$ before (since $2 r= {\cal{V}}  -
 {\cal{U}} $ and $r=0$ at the apex of $H$).

Fig.~(\ref{star}) gives a sketch of the exterior R and interior I of the
star in the orthogonal coordinate grid $u,v$ which have their usual
meaning in R as functions of $t,r$ eq.~(\ref {1.4}) and where in I we have
\begin{eqnarray}
2r &=&  {\cal{V}} - {\cal{U}}  = \lambda v + {B \over 4M} e^{-u/4M} +
constant \nonumber\\
2T &=&  {\cal{V}} + {\cal{U}}  = \lambda v - {B \over 4M} e^{-u/4M} +
constant^\prime
\label{fig4}
\end{eqnarray}
\dessin{1.000}{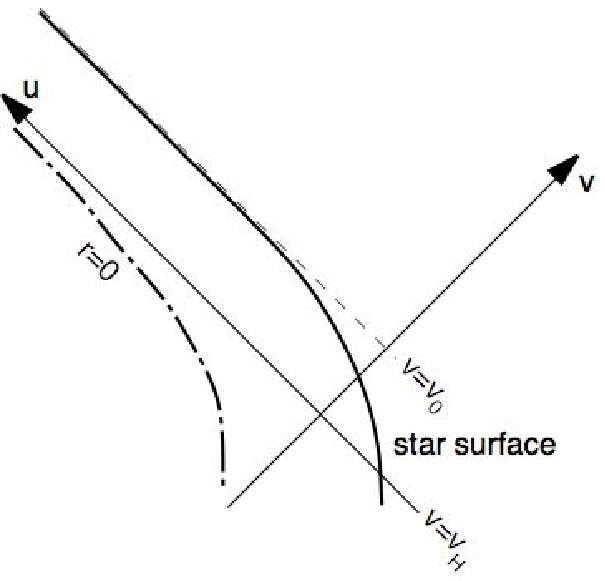}{star}{The geometry of a collapsing star in $u,v$ coordinates} 
One
should compare this drawing with the Penrose diagram Fig.(\ref{krusk}) and
compare the trajectory of light rays in the two drawings. One should also
compare the shell's trajectory eq.~(\ref{E.8b}) with the mirror trajectory eq.
(\ref{mirseven}) and remark that the Doppler shifts of eq.~(\ref{E.9}) and
eq.~(\ref{mirsix}) are identical.
Therefore, we shall find also 
a steady thermal flux (Hawking radiation) 
in the collapsing situation.

\section {Hawking Radiation}\label{hawking}

As in the preceding chapters, quantization proceeds through the construction of
modes, here solutions of $\square \psi = 0$. In the simple shell
model the d'Alembertian is a simple operator in the inside and outside
domains. Begin with the outside (Schwarzschild space) wherein from
eq.~(\ref {1.1})
 \begin{eqnarray}
\square \psi &=& {1 \over \sqrt{g}} \partial_\mu\  g^{\mu\nu}\!\sqrt{g}
\partial_\nu \ \psi \nonumber\\
&=&{1\over (r-2M)} \left[
-{\partial^2 \over \partial t^2} + 
{\partial^2 \over \partial r^{*2}}
+ (1- {2M \over r})({2M \over r^3} - {{\vec L}^2(\theta, \varphi) \over r^2})
\right]\  r \psi
\nonumber\\
\label{HR1}
\end{eqnarray}
Here $\vec L$ is the angular momentum operator. Since it commutes
with the d'Alembertian we pass over immediately to states of fixed $l$, i.e. 
modes are written as $ \psi_l=(\sqrt{4 \pi}
 r)^{-1} \varphi_l(r) Y^m_l(\theta ,
\varphi)$ where
\begin{equation}
\left( -{\partial^2 \over \partial t^2} + 
{\partial^2 \over \partial r^{*2}}
+ V_l(r)
\right)  \varphi_l=0
\label{HR2}
\end{equation}
with
\begin{equation}
 V_l(r)
= (1 - {2M\over r})({2M \over r^3} + {l(l+1)\over r^2})
\label{HR3}
\end{equation}

Similarly in the inside region one has
\begin{equation}
\left(-{\partial^2 \over \partial T^2} + 
{\partial^2 \over \partial r^2}
+ {l(l+1)\over r^2}
\right)  \varphi_l=0
\label{HR2BB}
\end{equation}

$V_l(r)$ plays the role of a centrifugal barrier, present even for
s-waves, where it gives a positive potential energy bump at
$r = 8M/3$. It has been shown by numerical calculation that about 
$90\% $ of Hawking radiation is in s-waves \cite{page},\cite{Sanch}. 

Therefore, in this section, we shall 
restrict ourselves to s-waves.
 As in Minkowski space, s-wave
modes vanish at the origin
$ \varphi_{l=0} (r=0) = 0$ (since $\psi = \varphi /r$)
 and define a one dimensional problem. 
Hawking
radiation is concerned  with the outgoing
reflected part. 
Because of the existence of $ V_{l=0} (r)$ in eq.~(\ref {HR2}),
low energy modes will be reflected back and the problem becomes a usual quantum
problem of finding the transmission coefficient. Once more for conceptual
purposes this complication is irrelevant and 
we shall put $ V_{l=0} (r) =0$. 
Nevertheless, at the appropriate place we shall mention the
modifications that ensue when its effects are included. In fact, in the last
analysis though this barrier is a nuisance 
for mathematics it will prove to be of
benefit since one will not have to 
confront the infra-red problem associated with
myriads of soft Hawking photons. Only frequencies $ \la \geq O ( 1/2M)$  are
passed. So the scrupulous reader may waive his qualms in the
knowledge that what follows is quite rigorous for sufficently large
$\la$.

With the above mentioned simplifications the s-waves 
obey
 \begin {eqnarray}
\left( {\partial ^2 \over \partial t^2} - {\partial ^2 \over \partial r
^{*2}} \right) \varphi =
{\partial  \over \partial u}{\partial  \over \partial v}\varphi
=0 &\quad& {\rm in\ R}
\label {HR.5} \\
\left( {\partial ^2 \over \partial T^2} - {\partial ^2 \over \partial
r ^{2}} \right) \varphi = {\partial  \over \partial  {\cal{U}} }
{\partial  \over \partial  {\cal{V}} }\varphi
=0 &\quad& {\rm in\ I}
\label {HR.6} 
\end {eqnarray} 
\noindent From eq.~(\ref {HR.5})  and eq.~(\ref {HR.6}) the solutions are the sum of an
ingoing and an outgoing piece
\begin{equation}
\begin {array}{cclcc}
\varphi& =& X(v) + \Xi(u) &\quad& {\rm in\ R}
\label{HR.7}\\
\varphi &=& \chi( {\cal{V}} ) + \xi( {\cal{U}} ) &\quad& {\rm in\ I}
\label{HR.8}
\end{array}
\end{equation}
where continuity on the star's surface 
imposes\footnote{In a realistic four dimensional model there will be some
scattering in the star (mixing $U$ and $V$ modes) at low frequencies.
But at the exponentially large frequencies encountered in Hawking radiation
the star is completely transparent} that $\chi( {\cal{V}} ) = X(v( {\cal{V}} ))$
and $\xi( {\cal{U}} ) = \Xi(u( {\cal{U}} ))$. 
Moreover, since $\varphi$
must vanish at ${r={\cal{V}}-{\cal{U}}=0}$,
one finds
that everywhere $\varphi$ takes the form
\begin {eqnarray}
\varphi &=& \chi( {\cal{V}} ) - \chi( {\cal{U}} ) 
\label{HR.9}
\end{eqnarray}
Before discussing in detail the form of the modes, 
we note that 
eq.~(\ref {HR.9}) 
already contains information about the division of modes into
the producing (class I) and non producing classes (class II)
as in chapters 1 and 2. In the distant
past, before collapse begins, we 
set up vacuum on  ${\cal I}^-$; this is in fact
Minkowski vacuum since the space is flat on  ${\cal I}^-$. The modes propagate
from  ${\cal I}^-$; hence they are 
of the form $e^{-i \omega v}$ (once more we are
thinking of a broad wave packet). They diminish 
in radius, and either penetrate or
do not penetrate the star's surface in there inward journey according to the
value of $v$ (See Fig.(\ref{collapse})). Hawking 
radiation comes about from modes in
the class that penetrates into the star, but not 
all of them. Only those that reflect
in a finite Schwarzschild time  ($v<v_H$ in Fig.(\ref{star})) 
thereby picking up a
dependence on $u$ which gives rise to the radiation. 

The asymptotic requirement that there be vacuum on ${\cal I}^-$
fixes that the incoming modes be of the form
$\varphi_\omega^{in}= \chi_\omega (v) = e^{-i\omega v}/ \sqrt {4 \pi \omega}$. 
We shall be concerned with the behavior of the modes in the vicinity
of the light cone $H$ that generates the horizon when
${\cal{U}}(u)$ reaches its asymptotic 
behavior (eq.~(\ref{E.10})). Hence we need the
relation between $v,u$ and $ {\cal{V}} , {\cal{U}} $ in the vicinity of $H$
 only. More precisely we
need only the function $v( {\cal{V}} )$ in the vicinity of 
$ {\cal{V}} =0$ when $H$ is an infalling light cone since 
we have already the function ${\cal{U}}(u)$. 
 Near $ {\cal{V}} =0$ the function $ {\cal{V}}  (v)$ can be approximated
linearly $v - v_H = \kappa  {\cal{V}}  + O( {\cal{V}} ^2)$  where $\kappa$ is an
irrelevant constant which depends on the shell's speed at $ {\cal{V}} =0$.
Hence in the region where $H$ is an infalling light cone
the modes take the forms
\begin{equation}
\varphi_\omega ^{in}=
{e^{-i \omega v}
 \over \sqrt{4 \pi \omega}}
= {e^{-i \omega (\kappa  {\cal{V}} +v_H)} 
\over \sqrt{4 \pi \omega}} 
\end{equation}
We have used the fact 
that one may coordinatize the whole space with either $v$ or ${\cal{V}}$ 
and used the 
continuity at the star's surface.
After reflection at $r=0$, $H$ is an outgoing light cone and the in modes  
read 
\begin{equation}
\varphi_\omega ^{in}
 =-{e^{-i \omega (\kappa  {\cal{U}} +v_H)} 
\over \sqrt{4 \pi \omega}} 
 =  -
{e^{i \omega ( Ke^{-u/4M} + v_H)}
 \over \sqrt{4 \pi \omega}}
\label{HR.11}
\end{equation}
where $K(= \kappa  B 4M$, see eq.~(\ref{E.10})) is a positive constant. 
Note that the approximation \begin{equation}
{\cal{U}} =- B 4M e^{-u/4M}= BU
\label{HR.11b}
\end{equation}
 is rapidly excellent
since the corrections are of the order $e^{-u/2M}$. 
Therefore the asymptotic form eq.~(\ref{HR.11}) is 
also rapidly valid. In this asymptotic regime, 
the Hawking radiation is independent of $K$ because $K$ 
can be reabsorbed into a redefinition of the origin of $u$.
For convenience we choose $ K = 4M $ and drop the phase
$-e^{-i \omega v_H}$. 

The rest is copying out the results of Chapter 2. We repeat the
salient facts. Once more because the
Bogoljubov transformation does not mix $U$'ness and $V$'ness, 
each sector may be
handled independently. Concerning the $V$ modes, at this point, nothing need be
said, the modes being of the form $e^{-i \omega v}$ throughout. The in-vacuum
stays equal to the out-vacuum i.e. the vacuum defined by single particle
states in R.

The physics under present scrutiny is encoded in the overlap between
the $u$ part of the scattered in-modes 
given in eq.~(\ref{HR.11})
 and the out-modes $\varphi_\lambda ^{out}=e^{-i\lambda u}/
\sqrt{4 \pi \lambda}$:
\begin{equation}
\varphi_\omega^{in} = \int_0^\infty \! d \omega \left[
\alpha_{\la \om} \varphi_\lambda ^{out} + \beta_{\la \om}
\varphi_\lambda ^{out *}
\right]
\label{twotwobh}
\end{equation}
\begin{eqnarray}
\alpha_{\la \om} &=& \int^{+ \infty}_{-\infty} 
\! du\ \varphi_{\la}^{out *}(u) (-i\! \lr{ \partial_u})
\varphi_\omega^{in} ({{\cal{U}}(u)})
\nonumber\\
 \beta_{\la \om} &=& \int^{+ \infty}_{-\infty}
 \! du\ \varphi_{\la}^{out} (u) 
(-i\! \lr{ \partial_u})\varphi_\omega^{in} ({{\cal{U}}(u)})
\label{alphabh}
\end{eqnarray}
Since ${{\cal{U}}(u)}$ is correctly approximated by $BU$ only when the 
exponential corrections $e^{-u/2M}$ are negligible, there is no universal
form for $\alpha_{\la \om}$ and $\beta_{\la \om}$ when the Doppler shift is
small i.e. when $\om \simeq \la$, hence there is no steady flux in this
early stage. On the contrary, when the Doppler shift becomes  exponential,
the corrections fade out, and one does find a universal behavior
for $\alpha_{\la \om}$ and $\beta_{\la \om}$ when $\om >> \la$. This is due to the 
finite interval in $u$ in which these overlaps acquire their value.
Hence, for $\om >> \la$, one can replace the expressions for 
$\alpha_{\la \om}$ and $\beta_{\la \om}$ by the following expressions,
thereby making contact with the Rindler--Minkowski Bogoljubov transformation, 
\begin{eqnarray}
\alpha_{\la \om} &=& {1 \over 2\pi} { \sqrt{ \lambda} \over \sqrt{\omega}}
\int^{+ \infty}_{-\infty}
 du e^{i \lambda u} e^{i \omega 4M e^{-u/4M}} \nonumber\\
 &=& { 4M \over 2\pi} { \sqrt{ \lambda} \over \sqrt{\omega}} \int^0_{-\infty}
dU (-4MU)^{-i \lambda -1} e^{-i \omega U} \nonumber\\
 &=& {4M \over 2 \pi } \sqrt{\la \over \om}
\Gamma(-i4M \la)
\left( {a \over \om} \right)^{-i4M \la}
e^{\pi 2M \la }\nonumber\\ 
\beta_{\la \om} &=& 
 e^{- 4M\la} \alpha_{\la \om}^*
\label{alphabh2}
\end{eqnarray}
where we have used eqs (\ref{twotwo}).
Hence one finds 
a
steady thermal flux at the Hawking temperature $T_H= 1/ 8\pi M = 1/\beta_H$
since, for all $\om >> \la$, one has
\begin{equation}
\vert {\beta_{\la \om} \over \alpha_{\la \om}} 
\vert ^2 = e^{-8\pi M \lambda}
\label{alphabh2B}
\end{equation}
We note that the transitory regime can nevertheless be handled
 analytically in
some specific cases,
see for instance Appendix \ref{sphradd} and reference
\cite{MaPa}, and one explicitly verifies that there
is an early stage without significant flux which, after  a certain $u$-time, tends
exponentially fast to the asymptotic behavior.
One can then verify that modes with $l>0$ do not contribute significantly 
to Hawking radiation. This is seen by comparing the Hawking temperature with the
height of the centrifugal barrier in eq.~(\ref{HR3}).

Let us recall how a flux arises.
As in section 2.4, 
for a given $ \omega$ and $
\lambda$, the integrals eq.~(\ref{alphabh2}) arise from
the regions around the saddle $u^*$ at (see eq.~(\ref{threeeight}))
\begin{equation}
\mbox{ Re}(u^*(\omega, \lambda))
= 4M \ln \omega/\lambda
\label{reson}
\end{equation}
With Gerlach \cite{Gerl}, we call this saddle time the "resonance" time. 
The width of the significant region around
$u^*$ is independent of $ \omega $ and given by $\simeq \sqrt{4 M / \lambda}$
thereby justifying the replacement 
of eq.~(\ref{alphabh}) by eq.~(\ref{alphabh2}) for $\om >> \la$.
This independence of the width allows for the derivation of an exact
asymptotic formula for the rate i.e.  for the flux. First pick an
interval $ u_1  \leq u \leq
u_1 + \Delta u $ with $ u_1$ sufficiently large so that the 
correction term in eq.~(\ref{E.10}) can be neglected
and $\Delta u >> \sqrt{4 M /
\lambda}$.  We
wish to calculate the flux in R due to quanta of frequency $\lambda$. In this
interval the frequencies $\omega$ which contribute to 
$<a^\dagger_\la a_\la > $ (i.e. the  number of quanta of frequency
$\lambda$) satisfy
\begin{equation}
\lambda e^{u_1/4M} \leq \omega \leq \lambda
e^{(u_1+\Delta u) /4M}
\label{reson2}
\end{equation}
 Whence the rate is given, following eqs (\ref{extrab}, \ref{threenineb}), 
\begin {eqnarray}
\lim_{\Delta u \to \infty} 
<a^\dagger_\la a_\la >{1 \over \Delta u}  &=&
\lim_{\Delta u \to \infty}
{1 \over \Delta u} \int_{\lambda e^{u_1/4M}}^{\lambda e^
{(u_1+\Delta u) /4M}} d\omega \vert \beta_{\la \om} \vert ^2 \nonumber\\
&=&
\lim_{\Delta u \to \infty} 
{4M \over \Delta u} \int_{\lambda e^{u_1/4M}}^{\lambda e^
{(u_1+\Delta u) /4M}} {d\omega \over \omega}  
{1 \over e^{\beta_H \lambda}-1} { 1
\over 2 \pi }\nonumber\\
&=& (e^{\beta_H \lambda}-1)^{-1} {1 /2 \pi}
\label {1.36}
\end {eqnarray}
Since at fixed $r$, one has $ \Delta u = \Delta t$, 
$ <a^\dagger_\la a_\la > 
/ \Delta u$ indeed represents a particle flux 
per unit time. In this way the $\delta(0)$ which would come up by applying
naively eq.~(\ref{TRRRBBB}) is replaced by $\Delta u /2 \pi$. (This point has
been the subject of some misunderstanding in the literature. There is no
infra-red divergence in the number flux at finite $\la$! To define the total
number flux one should take into account the cutoff provided at small $\la$ by
the potential barrier.)
Similarly the
total energy flux is 
\begin {eqnarray}
< T_{uu} > = 
\int _{0} ^{\infty} \! {\lambda d \lambda \over 2 \pi}
( e^{
\beta_H \lambda} - 1 ) ^{-1} = { \pi \over 12} \beta_H^{-2} \label {1.38}
\end {eqnarray}

These results overestimate the flux since they do not take into account the
suppression of the emission of low frequency quanta $ \lambda \leq
\beta_H^{-1}$ due to the s-wave repulsive barrier. 

Unruh \cite{Unru1} has given a nice mnemonic device to represent these
results. Namely the steady state of Hawking evaporation (in the early stages
when the decrease in $M$ due to the radiation is neglected) is simulated by a
vacuum state - U(nruh) vacuum. This state is a cross between Schwarzschild
vacuum (also called Boulware vacuum \cite{Boul1}) and Kruskal vacuum (also called
Hartle-Hawking vacuum \cite{HaHa}). Outside the star, U-vacuum is vacuum of
Schwarzschild $v$
 modes (i.e. $e^{-i \omega v}/ \sqrt{4 \pi \omega}$) but the
Kruskal vacuum of U modes (i.e. $e^{-i \omega U}/ \sqrt{4 \pi \omega}$). 
This is simply
because in the collapsing situation the $v$ modes are $ e^{-i\omega v}$  
and the $u$ 
modes $ e^{i \omega  {\cal{U}} }$ (up to phases and irrelevant constants). The 
matching
conditions on the surface of the star have universally
related $ {\cal{U}} $ to 
$U$, eq.~(\ref{HR.11b}), when the shell is close to the horizon $H$.
This is the crux of Unruh's beautiful
isomorphism. 

One must nevertheless bear in mind the following
restriction of the Unruh mnemonic.
The reference to the Kruskal character of the $u$ part of the in modes is
after their reflection from $ r = 0 $. There is an absolute frame of
reference in this problem which is given by the movement of the star's
surface. This precludes the possibility of making arbitrary boosts 
(in Kruskal coordinates eq.~(\ref{1.5})) so as to
change the character of $ \omega, \lambda$ mixing in eq.~(\ref{reson}). 
Such boosts correspond to
translations in $t$, so that an event at $ t = 0 $
with $ \omega = \lambda$ at resonance in the boosted frame could result say in
$ \omega = \lambda e^{t_1/4M}$. The Unruh isomorphism is carried out in a
fixed frame e.g. the star at rest until $ t = t_0 $. 
The Gerlach resonance condition $ \omega / \lambda = e^{u^*/4M}$ 
which describes the
Doppler shift at and around $u^*$ cannot be boosted away.
This restriction will play an important role upon confronting the 
consequences of the very high frequencies $\om$
entering into the Bogoljubov coeficients eq.~(\ref{alphabh2})
as seen from eq.~(\ref{reson}).

A contemporary and elegant derivation of the concept of U-vacuum is
due to Hawking \cite{Hawk3} and Damour and Ruffini \cite{DaRu}.
Their method is the black hole
analog of the technique illustrated in eqs. (\ref {Unm}, \ref {threeoneb}). 
It is of some interest to sketch their method since it
makes contact with the tunneling methods of Section 1.2.

In  Eddington Finkelstein coordinates eq.~(\ref  {1.7}) 
the d'Alembertian near the
horizon oustide the star is $ - {\partial_x} ( {x \over 2 M}
{\partial_x} + 2\partial_v)$ where $ x = r - 2M$. The mode
equation is \begin {eqnarray}
- {\partial\over \partial x} [ {x \over 2M} {\partial \over \partial x} - 2
i \lambda] \chi_\lambda (x) e^{-i\lambda v} = 0 \label {1.44}
\end {eqnarray}
having solutions $ \chi_\lambda (x) = $ const. 
($v$-modes) and the $u$-modes 
\begin {eqnarray}
\chi_\lambda(x) =  A \left[ \theta (x) {(x)^{4 i M
\lambda} \over \sqrt {4 \pi \lambda}} \right] 
+ B \left[\theta (-x) {(- x )^{4i M \lambda}\over \sqrt {4 \pi \lambda}}
\right] 
\label {1.45}
\end {eqnarray}
One checks that for $ x> 0$, $e^{ i\lambda v} x^{-4 i M \lambda}( = e ^{i
\lambda u} )$ is indeed a $u$ mode in R. Completeness in the complete space
spanned by the  Eddington Finkelstein coordinates (collapsing star) requires the term in $ \theta
(- x)$ as well. We now require that in this complete space the modes be
positive Kruskal frequency. Since at $v$ fixed, $ \partial_x $ is
pure light like (it is in fact $-2 \partial_U  $)
$ \chi_\lambda
(x) $ must enjoy upper half analyticity in $x$. 
The normed modes
are thus
\begin{eqnarray}
\chi_\lambda = 
{e^{+\beta_H \lambda/2} \over \sqrt{ 2 \mbox{sinh} \beta_H \lambda}}
\left[
\theta(x) 
{(x)^{4 i M
\lambda} \over \sqrt {4 \pi \lambda}} \right] + 
{e^{-\beta_H \lambda/2} \over 
\sqrt{ 2 \mbox{sinh} \beta_H \lambda}}
\left[\theta (-x) {(- x )^{4i M \lambda}\over \sqrt {4 \pi \lambda}}
\right] 
\mbox{ $\lambda >0$}\nonumber \\
\label{unruhmi}
\end{eqnarray}
as in eq.~(\ref{Unm}). 
The modes $\chi_\lambda$ are simply the 
rewriting of the in-modes $\varphi_{\omega}^{in}$ which 
diagonalize the Bogoljubov transformation~(\ref{twotwobh}). 
Indeed eq.~(\ref{unruhmi}) shows that the modes $\hat \varphi_\la = e^{-i\la v}
\chi_\la$ ($-\infty < \la < + \infty$)  (hereafter called Unruh modes) can be written
in terms of the  Schwarzschild modes $\varphi_\la^{out}$ as
\begin{eqnarray}
\hat \varphi_\la &=& \alpha_\la \varphi_\la^{out} 
+ \beta_\la \varphi_\la^{out F *} \quad \la>0 \nonumber\\
\hat \varphi_{-\la} &=& \beta_\la \varphi_\la^{out *} 
+ \alpha_\la \varphi_\la^{out F }\quad \la>0 \label{unruhmii}
\end{eqnarray}
where we have introduced the modes \begin{equation}
\varphi_\la^{out F }
= (1/\sqrt{4 \pi \la})e^{-i\la v} (-x)^{-4 i M \la}\theta (-x)
\label{unruhmiii}\end{equation}

To make contact with Section 1.2, we note the similarity between the
differential equation for $ \chi_\lambda (x)$:  $(x \partial_x -2
i 4M \lambda ) \chi_\lambda (x) = 0 $ 
and eq.~(\ref{equchi}). The analogy goes quite
far. For example we take the Fourier transform of $ \chi_\lambda (x)
(=\tilde \xi_\lambda (p))$ we get $ ({\partial_p} p -2 i
\lambda) \tilde \xi_\lambda (p) = 0$. The modes $ \tilde \xi_\lambda ( p )
$ having been obtained by integration over all $x$
 are complete and in fact can
serve as a complete set of in-basis states
\cite{PaBr}. Their Fourier transform then gives
them as a linear combination of out states.
Each of these latter live on one
side or the other of $H$,  the horizon line $ x = 0 $.

This is then precisely the same 
mathematical mechanism used in eq.~(\ref{vchi}) to go from the
in-state (proportional to $ \theta (u)$) to the 
linear combination of states defined
in terms of the conjugate momentum to $ u (=v) $, a combination of
$\theta(v)$ and $ \theta (-v)$. 
The classical exponential approach in a given region
of space-time to a horizon is translated in each case to a quantum
formalism of this type. 
However, in the
black hole case, the transcription of the modes to a manifest tunneling is
ambiguous. We may write the operator $ ix \partial _x $ as $
[\Pi^2 - \xi^2]$ where $ \Pi = {1 \over \sqrt{2}} (iA {
\partial_x} + A^{-1}x )$ and $ \xi = {1 \over \sqrt{2}} (iA {
\partial_x} - A^{-1}x )$  with $ [ \Pi, \xi ] = -i $. But the constant $A$ is
arbitrary so tunneling in $ \xi$ is not only non local in $x$ but occurs at
arbitrary scales. 
Nevertheless, the above highlights that the collapse has produced pairs
living on both sides of $x$
 and that these pairs are the realization of vacuum
fluctuation which have "tunneled" into reality. The pair
formation is illustrated in Figs (\ref{pairpen}) and (\ref{pairedf})  in a
Penrose diagram and in
 Eddington Finkelstein coordinates. To this end we take into account the fact
when extrapolated backward in time they bounce off $ r = 0 $. Of course the $
\theta (U)$ piece is never seen at finite Schwarzschild time, but only its
backward reflected piece. As the wave progresses from $ {\cal I}_-$ , this
partner of the Hawking photon never manages to reflect, but,
at $t \to \infty$, simply crowds into
the horizon line extended into the origin.
\dessin{1.000}{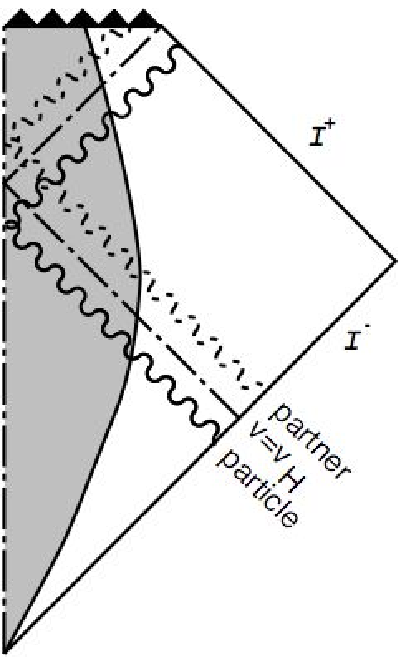}{pairpen}{The classical trajectory (stationnary phase) of a pair
produced in the geometry of a collapsing black hole represented on a Penrose
diagram. Only the member of the pair which reaches ${\cal I}^+$ is on shell,
the partner falls into the singularity.} 
\dessin{1.000}{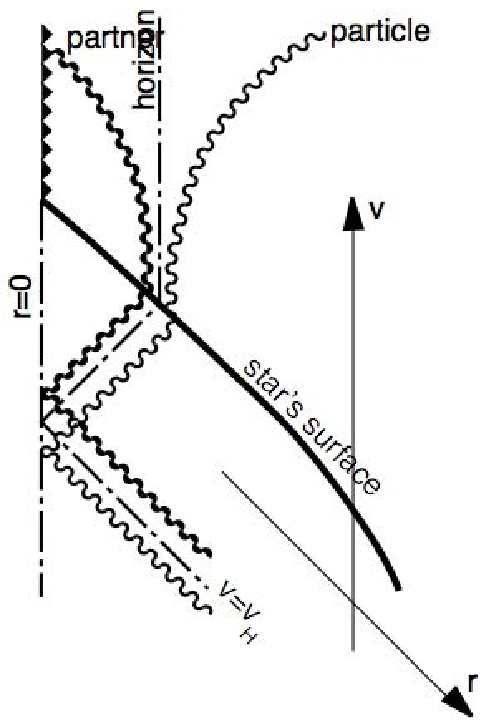}{pairedf}{The same as in the previous figure (3.5)
 but in Eddington Finkelstein coordinates.} 
\dessin{1.000}{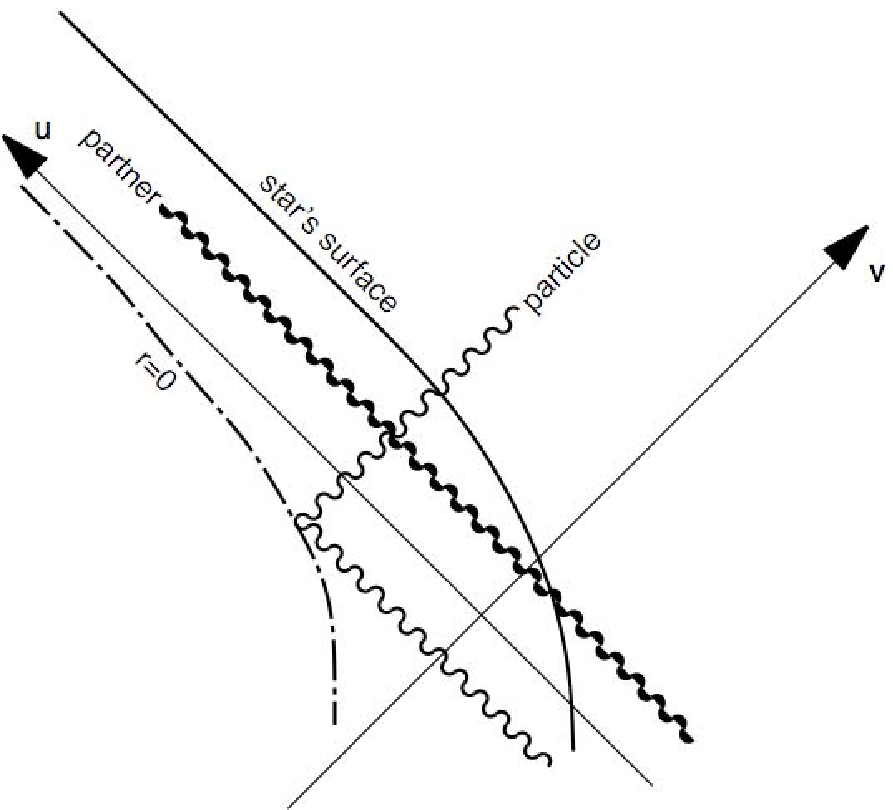}{pairuv}{Pair
formation drawn in $u,v$ coordinates.}

To compare with the accelerated mirror of Section 2.5 
we also have drawn in Fig.(\ref{pairuv})
the two pieces of $\hat \varphi_\lambda (U)$ in the extended $u,v$ system.
\par From these pictures and the above analysis it is seen that the conversion of
fluctuations into particles, unlike in Chapter 1 (Fig.~(\ref{F4})),
does not occur
over a well defined space-time domain. This 
is due to the absence of a scale, and it is
reflected in the arbitrariness of the parameter $A$ which appeared in the
definition of $\Pi$ and $\xi$ in the above paragraph (we note in passing that
the introduction of mass does not help since the frequencies of the modes
within the star rapidly become large compared to any known mass). Nevertheless,
though tunneling seems inappropriate as a tool in black hole physics (as it
appears in the present formulation) one should not lose sight of its
conceptual content. In chapter 1, particle production could have been
understood either in terms of tunneling of modes or of their backward
scattering in time according to the gauge choice. It is therefore useful to
define a generalized tunneling concept, the conversion of a mode from a domain
of virtuality (i.e. as a vacuum fluctuation) to a domain of realization
(i.e. measurable as a quantum of excitation, in a counter for example). The
process, in general, is caused by the imposition of an external field. In black
hole evaporation the process induced by the gravitational field of the
collapsing star is summarized succintly: virtuality on  ${\cal I}^-$ induces
reality on  ${\cal I}^+$. The precise formulation of this aspect of the problem
is the subject of Section \ref{VFHR}.

We also note here that we have come upon a rather unexpected fragility of the
theoretical foundations which have been set to work thus far. Is it reasonable
to expect modes to remain free and unmixed in their infinite excursion from 
 ${\cal I}^-$ to  ${\cal I}^+$? 
Indeed, eq.~(\ref{reson}) tells us that after a time of the order of $\Delta
u = M \mbox{ln} M$ the frequencies $\om$ of the vacuum fluctuations
converted upon resonance into Hawking quanta  exceed the Planck frequency.
More of this in Section \ref{troub}.


\section{Renormalized Energy Momentum Tensor in Unruh Vacuum.}\label{Tmunuren} 

One of the approaches used to address the back reaction to black hole
evaporation consists in solving the semi-classical equations \begin{equation}
G_{\mu\nu} =  8 \pi \langle T_{\mu\nu} \rangle \label{T1} \end{equation} wherein
the mean value of the energy momentum tensor is taken as source of Einstein's
equations. The fundamental assumption is that  the source terms are $ \langle
T_{\mu \nu} \rangle $ i.e. that the average solution is the solution of the
equations using the average source. When the fluctuations are large, or when one
wishes to evaluate the importance and the consequences of the fluctuations, one
has to analyze other matrix elements than the mean appearing in eq.~(\ref{T1}).
This is why we shall study in Section \ref{VFHR} non diagonal matrix elements of
the operator $T_{\mu \nu}$.

The mean energy momentum tensor which appears on the r.h.s. of eq.~(\ref{T1}) is
formally infinite and must be regularized  and renormalized before starting to
solve eq.~(\ref{T1}). In this section we deal with this part only, in the next 
section we shall treat the gravitational back reaction in this semi-classical
approximation.

Until this time, the complete four dimensional computation of $\langle
T_{\mu\nu}\rangle$ has not yet been carried out.
What has been achieved is a good approximation to  $\langle
T_{\mu\nu}\rangle$ in the static spherical symmetric case \cite{How}. In
the evaporating situation no analytical expression for the mean energy
momentum tensor has been obtained. However we recall that most of Hawking
radiation is due to s-waves and that, for the s-waves, the problem
simplifies a lot if one drops the residual centrifugal barrier.
(Indeed one ends up with a conformally invariant two dimensional theory.)
Hence one expects that the essential properties of the four dimensional theory
are recovered from the effective two dimensional one.

We shall therefore continue with this truncated theory
and calculate analytically $\langle T_{\mu\nu} \rangle$ in the Unruh vacuum
following ref.\cite{DFU}.
 Since the modes in the two dimensional model are rescaled by a factor
of $\sqrt {4 \pi} r$ as compared with the original field (see eq.~(\ref{HR2})),
the fluxes should be rescaled by $4 \pi r^2$
\begin{equation}
\langle T_{\mu\nu}^{4 D}\rangle = {1 \over 4 \pi r^2} \langle T_{\mu\nu}^{2 D}\rangle
\label{rescaled}
\end{equation}
This equation applies for $\mu, \nu = U, V$ and we shall see that Bianchi
indentities imply $\langle T_{\theta \theta}\rangle = 0$.

The renormalization of the energy momentum tensor has already been carried out,
in two dimensions,
 in the case of a flat background geometry in Section \ref{mirro} and we recall
the result, eq. ( \ref{A9})  (extended to the  V modes as well)  
\begin {eqnarray} \langle T _ {UU}\rangle _ {fg}
- \langle T _ {UU} \rangle _0 & =&  {1 \over 12 \pi} f^{\prime {1/2}} \partial
_{U} ^{2} f ^{\prime {-1/2}} \nonumber \\ \langle T _ {VV}\rangle _{fg}- \langle
T _{VV}\rangle _ 0 & = & {1 \over 12\pi} g^{\prime {1/2}} \partial _{V}^{2}
g^{\prime {-1/2}} \label {TT2} \end {eqnarray} Here $\langle T_{UU}\rangle _0 $
means Minskowski vacuum expectation value and $U$ means inertial Minkowski
coordinate. $ \langle \quad \rangle _{fg} $ means the average with respect to
the vacuum defined by the modes  $ e ^{-i \omega f (U)} /\sqrt{4 \pi \omega}$
and  $ e ^{-i \omega g(V)} /\sqrt{4 \pi \omega}$. An important property of
eqs~(\ref{TT2}) is that they can be inverted so as to express $\langle T_{ff}\rangle$
and $\langle T_{gg}\rangle$ in terms of the inverse functions $U(f)$ and $V(g)$:
all one does is flip the sign and replace $f,g$ by their inverse functions, as is
required by the reciprocity of these relations.  

In a curved background we shall generalize the subtraction (\ref{TT2}) by
subtracting from  $\langle T _ {\mu\nu}(x)\rangle $ the value  $\langle
I(x)\vert  T_{\mu\nu}(x)\vert I(x) \rangle$ calculated from the inertial modes
at $x$ \cite{MPB2},  i.e. those modes which most resemble Minkowski modes at
$x$. 
 This is taken on the principle that flat space is a solution, i.e. that the
expectation value of $T_{\mu\nu}$ in Minkowski vacuum is zero. The natural
generalization is to pose that in the local vacuum $\vert I(x) \rangle$ of local
inertial modes,  $\langle I(x)\vert  T_{\mu\nu}(x)\vert I(x) \rangle_{ren} =0 $.
For an arbitrary state one then postulates that $ \langle T_{\mu \nu} (x)
\rangle _{ren}/ 4\pi r^2 $ is the gravitational source  where  \begin
{eqnarray}    \langle T _ {\mu \nu} (x) \rangle _{ ren} =  \langle T_{\mu
\nu}(x) \rangle - \langle I(x)\vert  T_{\mu\nu}(x)\vert I(x) \rangle \label {T2}
\end {eqnarray}  Furthermore, one must implement this difference of infinities
with a regularization scheme which as in the previous chapters is taken to be
the split point method e.g.  
 \begin{eqnarray} \langle T_{UU} \rangle = \lim_{U \to U^ \prime}   \partial_U
\partial _{U ^\prime} \langle \phi (U) \phi (U^ \prime) \rangle \label {T3}
\end{eqnarray}

To compute $ \langle  T_{\mu\nu} \rangle_{ren}$ we first need to construct
inertial coordinates about $x$. To this end, we express the spherically
symmetric geometry, in a "conformal" gauge $ ( g_{UV}=C, g_{UU} = g_{VV} = 0 $
everywhere)  \begin {eqnarray}  ds^2 = - C(U,V) dU dV + r^2(U, V) d\Omega^2
\label {Ttwo} \end {eqnarray} 
The residual reparametrization invariance $U \to
U'(U), V \to V'(V)$ is fixed by requiring that the state of the field, in which
one wants to compute  $ \langle T_{\mu\nu} \rangle_{ren}$ is vacuum with respect
to the modes $e^{-i\om U}, e^{-i\om V}$.  In this way the conformal factor
$C(U,V)$ encodes the vacuum state of the $\phi$ field as in the flat exemple eq.
(\ref{A8}).

The inertial set $\hat U, \hat V$ based on the $U,V$ set is \begin{eqnarray}
\hat U - \hat U_0 & = & \int _{U_0} ^{U} \quad {C(U ^\prime, V_0) \over C (U_0,
V_0)} dU^\prime \nonumber \\ \hat V - \hat V_0 & = & \int _{V_0} ^{V} \quad
{C(U_0, V^ \prime) \over C(U_0, V_0)} \quad dV^\prime \label {T4}  \end{eqnarray}
 whereupon \begin{eqnarray} ds^2 = - {C(U,V) C^2 ( U_0, V_0) \over C(U_0, V)
C(U, V_0)}  d \hat U d \hat V + r^2(U, V)d\Omega^2
 \label{T5} \end{eqnarray} with $  U, V$ in eq.~(\ref{T5}) functions of $\hat U, \hat
V$ obtained by inversion of eqs~(\ref{T4}).  The coordinates $\hat U, \hat V$ are
affine parameters along the radial  light--like geodesics $U=U_0$ and $V=V_0$
which pass through $x=(U_0,V_0)$. Hence they are the inertial set at $x$ which
we shall use to define the local vacuum $\vert I(x) \rangle$ since they
constitute the most inertial parametrization of the light like geodesics
$U=U_0$, $V=V_0$ (in particular the Christoffel symbols vanish at $x$ in
coordinates $\hat U, \hat V$). One may now apply eqs~(\ref{TT2}) in terms of the
function $ U(\hat U)$. Lest the reader have no misunderstanding of the approximate 
status of the following computation, it comes from mode analysis, not from the form of the metric.
 We consider
only the contribution of the s-waves and in addition drop the residual barrier so as
to make the effective field theory conformal in the $U,V$ sector.

One identifies the Minkowski coordinates
of eqs~(\ref{TT2}) which defined the subtraction,  with the coordinates $\hat U, \hat
V$ of eq.~(\ref{T4}) and similarly one identifies the functions $f,g$, which defined
the state of the field, with the functions $  U(\hat U), V( \hat V)$  whereupon 
\begin{eqnarray} \langle T_{ UU} \rangle _{ren} & = &  -{1 \over 12 \pi} C^ {1/2
}\partial_{U}^{2} C^{-1/2}  \nonumber \\ \langle T_{VV} \rangle _{ren} & = & 
-{1 \over 12 \pi} C^{ 1/2} \partial_{V} ^{2} C^{-1/2}  \label {T6} 
\end{eqnarray} where we have used the reciprocity mentioned after eqs~(\ref{TT2}).

Moreover, in addition to eqs~(\ref{T6}),
 the trace of the energy momentum tensor ${\rm tr} \ T_{\mu\nu} = 4  T_{UV} /C =
m^2 \phi^2$ no longer vanishes even though classically it vanishes for a
massless scalar field.  This is because $C$ is a function of both $U$ and $V$
contrary to $f$ in eqs~(\ref{TT2}). Indeed,  requiring that the renormalized
energy momentum tensor be conserved
 \begin{equation}T^V_{\ U;V} ={1 \over C}T_{UU,V} + (C^{-1} T_{UV}),_U=0
\label{TT5}\end{equation} and the similar equation for $T^U_{\ U;V}$ and
requiring also that $T_{UU}$ and $T_{VV}$ be given by eqs~(\ref{T6}) imposes a trace
given by  \begin{equation} \langle T \rangle_{ren} = 2 g^{UV}  \langle T_{UV}
\rangle _{ren} = {1 \over 24 \pi} R \label{trace} \end{equation} where  $R=
{\square}\ln C = 4 C^{-1} \partial_U\partial_V \ln C$ is the
 two dimensional curvature.  Equation~(\ref{TT5}) is the energy conservation in two
dimensions but we emphasize that
it is also valid in four dimensions if one divides the two dimensional
fluxes by $4 \pi r^2$ and if one sets $T_{\theta \theta}
$ to zero. 
Then our $\langle T_{\mu \nu} \rangle _{ren}$, divided by $4
\pi r^2$,
 is a legitimate source for the Einstein equations in four dimensions.

One sees that the result $\langle T \rangle_{ren}=0$ when $m^2=0$ is erroneous
because one has not taken into account  the change in the local vacuum upon
changing the point $ x= (U_0,V_0)$ which leads to the non-vanishing of
the first term of the r.h.s. of eq.~(\ref{TT5}). More refined regularizations
leads to the result directly \cite{DFU} \cite{BD}. One can also interpret the
origin of a non vanishing trace by remarking that the presence of a curvature
within the Compton length of a vacuum fluctuation causes the existence of a term
$ O(R/m^2)$ which must be subtracted to keep the one loop approximation to the
effective gravitation action finite.

Note also that the 
 quantum trace given in eq.~(\ref{trace}) is  a pure geometric quantity, the 
same for all states\footnote{
This is true even in the full four dimensional theory wherein the trace 
anomaly is a quadratic form of the Riemann 4-tensor.}. Indeed 
the difference in energy between two states (of
which eqs~(\ref{TT2}) is a particular example) is given by conserved $\Delta
T_{UU}$ and $\Delta T_{VV}$, hence with $\Delta T_{UV}=0$.

The above procedure of subtracting $\langle I \vert T_{\mu\nu} \vert I\rangle$
can in principle be applied in the original four dimentional
theory and be generalized to fields with mass (see ref.\cite{Massar}).
 A number of alternative procedures have been devised
for renormalizing the energy momentum tensor (see ref.\cite{BD} for a review).
Happily they are all equivalent: indeed Wald \cite{Wald2} has shown that if the
renormalized energy momentum tensor satisfies  some simple and natural
conditions it is completely fixed (see nevertheless ref.\cite {OttewillBrown}). 

 We now apply eqs (\ref{T6}) to our problem. As a warm-up let us start with Boulware
vacuum. This is the case one would have if the star were eternally static at
some fixed radius $ (= R_0)$ greater that its Schwarzschild radius $ (= 2M)$ and
we shall have in mind the case $ {R_0 - 2M} << 2M$.

In the Schwarzschild region, the modes $ e^{-i \lambda u}/ \sqrt{4 \pi \lambda}$
and $ e^{-i \lambda v}/ \sqrt{4 \pi \lambda}$ define the Boulware ($B$) vacuum
(e.g. usual vacuum at  $r=\infty$) . The conformal factor $C$ is  $  (1 - 2M /
r) $ in the Schwarzschild coordinates $u,v$ defined in eq.~(\ref{1.4}). Since $
C(u,v) = C(v-u) $ we see immediately from eqs (\ref{T6}) that $ \langle T_{uu}
\rangle_B = \langle T_{vv} \rangle_B $ thereby implying no flux, $ \langle
T_{rt} \rangle_B = 0 $,
 as it should be for a static situation. Moreover since $\lim_{ r \to \infty }
\partial_r C (r) = 0  $ we have $ \langle T_{uu}\rangle _{ B} = \langle T_{vv}
\rangle _{B} = 0$ at $ r = \infty $. This is  as it should  be because the
 Schwarzschild metric  is asymptotically flat and because the Schwarzschild 
modes are identical to the usual Minkowski modes for large $r$.

It is also very easy to obtain $ \langle T_{uu} \rangle_B$ in  the vicinity of $
r = 2M $. Indeed, near $ r = 2M $, \begin{equation}
 C(r) \simeq - e^{r^*/2M} = -e^{(v-u)/4M} \label{cuv} \end{equation}
 so that from eqs~(\ref{T6})  \begin{equation}
  \lim_{r \to 2M} \langle T_{uu} \rangle_B = \langle T_{vv} \rangle_B = - (1/12
\pi) (1/6 4 M^2) =- (\pi/12) T_{H}^{2} \label{minus} \end{equation} i.e. minus
the asymptotic Hawking flux eq.~(\ref{1.38}). The origin of this negative energy
density lies on the Rindler character of the Schwarzschild geometry expressed in
the Schwarzschild coordinates $u,v$ in eq.~(\ref{cuv}). Indeed, the Minkowski
metric expressed in Rindler  coordinates (eq. (\ref{onesix})) reads
\begin{equation} ds^2= -dU dV = - e^{a(v-u)} du dv. \end{equation} Thus the
specification of being in Boulware vacuum  becomes, near $r=2M$, equivalent
 to being in Rindler vacuum in Minkowski space (see eqs. (\ref{threethree},
\ref{trindl})).  

It is to be observed that $\langle T_{uu} \rangle_B$ being finite on the future
horizon $u \to \infty$ ($ U \to 0$) leads to singular values of $\langle T_{UU}
\rangle_B$ on $H$: $\langle T_{UU} \rangle_B = (du / dU )^2 \langle T_{uu}
\rangle_B = (4M/U)^{2} \langle T_{uu} \rangle_B$. Similarly $\langle T_{VV}
\rangle_B$ blows up on the past horizon ($V=0$) which exists in the complete
Schwarzshild space. If this situation would pertain to the collapsing case (i.e.
to the true physical state of affairs) one would arrive at a catastrophic
situation in that as $U \to 0$, the values of $\langle T_{UU} \rangle$ would
tend to $-\infty$; and it is $\langle T_{UU} \rangle$ which is close to that
measured by an inertial observer near the horizon (since $ds^2 = - e^{-1} dU dV$
on the horizon, see eq.~(\ref{1.6})). The accomodation to this singular behavior
is a remarkable feature of black hole evaporation.

Indeed, in the collapsing case, in the Unruh vacuum, whereas the $v$-modes remain
$ e ^{-i \lambda v}/ \sqrt{4 \pi \lambda}$ in the outer Schwarzschild region
 the $u$-modes rapidly behave like $ e^{-i\omega U}/ \sqrt{4 \pi
\omega}$, see eqs.  (\ref{HR.11}, \ref{HR.11b}) and the discussion after eq.
(\ref{reson2}). Without  calculation we see from eqs~(\ref{T6}) that near the horizon
$\langle T_{uu}\rangle_U$ vanishes (where the subscript $U$ refers to Unruh
vacuum). Indeed space is regular near the horizon and the inertial modes differ
only slightly from the Kruskal modes. Therefore $ \langle T_ {UU}({r = 2M})
\rangle _U $ is finite and $\langle T_{uu}\rangle_U =({U/ 4M})^2 \langle
T_{UU}\rangle _U$ vanishes quadratically on the horizon $U=0$. The former
singularity is thus  obliterated. The quadratic vanishing is necessary for
having no singularity for a free falling observer since  the combined effect of
the gravitational  and  Doppler shifts already encountered in eq.
(\ref{g00}) is always present for any inertial trajectory crossing the future
horizon.  In refering once more to the isomorphism between this situation and 
the Minkowski-Rindler case, we see that the $u$ part of the state in the 
U-vacuum behaves like a regular Minkowski state on the horizon $U=0$. 

One can now either proceed by calculation to find $ \langle T_ {uu} \rangle _{U}
$,  using eq.~(\ref{T1}) to go from Boulware to Unruh vacuum (with $ g(v) = v; $ $
f(u) = U(u) = - 4M e^{-u/4M}$
 see eqs. (\ref{1.5}, \ref{HR.11b})), or easier yet (and perhaps more physical)
appeal to the equation of conservation (\ref{TT5}). Indeed $\langle T_{uu} \rangle
_{U} $ differs from $\langle T_ {uu} \rangle _{B}$  by a function of $ u$ only.
But this function of $u$ must be a constant because we are in a steady state
characterized by a rate eqs. (\ref{reson}, \ref{reson2}).  Whence
\begin{eqnarray} \langle T_{uu}(r) \rangle _U = \langle T_{uu} (r) \rangle _{B}
- \langle T_{uu}(2 M) \rangle _{B}
 \label {4.54}  \end{eqnarray} i.e. the constant is fixed by $ \langle T_{uu}
\rangle _{U} = 0 $ at $ r = 2M$.  Hence, from eq.~(\ref{minus}), the constant
flux  $ T^r_{\ t}  = T_{uu} - T_{vv}$ is  given, in Unruh vacuum,
 by $\langle T_{uu}(r=\infty) \rangle _{U} = (\pi/12)T^2_H$ which is the thermal
flux at the Hawking temperature as in the mode analysis of Section \ref{hawking}.

The behavior of  $ \langle T_{\mu\nu} \rangle $ for finite $r$ is obtained from 
eqs~(\ref{T6})  with $C = (1 - 2M / r)$ and $  \partial _{r^*} = (d r / d{r}^{*}) 
\partial_r$: \begin{eqnarray} \langle T_{vv} \rangle_U&= &\langle T_{vv}
\rangle_{B} =  \langle T_{uu} \rangle_{B} \nonumber\\&=&{\pi \over 12}T_H^{2} (
 {48 M^4 \over r^4}-{32 M^3 \over r^3}  )\label {T9} \end {eqnarray} and
\begin{eqnarray} \langle T_{uu}\rangle_{U} = {\pi \over 12}T_H^{2} (1 - {2 M\over
r})^2 (1 + {4 M \over r} + {12 M \over r^2}) \label {T10} \end{eqnarray}
wherein the quadratic vanishing at $r=2M$ and the asymptotic behavior are 
displayed.

All this is without back reaction i.e. without taking into account the decrease
of $M$ in time. But the  energy conservation (the Einstein equations at large
$r$) dictates that there is a necessary backreaction wherein
 \begin{eqnarray} {dM \over dt}  =  \langle T_{rt}\rangle _{U}
 = - {\pi \over 12} \left( {1 \over 8 \pi M} \right)^2 \label{T13} \end{eqnarray}
We recall that the flux $T_{rt}$ should be integrated over the sphere and that
our $T_{\mu \nu}$ has been rescaled by  $4 \pi r^2$, therefore eq.~(\ref{T13}) is
the usual four dimensional  expression for the mass loss. Upon correcting for
the existence of the potential barrier and adding the contribution of the higher
angular momentum modes $l\geq 1$ one finds 
 \begin{equation} {d M \over dt} = - \xi {1 \over M^2} \label{T14} \end{equation}
where $\xi $ depends on the spin of the radiated field \cite{page}.  Then, if
one assumes that at later times the rate of evaporation is given by the  same
equation, the decay time for complete evaporation is $ \tau_{decay} = M^3 /3
\xi$ in Planckian units.

In eq.~(\ref{T13}) the sphere over which one calculates can be situated at any
value of $r$ since the flux is conserved.  But the interpretation of the
integral changes with $r$. At large $r$ one has $\langle T^r_t \rangle_U =
\langle T_{uu} \rangle_U  $. This is a traditional positive energy outflow.
Instead,  near the horizon $\langle T^r_t \rangle_U = -\langle T_{vv} \rangle_B
$ since $\langle T_{uu} \rangle_U $ vanishes there.  How these two properties
contrive to modify the metric and describe a black hole with a slowly varying
mass parameter is the subject of the next section.


\section{The Semi-Classical Back Reaction}\label{semicc}

Up to this point in our primer we have concentrated on the
response of matter to a fixed background geometry, that
of the collapsing star. The dynamical response to this
time dependent geometry is the emission of energy to
infinity accompanied by an accommodation of the vacuum
near the black hole's horizon. 
Indeed 	at the horizon the
mean flux is carried by $\langle T_{vv} \rangle$ only. As
emphasized at the end of the previous section, the rate of change
of mass  is $
\langle T_{rt} \rangle\vert_{fixed \ r}$ where the
value of $r$ is arbitrary.
 Hence near the horizon the
description of the mass lost by the star is given in
terms of a halo of negative energy which accumulates
around the horizon.

The above discussion then suggests the shrinking of the
area of the horizon. But how does a significant reduction of
the area, hence a large change in the metric, affect 
$T_{\mu\nu}$, and how in turn do these
affect the metric? The complete answer to this question
(i.e. wherein one takes into account the 
full quantum properties of the operator $T_{\mu\nu}$) is 
the subject of present research and is far from
resolution -very far indeed in that its ultimate
elucidation may well entail (or lead to) the quantum
theory of gravity (see Section \ref{troub} in this regard).

Until the present time the only quantitative treatment
available is in the context of the semi-classical theory
i.e. gravity is classical and Einstein's equations are
driven by the mean value of $T_{\mu\nu}$ 
(see eq. (\ref{T1})). Since the latter
is a function of $g_{\mu\nu}$ and its derivatives, one
has highly non trivial equations to solve. In what
follows we shall report on what is known concerning this
enterprise. We first give  a semi quantitative
description of what happens. It is then followed by
more rigorous considerations and the presentation of 
the properties of the evaporating geometry obtained by numerical 
integration
of a
simplified model.

What are the features that one wishes to display? The
first concerns the validity of eq. (\ref{T13}) at later times. Is
the mass loss at time $t$ determined by $M(t)$, the mass
at that time, through Hawking's equation? In other words
is the system Markoffian? One finds that the answer is
yes, and that it is due to the continued negative energy
flux across the horizon. This results in an outside
metric which is determined by $M(t)$ and not by history.
The second question concerns the distribution of this
negative energy. Is it a halo, or is it distributed
homogeneously within the collapsing star so as to give
rise to a sort of "effective'' star of mass $M(t)$? One
finds the former. There is a halo which accumulates in a
region enclosed between the surface of the star and the
 horizon,  which we now explain on qualitative grounds.

In order to describe the geometry near the horizon it is
necessary to choose a coordinate system. A convenient
choice is the Eddington-Finkelstein set ($r,v$) 
introduced in eq. (\ref{1.7}).
The advantages of these
coordinates are: 

1) they cover both sides of the future
horizon; 

2) in the case of a purely ingoing light
like flux, the metric
\begin{equation}
ds^2 = - (1 - {2 M(v) \over r}) dv^2  + 2 dvdr+  r^2
d\Omega^2
\label{semia}
\end{equation}
where the only parameter is 
\begin{equation}
M(v)= \int^v dv T_{vv}
\label{semib}
\end{equation}
is an exact solution of Einstein's equations (the Vaidya solution). 
We remind the reader that we are working with 
energy momentum tensor rescaled by $4 \pi r^2$, see eq. (\ref{rescaled}). 
In the
evaporating situation near the horizon the flux is
entirely carried by $T_{vv}$ hence \ref{semia} is valid
in the vicinity of the horizon and can be used to
describe the physics there.

We are interested in the behavior of outgoing light
rays because the modes 
which give rise to Hawking radiation flow along these light rays. 
Let us first recall what values of $r$ a light ray,
reflected from $r=0$, visits without back reaction (i.e. with 
$M= M_0$). Photons which emerge from the
star at $r=2M_0 + \ge$ (with $\ge >0$) proceed to infinity
after adhering to the horizon $r=2M_0$ for a while 
(a $v$ lapse of order $M_0\ln(\ge/ 2M_0)$). Those
which emerge at $r=2M_0 - \ge$ also adhere to the horizon
for a while before falling into the singularity. These
trajectories describe the locus 
\begin{equation}
u=const=v-2 r^* = 
v- 2r - 4M_0 \ln \vert r/ 2M_0-1 \vert
\label{ueq}
\end{equation}
for $r$ near the horizon.
The main point is the existence of a horizon at
$r=2M_0$ which separates the out's from the in's, those
which escape from those which are trapped and crash into the singularity.

Now suppose that the mass decreases due to the absorbtion of
negative energy: $\langle T_{vv} \rangle \simeq -1/M^2$.
 Then the locus which separates the out's
from the in's is expected to shrink. Thus outgoing light
rays feel a slowly diminishing  gravitational potential
and some of those which previously were trapped may now
escape after having been sucked in for a while. The locus
where they cease to fall in and start to increase in
radius is called the apparent horizon. To describe this
quantitatively we consider the equation for outgoing light
rays when the mass is varying, in the metric \ref{semia}
\begin{equation}
{dr \over dv} = {1 \over 2} {r-2 M(v) \over r}
\label{semic}
\end{equation}
The apparent horizon is the locus where $dr/dv =0$, i.e.
\begin{equation}
r_{ah}=2M(v)
\label{rah}
\end{equation}
And it shrinks according to eq. (\ref{semib}) 
 i.e. $dr_{ah}/dv \simeq -1
/ r_{ah}^2$. However the apparent horizon does not separate
the geodesics which will ultimately fall into  the
singularity from the escaping ones since it is not an outgoing geodesic (for an evaporating
black hole, when $T_{vv}$ is negative, it is time like
 \cite{EH}). It 
is nevertheless very close to the event horizon (the light
ray which does separate the two classes). An estimate of the 
radius of the event horizon is
the inflexion point of the outgoing geodesics $d^2r/d^2v =0$: 
\begin{equation}
r_{eh}(v) = 2 M(v) + 8 M
dM/dv =  r_{ah}(v)  - O(1 / M)
\label{reh}
\end{equation}

 Therefore all the degrees of freedom which are in
the whole region $r<2M(v) -O(1/M)$ remain, in this semi-classical
treatment, inaccessible to the
outside observer.
This inaccessibility implies that the loss of mass does
not come from the evaporation of degrees of freedom from
inside to outside. Rather, it is due to the accumulation
of negative energy. 

The slow rate of change of $r_{ah}$ suggests that the processes which 
occur in this geometry are the same as in the case with no backreaction and
with $M$ equal to $M(v)$. For instance an important
property of the outgoing light rays in the
geometry \ref{semia} is that they are given, in good approximation,
 by the usual
formula, eq. (\ref{ueq}), with $M$ replaced by $M(v)$
\begin{equation}
v-2r -4M(v) \ln (r-2M(v)) =u
\label{semid}
\end{equation}
provided
\begin{equation}
M>>r-2M(v) >> 1/M
\end{equation}
where the restrictions come from the rate of change: $dM/dv  \simeq 1/M^2$.
Hence when this is valid the out-going geodesics are a scaled
replica of what happens without back reaction with $M$
replaced by $M(v)$. 

Based on the above one may conjecture that the radiation emitted at time $v$ is
also a scaled replica of the radiation emitted in the absence of back reaction
with $M=M(v)$. To understand why the evaporation is only controlled by $M(v)$
(and therefore why the past 
history of the black hole plays no role) and to have a full
appreciation of the processes which occur around the apparent horizon 
(where eq. (\ref{semid}) is not valid)
one must
resort to a more detailed analysis. This is done in the following paragraphs.
The upshot is that all that has been discussed qualitatively is correct to $O
(1/M)$.

[ Following Bardeen\cite{Bard} and York\cite{York} who
followed up the ideas of 
Hajicek and Israel \cite{HI} we shall use the metric 
\begin{equation}
ds^2=-e^{2 \psi}(1-2m(v,r)/r)dv^2 + 2 e^\psi dvdr 
+r^2 d\Omega^2
\label{c1}
\end{equation}
which describes a general spherically symmetric space-time.
In these coordinates Einstein's
equations are
\begin{eqnarray}
{\partial m \over \partial v} &=&  T^r_{\
v}\nonumber\\
{\partial m \over \partial r} &=& - T^v_{\
v}\nonumber\\
{\partial \psi \over \partial r} &=& 
T_{rr} /r
\label{c2}
\end{eqnarray}
Note that $\psi$ is defined only up to the addition of an arbitrary
function of $v$ corresponding to a reparametrization of the $v$ coordinate.
For simplicity we shall suppose that the right hand side 
of (\ref{c2}) is given by the two dimensional
renormalized energy momentum tensor discussed in section \ref{Tmunuren} (see
eq. \ref{T6} and \ref{trace} ). However the proof is general provided $M>>1$. 
The r.h.s. of \ref{c2} can be taken to be the full 4 dimensional renormalized
energy momentum tensor.

We shall proceed in three steps following \cite{Massar2}: first we
shall suppose that the renormalized energy momentum tensor resembles the
renormalized tensor in the absence of back reaction. We shall then show
that under this hypothesis the metric coefficients in
(\ref{c1}) are slowly varying functions of $r$ and $v$. Finally we shall solve
adiabatically the Klein Gordon equation in this slowly varying metric and
show that the renormalized tensor indeed possesses the
properties supposed at the outset thereby proving that the calculation is
consistent.

Our first task is to obtain estimates for $T_{\mu\nu}$ both far and near
the black hole. We begin with the former. We suppose that when $r$
is equal to a few times $2m$ (say $r=O(6m)$) there is only an outgoing flux
$T_{uu}(r>>2m)=L_H(u)$ where $L_H $ is the luminosity of the black
hole. This is
justified since in the absence of back reaction the other components of
$T_{\mu\nu}$ decrease as large inverse powers of $r$. For instance the trace
anomaly, in our 
2 dimensional problem decreases as $M/r^3$ in the static Schwarzschild
geometry. 
 We shall also suppose that $L_H$ is small 
(i.e. $L_H M^2 =O(1)$) and
varies slowly. Hence when $r>O(6M)$ an outgoing Vaidya metric is an exact
solution of Einstein's equations \begin{eqnarray} &ds^2 = -(1-2 M(u)/ 
r)du^2 - 2 du dr +r^2 d\Omega^2&\nonumber\\ &M(u) = \int^u \! du' \
L_H(u')&\label{c3} \end{eqnarray} 
The change of coordinates from (\ref{c2}) to (\ref{c3})
is obtained by writing the equation for infalling radial null geodesics in
the metric (\ref{c3}) as
\begin{equation}
F dv = du +  {2 dr\over 1 - 2 M(u)/r}\label{cIV}
\end{equation}
where $F$ is an integration factor. Upon using (\ref{cIV}) to change
coordinates from the set $(u,r)$ to $(v,r)$ one
finds that $e^{\psi}=F$ and $m(r,v)=M(u)$. 
Hence when $r>O(6M)$ the r.h.s. of (\ref{c2})
is given by
\begin{eqnarray}
T^r_{\ v}(r>O(6M)) &=& -e^\psi L_H\nonumber\\ 
-T^v_{\ v}(r>O(6M)) &=& 2 L_H/(1-2M/r)\nonumber\\
T_{rr}(r>O(6M))/r &=& 4 L_H / r (1-2M/r)^2\label{cc}\end{eqnarray}

We now estimate $T_{\mu\nu}$ near the horizon (i.e. near $r_{ah}=2m(r_{ah},v)$)
by assuming that the
energy momentum tensor measured by an inertial observer falling across the
horizon is finite and of order $L_H$. 
Near $r=r_{ah}$ one may neglect $g_{vv} = - e^{2 \psi} (1 - 2 m/r)$ and use
the reparametrization invariance of $v$ to choose $\psi(r_{ah},v)=0$
whereupon the metric becomes $ds^2 \simeq 2 dv dr + r^2 d \Omega ^2$. Hence
near $r_{ah}$, $r$ and $v$ behave like inertial light like coordinates (ie.
the proper time of an inertial infalling observer near the apparent horizon
is $cv + c^{-1} r$ where $c$ is a constant which depends on the precise
trajectory of the observer). That the energy momentum is of order $L_H$ near
the horizon is reexpressed as (for the components $T^v_v$ and $T_{rr}$)
\begin{eqnarray} T^v_v (r\simeq r_{ah}) &=& O(L_H)\nonumber\\ T_{rr} (r\simeq
r_{ah}) &=& O(L_H)\label{rahT} \end{eqnarray}
Where we have used the inertial character of the set $v,r$ near $r_{ah}$. In
addition we shall determine $T^r_v(r\simeq r_{ah})$ by making appeal to the
conservation of energy
\begin{equation}
T^r_{\ v,r} + T^v_{\ v,v} =0
\label{c8}
\end{equation}
where $T^v_{\ v,v} = O(L_{H,v})$. Integrating the conservation equation
from $r=2m$ to $r=O(6m)$ yields $T^r_v$ near the horizon in terms of its
value where (\ref{cc}) is valid. Putting everything together, near the horizon
we have  \begin{eqnarray}  {\partial m
\over \partial v} &=&-L_H e^\psi + O( m L_{H,v})\nonumber\\
{\partial m \over \partial r} &=&O(L_H)\nonumber\\
{\partial \psi\over \partial r}
&=&O({L_H / r})\label{c12}
\end{eqnarray}
As announced all metric coefficients vary slowly if $L_H$
is small and varies slowly. Integrating the equation for $\psi$ yields $e^\psi
\simeq r^{L_H}$ for all $r\geq r_{ah}(v)$. Hence 
$\psi$ can safely be neglected up to
distances $r = O(e^{1/L_H})$. From now on we suppose for simplicity of the
algebra that $\psi =0$.

In order to calculate the modes and $\langle T_{\mu\nu}\rangle_{ren}$ we must
first investigate the outgoing radial nul geodesics in the metric (\ref{c1})
with $\psi=0$. As mentioned in \ref{semid}, away from the horizon these geodesics
are a scaled replica of the geodesics in the absence of back reaction. Near
the horizon their structure is complicated by the distinction that has to be
made between the apparent and event horizon. We recall (equation \ref{rah}) 
that the
apparent horizon is the locus where outgoing geodesics obey $dr/dv =0$, therefore
solution of  $ r_{ah}(v) = 2 m (v, r_{ah}(v))$.

The event horizon $r_{eh}(v)$ is the last light ray which
reaches ${\cal I}_+$. It satisfies the equation of
outgoing nul geodesics
\begin{equation}
{dr_{eh} \over dv} = {1 \over 2} { r_{eh} - 2 m(r_{eh}, v)
\over r_H}
\label{c15}
\end{equation}
Setting $r_{eh} (v) = r_{ah}(v) + \Delta(v)$ one can rewrite
\ref{c15} in the form $\Delta(v) = 2(r_{ah}(v) + \Delta) (r_{ah})_{,v} +
\Delta_{,v}) + 2m(v, r_{eh} (v)) - r_{ah}(v)$. Solving recursively one
obtains an asymptotic expansion for $\Delta$ the first term of
which is $\Delta = r_{ah}(v) (r_{ah})_{,v} \simeq -1/M$ 
(see equation \ref{reh}). 

To obtain the trajectory of the outgoing nul geodesics we change
variables to the set $(v, x = r - r_{eh}(v))$ i.e. $x$ is the 
comoving distance from the
event horizon. In these coordinates the metric eq. (\ref{c1})
becomes
(using \ref{c15})
\begin{equation}
ds^2 = - {2 m(v,r_{eh} + x) x\over r_{eh}(r_{eh} + x)} dv^2
+ 2 dv dx + r^2 d\Omega^2
\label{c16}
\end{equation}
This metric resembles the Edington-Finkelstein metric in the absence of back
reaction in a crucial way. To wit $g_{vv}(x,v)$ vanishes on an outgoing null
geodesic, the event horizon $x=0$. Using this form for the metric it is now
straightforward to obtain the outgoing null geodesics. Indeed
when $x << r_{eh}$  the
equation for radial outgoing nul geodesics can be solved exactly to
yield an exponential approach to the horizon of the form
$\tilde 
v - 2 \ln x = f(u)$ where
\begin{equation} \tilde v = \int^v\!dv\
{2m(v,r_{eh}(v))\over r_{eh}^2(v)} \label{c18}\end{equation}
This motivates the following ansatz for the outgoing radial null
geodesics valid in all space time outside the collapsing star
\begin{eqnarray} \tilde 
v - 2 {x\over r_{eh}(v)} -2  \ln x 
+ \delta = \int ^u{du'\over 2 \tilde m(u')} + D
\label{c17}\end{eqnarray}
with $D$ a constant of integration.
This should be compared with the solution in the absence of
backreaction given in eq. (\ref{ueq}).
The function of $u$ on the r.h.s. of (\ref{c17})
has been written as $\int^u du'/\tilde m(u')$ for dimensional reasons. The quantity
$\delta$ is of order  $O(  L_H (Mx+ x^2) / M^2)$ for all $x$ (this is shown by
substitution of (\ref{c17}) into the equation for radial nul geodesics and
integrating the equation for $\delta$ along the geodesics $u=const$). The
function $\tilde m (u)$ is determined by requiring that the variable $u$ in
equation (\ref{c17}) be the same as  in the Vaidya metric equation (\ref{c3}).
The difference $M(u)-\tilde m(u)$ is then of order $O(M L_H)$. This is found by
using (\ref{c17}) to change coordinates from the set $(v,r)$ to the set
$(u,r)$ at the radius $r=O(6M)$ where (\ref{c3}) is valid.

Equation (\ref{c17}) is sufficient to prove  our hypothesis, to
wit that the flux emitted is $O(M^{-2})$ and that the energy momentum tensor is
regular at the horizon. Indeed, in our model,
 the flux emitted is given by eq. (\ref{TT2}):  
\begin{equation}
\langle T_{uu} (u,{\cal I}^+)\rangle=
(1/12 \pi) (d{\cal U}/du)^{1/2}\partial_u^2(d{\cal U}/du)^{-1/2}
\label{c50}
\end{equation}
  where the derivatives are taken at fixed $v$. The variable ${\cal U}$, defined
in eq. (\ref{E.1}), labels the outgoing geodesics as measured by an inertial
observer inside the star. The  jacobian $du/d{\cal U}$
is calculated by remarking that at the surface of the star the derivative at
fixed $v$ is 
$d{\cal U}/dr\vert_{v=vstar} = -2$. Hence  differentiating \ref{c17} yields
\begin{eqnarray}
du/d{\cal
U}\vert_v &=&(dr/d{\cal U})\vert_{v=vstar}( du/ dr)\vert_{v=vstar}\nonumber\\
&=& (1/2)4 \tilde m(u) /
x\vert_{v=vstar}= -4 \tilde m(u) / ({\cal U}-{\cal U}_{eh})
\label{dUdu}
\end{eqnarray}
Hence $\langle T_{uu} (u,{\cal I}^+)\rangle$ is equal to $(\pi /
12) T_H^2(u)$ where \begin{equation}
T_H(u) \simeq {1 \over 8 \pi \tilde m(u) }= {1\over 8 \pi M(u)} (1 +
O(L_H))
\end{equation}
is the Hawking temperature at time $u$ when the residual mass
is $M(u)$. 

The calculation of $\langle T_{uu}(v,x)\rangle_{ren}$ every place
(and not only on ${\cal I}^+$) is a slight generalization of the above
calculation. The renormalized energy momentum
tensor is given by 
\begin{equation}
\langle T_{uu} \rangle_{ren} =(d \hat U/ du)^2
(1/12 \pi) (d{\cal U}/d\hat U)^{1/2}\partial_u^2(d{\cal U}/d\hat U)^{-1/2}
\end{equation}
where  $\hat U(u, v)$ is the inertial coordinate introduced in equation
\ref{T4}. Since  $\hat U(u, v)$ is an affine parameter along
radial nul geodesics $v=const$ one obtains that $\hat U(u, v) = x(u,v)$. Hence
\begin{eqnarray}
d \hat U / d{\cal U}\vert_v &=&
d x / d{u}\vert_v d u / d{\cal U}\vert_v
\end{eqnarray}
which upon differentiating eq. \ref{c50} and using \ref{dUdu} is found to be
finite on the horizon. Hence as in section \ref{Tmunuren}  $\langle T_{uu}
\rangle_{ren}$ vanishes quadratically at the event horizon
${\cal U}={\cal U}_{eh}$. Thus the mean
$\langle T_{\mu \nu} (r,v)\rangle$ is the one computed without backreaction, in
the collapsing geometry of a star whose mass is $M(v)$ and this is valid
as long as $\partial_v M(v)  << 1 $.]

These conclusions  have been verified numerically in a  model \cite{PP}
wherein
the sources of the Einstein equations are a classical
infalling flux of spherical light-like dust (i.e. falling along $v=constant$)
and the energy momentum tensor given once more by eqs. (\ref{T6} and
\ref{trace}) properly rescaled see eq. (\ref{rescaled}).
Since this quantum energy momentum tensor is a local function
of derivatives of
the conformal factor $C$,
one is lead to a new set of differential equations
of the second order which describe dynamically the evolution of space-time.
The main advantage of this 
numerical integration is to provide the whole geometry from the distortion of 
Minkowski space-time due to the infalling shell up to the complete evaporation
of the collapsed object.

There is little point here in going through the algebraic formulation for
this model in view of the general considerations which have just been set
forth. We do wish however to point out one interesting feature of this work.
In view of the $T_{\mu\nu}$ used, the system of coordinates is that of eq.
\ref{Ttwo}. It is relevant to remark that there has been an 
independent mathematical approach to the black hole problem inspired by some
elements of string theory called the dilatonic black hole in 1+1 dimensions.
It resembles in form the abovementioned calculation. Indeed the dilatonic
scalar field $\phi(U,V)$ is played by $\ln r(U,V)$ 
\cite{CGHS}\cite{RST}\cite{SVV}\cite{PS}. 
The original hope was that the dilatonic black
hole model could be developed into a complete quantum mechanical theory.
Unfortunately these hopes have not been realized and at the present time our
knowledge is restricted to the semiclassical theory. Since we are now
possessed of a more physical semiclassical theory we shall not present this
work in this review.

We now present the numerical results in a series of figures.

\dessin{1.000}{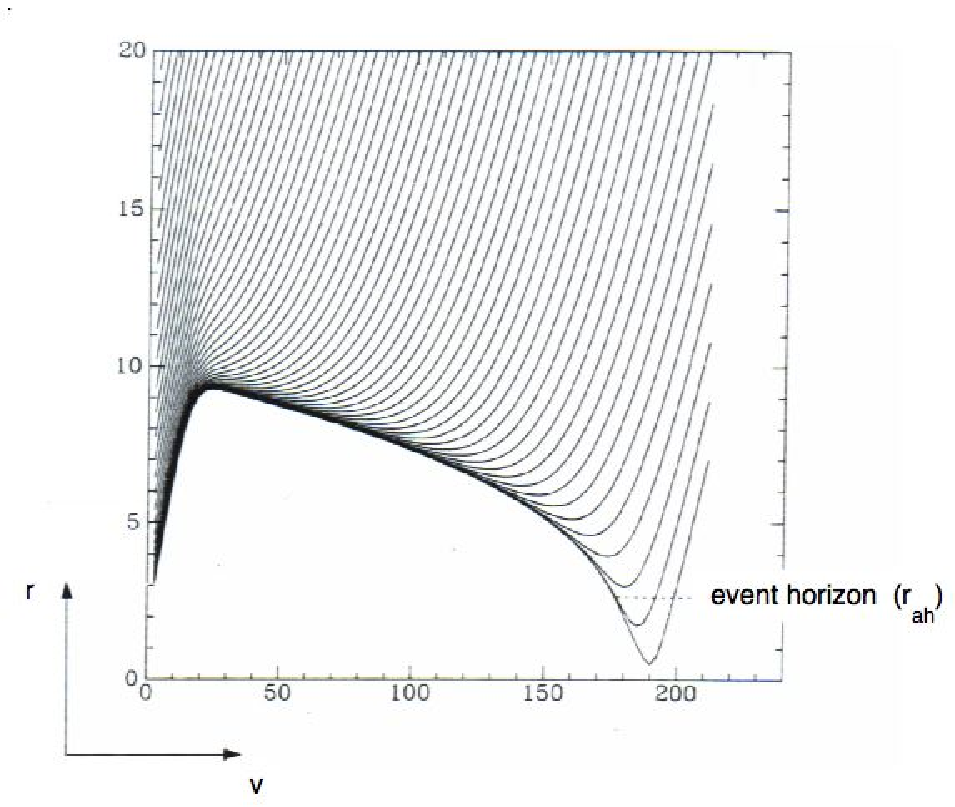}{PP3}{The outgoing nul geodesics ($u=const$) in the geometry of an
evaporating black hole, depicted in $r,v$ coordinates.}
Figure (\ref{PP3}) shows what happens in terms of the coordinates $r$ and $v$
(see eq. \ref{semia} and \ref{c1}) wherein the axis 
$v$ is drawn horizontally and
the axis $r$ vertically. The units are Planckian and $M_0$ is equal to eight.
The classical trajectory of infall is not designated but it is a packet
centered around the line $v=25$. The contour lines which are shown are
equally spaced $u=const$ outgoing light rays (that is equally spaced on
${\cal I}^+$, here represented essentially by $r=20$).
The event horizon appears as a thick line because it is where all these
outgoing geodesics accumulate. The geodesics which fall into the singularity
are not represented. They would lie in the white zone which lies beneath the
event horizon. All of these geodesics emanate from the point of reflection
$r=0$ (that is well within the star). The first ones continue to increase in
$r$ on their voyage to ${\cal I}^+$ whereas the later ones first increase in
$r$, then decrease in the evaporating geometry, then increase once more. 
The locus where they reexpand 
is the apparent horizon.
It is interesting to remark that
the paths of  photons in the white zone which begin their trajectory at $r=0$
and fall back towards $r=0$ resemble those of a closed Robertson Walker universe.
Indeed the maximum radius encountered by the outgoing light rays in this zone
is $2M_0$ and then the
radius diminishes according to the evaporation rate. At the apparent horizon
$r_{ah}=2M(v)$
one finds the throat which connects the interior region
to the external one. 
Thus at the
end of the evaporation, one has two macroscopic (smooth i.e.
wherein the mean geometry is
far from the Planckian regime) regions connected by
a throat of Planckian dimensions. This
is the situation that precedes the splitting of the geometry
into two
disjoined regions (universes) with a change of
topology \cite{HawkL}. From the semi-classical
scenario this seems unavoidable and confirms that information is
forever lost to the outside universe. 

In this theory there is a singular space like line $r=r_\alpha=O(1)$ (which is
not represented in figure \ref{PP3}). This line is singular in that  
the dynamical equations loose meaning at this radius. When the apparent horizon
reaches $r_{\alpha}$ (i.e. when the residual 
mass of the hole is $r_{\alpha}/2$),
 one has to stop the numerical integration. 
Thus the semiclassical model cannot
  describe or give any hint about the endpoint of the black hole evaporation.
This  is as it should be since at the end of the evaporation, 
the mean curvature reaches the Planckian
domain where the semiclassical treatment has no justification whatever.

Two other figures of interest are Figs (\ref{PP1}) and (\ref{PP2}) . These
represent the same geometry in other coordinates.

In Fig (\ref{PP1}) 
\dessin{1.000}{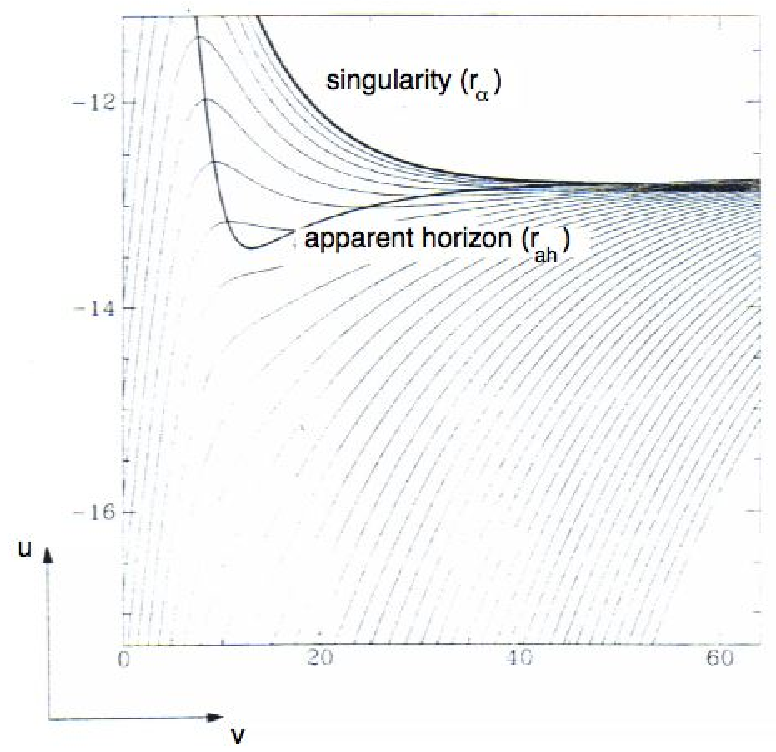}{PP1}{The $r=const$ lines in the geometry of an
evaporating black hole, depicted in ${\cal U}$,${ \cal V}$ coordinates.} 
we present
the geometry in the $\cal U, \cal V$ coordinates defined in eqs. (\ref{E.1}). 
These are the inertial coordinates in the Minkowski region
inside
the spherical infalling shell. We have represented the 
contour lines of $r=constant$ since they offer 
a convenient visualization of the evaporation 
process. 
Indeed, inside the shell, where there is no
matter  one has Minkowski space time in which $r=constant$ are
always time like straight lines. 
Outside the shell, in the absence of back reaction, one
would have  
Schwarzschild space with mass $M_0$ and  
the apparent horizon coincides with the static
event horizon at $r=2M_0$ which separates time like from space like
$r=constant$ lines.

As we have explained,
in the presence of back reaction, the evaporation process is accompanied by
the shrinking of the apparent horizon $r_{ah}(v)$. 
This horizon is the 
exterior boundary of the trapped region wherein $r=constant$ lines are
spacelike. The inner boundary of the trapped region (the other solution of
$\partial_v r =0$ at fixed ${\cal U}$) lies within the shell and is space like.
[Indeed because the flux $T_{vv}$ is positive here, due to classical matter
falling into  the black hole,
the apparent horizon is space like. In the evaporating situation $T_{vv}$ is
negative and the associated apparent horizon is time like.]
In this figure  we have also drawn the singular space-like line $r=r_{\alpha}$
which meets the apparent horizon $r_{ah}(v)$ at ${\cal {V}}\simeq 52$.

In the $\cal U, \cal V$ coordinates, the whole evaporation period is contained
in a tiny $\cal U $ lapse ($-14 < {\cal U} < -12.8$). 
This is why we have presented in Fig (\ref{PP2}) 
the evaporating geometry in the
$u,v$ coordinates. 
\dessin{1.000}{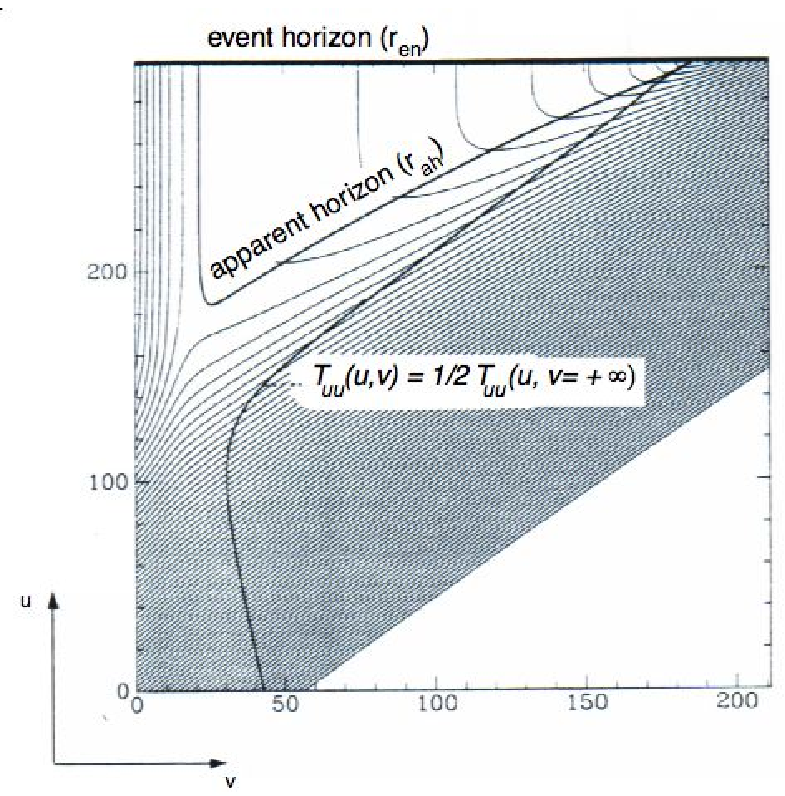}{PP2}{The $r=const$ lines in the geometry of an
evaporating black hole, depicted in $u$, $v$ coordinates. }
These later are the inertial ones at $r=\infty$ where
they are normalized by $v-u = 2r$. One sees the dramatic 
effect of the exponential jacobian eq. (\ref{dUdu})
relating $u$ to $\cal U$ which blows up the region between $r_{ah}$ and
$r_{eh}$. In the $u,v$ coordinates, the apparent horizon appears as an
almost static
line
[where static is defined as 
follows : if a mirror is put along
it, two infalling light rays separated 
by $\Delta v$ (i.e. by $\Delta t$ at fixed $r$)
will be reflected with the same $\Delta u$ (i.e. 
by the same $\Delta t$).] The property
of staticity is obviously satisfied by the $r=constant$ lines in a 
static geometry but strictly speaking no longer
in the evaporating situation where, as we have just described,
 $r=constant$  
passes from space-likeness back to
time-likeness. 
We also note that the  singular line at $r=r_{\alpha}$ 
is not present in this figure since
it is beyond the last u-line (i.e. the event horizon $r_{eh}(v)$) at 
which the 
apparent horizon meets this
singular line.

In Fig (\ref{PP2})
we have drawn an extra line which sits outside the trapped region.
This line designates the locus where $\langle
T_{uu}(r,u) \rangle$ reaches half of its asymptotic
value ($\langle T_{uu}(\infty,u)) \rangle$). 
Being well outside the trapped region, this proves that the flux 
is concerned with the external geometry only and is characterized
by the time dependent mass $M(u)$ and not by 
the whole interior geometry. This shows also
that the infalling matter is not at all affected by the evaporation process
since the infalling matter is in the causally inaccessible {\it{past}}
of the places
where the mean fluxes build up.

The semiclassical theory which we have presented in this section is a
mathematically consistent and well understood theory which predicts that
black holes evaporate following the law $dM/dt = -\xi M^{-2}$. However the
validity of the semiclassical theory even when the curvature is far from the
Planck scale is far from obvious, owing to the fact that very small distance
scales are invoked in order to derive the solution. Indeed the
jacobian $d{\cal U} / du = e^{-u/ 4 M(u)}$ used in obtaining the flux
at infinity in eq. \ref{c50} becomes exponentially small, ie one makes appeal
to the structure of the vacuum inside the star on exponentially small scales in
order to derive the radiation at later times. Another related problem is
that the distance between the event horizon and the apparent horizon is
$O(1/M(v))$ which for  macroscopic black holes is much smaller than the Planck
length.  [This distance is an invariant: it is the 
maximum proper time to go from one horizon to
the other. This is seen by writing the metric 
eq. (\ref{c1}) near the horizon in the form $ds^2=
2drdv+r^2d\Omega^2$ since we can drop
the term in 
$(r-2M)/ 2M dv^2$]. We shall discuss in more detail this problematic aspect
of the theory of black holes in section \ref{troub}.


\section{From Vacuum Fluctuations  
to Hawking Radiation}
\label{VFHR}

In the previous sections we have described the mean 
value of the energy momentum tensor and its effect on the
background geometry.  In this section we consider 
fluctuations. More precisely we
describe the  field configurations which evolve into a
particular Hawking photon using the
weak value formalism. This will provide us with the explicit history 
of  the creation of the particle out of vacuum. It will also provide us with the matrix
elements of $T_{\mu\nu}$ which play the r\^ole of the mean when calculating backreaction
effects to S matrix elements wherein the final state contains this specific pair.

As in the case of the accelerating mirror, we follow the history of the
vacuum fluctuation associated with the emission of this photon. What
follows is a resum\'e of Sections \ref{mirro}, \ref{weakacc} and the forthcoming
paragraphs.

Consider a created particle in a wave packet emitted at retarded
time $u_0$, of frequency $\la$ (where
$\la=O(M^{-1})$) with width $\Delta u= O(\la^{-1})$. From the Gerlach
resonance condition $ u^*({\om,\la}) = 4 M \ln (\om/\la)$, the
frequency $\om$ of the mode comprising the fluctuation within the
star that is converted into this Hawking photon is $\om = O(\la e^{u_0/
4M})$. This fluctuation is set up on ${\cal I}^-$ and is composed of
three parts. Let $v=v_H$ be the backward reflected light cone of the
horizon ${\cal{U}}=0$ (see Fig. \ref{collapse}). Then, as in Section \ref{mirro}, the
fluctuation, represented as a packet, straddles $v=v_H$. It is spread out
on either side of $v=v_H$ with a spread $\Delta v =O(\om^{-1})=O(\Delta u
e^{-u_0/4M})$. The part with $v>v_H$ has positive energy density. The
fluctuation with $v<v_H$ also has a positive energy hump, as well as an
oscillating broader distribution of energy. This latter is net negative
and the total energy of all these contributions vanishes as behooves a
vacuum fluctuation.

As in Section \ref{mirro}, the two different contributions, $\theta(\pm
(v-v_H))$ possess very different future destinies. The piece for $v-v_H<0$,
reflects off $r=0$, travels through the star and gets converted into a
Hawking photon. However unlike the mirror it is only at large radius
that space gets flat and that one  gets a true on mass shell quantum. Once the photon gets
out of the star, the oscillating net negative piece of the fluctuation becomes negligible
because of the time dependent Doppler shift $(d{\cal{U}}/du)^2$  
encountered in converting $T_{{\cal{UU}}}$ to $T_{uu}$.
 The
emerging photon is a net lump of positive energy, a propagating
outgoing photon. On the contrary, the piece of the fluctuation for
$v-v_H>0$ cannot get to the horizon in finite Schwarzschild time. It
approaches the horizon exponentially near the center of the star and
there it sits carrying net positive energy of $O(M^{-1}e^{u_0/4M})$.
However, as in the case of the accelerating mirror, the average energy
carried by all fluctuations is zero. Thus an observation of the
absence of a photon emitted at $u=u_0$ is associated with an ``anti--partner''
fluctuation carrying negative energy near the center of the star. So
on the average the fluctuations carry no energy. 
The problems raised by the gravitational back reaction to the exponentially large
fluctuations near the horizon will be discussed in Section \ref{troub}.

The detailed evaluation of the weak values of $T_{\mu \nu}$ for the effective
two dimensional model proceeds exactly as in the
accelerated mirror problem. Indeed the two dimensional
part of the black hole geometry 
depicted in the $u,v$ coordinate system (see Fig. \ref{star}) 
is almost identical to the
accelerated mirror problem (see Fig. \ref{ACCEL1}). The 
role of the mirror is played by the center
of the star $r=0$ (see the reflection condition eq. (\ref{HR.9}). 
The only difference is that
the conformal factor of the metric is trivial in the mirror problem
($ds^2= -dudv$) whereas it is nontrivial in the black hole problem. This
introduces some complications when computing the renormalized energy
momentum tensor  which we now address. But it does not affect the fluctuating
part of $T_{\mu \nu}$ since we are considering the simplified s-waves which are conformally invariant.

In Section \ref{Tmunuren}  the mean energy momentum tensor of the
truncated model was computed.
In order to calculate any matrix element (and not only the mean) of the
renormalized energy momentum operator 
we need a slightly more general formalism wherein the  renormalized
energy momentum operator is written as
\begin{eqnarray}
 T_{\mu\nu}(x)_{ren} =
 T_{\mu\nu}(x) - \elematrice{I(x)}{  T_{\mu\nu}(x)}{I(x)}I
\label{ww2}\end{eqnarray}
Here $ 
T_{\mu\nu}(x)$ is the bare energy momentum
operator, $\ket{I(x)}$ is the inertial vacuum, 
$I$ is the identity operator, $\elematrice{I(x)}{
 T_{\mu\nu}(x)}{I(x)}$is the expectation value of $T_{\mu\nu}$ in
the inertial vacuum which is conserved upon including the trace anomaly.
Note that in this expression the trace anomaly is entirely included
in the second term so it is state independent as required. A
convenient reexpression of eq.~(\ref{ww2}) which isolates the mean of
$T_{\mu\nu}$ from its fluctuating part is  \begin{eqnarray}
 T_{\mu\nu\ ren}
&=& : \! T_{\mu\nu}\!:
+ \left (\elematrice{0_{in}}{  T_{\mu\nu}}{0_{in}}
-\elematrice{I(x)}{  T_{\mu\nu}}{I(x)} \right)I\nonumber\\
&=& : \! T_{\mu\nu}\!:
+ \elematrice{0_{in}}{  T_{\mu\nu}}{0_{in}}_{ren}
\label{ww3}
\end{eqnarray}
where  
$\elematrice{0_{in}}{  T_{\mu\nu}}{0_{in}}$ is the expectation
value of $ T_{\mu\nu}$ in the Heisenberg vacuum $\ket{0_{in}}$ and $:\! 
T_{\mu\nu}\!:$ is the energy momentum
operator normal ordered with respect to this Heisenberg
vacuum. 

We are now in position to recopy the results of Section \ref{mirro}.
We first recall (see eqs~(\ref{unruhmi},\ref{unruhmii},\ref{unruhmiii}) that to each out
Schwarzschild mode
\begin{equation}
\varphi_\la^{out} = {1\over \sqrt{4 \pi \la}}\left(
\vert {v - v_H\over 4M}\vert^{-i\la 4M}\theta(v_H-v) - e^{-i\la u} \right)
\label{modeI}
\end{equation}
there corresponds a partner mode
\begin{equation}
\varphi_\la^{out F} = {1\over \sqrt{4 \pi \la}}\left(
( {v-v_H\over 4M})^{-i\la 4M}\theta(v-v_H) - e^{-i\la u_F} \right)
\label{modeII}
\end{equation}
where we have set for simplicity the constants which 
appear in eq.~(\ref{HR.11}) to  $K=4M$,
$B=1$, $\kappa=1$. The Unruh modes which are positive frequency on ${\cal I}^-$ are given by
\begin{eqnarray}
\hat \varphi_\la &=& \alpha_\la \varphi_\la^{out} 
+ \beta_\la \varphi_\la^{out F *} \quad \la>0 \nonumber\\
\hat \varphi_{-\la} &=& \beta_\la \varphi_\la^{out *} 
+ \alpha_\la \varphi_\la^{out F }\quad \la>0 \label{unruhmiiB}
\end{eqnarray}
To these modes are  associated the operators $a_\la^{out}$, $a_\la^{out F}$,
$\hat a_\la$. 

The Heisenberg state $\ket{0_{in}}$ is that anihilated by the $\hat a_\la$ operators. It can
be expressed in terms of  out states as
\begin{equation}
\ket{0_{in}} = {1 \over \sqrt{Z}} \prod_\la e^{{\beta_\la \over \alpha_\la}
a_\la^{out \dagger} a_\la^{out F\dagger}}\ket{0_{out}}
\label{wi}
\end{equation}
thereby exhibiting the correlations between the produced Hawking quanta
($a_\la^{out \dagger}$) and the partners ($a_\la^{out F\dagger}$). Following the
development of eqs~(\ref{labeli}) et seq.  one shows that  to each
produced hawking photon in a wave packet 
\beq \psi_i =
\int_0^\infty\! d\la\ \gamma_{i\la}
\varphi_\la^{out}\label{wiB}\feq 
there corresponds  a partner in the packet $
N_i^{-1}  \int_0^\infty\! d\la\ \gamma_{i\la}
\left({\beta_\la/\alpha_\la}\right) \varphi_\la^{out F}$ where $N_i$ is a normalization
factor given by $N_i^2= \int d\la\vert \gamma_{i\la}
({\beta_\la/\alpha_\la}) \vert^2$. Note that the wave function of the
partner does not have the same mode decomposition as $\psi_i$.

We can now compute the weak value of the energy momentum tensor, i.e. 
the energy correlated
to the creation of a photon in mode $\psi_i$.  \begin{equation}
\langle T_{\mu\nu}\rangle_w ={  \bra{0_{in}}\Pi T_{\mu\nu} \ket{0_{in}}
\over \bra{0_{in}}\Pi  \ket{0_{in}} }
\label{wii}
\end{equation}
As in Section \ref{mirro}, the projector $\Pi$ selects the
final state wherein one photon is produced in the mode $\psi_i$:
\beq
\Pi = I_{partners}\otimes
\int_0^\infty \!d\la \gamma_{i\la}a_\la^\dagger \ket{0_{out}}
\bra{0_{out}} \int_0^\infty \!d\la \gamma_{i\la}^* a_\la
\label{wiii}\feq
A calculation similar to that leading to eq. (\ref{Tn}) yields
for $T_{vv}=\partial_v\phi\partial_v \phi$ (a similar expression obtains
for $T_{uu}$) \beqa
\langle T_{vv}\rangle_w
&=& 
2{
\left(\int_0^\infty \! d\la\ 
(\gamma^*_{i\la}/\alpha_\la)  \partial_v \hat\varphi^*_\la\right)\left(
\int_0^\infty \! d\la^\p\ 
\gamma_{i \la^\p} (\beta_{\la^\p} / \alpha_{\la^\p}^2 )
 \partial_v \hat\varphi^*_{-\la^\p}\right)
\over 
\int_0^\infty \! d\la\ 
\vert \gamma_{i\la} \vert^2 ( \beta_\la /\alpha_\la)^2
}
 \nonumber\\ & &+ 
{\bra {0_{out}} :\!T_{vv} \!: \ket{0_{in}} \over \langle
{0_{out}}  \ket{0_{in}}  } 
+\bra {0_{in}} T_{vv} \ket{0_{in}}_{ren}
\label{wiiii}
\feqa
The second and third terms of eq.~(\ref{wiiii}) are background (they are independent of $\psi_i$).
The third term was the subject of Section \ref{Tmunuren} and the second is the difference
between Unruh and Boulware vacuum. Therefore in computing 
these weak values the background is that
of Boulware vacuum. This is the precise analogue of Section \ref{mirro} where the
background was Rindler vacuum. 

The first term (hereafter referred to as $\langle T_{\mu\nu} \rangle
_{\psi_i}$) describes the energy momentum of the vacuum fluctuation which will
become the Hawking photon $\psi_i$ and its partner. From eq. (\ref{wiiii})
it is apparent that the energy of this vacuum fluctuation vanishes.
Indeed the annihilation of the vacuum by the total energy
operator $\int_{-\infty}^{+ \infty} dv :\!T_{vv} \!: \ket{0_{in}} =0$ implies
that the integral of the first two terms on the r.h.s. of eq.~(\ref{wiiii}) 
vanish as in eq.~(\ref{vanish}). 

It is convenient to rewrite $\langle T_{vv} \rangle
_{\psi_i}$
in terms of out modes 
to obtain
\beqa
\langle T_{vv} \rangle_{\psi_i} &=&2\/\theta(v_H-v){
\left(\int_0^\infty \! d\la\ 
\gamma_{i\la} ( \beta_\la^2/ \alpha_\la^2) \partial_v \varphi^{out }_\la\right)\left(
\int_0^\infty \! d\la^\p\ 
\gamma_{i \la^\p}^* 
\partial_v \varphi^{out  *}_{\la^\p}\right)
\over 
\int_0^\infty \! d\la\ 
\vert \gamma_{i\la} \vert^2( \beta_\la^2/ \alpha_\la^2)
} \ +\nonumber\\
&\ & 2\/ \theta(v-v_H){\left(
\int_0^\infty \! d\la\ 
\gamma_{i\la}( \beta_\la/ \alpha_\la)\partial_v \varphi^{out  F}_\la\right)\left(
\int_0^\infty \! d\la^\p\ 
\gamma_{i\la^\p}^* ( \beta_{\la^\p}/ \alpha_{\la^\p})
\partial_v \varphi^{out F *}_{\la^\p}\right)
\over 
\int_0^\infty \! d\la\ 
\vert \gamma_{i\la} \vert^2 ( \beta_\la^2/ \alpha_\la^2)
}\nonumber\\
&&
\label{wv}
\end{eqnarray}
In this form it is apparent that the $\theta(v-v_H)$ piece is real and
positive whereas the $\theta(v_H-v)$ piece is not. It is complex and
oscillates in such a way that the total energy $\int^{\infty}_{-\infty}
dv T_{vv}$ vanishes.

Only the piece $\theta(v_H-v)$ is reflected in finite Schwarzschild time
and reaches ${\cal I}^+$. There it takes the form
\beqa
\langle T_{uu} \rangle_{\psi_i} &=&
2{
 \left(\int_0^\infty \! d\la\ 
\gamma_{i\la} (\beta_\la^2 /\alpha_\la^2) \partial_u \varphi^{out }_\la\right)\left(
\int_0^\infty \! d\la^\p\ 
\gamma_{i\la^\p}^* 
\partial_u \varphi^{out *}_{\la^\p}\right)
\over 
\int_0^\infty \! d\la\ 
\vert \gamma_{i\la} \vert^2 ( \beta_\la^2/ \alpha_\la^2)
}
\label{wvi}
\end{eqnarray}
It carries the Schwarzschild energy $\la_0$ of the post selected photon. Indeed one has
\beq
\int_{-\infty}^{+\infty}
du \langle T_{uu} \rangle_{\psi_i}
={ 
\int_0^\infty \! d\la\ \la
\vert \gamma_{i\la} \vert^2 \vert \beta_\la /\alpha_\la\vert^2
\over 
\int_0^\infty \! d\la\ 
\vert \gamma_{i\la} \vert^2 \vert \beta_\la /\alpha_\la\vert^2
} \simeq \la_0
\label{wvii}\feq
where $\la_0$ is the energy of the produced hawking photon.

We mention that one may in similar fashion postselect the
presence of any number  Hawking photons. In this case one
will find a decomposition of the weak value very similar to eq.~(\ref{wiiii}).
There will be a term corresponding to the Boulware energy  (the second
line of eq. (\ref{wiiii})) and a sum of terms of the form $\langle
T_{\mu\nu}\rangle_{\psi_i}$, one for each post selected photon. Thus  the background
contribution 
can be naturally separated from the fluctuating
contributions.

As in the electric field (see eq. (\ref{PEnew})), 
the physical relevance of the imaginary part
of $\langle T_{uu} \rangle_{\psi_i}$ can be seen by modifying slightly
the background geometry: $g_{\mu \nu} = g_{\mu \nu} + \delta g_{\mu \nu}$.
Then the change in the probability of finding on ${\cal I}^+$ the Hawking
photon selected by $\Pi$ (eq. (\ref{wiii})) is given by
\begin{eqnarray}
P_{g_{\mu \nu} + \delta g_{\mu \nu}} = P_{g_{\mu \nu}} \left(1 - \int d^4 x 
\sqrt{g} \delta g_{\mu \nu} 2 {\mbox Im} \left[\langle 
T_{\mu \nu}\rangle_w \right] \right)\label{physrela} 
\end{eqnarray}


\section {Thermodynamics of Black Holes}\label{thermm}

In order not to disrupt the continuity of the text we shall 
introduce the notion of black hole entropy by giving a thermodynamic
interpretation to the evaporation phenomenon which has been the subject
of this review until this point. The argument pursued will be heuristic
and based on analogy. Subsequently a more rigorous derivation based on a
true equilibrium situation, the eternal black hole will be presented.

A convenient pedagogical crutch to start with is the idealization 
used in the preceding chapters,
the 1+1 dimensional problem with unit transmission coefficient. We
emphasize at the outset that the notion (and the value) of the entropy of
the black hole itself which is deduced from this model, is by no means
contingent on the idealization used to get it.

The central result of the idealized model is that at a certain 
time 
a thermal flux emerges
as the collapsing star approaches its horizon. Then the mean energy emitted 
during $\Delta v$ (or $\Delta t$ at fixed $r$) is
given in eq.~(\ref{1.38}): $ <T_{uu}> \Delta t  =
{\pi \over 12}\beta_H^{-2} \Delta t$ and it is distributed
thermally (with $ \beta_H  = 8\pi M$. 
The probability for the simultaneous occurrence of quanta in a
state whose occupation numbers for frequency $\omega_i$ are $n_i$ is
\begin{equation}
P_{\{ n_i\}} = Z^{-1} e^{- \beta_H \sum n_i \omega_i}
\label{th0.1}
\end{equation}
We remind the reader that the density matrix eq.~(\ref{th0.1}) is a 
consequence of the Bogoljubov transformation
 eq.~(\ref{twotwobh}).
The normalizing factor $Z$ is likened to a partition function of a one
dimensional gas (c.f. eq.~(\ref{partf}))
whose volume is $\Delta t$, i.e.
\begin{eqnarray}
\ln Z &=& -\Delta t \int {d \omega \over 2 \pi} \ln (1 - e^{-\beta_H
\omega})
{-1}
\label{th0.2}
\end{eqnarray}
and
the mean number of quanta with frequency $\om_i$ is
\begin{equation}
\langle n_i \rangle = {\partial \ln Z \over
\partial \beta \omega_i} =
(e^{\beta_H \omega_i} -1)^{-1}
\label{th0.3}
\end{equation}
The fact that the photons are all outgoing does no injustice to the
application of usual statistical mechanics and thermodynamics defined by
the canonical ensemble.

The conditions under which
evaporation is taking place are isothermal to a very good approximation.
By that we mean there is a $\Delta t$ sufficiently small so that even
when a macroscopic number of photons evaporate in this time interval,
$\beta_H$ does not change. For example for $M\simeq 1$ solar mass
($ = 10^{38}$ Planck masses) during the time interval
that $10^{20}$ photons are emitted one has $\Delta M = O(10^{20}/M)$
and $\Delta \beta_H / \beta_H =
O(\Delta M / M) = O(10^{-56})$.

One may liken this quantum evaporation 
to the irreversible process in which a large
mass of liquid evaporates into a tiny amount of vapor in time $\Delta t$.
The whole system is insulated from its surroundings (i.e.
$E_{total}=constant$) with a little vapor occupying a  space above the
liquid. Then increase the volume a little bit. To make the illustration
more cogent
we neglect terms proportional to the chemical potential (after all
it doesn't cost anything to make a vacuum fluctuation).  Then the only increase
in entropy is due to the increase of volume of the vapor phase. The
liquid loses a "little'' energy $\Delta E$ and the vapor increases in
energy $\Delta E$ where "little'' means $\Delta E / E_{total} <<1$ and we
envision that almost all the energy is localized in the liquid.
Then the temperature change is negligible
($\Delta T / T = O(\Delta E / E)$). One says that the liquid is a
"reservoir''. The "reservoir limit'' for the change in its entropy is
$\Delta S = \Delta E / T$. [In thermodynamics one introduces a reservoir
to convert entropy considerations of a system to its free energy, 
the relevant function to describe isothermal processes undergone by a system
in contact with the reservoir].
\par From the above interpretation we are led to ascribe to the black hole an
entropy. In point of fact it was this remarkable insight of Bekenstein \cite{beken} that must
 have incited Hawking to attribute a temperature to the black hole
and hence evaporation.
During the time $\Delta t$ wherein a mass $\Delta M_{BH}$ 
($= - \Delta M_{gas} =-\sum_i
\langle n_i \rangle \omega_i$) is evaporated ( with $| \Delta M_{BH} |
/ M << 1$),
the process is effectively isothermal and the black hole acts as a
reservoir. Its change in entropy is thus
\begin{equation}
\Delta S_{BH} = \beta_H \Delta M_{BH} = 8 \pi M \Delta M_{BH} \simeq \Delta
(A/4)
\label{th0.4}
\end{equation}
where we have introduced the area of the horizon surface ($A =4 \pi (2
M)^2$). 

A noteworthy feature of black hole evaporation as compared to the liquid
vapor analogy is that no time is required to get the 
"vapor'' into an equilibrium state after it evaporates.
To complete this discussion, one may
calculate the increase in entropy occasioned by the evaporation. Using
eq.~(\ref{th0.1}) one has
\begin{eqnarray}
\Delta S_{total} &=& \Delta S_{BH} + \Delta S_{gas} = \beta_H(\Delta
M_{BH} + \Delta M_{gas}) + {\partial \ln Z \over \partial \Delta
t}\Delta t\nonumber\\
&=& 
{\partial \ln Z \over \partial \Delta
t}\Delta t = p \Delta t
\label{th0.5}
\end{eqnarray}
Recall that the pressure is given
by $\partial \ln Z / \partial V$ and in our case $\Delta V$ is the one
dimensional volume $\Delta t$.
Thus the entropy increase is $p\Delta V$ as for the analog liquid-vapor
example (with zero chemical potential). 

But how is that? We started with one state and we are now
calculating the increase in the total number of states. The answer
of course is that we are describing the gas by a density matrix
and have completely forgotten the correlation to the degrees of
freedom left within the black hole. 
It is therefore the extra trace
over these degrees of freedom which is responsible for the
increase of entropy $ \Delta S_{total}$ and this is a perfectly
legitimate description for the outside observer. 

It is thus possible to develop a phenomenological thermodynamics
for the outside observer which ascribes to the black hole an
entropy change which is $ \beta_H (M) \Delta M$. 
We may then calculate the 
entropy of the black hole by integrating this change
from 0 to $M$ as to obtain
\begin{eqnarray}
S_{BH}(M) - S_{BH}(0) = \int_{0}^{M} \beta_H(M')
dM'= \pi (2M)^2 = A /4 \label {T.8}
\end{eqnarray}
One may ask why $S_{BH}(M)$ is intrinsic to the black hole i.e. equal to
$log \Omega (M)$, where $\Omega$ is the number of degrees of freedom of the
black hole. After all the black hole was used as a reservoir only.
The answer is that the black hole even if it is a reservoir for the radiation
is at the same time in very good approximation a microcanonical ensemble unto
itself. This is because $M$ changes so slowly ($dM/dv \simeq - 1/ M^2$).
Thus $\beta_H = \partial_M log \Omega (M)$ is a valid microcanonical
expression for how the density of states varies with the mass. In usual statistical
mechanics, one turns the argument the other way around since one starts with
the density of states. The (temperature)$^{-1}$ is then introduced as the
logarithmic derivative of $\Omega$.

A possible  constant of integration in eq.~(\ref{T.8}) is 
a subject of much
debate \cite{CaEn}. If there is 
a remnant at the end of evaporation then $ S_{B H}
(0)$ is its entropy. Since we have as yet no way to think about
this remnant it is the better part of valor not to commit oneself as to its
value. If the remnant leaks away  then this leakage should be accompanied by
further entropy increase, in a model dependent way and in the sense
of a density matrix. The situation is analogous to that sketched in
Fig. (\ref{ACCELiii}) for a decelerating mirror. In principle one could measure the
correlations between these late leakage photons and the earlier radiated
ones. Then one would be able to check whether or not
there is a pure state with no change
in entropy. Thus the former entropy increase would have resulted
from neglecting the correlations.
If otherwise the remnant sinks through a singularity
these degrees of freedom  get lost  and the entropy
increase is given by eq.~(\ref{th0.5}), i.e. we are stuck with the density
matrix description and quantum mechanics applied to this problem has
become non unitarity as initially suggested by Hawking  \cite{Hawk3}. 
Nothing could be more interesting-or exasperating.
For further discussion, see Section \ref{troub}.

We mention that when one takes into account the transmission
coefficient, much of what has been said still survives. Eq.
(\ref{th0.1}) must be changed in that each factor $ e^{-\beta
\omega_i}$  must be multiplied by a transmission factor $\Gamma_i$, and one must include
all the angular momentum modes.
The value of $Z$ changes, but thermodynamics is retained in a
modified sense since for $ N >> 1$ (where $N= \sum_i n_i$), relative energy fluctuations
are still ${ \cal{O}} (1/ \sqrt N) $ because the distribution remains Poisson, 
and the process is still isothermal.
All that is modified is the numerical value of $\Delta S_{total}$. In
particular eq.~(\ref{th0.5}) is retained since $ -\partial \ln Z/
\partial \beta \vert_{\beta_H} = \Delta M_{gas}$ for the modified
ensemble as well. The entropy ascribed to the black hole and its
physical interpretation as described after eq.~(\ref{th0.5}), is
independent of the mechanical details. 
And one still has $\Delta S_{total} = p \Delta t >0$.

The idea of black hole entropy in terms of a reservoir is
reinforced by the analysis of the equilibrium situation which is
characteristic of the eternal black hole. This idea is once more
a seminal discovery of Hawking \cite{Hawk2}. We shall follow Hawking's idea,
but, unlike him, we shall also deduce the black hole entropy
itself.
 
In a closed box of volume $V$, it has been shown by Hawking
that for a sufficiently large energy, the energy becomes partitioned into 
a black hole surrounded by radiation in thermal equilibrium. We 
shall see below that there is a limit wherein 
essentially all the energy is in the black hole
whose radius is nevertheless 
much smaller than the radius of the box. Then, the only
r\^ole of the radiation is to furnish a temperature.
 In the limit envisaged, this latter is the Hawking temperature, $T_H$,
which value is then used to compute the entropy of the hole.

The relevant geometry which describes the static situation characteristic of 
the eternal black hole is the full Schwarzschild quadrant R ($r>2M$)
of Kruskal space.
This geometry is depicted in
Fig. (\ref{krusk}).
In the case
where almost all the mass is in the black hole the geometry in R
approximates to that of empty Schwarzschild space.
This space has both past and future horizons. For the case envisioned of
a black hole formed from a collapsing shell the past horizon is
not present (see Fig. (\ref{collapse})). Nevertheless, the 
use of the eternal black hole geometry is a 
correct mathematical idealization to be understood as follows. Outgoing photons
that issue from the black hole, in the course of its formation at late
stages are Kruskal in character (i.e. the modes used to describe Unruh
vacuum). Because the black hole is enclosed in a finite volume, these
modes get reflected off the walls and come back as Kruskal modes. So
after some time a stationary state gets established in which both the
incoming and outgoing modes (or pieces thereof) are of Kruskal character.
Note that all angular momenta participate in thermal equilibrium
since considerations of reversibility render useless any appeal to
the smallness of the transmission coefficient for the higher angular
momentum waves. In this equilibrium case, the hole is surrounded by a gas of
energy density proportional to $T_H^4$. So the s-wave truncation
makes no sense in this case.
The state is then set up in terms of the modes which traverse the whole
space. But the part of the space which is physically relevant is
restricted to that part of R which is bordered by the surface of
infalling matter.

Every stationary state of the photon gas in the eternal
geometry except the Hartle-Hawking
vacuum gives rise to a singular energy momentum tensor on at least one of
the horizons. The Hartle-Hawking vacuum  \cite{HaHa} is that constructed
from the quanta of Kruskal modes. It is the analog of Minkowski vacuum in
Minkowski space and R is the analog of the right Rindler quadrant.
One can understand the origin of the theorem on
singular energy density near the horizon by reference to eq.~(\ref{minus}) et seq.
where it was shown that in Boulware vacuum $ <T_{UU} > $ is
singular. To undo this singularity clearly requires a very
special condition (which, for the collapsing case, was shown to be
$ \langle T_{uu} (r\simeq 2M)\rangle = O [ ( r - 2 M)^2 ]$).

In order to prove this point, to see the thermal character of Hartle-Hawking
vacuum in $ R $, and in fact to construct the corresponding density matrix, it is
useful to make the euclidian continuation of the Schwarzschild geometry $ t \to it
$ (a construction which is possible owing to the staticity of the geometry). 
Forgetting angles the line
element is $ (1 - {2 M / r}) (d (it))^2 + ( 1 - 2
M/r)^{-1} dr^2 $. This has the form of a cigar which terminates at $ r = 2
M $ and extends as a cylinder of radius $ 4 M $ out to infinity. To
see this note that near $ r = 2 M $, the line element takes the
form \begin{eqnarray}
ds^2 \quad _{\stackrel {\longrightarrow}{r \to 2 M}} \quad \rho^2 d
\theta^2 + d \rho^2 
\label {T.9}
\end{eqnarray}
where $ \rho = 2 \sqrt{ 2 M (r - 2 M)}$ ; $ \theta = it/{4 M} $.
Equation~(\ref{T.9}) is the line element in the neighborhood of $r=2M$ written
in local euclidean polar coordinates. No conical singularity at that point
requires periodicity of $ \theta $ equal to $ 2 \pi $. On the other hand as
$ r \to \infty $ the metric goes over to the metric of a cylinder $ ( = (4
M) ^2 d \theta ^2 + d r^2) $, the periodicity in $ \theta $ remains equal
to  $ 2 \pi $.

Thus regular functions defined on this space are periodic in $ it
$ with period equal to $ 2 \pi (4 M) = \beta_H $. For large $ r $,
Schwarzschild $ t $ coincides with proper time, hence
$\beta_{H}^{-1}$ is temperature at large $ r $. And Green's
functions defined as Hartle-Hawking expectation values are regular.
A discussion of these Green's functions, in general, is given in ref.
 \cite{FuRu}. For the purpose at hand it suffices to note that
near the horizon the geometry is regular (eq.~(\ref{T.9})). In this
region Hartle-Hawking Green's functions are related to Boulware Green's
functions in the same way that in flat space Minkowski Green's functions
are related to Rindler Green's functions. Near the horizon the analogy is
exact with the acceleration replaced by $\rho^{-1}$.

Having displayed the properties at equilibrium, let us now 
construct the necessary conditions to  derive
the black hole entropy.
Our condition of negligible energy in the gas is 
\begin{eqnarray}
M >>V M^{-4} \quad \mbox{thus} \quad
M >> V^{1/5}
\label {T.11}
\end{eqnarray}
On the other hand we require that the volume occupied by the gas
be much greater than that occupied by the hole in order to validate
the estimate of its energy (i.e. $E_{gas} = V T^4$ as in absence of gravity) 
\begin{eqnarray}
V >> M ^3 
\label {0.12}
\end{eqnarray}
Equations (\ref{T.11}) and  (\ref {0.12}) are compatible if $ M^5 >> M^3 $ i.e.
$ M >> 1 $ . In that case the total energy at equilibrium is in very good
approximation 
\begin{eqnarray}
E = M + V \beta^{-4}_{H} \label {0.13}
\end{eqnarray}
We now use the fact that this equilibrium configuration should be derivable
from the variational principle of entropy 
$\delta S_{total} \vert_{E,V} = 0$. Indeed, by attributing an entropy
to the black hole ($S_{BH}$) and by taking the variation with the
energy repartition ($\delta M = - \delta E_{gas}$), one finds
\begin{eqnarray}
 \delta S_{total} \vert_{E,V}  &=& \delta M
 ({\partial S_{BH} \over \partial M} -
{\partial S_{gas} \over \partial E_{gas}} ) \nonumber\\ &=&
 \delta M
 ({\partial S_{BH} \over
\partial M} - \beta_H) = 0 
\label{th0.14} 
\end{eqnarray}
 whence the equality of the temperatures gives
\begin{eqnarray}
 {\partial S_{BH} \over \partial M} = 8 \pi M 
\label{th0.15}
\end{eqnarray}
In eq.~(\ref{th0.14}), we have used conventional 
canonical thermodynamics for the radiation (i.e. $dE_{gas} = T_H dS_{gas}$) 
and ascribed to it the temperature $T_H$ albeit that the total system
is microcanonical in which one phase (the black hole) acts as a reservoir. 
The above considerations are quite general and the use of
the photon gas was by way of illustration. Equation~(\ref{th0.14}) will be
true for any model of matter.
In this sense eq.~(\ref{th0.15}) is a very powerful result. Whatever
happens in the ultimate destiny of black hole physics one would be
loath to give it up. 

A critique (see ref. \cite{PKO}) of the notion of black hole entropy follows from a
closer inspection of the distribution of energy density in the 
Hartle-Hawking vacuum \cite{How}. 
Vacuum polarization effects in $ < T_{\ \mu} ^{\nu}
> $ ( which lead to negative energy density near the horizon) prevent a
clean split of total entropy into its matter and black hole components.
Therefore the value $ S_{BH} = A / 4 $ is to be considered as valid only
in the reservoir limit as discussed above. This does not mean that $
S_{total} $ does not exist but rather that in the general case
it cannot indeed be unambiguously divided into two terms (matter and
black hole) because gravitation is a long range force.

That the entropy is proportional to the area of the horizon 
has been based on the fact that the Hawking temperature is
proportional to $M^{-1}$. This in turn is related to the
imaginary period of the relevant Green's functions of
quantum fields in Schwarzschild space. A transfer of
dimensions has occurred from ${\rm mass}^{-1}$ to length (or
time since $c=1$) owing to the existence of $\hbar$. Thus it
would seem that the identification of entropy with area is
essentially quantum in character. But quite surprisingly, a
purely classical development already foretells a good bit.
On one hand, Hawking and collaborators have shown that in
the classical evolution of matter-gravitation
configurations, the area of the horizon always increases.
[This review is not the place to prove this important
classical topological theorem. The reader is referred to the
important monograph of Ellis and Hawking \cite{EH}.] This can be related to more
general results that the total entropy of black holes and matter always increases
(see  ref. \cite{Beck94}). On the other
hand, Bardeen, Carter and Hawking \cite{BCH}  have used a 
tool called the
Killing identity to investigate the role of the horizon's area
as an entropy, in purely classical
terms. 
The point of departure is an exact
geometrical identity satisfied by Killing vectors, upon which one grafts
Einstein's equations to relate the curvature, which appears
in the identity, to the energy momentum tensor. 
The version we shall present below is applicable to static
spherically symmetric systems (See also ref. \cite{CaEn} for a similar
derivation in the framework of the hamiltonian formalism). 
In addition we refer to efforts \cite{GibbonsHawking}, \cite{YorkB}
which identify the black hole
entropy as an action integral for the pure gravitational sector.
The value $ S_{BH} = A / 4 $ is then obtained from the
classical Einstein action taking due care of boundary terms. The
bearing of this result on the characterization  of the degrees
of freedom locked within the black hole (but coupled to the
external world) remains a subject of debate.

Spherically symmetric static systems can be described by the
line element
\begin{equation}
ds^2 = -e^{2 \phi(r)}dt^2 + e^{2 \Lambda(r)} dr^2 + r^2
d\Omega^2
\label{K.1}
\end{equation}

We assume that at large $r$ there is no matter present, so
Birkhoff's theorem applies and one has
\begin{equation}
e^{-2 \Lambda} = (1 - {2 M\over r})
\quad r \to \infty
\label{K.2}
\end{equation}
 Further we fix
completely the coordinates by imposing $\phi(r=\infty)=0$. 
$M$ is the total Keplerian mass measured from infinity. It is the sum of the
black hole and matter mass.  Indeed, the equation $R^0_0 - R/2 = 8 \pi T^0_0$ can be
integrated exactly to yield
\begin{eqnarray}
&e^{-2 \Lambda} = (1 - {2 m(r) \over r})&\nonumber\\
&m(r) = M - \int_r^\infty dr 4 \pi r^2 \rho (r)
\label{K.2b}
\end{eqnarray}

The  horizon of the  black hole is the radius   $r=2m_B$
at which $e^{-2 \Lambda}$ vanishes.
Regularity of the geometry then implies that in the vicinity
of the horizon  one has \begin{equation} e^{-2 \Lambda}
\simeq {r- 2 m_B \over 2 m_B} \quad ; \quad
e^{2 \phi} \simeq k^2 {r- 2 m_B \over 2 m_B}
\label{K.3}
\end{equation}
The value of $k$ differs from unity owing to the presence of
matter. It plays an important conceptual role in what
follows.

The relevant  geometrical (Killing) identity in this approach is
\begin{equation}
R^0_0 r^2 e^{\phi + \Lambda} = -\left ( e^{\phi - \Lambda} r^2 \phi^\prime
\right)^\prime
\quad ,\label{K.3b}
\end{equation}
a consequence of the staticity of the geometry . One then uses Einstein's
equation $R^0_0 = 8 \pi (T^0_0 - {1 \over 2} T)$.
We shall suppose that the matter is a perfect fluid
$T^\nu_\mu = (\rho + p) \delta^\nu_0 \delta^0_\mu - p \delta^\nu_\mu$
(introducing a difference between radial and tangential pressure does 
not change the final answer),
to give
\begin{equation}
{d \over dr} (e^{\phi-\Lambda} r^2 {d\phi \over dr}) = 4
\pi(\rho + 3 p )e^{\phi + \Lambda}\label{K.4}
\end{equation}
whereupon integration from $r=2m_B$ to $r=\infty$ yields
\begin{equation}
M - k m_B = 4 \pi \int _{2 m_B}^\infty (\rho + 3 p) r^2
e^{\phi + \Lambda} dr
\label{K.5}
\end{equation}
(This is an alternative way of writing $M$ as compared to $M
= m_B + 4 \pi \int_{2 m_B}^\infty\rho r^2 dr$; see eq.~(\ref{K.2b}).)

The  term $k m_B$ will be expressed henceforward as a term
proportional to the black hole area ($A=4 \pi (2 m_B)^2$)
\begin{equation}
k m_B = {\kappa \over 4 \pi } A\quad ; \quad \kappa = k/4 m_B
\label{K.6}
\end{equation}
The constant $\kappa$, called the surface gravity, has very
important physical significance. It is the gravitational
acceleration at radius $r$ measured at infinity, for example
from the tension in a string attached to a test mass located
at $r$ \cite{Pagepr}. To see this, suppose the test mass, $\mu$,
initially at rest, is dropped by a static observer at $r$. It
picks up kinetic energy $T$ (as measured by a static observer
at $r+\delta r$) in an infinitesimal distance $\delta r$ (ie.
proper distance $\delta l = e^{\Lambda} \delta r$) equal to
$\mu (d\phi / dr) \delta r$. When measured at $r=\infty$,
this energy is redshifted to $\mu e^{\phi}\phi^\prime
\delta r$. Equating it to a work term $F_\infty \delta l$
one finds $F_\infty = \mu e^{\phi-\Lambda} \phi^\prime$.
For $r$ at the horizon this gives $F_\infty / \mu =\kappa$.
It is noteworthy that in the absence of matter (with
$e^{2\phi}=e^{-2 \Lambda}=(1 - 2 m_B /r)$) one has
$F_\infty = \mu m_B / r^2$.
Consistency with eq.~(\ref{K.4}) when $\rho=p=0$ is an
essential point. Were Newton's law for the force other than
$r^{-2}$, the repercussions would be serious indeed.

The reason why we have belabored the physical
interpretation of $\kappa$ is that in the euclidean
continuation of the metric near the horizon, eq.~(\ref{K.3})
gives rise to the polar coordinate representation of flat
space which for Schwarzschild space is given by eq.~(\ref{T.9}) 
with $\theta = i t/(4M)$. The only difference when $k\neq
1$ is $\theta = k(i t ) /(4 m_B) = \kappa (it)$. When
transcribed into quantum mechanics this period in proper euclidean
 time
is transformed into the inverse Hawking temperature
$\beta_H = 8 \pi m_B / k = 2 \pi / \kappa$.

To see that the area has to do with an entropy one compares two
static solutions by varying external parameters $\rho$, $p$,
$m_B$, etc in such  manner as  to be consistent with Einstein's
equations. The steps require a bit of algebra and is
relegated to a parenthesis. We first quote the result
\begin{equation}
\gd M = {\kappa \over 8 \pi} \gd A + \int_{2 m_B}^\infty
(\tilde T d \gd S_{matter} + \tilde \mu d \gd N_{matter})
+p_B k A \gd(2 m_B)
\label{K.7}
\end{equation}
Here $\tilde T$ and $\tilde \mu$ are the local temperature
and chemical potential, scaled correctly by the red shift
(Tolman scaling) $\tilde T = e^\phi T(r)$, $\tilde \mu =
e^\phi \mu(r)$. $\gd S_{matter}$ and $\gd N_{matter}$ are
the local entropy and particle number of matter. There are
no pressure terms at infinity since  the space is
asymptotically flat but we have retained the work term due to the pressure $p_B$
occurring near the horizon. This term doesn't appear in the original formula of
Bardeen, Carter, Hawking who assumed that $\rho$ and $p$ vanish at the horizon.
However it becomes relevant for instance when we compare geometries of black
holes in a box surrounded by radiation in a Hartle-Hawking state. If
instead of the model considered here above one  considers a black hole and
surrounding matter enclosed in a finite volume having walls, the energy of
neighboring configurations have to differ by a term like $-p \delta V$ but also by
a term (model dependent) taking into account the stress in the wall.

[The difference between two neighboring solutions are
characterized by $\gd M$, $\gd m_B$, $\gd \kappa$, $\gd
\rho$, $\gd p$, $\gd \phi$ and $\gd \gL$, quantities that
cannot be all independent because of the Einstein field
equations. Starting from eq.~(\ref{K.5}) we obtain
\begin{equation}
\gd M = {1 \over 4 \pi} \gd ( \kappa A) +
4 \pi \gd \int_{2 m_B}^\infty
(\rho + 3 p) r^2 e^{\phi + \gL} dr
\label{par01}
\end{equation}
The variation of this last integral is the main task of the
calculation. This is most easy done by splitting it as a
sum of two parts  using $-R = 8 \pi (-\rho +
3p)$
\begin{eqnarray}
4 \pi 
 \int_{2 m_B}^\infty
(\rho + 3 p) r^2 e^{\phi + \gL} dr
&=&  \int_{2 m_B}^\infty - {R \over 2} e^{\phi + \gL} r^2 dr
\nonumber\\
&\ &+ 8 \pi  \int_{2 m_B}^\infty \rho e^{\phi + \gL} r^2 dr
\label{par02}
\end{eqnarray}
The first term, hereafter denoted $I_1$, on the right hand
side is the gravitational action. Its variation must take
into account boundary terms which arise both because $m_B$
varies and by integration by parts to yield
\begin{eqnarray}
\gd I_1 &=& e^{\phi + \gL} r^2 R\!\vert_{2 m_B} \gd m_B +
(\lim_{r\to\infty} -\lim_{r\to 2m_B})
e^{\phi - \gL} r^2 ( \gd \phi^\prime + \phi^\prime \gd \phi -
({2 \over r} + \phi^\prime)\delta\gL )\nonumber\\
&\ &+ \int_{2m_B}^\infty (G^0_0 \gd \phi + G^r_r \gd \gL )
e^{\phi + \gL} r^2 dr
\label{par03}
\end{eqnarray}
To prepare the evaluation of the variation of the second
term, let us recall the  thermodynamic relation
\begin{equation}
\gd E = -p\gd V + T \gd S + \mu \gd N
\label{par04}
\end{equation}
which in terms of local densities on 3-surfaces $t=constant$
becomes
\begin{eqnarray}
\gd ( \rho 4 \pi e^{\gL} r^2 dr) &=&
-p \gd (  4 \pi e^{\gL} r^2 dr) 
+ T \gd ( s 4 \pi e^{\gL} r^2 dr)
+ \mu \gd ( n 4 \pi e^{\gL} r^2 dr)\nonumber\\
&=&- p 4 \pi e^\gL r^2 \gd \gL dr + T d\gd S + \mu d \gd N
\label{par05}
\end{eqnarray}
Acordingly we obtain
\begin{eqnarray}
\gd I_2 &=& - 8 \pi \rho e^{\phi + \gL} r^2 \vert_{2 m_B} \gd
(2 m_B)
\nonumber\\
& &+2 \int_{2m_B}^\infty e^{\phi + \gL} 4 \pi \rho r^2 \gd
\phi dr\nonumber\\
& &+ 2 \int_{2m_B}^\infty e^{\phi } \left[ T d \gd S +
\mu d \gd N) - p 4 \pi r^2 e^{\gL}\gd \gL dr\right]
\label{par06}
\end{eqnarray}
The variation of $\kappa A$ results from the variation of
both $m_B$ and $\phi$ and $\gL$: \begin{eqnarray}
{1 \over 4 \pi} \gd (\kappa A ) &=&
\gd ( e^{\phi - \gL} r^2 \phi^\prime)\nonumber\\
&=& ( e^{\phi - \gL)} r^2 \phi^\prime )^\prime
\!\vert_{2 m_B} \gd (2 m_B ) \nonumber\\
&\ &+ e^{\phi - \gL} r^2 (\gd \phi^\prime + \phi^\prime (\gd
\phi - \gd \gL ) )\!\vert_{2 m_B}
\label{par07}
\end{eqnarray}
Putting all together and using the Einstein equations $G_0^0 =
8\pi \rho$, $G^r_r = 8 \pi p$ and the expression of ${R/
2} = {1 / r^2} - e^{-2 \gL} ( {1 \over r^2} +
\phi^{\prime\prime} + \phi^{\prime 2} -\phi^\prime \gL^\prime
+ {2 \phi^\prime / r} - {2 \gL^\prime /r})$ 
and the asymptotic behavior of the metric components $e^{\phi-\gL}=1$, $\gd
\phi^\prime = \gd M/r^2$, $\gd \phi = -\gd \gL = -\gd M/r$
and the condition defining the radial coordinate at the horizon $e^{-2 \gL} (\gd
\gL + \gL^\prime \gd(2m_B))=0$ we obtain
\begin{eqnarray}
\gd M &=& k \gd m_B + \int_{2m_B}^\infty
( \tilde T d \gd S + \tilde \mu d
\gd N )
-4 \pi e^{\phi + \gL} \rho r^2 \!\vert_{2 m_B} \gd(2 m_B)
\label{par08}
\end{eqnarray}
where $\tilde T = e^{\phi }T$ and $\tilde \mu = e^\phi \mu$
are the Tolman temperature and chemical potential.
On the horizon, staticity implies $R_{00}=0$, ie. $\rho + p=0$ (the same can be
deduced from the regularity of $\sqrt{-g}\!\vert_{2 m_B} = k 4 m_B^2$), so the
last term can be translated into a term giving the work (measured at infinity) due
to the pressure at the horizon when the volume of the exterior of the black hole
varies.]

 In eq.~(\ref{K.7}) it
is difficult to resist setting the black hole entropy
equal to $A/4$ since $\kappa/ 2 \pi$ is 
the periodicity of the euclidianized time hence the 
temperature. How is it that the classical theory
anticipates quantum mechanics? 
To this end, we cite the efforts of Gibbons and
Hawking and Brown and York referred to above, based as they are on the classical
euclidean gravitational action with imaginary time of period $\beta$.
 According to the tenets of quantum statistical
theory, one converts action to entropy universally through
division by $\hbar$. It is only through quantum mechanics
that entropy acquires an absolute sense. The term $\kappa dA$
in eq.~(\ref{K.7}) is derivable from the difference between the 
classical actions of gravity
which arises owing to different matter configurations. From the above
identification of $\kappa /2 \pi $ with $T_H/\hbar$, one rewrites
$\kappa dA /8 \pi$ as $T_H dS_{BH}$ hence $dS_{BH} = dA/ 4\hbar$.
 Thus it seems that the classical Killing indentity "derivation" of
entropy has no thermodynamical interpretation without quantum mechanics
(See ref. \cite{EMP}).


\section{Problems and
Perspectives}\label{troub}

As elegant as is the semi-classical theory of black hole 
radiation, it is fraught with severe conceptual problems.

The most essential 
is concerned with 
the consequences of the exponential increase of
the energy densities of the vacuum fluctuations inside
the star and about the horizon which are converted into
Hawking photons \cite{THooft}, \cite{Jacobson1}. 
This is due to the Doppler shift,
eq. (\ref{reson})
\begin{equation}
\omega
= \lambda e^{u/4M}
\label{reso}
\end{equation}
as the surface of the star approaches the horizon ($u \to
\infty$). Therefore the Planck scale $\omega=O(1)$ 
 is reached very early in the history of the evaporation
for a 
characteristic value of $\la$ of $O(M^{-1})$) 
after a time $u = O(M \ln M)$. This should be compared to the
lifetime of the black hole $O(M^3)$. 
Thus after the emission of a few
photons ($\Delta n = O(\ln M)$), Hawking radiation is
concerned with the conversion of ``transplanckian''
vacuum fluctuations (of frequencies $\omega)$
into ``cisplanckian'' photons (of frequencies $\la$).
 The point to emphasize is the mixing of radically different
energy scales which is occasioned by 
the exponential growth of the
Doppler shift. This is
in contradistinction  to more conventional physics 
wherein different scales remain separate 
(e.g. atomic versus nuclear).

We recall that the observation of a photon near
${\cal I}^+$ implies the existence of a particular
localized vacuum fluctuation all the way back on ${\cal
I}^-$. The energies involved in this  vacuum fluctuation
are of the order of $\omega$.
 Thus there is an implicit assumption of infinite
mean free path of modes. This is implied ab initio
through the use of free field theory ($\square \phi =
0$). Whilst there is nothing wrong with that insofar as
elementary particle interactions are concerned (their
scale being of $O($Gev or Tev$)$ and presumably
being asymptotically free), 
 it is most probably incorrect since the gravitational
interactions really get in the way. In fact, if anything,
the existence of the Planckian mass scale would inevitably
tend to
stronger forces at higher energies (in local field theory at least,
but maybe not for string theory). 
Be that as it may, if
one uses the Newtonian law of gravitation as a guide, two
spherically symmetric shells of mass $m$ at radius $r$
will have a gravitational interaction energy which
exceeds the mass $m$, for $m=O(1)$ and $r =O( 1)$. From
the above considerations, 
it is seen that these scales are attained
at the threshold of entry into the transplanckian region ($\om=O(1)$, 
$u=O(M\ln M)$)
as the fluctuation approaches the center of the star. The
assumption of free field theory is thus \`a priori
completely inadequate.

Whatever is the accommodation of the transplanckian fluctuations
within the star, there is another vexing short distance problem
which comes about from the fluctuations of the position of the
apparent horizon due to the emission or non emission of a Hawking
photon. In the
semi-classical theory the change in radius of the apparent horizon
is due to $\langle T_{vv} \rangle_{ren}$ only,
since $\langle T_{uu} \rangle_{ren}=0 $ on the horizon.
 However the post-selected presence (absence) of a photon 
around $u_0$ 
is contiguous with the contraction (expansion) of the apparent
horizon with respect to the mean evolution.
More precisely, when the
Hawking photon of frequency  $\la =M^{-1}$ (measured on $\cal I ^+$)
 is within $M^{-1}$
of the horizon (i.e. $r-2M$ at fixed $v$ is 
less than $
M^{-1}$) 
the classical geometric concept of the apparent 
horizon (the locus where $\partial
_v r\vert_u = 0$) 
loses meaning since its radius fluctuates by $O(M^{-1})$.
So the approximation of a free field in a fixed background seems to break down
near the horizon. Thus how can we be sure that the result of the semiclassical
theory wherein
there is no coupling between $u$ and $v$ modes
corresponds to the true physics near the horizon ?
Might there not arise crucial 
correlations which are absent in the semi-classical
theory $(\langle T_{uu}T_{vv} \rangle - 
\langle T_{uu} \rangle
\langle T_{vv} \rangle \neq 0$) ?
 Then putting the whole blame on
$\langle T_{vv} \rangle$ for the change of the geometry near the horizon
could be quite misleading. The mechanism of evaporation
might then be closer to pair creation near the horizon.

Does this mean that everything contained in the semi classical approximation
is
irrelevant\cite{Jacobson2}? Probably not. For one thing it is very
unlikely that Hawking radiation does not occur. 
One must distinguish between the theory of vacuum fluctuations based on 
free field
theory (given in Section \ref{VFHR}) and Hawking radiation as a mean 
theory (see
Section \ref{semicc}).
In the first, expectation values of $\langle T_{\mu\nu} \rangle$ in
the in-vacuum wash out the fluctuations and provide a dynamical origin 
of the Hawking flux. So one can refer to the derivation of Hawking 
radiation in a strong or weak sense, i.e. as derived from free field
theory
or as some effective theory giving rise to a similar 
$\langle T_{\mu\nu} \rangle$.
In support of the existence of such a effective theory, one needs but appeal
to the regularity of $\langle T_{\mu\nu} \rangle$ on the horizon(s).
Indeed both in the collapsing and eternal situations (Sections \ref{Tmunuren} 
and \ref{thermm})
it has been emphasized that one can envisage Hawking radiation as a
response to an incipient singularity of the Boulware vacuum at the 
horizon(s) in such a manner as to erase that singularity.
(To see the connection between the eternal and collapsing case imagine punching
a small hole in the surface of a recipient containing the ``eternal'' black
hole. It then becomes ``ephemeral''. Radiation leaks out and this is neither
more nor less than Hawking radiation albeit with a small transmission
coefficient. In fact the potential barrier which stops
s-waves with energy smaller than $O(1/M)$ and almost all higher
angular momentum modes plays already the role of the small hole since
one has almost a thermal equilibrium behind the 
barrier\cite{Candelas}).
Such a general consideration might well be inherent in the complete
theory wherein one accounts for the gravitational quantum back reaction.
This latter could give rise to violent fluctuations at Planckian
distances which nevertheless leave a regular mean to drive a classical
background geometry at larger scales.

How might we then envisage the fluctuations ?
Almost certainly one would expect that the reduction of
the problem to free s-wave modes is incorrect. Rather
modes will interact mixing angular momenta out to
distances where the notion of a free field is legitimate
-sufficiently outside the horizon so that the Doppler
shift does not entail transplanckian frequencies when the
Hawking photon is extrapolated backwards in time. Within
the interior region one would expect a ``soup''.
And it is to be noted that this soup is
present in Minkowski space, since it is inside the star.

It is interesting to speculate on how one might describe this situation. After
all, even if there is an ungainly soup, it nevertheless fluctuates in a
systematic way if one imposes spherical symmetry and translational symmetry in
time. The fluctuations can then be sorted out according to angular momentum and
frequency; perhaps they must be endowed with a lifetime as well. These things
are quasi particles. If Hawking radiation exists then we know for sure that as a
quasi particle passes through the horizon region it gets converted from
``quasi'' to the free field fluctuations which we have treated in this review.
So it should be possible to come to grips with this problem somewhere in the
middle region. It is not impossible that one will be able to prove that the
Hawking radiation develops out of this Planckian nether nether land and that
the true transplanckian fluctuations are irrelevant
\footnote{In a recent numerical calculation in a model which is analogous to the black hole, Unruh \cite{xxx} has shown that a severe modification of the dispersion relation $\omega(k)$ for $k$ greater than some threshold value in no way affects the thermal spectrum of Hawking emission.}. After all these latter
don't seem to bother us very much here and now so there is some hope that they
are no nuisance there either. They may even result, as suggested above, in a
Planckian spread of the region between the apparent and real horizon since they
incorporate gravity within them.

Another very disturbing problem 
arises in the semi classical
theory. This is the so called unitarity issue\cite{Hawk3}.
In usual evaporation of an isolated system into vacuum,
correlations
are always present. At the early stages an evaporated molecule
gives a kick back to the unevaporated mass causing
correlations between what has and what has not
evaporated. At later stages these correlations get
transferred to correlations among the evaporated
molecules\cite{page93}.
If one were in a pure quantum state, it is
these correlations which encode its purity. Of course
one says evaporation is accompanied by entropy
increase, hence increase in the number of states. But we
attribute this to the coarse graining that is  implicit
in the definition of entropy.

In black hole evaporation, in the semi classical theory,
the correlations 
are in nature
similar to that encountered in usual evaporation in its early stages.
Reference to Section \ref{VFHR}
shows that each Hawking photon leaves behind a
``partner'', a field configuration in vacuum of a
specified local character within the star. As the
evaporation proceeds these correlating configurations
build up.  However in the semi classical theory these
configurations never get out (they are shut up in the closed
 geometry which devellops during the evaporation see Section \ref{semicc})
 and the Hawking
radiation contains no information on the quantum state of
the star. One does not recover unitarity at $r=\infty$.

A few 
options (see for instance the
review article of Preskill\cite{Presk})
seem available on how to confront this situation: 
\begin{itemize} 
\item 
As suggested by the semi classical theory, these inside
configurations are forever lost to the outside observer
(they could end up in the singularity, or in an infinitely
long lived remnant). Unitarity 
is truly violated as originally claimed Hawking\cite{Hawk3}. 

\item The semi classical theory fails at Planckian size black hole, evaporation
stops and a finite long living 
remnant forms whereupon the correlations between the Hawking quanta,
their partners and 
the star's matter are recovered at $r=\infty$ to reconstruct purity.
We recall that for the accelerating mirror (Section \ref{mirro})
the correlations to the partners are completely recovered upon decelerating the mirror. However these considerations cannot be applied directly
for the black hole since, for purity, the degrees of freedom of the star should be recovered as well.

\item 
The semi-classical picture is all wrong in
this regard and the correlations occur 
outside the horizon. 
All the information about the state
of the star leaks out in the radiation\cite{THooft}\cite{THooft2}\cite{SVV}.
 This option however
necessitates either a violation of causality or a
fundamental revision of the concept of background geometry
at scales large compared to the Planck scale. Some authors entertain the
thought that different backgrounds are appropriate for different 
observers\cite{STUg},
e.g. Schwarzschildian or free falling observer. The backgrounds would be
``post-selected'' by the observer. \end{itemize}

In the opinion of the authors it is futile to confront
the unitarity issue without some clear ideas about the
transplanckian issue i.e. quantum gravity.
Indeed the unitary problem cannot be settled without a deep understanding of
the dynamical origin of black hole radiation.

We also wish to
point out that in this quest the fact that the entropy of
a black hole is proportional to its area may play a vital
role. How is it that the entropy of a black hole is equal
to the number of Planckian cells 
 that are necessary to
pave its surface? The horizon seems to block out the
cells which lie deeper than a Planck length within the
hole. Is this related to the expected scenario that
emission will occur at the surface outside the apparent 
horizon i.e. where a quantum
fluctuation begins to belie its presence?

Such are the problems that one must face. Whether their
solution will lead to the quantum theory of gravity or 
the inverse is a moot point. And this primer is certainly
not the place to speculate any further on the question.

No doubt there will turn up further stormy weather to
stir up the already troubled waters that must be
traversed on this journey to terra incognita. Nevertheless
we wish the reader at least some fair weather. Good Luck
and Bon Voyage.


\appendix
\chapter{  Bogoljubov Transformation. }\label{appbog}

In superfluid helium at rest a macroscopic number of particles
occupy the zero momentum state $ <a_0^{\dagger}a_0 > = N_0 $ and
$ N_0 / N $ is a finite fraction as well as $ N / V $ where $V$ is
the volume, $N$ the total number. The thermodynamic limit is $ N
\to \infty, V \to \infty $ with $ N_0 / N $ and $ N / V $ fixed.
The commutation relation $ [a_0, a_{0}^{+}] = 1 $ is then a
negligible consideration when considering operators containing $
a_0$ and $ a_{0}^{+} $ as products that multiply  the typical
unperturbed states which make up the vacuum (ground state), these
unperturbed states being eigenfunctions of $ < n_k >$.

The hamiltonian of interacting particles is (with $  \xi_k = 
k ^{2}/2m)$

\begin{eqnarray}
H & = & H_0 + V \nonumber \\
H_0 & = & \sum_k  n_k  (\xi_k -\mu) \nonumber \\
V & = & {1 \over 2} \sum_{ k_1, k_2, k_3, k_4 } v(k_1, k_2; k_3,
k_4) \quad \delta _{k_1 + k_2, k_3 + k_4} 
	a_{k_1}^{+} a_{k_2}^{+} a_{k_3} a_{k_4} 
 \nonumber \\
\label {A.1} 
\end{eqnarray}

The interaction potential has matrix elements of $ O(1/V^2) $
since the unperturbed states in a box are $ e^{ikr}/ \sqrt V $.

Each $ {\sum_k} $ is of $ O (V) $ so each of the terms in
eq. (\ref{A.1}) is $ O(N)$, $ \mu $ is a chemical potential put in for
convenience so that one can allow $ N $ to fluctuate albeit such
that $ <\Delta N^2 > / < N >^2 = O ( 1/N) $.

The unperturbed ground state has $ N_0 = N$,  $\mu = 0 $, and
$ E = 0 $. The idea of Bogoljubov \cite{Bogo} was to develop a perturbation
theory in the small number $ [ (N_0/N) - 1 ] $. So the technique
is to keep terms in leading order in $ \sqrt {N_0} $ in eq. (\ref{A.1}) 
where one counts $ a_0 = \sqrt {N_0} e^{i \varphi} $, $a_{0} ^{+} =
\sqrt {N_0} e^{- i \varphi} $ in accord with neglect of the
commutator. The leading orders are then $ O(N_{0}^{2}) $ and $ O
(N_0) $. One returns to terms of $ O(\sqrt {N_0}) $ and $ O (1) $
in a standard perturbative procedure as a subsequent step.

Thus to $ O(N_0) $, the perturbation V becomes

\begin{eqnarray}
V = { N_0 \over 2}  \sum_k &&\left[ v(k, - k; 0,0)( e^{ 2 i \varphi} a_{k}^{+}
a_{-k}^+ + e^{- 2 i \varphi} a_k a_{-k} ) \right. \nonumber \\
&&\left. +v(0, k; 0, k)
 a_{k}^{+} a_k  + v(k, 0; 0,k) a_{k} a_k^+ \right]
\label{A.2} 
\end{eqnarray}
The third and fourth terms of eq. (\ref{A.2}) are standard Hartree Fock
single particle energies and may be absorbed into $ H_0 $

\begin{eqnarray}
H_0 & = & \sum _k n_k ( E_ k - \mu) \nonumber \\
E_k & = & \xi _k + [ v (0, k; 0, k) + v (k, 0; 0, k)] N_0/2
\label {A.3}
\end{eqnarray}

The result is a quadratic hamiltonian. This is diagonalized in the
following (Bogoljubov) transformation

\begin{eqnarray}
b_k & = & \alpha_k a_k + \beta_k a_{-k}^{\dagger} \nonumber \\
b_{k} ^{\dagger} & = & \alpha _{k} ^{*} a_{-k}^{\dagger} + \beta
_{k}^{*} a_k \label {A.4}
\end{eqnarray} 
where the phase of $ \alpha $ is $ e^{-i\varphi} $ and of $
\beta $ is $ e ^{+i \varphi} $ and 
\begin{eqnarray}
E_k \vert \alpha_k \beta_k \vert =( \vert \alpha _k \vert^2 +
\vert \beta_k \vert ^2) \vert V_k \vert \label {A.5}
\end{eqnarray}
with $ \vert V_k \vert = \vert v(k, -k; 0, 0) \vert$. Canonical commutation
relations for $ b_k $ gives $ \vert \alpha_k \vert ^2 - \vert \beta _k
\vert ^2 = 1 $.

Thus the unperturbed ground state is not an eigenstate 
of the total hamiltonian $H$ but is unstable. In particular, had we chosen this
state as the initial state it  would evolve to the true ground state through
emission of
 $ k,-k $ pairs. We leave  to the reader the pleasure to confirm the
Nambu Goldstone theorem for this case 

\begin{eqnarray}
[ H - < \Omega \vert H \vert \Omega > ] = \sum \omega _k b_{k}
^{\dagger} b_k \label {A.6}
\end{eqnarray}
where $ \lim_{k\to 0}\omega _k = C \vert k \vert +
O(k^2)$. He may also show that $ b_{k} ^{\dagger} \vert 
\Omega \rangle$
corresponds to the creation of a longitudinal density fluctuation
in this approximation, i.e. a phonon.


\chapter{ 
Functional Integral Technique. }\label{functint}

We use standard field theoretical techniques. The uninformed
reader will find an account in refs. \cite{Schw}\cite{BD}\cite{PaBr0}.
The in	vacuum to out
vacuum amplitude is

\begin{eqnarray}
e^{iW} = \int {\cal D} \phi\  e^{i S (\phi)} = \langle out, 0, \vert 0, in\rangle 
\label {A.7} 
\end{eqnarray}

\noindent where $ \phi$ is a complex scalar field, $ S(\phi) $ is the
action to go from initial to final configurations [ which we take
to be a quadratic form: free field theory in the presence of an
electromagnetic and/or gravitational field ] in time $ t $ (which
tends to $ \infty $ ) and the $\rm mass^2 $ has a
small negative imaginary part. The mass dependence of $ S $ is of
the form $ \int -m^2 \phi \phi^*\/d^d x $ (with $ \sqrt g = 1) $ so that

\begin{eqnarray}
{\partial W \over \partial m^2}& =& -
 \int d^d x \langle out, 0, \vert 
 \phi^2 \vert0, in\rangle  = - \int G_F (x, x) d^d x \nonumber\\
& = & - {\rm tr} G_F \quad ,\label {A.8}
\end{eqnarray}
where $ G_F (x, x^{\prime}) $ being the Feynman propagator to go from $
x^{\prime}$ to $ x $. In terms of the heat kernel $K$ one has

\begin{eqnarray}
G_F (x, x^{\prime}) = \int_{0} ^{\infty} \!ds\  e^{- i m^2 s} K(x,
x^{\prime}; s) \label {F.9}
\end{eqnarray}
where
\begin{eqnarray}
(\Dalamb + i {\partial \over \partial s}) K = i \delta
(s)\delta(x-x^{\prime}) \label {F.10} 
\end{eqnarray}

In eq. (\ref{A.8}) we have written $ \partial W/ \partial m^2 $ as a trace.
Clearly this is formal and one has to watch one's step on the
measure. The following steps are valid because the passage from $
x$ representation to $u$ representation is unitary \cite{PaBr0}.
Putting it all together we have upon integrating eq. (\ref{A.8}) 
over $
m^2$ 

\begin{eqnarray}
W = -  \int _{0} ^{\infty} {ds \over s } e^{-im^2 s}  {\rm tr}
K{(s)} \label {F.11} 
\end{eqnarray} 

We now specialize to the case of constant electric field in the
gauge $ A_x = 0 $, $ A_t =  E x $. As in Chapter \ref{ELEC},
we can label the modes by $
\omega$ and work in the $u$ representation, whereupon
\begin{eqnarray}
W = - {i} \sum_\omega \int _{0}^{\infty}{ds \over s } e^{-im^2 s} \int
K_\omega (u, u; s) du \label {F.12}
\end{eqnarray}
where $ K_\omega (u,u_0, s)$ obeys
\begin{eqnarray}
[ u \partial_u + {1 \over 2} - \partial_\tau] K_\om = \delta(\tau)
\delta(u - u_0) \label {F.13} 
\end{eqnarray}
and $ \tau = 2E\ s $, and we used eq. (\ref{equchi})
 with $i \varepsilon $
replaced by $ \partial/\partial \tau.$
The solution is
\begin{eqnarray}
K_\om(u, u_0; \tau) = \theta (\tau) \delta (u e^{\tau} - u_0)
e^{\tau /2} \label {F.14}
\end{eqnarray}
to give from eq. (\ref{F.11}) 
\begin{eqnarray}
{\rm Im}W & = & - {\rm Im}\ {i}\sum_ \omega \int_{0} ^{\infty} {d \tau \over \tau}
e^{-im^2 \tau/2E} \int du  \delta (u e^ {\tau} - u) e^{\tau/2}
\nonumber \\ 
& = & - {\rm Im}{i\over 2 }\sum _\omega \int _{0} ^{\infty} {ds \over s }
{e^{-im^2 s} \over \mbox {sinh}Es} 
\nonumber \\
& = &{1\over 2} \sum _\omega \mbox{ln} (1 + e^{-\pi m^2/E}) \label {F.15}
\end{eqnarray}
where the last equality is obtained by picking 
up the poles (with $ m^2 = m^2-i\epsilon)$
recovering therefore
the Schwinger formula eq. (\ref {schwfor})
since $\sum_ \omega = ELT/2 \pi $.

When the path parameter $ \tau $ is expressed in terms
of proper time these poles on the imaginary axis are related to
multiple excursions through the tunneling region. For example, Born
approximation (WKB) for the tunneling amplitude corresponds
to the pole at $  Es = i \pi $ and represents one excursion back
and forth $ (\Delta \tau_{proper} = 2 \pi /a $ in the movement of
the wave packets of Section 2.3).

One may perform similar tricks to put into evidence the
instability of Schwarzschild vacuum in the complete space spanned
by the Eddington Finkelstein coordinates. Here the Schwinger
counting parameter $ s $ is once more proportional to proper
time. The complete analysis is considerately more tedious and
complicated than the above but the essential features are the
same. We  refer the reader
to ref. \cite{PaBr0} for details.

In this case it turns out that the convenient variable that
describes the effective motion of a packet is $ p $, the momentum
conjugate to the Eddington Finkelstein
 coordinate $x= r - 2 M $ at fixed $v$. There are three
classes of paths which contribute to $ W $, those in which initial
and final momenta have the same sign and that in which it goes from
negative to positive momentum. Upon taking the trace initial and
final momenta are set equal, so this operation requires a
careful limiting procedure. One finds that it is the third class
that encodes the instability and that the time to execute this
movement is $ \Delta v = i 8 \pi M = i \beta_H $ i.e. $ \beta_H
$ is the imaginary time to go from $p$ to $- p$ and back. One finds 
\begin{eqnarray}
{\rm I m} W_{BH} = - {T \over 4 \pi } \int _{0} ^{\infty} \!d \omega
\ \ln [ 1 - e^{- \beta_H \omega} ] \nonumber\\
\end{eqnarray}
precisely the one dimensional partition function. In this manner
$ \beta_H \omega $ is indeed interpretable as the action for a
Hawking photon to tunnel out into existence.


\chapter{
Pre- and Post-Selection, Weak Measurements. }\label{weak}

Pre- and post-selection consists in specifying both the initial and the final
state of a system (denoted by $S$
in the sequel). 
Pre and post selection is not an unusual procedure in
physics. For instance when dealing with transition amplitudes, scattering
amplitudes, etc... one is performing pre and post selection. 

In the first part of this appendix we shall implement post-selection in a 
rather formal way by acting on the state with projection operators which select
the desired final state(s) following the treatment of
\cite{MaPa}. This generalizes the approach of \cite{Ahar}.

In the second part of this appendix we show 
how  post-selection may be realized operationally following the rules of
quantum mechanics by coupling to $S$
an additional system in a metastable
state   (the "post selector" $PS$) which will make a transition only if the
system is in the required  final state(s).
The weak value of an operator obtained in this manner changes as time goes by
from an asymmetric form to an expectation value, thereby making contact with
more familiar physics. This extended formalism finds important application when
considering the physics of the accelerated detector since the accelerated
detector itself plays the role of post selector. In this way one can study the 
EPR correlations between the state of a uniformly accelerated detector and the
radiation field, thereby clarifying and generalizing the results of
\cite{UnWa}, \cite{AuMu}, \cite{Grow}.

\section{Weak Values}
The approach developed by
Aharonov et al.\cite{Ahar} for studying pre- and post-selected
ensembles
consists in performing at an intermediate time a "weak measurement"
on $S$. In
essence one studies the first order 
effect of $S$ (ie. the back reaction) onto an
additional system
taken  by Aharonov et al. to be the measuring device. But the formalism is
more general. Indeed
when the first order (or weak-coupling)  approximation is
valid, the backreaction takes a simple and universal form governed by  a
c-number, the "weak value'' of the operator which controls the interaction.

The system to be studied is in the state $\ki$ at
time $t_i$ (or 
 more generally is described by a density matrix $\rho_i$).
The unperturbed time
evolution of this pre-selected state
 can be described by the following density matrix
\begin{equation} \rho_S(t) = U_S(t,t_i)\ki \bi U_S(t_i,t) \label{weaki}
\end{equation}  where $U_S = \exp (- i H_S t)$ is the time evolution operator
for the system $S$. The post-selection at time $t_f$ consists in specifying
that the system belongs to a certain subspace, ${\cal H}_S^{0}$, of
${\cal H}_S$.
Then the probability to find the system in
this subspace at time $t_f$ is \begin{equation}
P_{\Pi_S^{0}} = Tr_S \Bigl [ \Pi_S^{0} \rho(t_f) \Bigr] = Tr_S \Bigl[
\Pi_S^{0} U_S(t_f,t_i)\ki \bi U_S(t_i,t_f) \Bigr] \label{weakii}
\end{equation}   where
$\Pi_S^{0}$ is the projection operator onto ${\cal H}_S^0$ and $Tr_S$ is
the trace over the states of system S. 
In the special cases wherein the specification 
of the final state is to be in a pure state $\ket{\psi_f}$ (ie. 
$\Pi_S^0=\ket{\psi_f}\bra{\psi_f}$)
then the probability is simply given by the overlap 
\begin{equation}
P_{f}= \vert \bra{\psi_f} U_S(t_f,t_i)\ket{\psi_i}\vert^2
\end{equation}

Following Aharonov et al. we introduce an additional system, called the "weak
detector" ($WD$), coupled to $S$. 
The
interaction hamiltonian between $S$ and $WD$ is taken to be of the form
$H_{S-WD} (t) = \e f(t) A_S B_{WD}$ where $\e$ is a coupling constant,
$f(t)$ is a c-number function, $A_S$ and $B_{WD}$ are hermitian operators acting
on $S$ and $WD$ respectively.

Then to first order in $\e$ (the coupling is weak), the evolution of the coupled
system $S$ and $WD$ is given by
 \begin{eqnarray} 
\rho(t_f)& =&\ket{\Psi (t_f)}\bra{\Psi(t_f)}\nonumber\\
\mbox{\rm where}&&\nonumber\\
\ket{\Psi(t_f)}& =& \Bigl [ 
U_S(t_f,t_i)U_{WD}(t_f,t_i) - i\e \int_{t_i}^{t_f}\! dt \
 U_S(t_f,t)U_{WD}(t_f,t)  f(t) A_S B_{WD}\times \Bigr. \nonumber\\ 
&& \Bigl.
  U_S(t,t_i)U_{WD}(t,t_i) \Bigr] \ki \ket{WD}
\label{weakiii}
 \end{eqnarray} 
where $U_S$ and $U_{WD}$ are
the free evolution operators for $S$ and  $WD$ and  $\ket{WD}$ is the initial
state of  $WD$. Upon post-selecting at $t=t_f$ that $S$ belongs to the
subspace ${\cal H}_S^{0}$ and tracing over the  states of the
system $S$, the reduced density matrix describing the $WD$ is obtained
\begin{eqnarray}
 \rho_{WD}(t_f) &=& Tr_S
 \Bigl [ \Pi_S^0 \rho(t_f)
\Bigr]\end{eqnarray}
In
the first order approximation in which we are working it takes a very simple
form \begin{eqnarray}
 \rho_{WD}(t_f)&\simeq &
P_{\Pi_S^{0}}\ket{\Psi_{WD}(t_f)}
\bra{\Psi_{WD}(t_f)}\nonumber\\ 
\mbox{\rm where}&&\nonumber\\
\ket{\Psi_{WD}(t_f)}&=&
\left[ U_{WD}(t_f,t_i) -
 i\e
\int_{t_i}^{t_f}\! dt \ U_{WD}(t_f,t)  f(t) A_{Sweak}(t)B_{WD}U_{WD}(t,t_i)
 \right] \ket{WD}\nonumber\\
&&\label{weakiv}
 \end{eqnarray} 
where $P_{\Pi_S^{0}}$ is the probability to be in subspace ${\cal H}_S^{0}$
and
  \begin{equation} A_{S weak}(t) = {Tr_S
\Bigl [ \Pi_S^{0} U_S(t_f,t) A_S U_S(t,t_i) \ki\bi U_S(t_i,t_f) \Bigr]
\over Tr_S \Bigl [ \Pi_S^{0} U_S(t_f,t_i)\ki\bi U_S(t_i,t_f) \Bigr]}
\label{weakv}
\end{equation}
 is a c-number called  the weak value of $A$. If
one specifies completely the final state, $\Pi_S^{0} =
\ket{\psi_f}\bra{\psi_f}$  then
the result of Aharonov et al.
obtains:
\begin{equation} A_{S weak}(t) = {\bra{\psi_f}
U_S(t_f,t) A_S U_S(t,t_i) \ki  \over \bra{\psi_f} U_S(t_f,t_i) \ki}
\label{weakvi}
 \end{equation}
The principal feature of the above formalism is its independence on the
internal structure of the $WD$. The first order backreaction of $S$
onto $WD$ is
universal: it is always controlled by the c-number $A_{S weak}(t)$,
the ''weak
value of $A$". Therefore
if $S$ is coupled to itself by an
interaction hamiltonian, the backreaction
will be controlled by the
weak value of $H_{\rm int}$
in first order perturbation theory.
For instance the modification of the probability that the
final state belongs to ${\cal H}_S^{0}$ is
given by the imaginary part
of $H_{{\rm int}\ weak}$.
Indeed 
\begin{eqnarray}
P_{\Pi_S^0}^\prime &=& Tr_S \left[
\Pi_S^0 ( 1 - i \int\! dt\  H_{\rm int} )
\rho_i ( 1 - i \int\! dt\  H_{\rm int} )\right]\nonumber\\
&=&P_{\Pi_S^0} (1 - 2 { \rm Im} H_{{\rm int}\ weak})
\end{eqnarray}

The weak value of $A$ is complex. By  performing a
series of
measurements on $WD$
and by varying the coupling function $f(t)$, the real and imaginary part of
$A_{S weak}$ could in principle be determined. Here the word ''measurement"
must be understood in its usual quantum sense: the average over repeated
realizations of the same situation. This means that the weak value of $A_S$
should also be understood as an average. The fluctuations around $A_{S weak}$
are encoded in the second order terms of eq.~(\ref{weakiii}) which have been
neglected in eq.~(\ref{weakiii}).

To illustrate the role of the real and imaginary parts of $A_{S weak}$, we
recall the example of Aharonov et al consisting of a  weak
detector which has one degree of freedom $q$,
 with a gaussian initial state
$\scal{q}
{WD}=e^{-q^2/
2\Delta^2}$,$-\infty < q < + \infty$.
 The unperturbed hamiltonian of $WD$ is taken to vanish
(hence $U_{WD}(t_1,t_2) = 1$)
 and the interaction hamiltonian is  $H_{S-WD} (t) =\epsilon
\delta(t-t_0) p A_S$ where $p$ is the momentum conjugate to $q$. Then after
the post-selection the state of the $WD$ is given to first order by
\begin{eqnarray}
\scal{q}{WD(t_f)} &=&\left(1-i\epsilon pA_{S weak}(t_0) \right)
e^{-q^2/ 2\Delta^2}\nonumber\\ &\simeq& e^{-i\epsilon pA_{S weak}(t_0)} e^{-q^2/
2\Delta^2}\nonumber\\ &=&e^{-\left(q- \epsilon A_{S weak}(t_0)\right)^2/
2\Delta^2}\nonumber\\ &=& e^{-\left(q- \epsilon {\rm Re }  A_{S
weak}(t_0)\right)^2/ 2\Delta^2} e^{+i\epsilon q {\rm Im}A_{S weak}(t_0)/
\Delta^2} \label{weakviii}
 \end{eqnarray}
 The real part of $A_{S weak}$ induces a
translation of the center of the gaussian, the imaginary part
a change in the
momentum. Their effect on the $WD$ is therefore measurable.
The validity of the first order approximation requires 
$\e A_{S weak} / \Delta <<1$.

It is instructive to see how unitarity is realised in the above formalism.
Take
 $\Pi^j_S$ 
to be  a 
complete orthogonal set of projectors acting on the
Hilbert space of $S$. Denote by $P_j$ the probability that
the final state of the system belong to the subspace spanned by $\Pi^j_S$ and
by $A_{S weak}^j$ the corresponding weak value of $A$. Then the mean value of
$A_S$ is  \begin{equation} \bi A_S \ki =
\sum_j P_j A_{S weak}^j  \label{weakix} 
\end{equation} 
Thus the mean
backreaction if no post-selection is performed is the average over the
post-selected backreactions (in the linear response approximation). 
Notice that the imaginary parts of the weak
values necessarily cancel since the l.h.s. of eq.~(\ref{weakix}) is
real. Equation (\ref{weakix}) is the short cut used in the main text to
obtain with minimum effort the weak values.

\section{Physical Implementation of Post Selection}
Up to now the postselection has been implemented by projecting
by
hand
the state of
the system onto a certain subspace ${\cal H}^{0}_S$.
Such a projection may be
realised operationally by introducing
an additional
quantum
system, a ''post-selector" ($PS$),
coupled in such a way
 that it will make a transition if and only if the system $S$
is in the required final state. Then by considering
only that subspace of the
final states in which $PS$ has made the
transition, a post-selected state is specified. This
quantum description of the
post-selection
is similar in spirit to the measurement theory developed in ref.
\cite{Vonn}: by introducing explicitly the measuring device in the
hamiltonian the collapse of the wave function ceases to be a necessary
concomitant of measurement theory. As we have mentioned, this formalism is
the basis for a general treatment of the energy density correlated to
transitions of an accelerated detector.

We shall consider the very simple model of a $PS$ having two states,
initially in the ground state, and coupled to the system by an interaction
of the form \begin{equation}
H_{S-PS} = \la g(t) ( a^\dagger Q_S + a Q_S^\dagger)
\label{weakx}
\end{equation}
where $\la$ is a
coupling constant, $g(t)$ a time dependent function, $a^\dagger$  the operator
that induce transitions from the ground state to the exited state of the
$PS$, $Q_S$ an operator acting on the system $S$. The
postselection is performed at $t=t_f$ and consists in finding the $PS$ in the
exited state.

For simplicity we shall work to second order in $\la$ (although in principle
the interaction of $PS$ with $S$ need not be weak).
The wave function of the combined system $S+WD+PS$ is in interaction
representation to order $\epsilon$ and order $\lambda^2$
\begin{eqnarray}
\lefteqn{{\cal T} e^{-i \int\! dt \ H_{S-WD}(t) + H_{S-PS}(t)}
\ket{\psi_i}\ket{WD}\ket{0_{PS}}=}\nonumber\\
& & \left[
1 -i \int dt \left( H_{S-WD}(t) + H_{S-PS}(t) \right)\right.
\nonumber\\
& & \ -{1 \over 2}
\int dt \int dt^\prime
{\cal T}\left[ H_{S-PS}(t)  H_{S-PS}(t^\prime) \right]
-
\int dt \int dt^\prime
{\cal T}\left[ H_{S-WD}(t)  H_{S-PS}(t^\prime) \right]
\nonumber\\
& & \left.\ 
+{i \over 2}
\int dt \int dt^\prime\int dt^{\prime\prime}
{\cal T}\left[ H_{S-PS}(t)  H_{S-PS}(t^\prime) 
 H_{S-WD}(t^{\prime\prime})\right]
\right]
\ket{\psi_i}\ket{WD}\ket{0_{PS}}\nonumber\\
& \strut & \label{weakxi}
\end{eqnarray}
where $\ket{0_{PS}}$ is the ground state of $PS$ and ${\cal T}$ is
the time ordering operator. 
The probability of being in the excited state at $t=t_f$ at order $\la^2$ is
\begin{equation}
P_{excited}
= \la^2  \bra{\psi_i} \int dt g(t) Q^\dagger_S
\int dt^\p g(t^\p) Q_S \ket{\psi_i} 
\end{equation}
Upon imposing that the
$PS$ be in its excited state at $t=t_f$ the
resulting
wave function  is, to order $\epsilon$ and $\la^2$,
\begin{eqnarray}
&&\quad \left [
-i \int dt \la g(t) Q_S(t)  \right.
\nonumber\\
&& \left. -
\int dt \int dt^\prime
{\cal T}
\left[\epsilon f(t) A_S(t) B_{WD}(t) \la g(t^\prime) Q_S(t^\prime)
\right]
\right]
\ket{\psi_i}\ket{WD}a^\dagger\ket{0_{PS}}\nonumber\\
\label{weakxii}\end{eqnarray}
Making a density matrix out of the state (\ref{weakxii}), 
tracing over the states of $S$ and $PS$
yields the reduced density matrix $\ket{\Psi_{WD}}\bra{\Psi_{WD}}$ of $WD$ to
order $\epsilon$ where
\begin{equation}
\ket{\Psi_{WD}}=\left[
1 - i \epsilon \int dt_0 f(t_0) B_{WD}(t_0)
A_{S weak}^{excited}(t_0) \right] \ket{WD}
\label{weakxiii}
\end{equation}
and
\begin{equation} A_{S weak}^{excited}(t_0) =
{ \bi \int dt g(t) Q_S^\dagger (t)  \int dt^\p
g(t^\p) {\cal T } \left[ A_S(t_0) Q_S(t^\p) \right]
\ki \over \bi \int dt g(t) Q_S^\dagger (t) \int
dt^\p g(t^\p) Q_S(t^\p) \ki} \label{weakxiv} \end{equation}
Note how the weak value of $A_S$ results from the quantum mechanical
interference of the two terms in eq.~(\ref{weakxii}). 

There are
several important cases when the time ordering in eq.~(\ref{weakxiv}) 
simplifies.
If $g(t)$ is non vanishing only after $t=t_0$ then $A_W$ takes a
typical (for a weak value) asymmetric form 
\begin{equation} A_{S weak}^{excited}(t_0) =  {
\bi \int dt g(t) Q_S^\dagger (t)
 \int dt^\p g(t^\p) Q_S(t^\p)  A_S(t_0) \ki \over \bi \int dt g(t) Q_S^\dagger
(t) \int dt^\p g(t^\p) Q_S(t^\p) \ki} \label{weakxv} \end{equation}
If in addition $g(t)=\delta(t-t_f)$, $t_f >t_0$ and $Q_S=\Pi_S^0$,
eq.~(\ref{weakv}) is recovered using $ (\Pi_S^0)^2 = \Pi_S^0$. This
is expected since in this case the post--selector has simply gotten
correlated to the system in the subspace ${\cal H}^0_S$

If on the other hand $g(t)$ is non vanishing only before $t=t_0$ then
the time
ordering operator becomes trivial once more and eq.~(\ref{weakxiv}) takes the
form \begin{equation} A_{S weak}^{excited}(t_0) =  { \bi \int dt g(t)
Q_S^\dagger (t) A_S(t_0)
 \int dt^\p g(t^\p) Q_S(t^\p)   \ki \over \bi \int dt g(t) Q_S^\dagger (t)
\int dt^\p g(t^\p) Q_S(t^\p) \ki} \label{weakxvi}
\end{equation}
This is by construction the
 expectation value of $A_S$ if the $PS$ has made a transition. 
It is
necessarily real contrary to  eq.~(\ref{weakxv}) when
the weak measurement is performed before the 'collapse' induced by 
the post-selection.

Finally, the weak value of $A_S$ if the $PS$ has not made a
transition can also be computed. 
Once more the two cases discussed in eqs (\ref{weakxv}) and (\ref{weakxvi})
are particularly simple: if $g(t)$ is non vanishing only after $t=t_0$ one
finds \begin{eqnarray}
A_{S weak}^{deexcited}(t_0) &=&{1 \over 1 - P_{excited}  }\Biggl(
\bra{\psi_i}A_S\ket{\psi_i} \nonumber\\
& &-\la^2 {\rm Re}  \left[
\bi \int dt g(t) Q_S^\dagger (t)
 \int dt^\p g(t^\p) Q_S(t^\p)  A_S(t_0) \ki\right]\Biggr)\nonumber\\
&&  \label{weakxvii}
\end{eqnarray}
On the other hand if $g(t)$ is non vanishing only before $t=t_0$ one finds
\begin{eqnarray}
A_{S weak}^{deexcited}(t_0) &=&{1 \over 1 - P_{excited}  }\Biggl(
\bra{\psi_i}A_S\ket{\psi_i} \nonumber\\
& & -{1 \over 2} \la^2 {\rm Re} \left[
\bi A_S(t_0) \int dt 
 \int dt^\p {\cal T} g(t) Q_S^\dagger (t) g(t^\p) Q_S(t^\p)   \ki\right]
\Biggr) \nonumber\\
\label{weakxviii} \end{eqnarray}
These are related to the mean value of $A_S$
and to eq.~(\ref{weakxiv}) through the unitary relation eq.~(\ref{weakix}):
if $g(t)$ is non vanishing only after $t=t_0$
\begin{eqnarray}
 P_{excited}A_{S weak}^{excited}(t_0)
+ (1 - P_{excited} )A_{S weak}^{deexcited}
= \bi A_S \ki \label{weakwviiii}
\end{eqnarray}
if $g(t)$ is non vanishing only before $t=t_0$
\begin{eqnarray}
\lefteqn{ P_{excited}A_{S weak}^{excited}(t_0)
+ (1 - P_{excited} )A_{S weak}^{deexcited}\ =}\nonumber\\ 
& &
\bi {\cal T} e^{i\int \! dt \ H_{S-PS}}
A_S 
{\cal T} e^{-i\int \! dt \ H_{S-PS}}
\ki
\ =
\nonumber\\ 
& &\bi A_S \ki 
  +\la^2 
\bi \int dt g(t) Q_S^\dagger (t) A_S(t_0)
\int dt^\p g(t^\p) Q_S^\dagger (t^\p)\ki \ -\nonumber\\
& &  -\la^2 {\rm Re } \bi A_S(t_0) \int dt
\int dt^\p {\cal T} g(t) Q_S^\dagger (t) g(t^\p) Q_S^\dagger (t^\p)\ki 
\label{weakwxx}
\end{eqnarray}
where the right hand side is the average value of $A_S$ before
eq.~(\ref{weakwviiii}) and after eq.~(\ref{weakwxx}) the detector has interacted with $S$.


\chapter{S-wave Hawking Radiation for a General Collapsing Spherical
Star Without Back Reaction
 }\label{sphradd}

We shall consider a collapsing sphere of matter with a well
defined surface. Exterior to the surface the geometry is
Schwarzschild, parametrized by a fixed mass M. This part of
the space shall be coordinatized by advanced Eddington-Finkelstein
 coordinates ($ v, r,\theta, \varphi $) defined in eq.~({\ref{1.7}).
 Interior to the star we
shall use a set ($ T, X, \theta, \varphi $) in terms of which the length interval for
a general dynamic spherically symmetric space is

\beq  ds^2 = -a^2 (T, X)\  [dT^2-dX^2 ] + b^2 (T, X)\  d\Omega^2
\label{AHwrad1} \feq

\noindent with $  d \Omega^2 = d \theta^2 + \sin^2\ \theta d
\varphi^2 $. This system is used in the  numerical calculation
of $\langle T_{\mu \nu} \rangle$ that was reported in 
Section \ref{Tmunuren} (with $U, V = T \mp X$). In that case it was
used to cover the whole space so as to account for the back
reaction outside the star as well. Since we are not taking
into account the back reaction in the present case, we use
eq.(\ref{AHwrad1}) in the inside region only. The curvature stemming from
the metric components $a$, $b$ is driven by classical sources.

The coordinate $X$ is a radial coordinate, but areas of
spheres are $ 4 \pi b^2$. Thus at a given point of coordinates ($T,\ X$) on the
star's surface one has the identification

\beq r = b (T, X) \label{AHwrad2}\feq

\noindent We shall choose the origin of $X$ so that $r = 0$
coincides with $X = 0$.

Hawking radiation follows from the combined gravitational and
Doppler shifts that a photon experiences on its voyage from
${\cal I}^- \mbox{\rm to } {\cal I}^+$. It begins on ${\cal I}^-$ as a
packet (superposition) of modes $ e^{-i\omega v}$ and end up on $ {\cal I}^+$ as a
packet of modes $ e^{i \omega f(u)}$. We consider a particular radial packet
that emanates from a point on $ {\cal I} ^-$ denoted by $ v =
{\hat u} $. The function $f(u)$ varies according to the value of
$ \hat u$ so we will call it $ \hat u (u)$. The shift of
frequency on $ {\cal I} ^+$ is

\beq \omega (u)\big|_{{\cal I}^+} = \omega (d {\hat u} / d
u)\qquad . \label{AHwrad3}\feq

\noindent  One may deduce this, for example, by the
conservation of the number of oscillations in a given segment
of wavefront $( =  \omega d v = \omega d {\hat u} =
\omega \vert_{{\cal I}^+} d u)$

There are three important points which the ray visits. These
are:

\begin{enumerate}
\item $P_I\equiv ({\overline X},{\overline T})$ where the ray penetrates the
star. This labels the sphere which is the intersection between
the incoming lightcone $ v = \hat u$ and the star's surface.
\item $P_C$ where reflection on the axis $X = 0$ occurs.
\item $P_O\equiv ({\tilde X}, {\tilde T})$, where the ray leaves the
star.
\end{enumerate}

\noindent One has
\beq {\tilde T} - {\tilde X} = {\overline T} + {\overline X} \label{AHwrad4} \feq
\noindent since the coordinate time interval $ (= {\tilde T}
- {\overline T}) $ is equal to the coordinate distance toward
in the star $ (= {\tilde X} + {\bar X}) $.

We now must specify the surface of the trajectory. In any
given model this specification is, of course, correlated to
the metric coefficients $a$, $b$ along the trajectory, but there
is no need to keep track of this in the general analysis. We
parametrize the trajectory of the surface in each of the
systems according to

\begin{eqnarray}
v &=& V_{\mbox{\rm s}} (r)\quad ,  \nonumber \\
X &=& \Xi_{\mbox{\rm s}} (T)\quad .  \label{AHwrad5}
\end{eqnarray}
\noindent Then the first intersection point $P_I$ is given in terms
of the ray $ {\hat u}$ by the solution of

\begin{eqnarray}
{\hat u} &=& V_{\mbox{\rm s}} ({\overline r})\nonumber \\
{\overline r} &=& b ( {\overline T}, \Xi_{\mbox{\rm s}} ({\overline
T}) \nonumber \\
{\overline X} &=& \Xi_{\mbox{\rm s}} ({\overline T}) \label{AHwrad6}
\end{eqnarray}

\noindent The second intersection point is obtained from eq.(\ref{AHwrad4})
\beq {\tilde T} - \Xi_{\mbox{\rm s}} ({\tilde T}) = {\overline T} +
\Xi_{\mbox{\rm s}} ({\tilde T}) \label{AHwrad7} \feq
\noindent The wave then propagates out to $ {\cal I}^+$
according to $ e^{i \omega {\tilde U} }$ where $ {\tilde U} =
{\tilde T} - {\tilde X}$. The exterior coordinate $u$ is given
by
\beq u = V_{\mbox{\rm s}} ({\tilde r}) - 2 r^{\star} ({\tilde r}) \label{AHwrad8} \feq
\noindent where $ {\tilde r} = b ({\tilde T}, \Xi_{\mbox{\rm s}} ({\tilde
T}))$. Inverting this last relation gives $ {\tilde T}
({\tilde r})$. One then computes $ {\tilde X}({\tilde r})$
from eq.(\ref{AHwrad2}) hence $ {\tilde U}({\tilde r})$. Together with eq.(\ref{AHwrad8})
this chain then gives the required relation between $u$ and ${\tilde U}$.
This matter is the sought function ${\tilde U}= {\hat u} (u)$. In this
way it is seen that the phase of the outgoing wave gives a
sort of `` X-ray picture " of the interior of the star.

To perform these various inversions is an arduous task. At the
end of the section we shall present  the results of a calculation
wherein the collapsing star consists of dust. As announced in Section
\ref{semicc}  almost all of Hawking radiation occurs in
a small interval where the point $P_O \equiv ({\tilde X}, {\tilde T}) $ is
near the horizon, ${\cal H}$. As in Section \ref{hawking} we shall
linearize the equation of motion of the star's trajectory for
such points. Similarly, the point $P_I \equiv  ({\overline X}, {\overline
T}) $ is near the extension of ${\cal H}$ into a past lightcone (the
last null rays that are reflected into a future lightcone expanding up to the
asymptotically flat infinity). See Fig.~(\ref{collapse}b). We shall denote the
intersection of the star's surface with ${\cal H}$ as $O\equiv (X_O, T_O) $ and
the intersection with the backward extension of ${\cal H}$ as $I\equiv(X_I, T_I)$.
 The linearized forms of the trajectory of the surface of the star for points
near $O$ and $I$ are
\begin{eqnarray}
v - v_O &=& k_O (r - 2 M) \quad ,\nonumber \\
v - v_I &=& k_I (r - r_I) \quad ,  \label{AHwrad9}
\end{eqnarray}

\noindent which in interior coordinates  will be written
\begin{eqnarray}
X - X_O &=& \beta_O (T - T_O)\quad ,  \nonumber \\
X - X_I &=& \beta_I (T - T_I) \quad . \label{AHwrad10}
\end{eqnarray}

The relation between $ \beta_O $ and $ k_O$ (or $ \beta_I $
and $ k_I$) is obtained by equating the expressions  of proper time intervals
$(= ds^2)$  on the star's surface at the intersection
points $O$ and $I$, when expressed in both coordinate
systems. Thus

\beq (1 - {2 M \over r})\; dv^2 - 2 dr dv = a^2 [ d T^2 - d X^2
] \label{AHwrad11} \feq

\noindent with all differentials taken along the trajectory of
the star's surface. In such differentials we also have

\beq dr = b^\prime \; d X + {\dot b}\; d T \qquad .\label{AHwrad12} \feq

\noindent Dots are derivatives with respect to $T$ and primes derivatives 
with respect
to $X$. Using eqs.(\ref{AHwrad9}, \ref{AHwrad10}, \ref{AHwrad11}, \ref{AHwrad12}) gives
at $X = X_O$ 
\begin{eqnarray}
[ ( 1 - {2 M \over r_I}) \; k_I ^2 -  2 k_I ] [ {\dot b}_I +
b_I ^\prime \beta_I ]^2 &= &a_I (1 - \beta_I ^2 ) \label{AHwrad13} \\
-  2 k_O [ {\dot b}_O + b_O ^\prime \beta_O ] ^2& =& a_O (1 -
\beta_O^2)  \label{AHwrad14} 
\end{eqnarray}
\noindent Here $  a_O, b_O, (a_I, b_I)$ denotes the values of the metric components
$a$ and $b$ at points $O$ and $I$ respectively.

Let us now track the various chain of variables by following
the ray backwards in time. Near $O$ we have on the suface point
$ (\tilde X, \tilde T)$ the value of $u$ given by

\beq  u \simeq v_0 - 4 M - 4 M ln {\tilde r - 2 M \over 2 M}
\label{AHwrad15} \feq
\noindent where
$$  {\tilde r} = b ( {\tilde T}, \Xi_{\mbox{\rm s}} ({\tilde T}) \cong
2 M + \dot b_O ({\tilde T} - T_O) + b^\prime _O \dot \Xi_{\mbox{\rm s}} ({\tilde
T} - T_O) \nonumber $$
\noindent so that
\beq  {\tilde r} - 2 M \cong ({\dot b}_O + b^\prime _O \;\beta_O) (
{\tilde T} - T_O) \label{AHwrad16} \feq
\noindent And from eq.(\ref{AHwrad7}) we have
\beq  {\tilde T} - T_O = {\overline T} - T_I + ( \Xi_{\mbox{\rm s}}
({\overline T}) - \Xi_{\mbox{\rm s}} (T_I)) + (\Xi_{\mbox{\rm s}} ({\tilde T}) -
\Xi_{\mbox{\rm s}} (T_O)) \nonumber \feq
\noindent to give after linearization

\beq  ({\tilde T} - T_O) \cong {1 + \beta_I \over I - \beta_O}
({\overline T} - T_I) \label{AHwrad17} \feq

\noindent Equations (\ref{AHwrad15}, \ref{AHwrad16}, \ref{AHwrad17}) then permit
one to express $u$ in terms of $( {\overline T} - {\overline T}_I)$. We
must go one step further back and express $ ({\overline T} -
T_I)$ in terms of $ v - v_I$. For this we use 

\beq  {v - v_I \over k_I} = r - r_I = b^\prime _I ( {\overline X} -
X_I) + {\dot b}_I ({\overline T} - T_I) = (b^\prime_I \beta_I + {\dot
b}_I) ({\overline T} - T_I)\quad . \label{AHwrad18} \feq

\noindent Substituting for $ ({\overline T} - T_I)$ the value $
(v - v_I)$ of eq.~(\ref{AHwrad18}) and proceeding the chainwise through eqs.
(\ref{AHwrad17}, \ref{AHwrad16}, \ref{AHwrad15}) and using the 
relationships eqs.
(\ref{AHwrad13})
 and (\ref{AHwrad14})
yields the final result
\beq u = v_O - 4M - 4M \ln 
\Big|   {\sqrt{(1 - {2M \over r_1}) k^2 _I - 2 k_I} \over k_I
\sqrt{-2 k_O}} \; {a_O \sqrt{1 + {\beta}_O \over 1 - {\beta}_O}
\over a_I \sqrt{1 - {\beta}_I \over 1 + {\beta}_I}} \;
 {\hat u - v_I \over 2M}\Big| \label{AHwrad19} \feq

\noindent for the coordinate $u$ of the outgoing ray in the
exterior space which began on $ {\cal I}_-$ at the point $ 
v = {\hat u}$.
\par\noindent Equation (\ref{AHwrad19}) has a physical interpretation when one
writes the total shift as a product of three shifts (denoted
 $D_1$, $ D_2$ and $ D_3$).The factor  $D_1 $ is 
the shift produced in the voyage
from $ {\cal I}^-$ to $I$, $ D_2 $ from $I$ to $O$ and $  D_3 $ from
$O$ to $ {\cal I}^+$. The identifications are
\begin{eqnarray}
  D_1 & = & {{k_I} \over {\sqrt {(1 -{2M \over r_I})k_I^2 - 2 k_I}}}
\nonumber \\
 D_2 & = & {a_I \over a_O} {\sqrt {{1 + {\beta}_O} \over 1 -
\beta_O}}{\sqrt {{1 + {\beta}_I} \over 1 -
\beta_I}}\nonumber \\
 D_3 & = & \sqrt {-2 k_O} \; {(r_O - 2M) \over 4M} \nonumber
\end{eqnarray} 
\noindent It is of course $  D_3 $ that is the important
shift that gives rise to the steady state Hawking radiation.

To illustrate this discussion and to exhibit how the Hawking flux reaches 
its asymptotic value we consider a star consisting
 a cloud of dust in parabolic collapse. The trajectory of the
surface of the star is 
\beq  v = 2 M \left[ {5 \over 3} - {2 \over 3} ({r \over 2M})^
{3/2} + { r \over 2 M} - 2 \sqrt {{r \over 2M}} + 2 \ln \left (
{1 + \sqrt {{r \over 2 M}} \over 2} \right) \right] \nonumber \feq
\noindent so that  $v = 0$ at $r = 2M$. The interior metric is a
Roberston-Walker one, given by
\beq  
d s^2 = \left( {M\over 2 X _{\star}^3} \right)^2
T^4 \; [ - d T^2 + d X^2 ] + X^2 \left({ M \over 2 X
_{\star}^3} \right)^2 T^4 \; d \Omega^2 \nonumber 
\feq
\noindent where M is the mass of the star, and $  X =
X_{\star}$ the equation of motion of its surface. (The value of
$  X_{\star}$ fixes the density in the star).

The figures (\ref{fluxinf}) and (\ref{fluxsurf}) show the outgoing flux  
( times $ 24 \pi \ (2M^2)$) measured at infinity ($\Phi_{\infty}$) and
by  a free falling observer on the surface of the star ($\Phi_{s}$).
The fluxes are calculated using the two dimensional energy momentum 
tensor discussed in Section \ref{Tmunuren}.
Their 
analytic expressions (where we have choosen $ X_{\star} = 10M$) are:
\beqa
\Phi_{\infty}&=&{1\over {24\pi\ 4M^2}}{{18{\rho}^{3/2}+27{\rho}-21{\rho}^{1/2}
-15}\over{{\rho}^4(8+24{\rho}^{1/2}
+26{\rho}+12{\rho}^{3/2}+2{\rho}^2}}\\
{{t_{\infty}}\over {2M}}&=& {5 \over 3} - {2 \over 3} (\rho)^
{3/2} + {\rho} - 2 \sqrt {\rho} + 2 \ln \left(
{1 + \sqrt {{\rho}} } \right)\nonumber\\
& &-2\left[\rho+\ln(\rho-1)\right]\label{D24}\\
\Phi_{s}&=&{1\over
{24\pi\
4M^2}}{{-(4{\rho}^{3/2}+24{\rho}+39{\rho}^{1/2}+15)}\over{{\rho}^4(8+24{\rho}^{1/2}
+26{\rho}+12{\rho}^{3/2}+2{\rho}^2}}\\  
{{\tau_{s}} \over {2M}}&=&{2\over 3}(1-\rho^{3/2})
\feqa
where the parameter $\rho=r/2M$ represents the radial Eddington-Flinkenstein
coordinate of the surface of the star and $\tau_{s}$ the proper time measured
along it, $t_{\infty}$ being the minkowskian time at infinity.
The Hawking flux attains its asymptotic value at infinity when the last term 
in eq.~(\ref{D24}) becomes dominant, ie. when the approximation 
eq.~(\ref{AHwrad19}) is
valid.

\dessin{1.000}{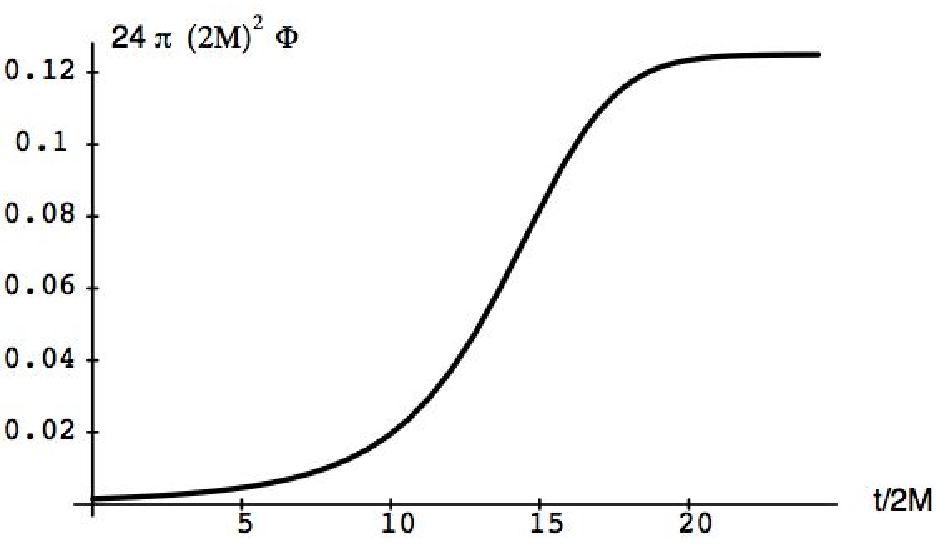}{fluxinf}{Hawking flux at infinity as a function of time. 
The background geometry
is that of a collapsing cloud of dust.}
\dessin{1.000}{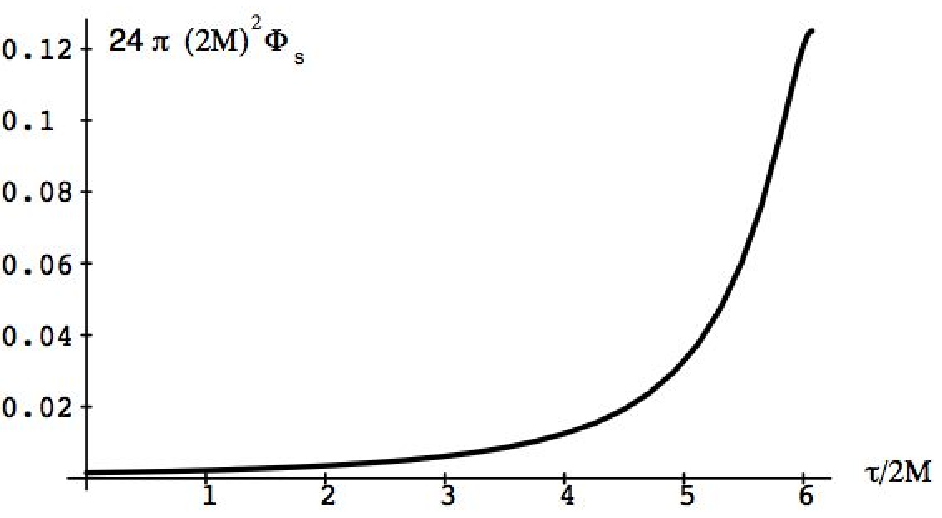}{fluxsurf}{Outgoing flux seen by an observer 
on the surface of the star as a function of
his proper time.}


{\huge Note Added in Proof}

In connexion with the remark at the end of Section 1.3 concerning the thermal
distribution of pairs produced in an external field and in connexion with the analysis of
the mean energy emitted by a uniformly accelerated detector we would like to mention the
work of Nikishov[Ni] (see also Myrhvold[My]) who anticipated the Unruh effect by an
analysis of the photons emitted by accelerated electrons.

As mentioned in footnote 10 (pages 153 and 154), upon taking into account recoil effects
the properties of the emitted fluxes are drastically modified. One can show that the
correlations which encode the weak values are also modified. The result is that due to
recoil effects the weak value Eq.~(2.149) vanishes on ${\cal I}^-$. Upon taking into
account quantum gravitational effects in the black hole problem (see the end of Section
3.5), such recoil effects may play an important r{\^o}le and could for instance modify
the properties of the weak values. Indeed we emphasize that the peculiar properties of
$\langle T_{vv} \rangle_{\psi_i}$ (Eq. (3.106)) result from the (unjustified) assumptions
of a free field theory evolving in a given classical background geometry. We hope 
that the study of the weak value $\langle T_{vv} \rangle_{\psi_i}$  can be used to
investigate the validity of both assumptions.

In order to understand the r{\^o}le of the transplanckian frequencies that are involved
in the emergence of Hawking quanta, Unruh showed in a recent paper[93] (see footnote
13, page 201), through a numerical analysis that Hawking radiation is unaffected by a
truncation of the free field spectrum at the Planck scale. We have investigated and
extended his result in Ref.~[BMPS95] where we show analytically how the appeal to
transplanckian frequencies can be avoided whilst retaining the thermal spectrum of emitted
particles.



\begin{thebibliography}{999}


\bibitem{Houches}{\em Black Holes}, Les Houches Summer School 1972, ed. C. and
B. DeWitt, Gordon and Breach, New York (1973)

\bibitem{NATO}{\em Black Hole Physics}, NATO ASI Series C, Vol. {\bf 364}, ed. V. de
Sabbata and Z. Zhang, Kluwer Academic Publishers, Dordrecht (1992)

\bibitem{Ahar} Aharonov Y., Albert D.,  Casher A. and Vaidman L., Phys. Lett.
A {\bf 124} 199 (1987),\\  
Aharonov Y. and Vaidman  L.,  Phys. Rev. A {\bf 41} 11 (1990),\\
Aharonov Y., Anandan  J.,  Popescu S. and Vaidman  L.,
     Phys. Rev. Lett. {\bf 64}, 2965 (1990)

\bibitem{AuMu}Audretsch J. and M\"uller R., Phys. Rev. D {\bf 49} (1994)
 4056; Phys. Rev D {\bf 49} (1994) 6566; 
Phys. Rev A {\bf 50} (1994) 1755

\bibitem{BaVo}Balazs N.L. and Voros A., Ann. Phys.(N.Y.) {\bf 199} (1990)
123

\bibitem{Bard} Bardeen J. M., Phys. Rev. Lett. {\bf 46} (1981) 382

\bibitem{BCH}Bardeen J. M., Carter B. and Hawking S.W., Commun. Math. Phys. {\bf 31}
(1973) 161

\bibitem{beken} Bekenstein J. D., Phys. Rev. D {\bf7} (1973) 2333; Phys. Rev D {\bf
9} (1974) 3292 

\bibitem{Beck94} Bekenstein J. D.,  {\em Do We Understand Black Hole Entropy ?}, Plenary talk
at Seventh Marcel Grossman meeting at Stanford University
(1994) preprint gr-qc/9409015

\bibitem{BD}Birrel N. D. and Davies P. C. W., {\em Quantum Fields in 
Curved Space} Cambridge University Press, Cambridge (1982)

\bibitem{BochN}Bloch F. and Nordziek A., Phys. Rev. {\bf 52} (1937) 54

\bibitem{Bogo}Bogoljubov N.N., Zh. Eksp. Teor. Fiz.,{\bf  34} (1958) 58 (Sov. Phys.
JETP, {\bf 7 }(1958) 51)

\bibitem{Boul1}Boulware D. G., Phys. Rev. D {\bf 11} (1975) 1404; Phys. Rev. D {\bf
12} (1975) 350

\bibitem{Boul2}Boulware D. G., Annals of Physics {\bf 124} (1980) 169

\bibitem{BPS} Brout R.,  Parentani R. and Spindel Ph. Nucl. Phys. B353 (1991) 209

\bibitem{BMPPS}Brout R., Massar S., Popescu S., Parentani R. and Spindel Ph., {\em
Quantum Source of the Back Reaction on a Classical Field}, preprint 
ULB-TH 93/16 UMH-MG
93/03 (1993) hep-th/ 9311019

\bibitem{OttewillBrown}Brown M. R. and Ottewill A. C., Proc. R. Soc. (London) A 389
(1983) 379

\bibitem{YorkB}Brown J. D. and York J. W., Phys. Rev. D {\bf 47}(1993) 1420

\bibitem{CGHS} Callan C. G., Giddings S. B., Harvey J. A. and Strominger A.,
Phys. Rev. D {\bf 45} (1992) 1005


\bibitem{Candelas} Candelas P., Phys. Rev. D {\bf 21} (1980) 2185

\bibitem{Carlitz} Carlitz R. D. and Willey R. S., Phys. Rev. D {\bf 36}
(1987) 2327 and Phys. Rev. D {\bf 36}
(1987) 2336

\bibitem{Cart}Carter B., in {\em Gravitation and Astrophysics}, 
Carg\`ese Summer School
1986, ed. B. Carter, J.B. Hartle, Plenum Press, New York (1986)

\bibitem{CaEn} Casher A. and Englert F., Class. Quantum. Grav. {\bf 10} (1993) 2479


\bibitem{Chan}Chandrasekhar S., {\em The Mathematical Theory of Balck Holes},
Oxford University Press, New York (1983)

\bibitem{CoMo}Cooper F. and Mottola E., Phys. Rev. D {\bf 40} (1989) 456
\\
Kluger Y., Eisenberg J.M., Svetitsky B., Cooper F. and Mottola E., 
Phys.
Rev. Lett. {\bf 67} (1991) 2427;  Phys. Rev. D {\bf 45} (1992) 4659

\bibitem{CHKMPA}Cooper F., Habib S., Kluger Y., Mottola E., 
Paz J. P. and Anderson P. B., Phys. Rev. D {\bf 50} (1994) 2848


\bibitem{DaRu}Damour T. and Ruffini R., Phys. Rev. D {\bf 14} (1976) 332

\bibitem{Davi}Davies P.C.W., Proc. R. Soc. London A {\bf 35}1 (1976) 129

\bibitem{DaFu} Davies P. C. W. and Fulling S. A., Proc. R. Soc. 
London {\bf A 356} (1977) 237


\bibitem{DFU}Davies P. C. W., Fulling S. and  Unruh W.G., Phys. Rev. D {\bf 13} (1976)
2720

\bibitem{Dewi}DeWitt B. S., in {\em General Relativity}, eds. S.W. Hawking
and
W. Israel (Cambridge: Cambridge Univerity Press) (1983)

\bibitem{Eddi}Eddington A., Nature {\bf 113} (1924) 192

\bibitem{EMP} Englert F., Massar S. and Parentani R.,
 Class. Quantum. Grav. {\bf 11} (1994) 2919

\bibitem{Fink}Finkelstein D., Phys. Rev. {\bf 110} (1958) 965

\bibitem{Full}Fulling S. A., Phys. Rev. D {\bf  7} (1973) 2850 

\bibitem{FuDa}Fulling S. A. and Davies P. C. W., Proc. R. Soc. 
London {\bf A 348} (1976) 393

\bibitem{FuRu}Fulling S. A. and Ruijsenaars S. N. M., Phys. Reports {\bf 152}, 3
(1987) 135



\bibitem{Gerl}Gerlach U., Phys. Rev. D{\bf 14} (1976) 1479

\bibitem{GibbonsHawking} Gibbons G. W. and Hawking S. W., Phys. Rev. D {\bf
15} (1977) 2738

\bibitem{MTW} Misner C.W., Thorne K.S. and Wheeler J.A., {\em Gravitation}, Freeman, 
San Francisco (1973)

\bibitem{Grow}Grove P., Class. Quantum Grav.{\bf  3} (1986) 801

\bibitem{HI} Hajieck P. and Israel W., Phys. Lett. A {\bf 80} (1980) 1

\bibitem{HaHa}Hartle J.B. and Hawking S.W., Phys. Rev D {\bf 13} (1976) 2188

\bibitem{Hawk}Hawking S.W., Nature {\bf  248} (1974) 30;
Commun. Math. Phys., 43 (1975) 199

\bibitem{Hawk2} Hawking S.W., Phys. Rev. D {\bf 13} (1976) 191

\bibitem{Hawk3} Hawking S.W., Phys. Rev. D {\bf 14}  (1976)  2460 

\bibitem{EH} Hawking S.W. and Ellis G. F. R., {\em The Large Structure of Space Time} 
Cambridge University Press, England (1973)

\bibitem{HawkL}  Hawking S.W. and Laflamme R., Phys. Lett. {\bf B209} (1988) 39

\bibitem{How} Howard K. W., Phys. Rev. D {\bf 30} (1984) 2532

\bibitem{HeEu}Heisenberg W. and Euler H.,  Z. Phys. {\bf  98} (1936) 714

\bibitem{THooft} 't Hooft G., Nucl. Phys. B {\bf 256} (1985) 727

\bibitem{THooft2} 't Hooft G., 
{\em Horizon Operator Approach to Black Hole Quantization} 
preprint THU-94/02 (1994) gr-qc/9402037 


\bibitem{Jacobson1} Jacobson T., Phys. Rev. D {\bf 44} (1991) 1731

\bibitem{Jacobson2} Jacobson T., Phys. Rev. D {\bf 48} (1993) 728

\bibitem{KaUm}Kamefuchi S. and Umezawa H., Il Nuovo Cimento {\bf  XXXI}, 2 (1964) 429

\bibitem{Klein} Klein O., Z. Phys. {\bf 53} (1929) 157

\bibitem{Krus}Kruskal M., Phys. Rev. {\bf 119 }(1960) 1743

\bibitem{Lee}Lee T. D., Phys. Rev. {\bf  45} (1954) 1329

\bibitem{Nien}Nielsen H. D. and Nimomya M., Phys. Lett. 
B {\bf 130}, 389 (1983)


\bibitem{NoFr}Novikov I. and Frolov V., {\em Physics of Black Holes}, Kluwer
Academic Publishers, Dordrecht (1989)

\bibitem{Massar} Massar S., Int. J. Mod. Phys. D {\bf 3} (1994) 237

\bibitem{Massar2}  Massar S., {\em The semi classical back reaction to black
hole evaporation} preprint ULB-TH 94/19, gr-qc/9411039

\bibitem{MaPa}Massar S. and Parentani R., {\em From Vacuum Fluctuations to
Radiation: Accelerated Detectors and Black Holes}, preprint ULB-TH 94/02, gr-qc/9404057

\bibitem{MPB}Massar S., Parentani R. and Brout R., Class. Quantum Grav.
{\bf 10} (1993) 385

\bibitem{MPB2} Massar S., Parentani R., and Brout R. Class. Quantum Grav.
{\bf 10} (1993) 2431

\bibitem{Meye}Meyer Y., {\em Wavelets, Algorithms and Applications}, SIAM,
Philadelphia (1993)

\bibitem{Mul}M\"uller B, Greiner W. and Rafeski J., Phys. Lett. A63
(1977) 181 



\bibitem{page} Page D. N., Phys. Rev. D {\bf 13} (1976) 198

\bibitem{page93} Page D. N., Phys. Rev. Lett. {\bf 71}  (1993) 1291 

\bibitem{Pagepr} Page D. N., private communication

\bibitem{Pare}Parentani R., Class. Quantum Grav. {\bf 10} (1993) 1409

\bibitem{Par}Parentani R., in preparation

\bibitem{PaBr0}Parentani R. and Brout R., Nucl. Phys. B388 (1992) 474


\bibitem{PaBr}Parentani R. and Brout R., Int. J. Mod.
Phys. D {\bf 1 }(1992) 169

\bibitem{PP} Parentani R. and Piran T.,  Phys. Rev. Lett. {\bf 73} (1994) 2805 

\bibitem{PKO} Parentani R., Katz J. and Okamoto I., 
{\em Thermodynamics of a Black Hole in a Cavity}, preprint (1994) gr-qc/9410015

\bibitem{PS} Piran T. and Strominger A., Phys. Rev. D {\bf 48} (1993) 4729


\bibitem{Presk} Preskill J.,{\em Do Black Holes Destroy Information?},
preprint CALT-68-1819,hep-th/9209058

\bibitem{epr} Pringle L. N., Phys. Rev. D {\bf 39} (1989) 2178

\bibitem{RSG}Raine D., Sciama D. and Grove P., Proc. R. Soc. A {\bf 435} (1991)
205

\bibitem{Rind}Rindler W., Am. J. Phys.{\bf  34} (1966) 1174

\bibitem{RST} Russo J. G., Susskind L. and Thorlacius L., Phys. Rev. D {\bf 47}
(1993) 533


\bibitem{Sanch} Sanchez N., Phys. Rev. D {\bf 18} (1978) 1030

\bibitem{SVV} Schoutens K., Verlinde E. and Verlinde H., Phys. Rev. D {\bf 48}
(1993) 2670 


\bibitem{SCD}Sciama D. W., Candelas P. and Deutsch D., Adv. Phys. {\bf 30} (1981)
327

\bibitem{Schw}Schwinger J., Phys. Rev. {\bf 82} (1951) 664

\bibitem{Step}Stephens C., Ann. Phys. (N.Y.) {\bf 193} (1989) 255

\bibitem{Spin}Spindel Ph., in {\em The Gardner of Eden} ed. P. Nicoletopoulos, J. Orloff,
 Physicalia Mag. {\bf 12} Suppl. (1990) 207 

\bibitem{STUg} Susskind L., Thorlacius L. and Uglum J., {\em 
The Stretched Horizon and Black Hole Complementarity}
SU-ITP-93-15 (1993) hep-th/9306069 


\bibitem{TPM}Thorne K.S., Price R.H., and Macdonald D.A., {\em Black Holes: The
Membrane Paradigm}, Yale University Press, New Haven (1986)

\bibitem{Unru1}Unruh W. G., Phys. Rev. D {\bf  14} (1976) 870

\bibitem{Unru2}Unruh W. G., Phys. Rev. D {\bf 46} (1992) 3271

\bibitem{Unru3}Unruh W. G.,{ \em  Dumb Holes and the Effects of High Frequencies on Black Hole Evaporation} gr-qc/9409008

\bibitem{UnWa}Unruh W. G. and Wald R. M., Phys. Rev. D {\bf 29} (1984) 1047



\bibitem{Vonn}von Neuwmann J., {\em Mathematical Foundations of Quantum
Mechanics}, Princeton University Press, Princeton (1955)

\bibitem{York} York J. W., Phys. Rev. D {\bf 28} (1983) 2929;
in {\em Quantum Theory of Gravity} ed.  S. M. Christensen, Adam Hilger,Ltd. Bristol (1984)
135

\bibitem{Wald}Wald R.M., {\em General Relativity}, Chicago University Press,
Chicago (1984)

\bibitem{Wald2} Wald R.M., Commun. Math. Phys., {\bf 54} (1977) 1;
Ann. Phys. (NY) {\bf 110} (1978) 472;
Phys. Rev. D {\bf 17} (1978) 1477


\bibitem{WiWa}Whittaker E. T., Watson G. N., {\em A course of modern analysis,
 $4^{th}$ edition}, chap. 16, Cambridge Univ. Press, Cambridge, England (1973)

\bibitem{Wil} Wilczek F., {\em Quantum purity at a Small Price: Easing a Black Hole
Paradox}  hep-th/9302096

\bibitem {BMPS95} Brout R., Massar S., Parentani R., Spindel Ph.,  Phys.Rev.D{\bf 52} (1995) 4559 

\bibitem {My} Myrhvold  N.P., Ann. of Phys. 160 (1985) 102.

\bibitem {Ni} Nikishov A.I., Sov. Phys. JETP 32 (1971) 690.

\end{thebibliography}
\end{document}